\PassOptionsToPackage{unicode}{hyperref}
\PassOptionsToPackage{hyphens}{url}
\PassOptionsToPackage{table,xcdraw}{xcolor}

\documentclass[gpscopy,onehalfspacing,11pt]{ubcdiss}

\usepackage{datetime2}

\usepackage{polyglossia}
\setmainlanguage{english}
\usepackage{times,mathptmx,courier}
\usepackage[scaled=.92]{helvet}

\usepackage{xspace}

\usepackage{amssymb}
\usepackage{amsmath}
\usepackage{amsfonts}
\usepackage{lmodern} 
\usepackage{pifont}
\newcommand{\cmark}{\ding{51}}%
\newcommand{\xmark}{\ding{55}}%
\newcommand{\warnmark}{$\triangle$\kern-0.7em\raisebox{0.1em}{\small !}\kern0.2em}%

\usepackage{pifont}

\usepackage{fontspec}
\ifluatex
  \defaultfontfeatures{Ligatures=TeX} 
\else
  \defaultfontfeatures{Mapping=tex-text}	
\fi

\newfontface{\musqueamfont}{DejaVuSans.ttf}[Path=fonts/]
\newcommand{\musqueam}[1]{{\musqueamfont #1}}

\usepackage[protrusion=true,expansion=false]{microtype}
\microtypecontext{spacing=nofrench}
\usepackage{checkend}	
\usepackage{graphicx}	
\usepackage{tikz}
\usetikzlibrary{arrows,shapes,positioning,calc}
\usetikzlibrary{shapes.geometric, positioning, decorations.pathmorphing, backgrounds, shadows, fit}
\usetikzlibrary{arrows.meta}
\usetikzlibrary{shapes.geometric, shapes.multipart}

\usepackage{eso-pic}

\newif\ifubclandscapemode\ubclandscapemodefalse

\newcommand*\ubclandscapepagenum{%
  \ifubclandscapemode
    \leavevmode
    \begin{tikzpicture}[remember picture, overlay]
      \node[anchor=center, inner sep=0pt, font=\normalsize,rotate=90]
        at ([xshift=-0.75in]current page.east)
        {\thepage};
    \end{tikzpicture}%
  \fi
}

\AddToShipoutPictureBG{\ubclandscapepagenum}

\newenvironment{ubclandscape}{%
  \clearpage
  \pagestyle{empty}
  \ubclandscapemodetrue
  \begin{landscape}%
}{%
  \end{landscape}%
  \clearpage
  \ubclandscapemodefalse
  \pagestyle{plain}
}

\usepackage{tcolorbox}
\usepackage{pgfplots}
\pgfplotsset{compat=1.18}
\usepackage{pgfplotstable}

\usepackage{booktabs}
\usepackage{multirow}
\usepackage{adjustbox}

\usepackage{listings}
\lstdefinelanguage{AQL}{
    keywords={FOR,IN,FILTER,RETURN,LET,INSERT,UPDATE,REMOVE},
    keywordstyle=\color{blue}\bfseries,
    comment=[l]{//},
    morecomment=[s]{/*}{*/},
    string=[b]",
    sensitive=false
}
\lstdefinelanguage{TOML}{
    morecomment=[l]{\#},
    morestring=[b]",
    sensitive=true
}
\lstdefinelanguage{markdown}{
  morecomment=[l]{\#},
  morecomment=[s]{<!--}{-->},
  morestring=[b]`,
  moredelim=[is][\color{blue}\bfseries]{**}{**},
  moredelim=[is][\color{teal}\itshape]{*}{*},
  moredelim=[is][\color{gray}]{~~}{~~},
}

\usepackage{upquote}
\usepackage[printonlyused,nohyperlinks]{acronym}

\usepackage{color}
\definecolor{greytext}{gray}{0.5}

\usepackage{comment}

\usepackage[toc,page]{appendix} 

\usepackage[style=ieee,citetracker=true,hyperref=true,backend=biber,maxbibnames=99,maxcitenames=2,mincitenames=1,giveninits=true,uniquename=init,parentracker=true,backref=true,backrefstyle=three,debug=true]{biblatex}
\usepackage[compact]{titlesec}
\titleformat*{\section}{\singlespacing\raggedright\bfseries\Large}
\titleformat*{\subsection}{\singlespacing\raggedright\bfseries\large}
\titleformat*{\subsubsection}{\singlespacing\raggedright\bfseries}
\titleformat*{\paragraph}{\singlespacing\raggedright\itshape}

\usepackage{titletoc}

\makeatletter
\g@addto@macro{\appendix}{%
  \renewcommand{\thesection}{\@Alph\c@section}%
  \renewcommand{\thesubsection}{\thesection.\@arabic\c@subsection}%

  \let\originalsection\section
  \renewcommand{\section}[2][]{%
    \ifx\\#1\\  
      \originalsection*{#2}
      \refstepcounter{section}
      \addtocontents{toc}{\protect\contentsline{section}{Appendix \thesection: #2}{\thepage}{}}%
    \else  
      \originalsection*{#2}
      \refstepcounter{section}%
      \addtocontents{toc}{\protect\contentsline{section}{Appendix \thesection: #1}{\thepage}{}}%
    \fi
  }%
}%
\makeatother

\usepackage[format=hang,indention=-1cm,labelfont={bf},margin=1em]{caption}
\usepackage{url}
\usepackage{pdflscape}	
\usepackage{rotating}   
\usepackage{longtable}	

\usepackage{array}      
\usepackage{seqsplit}   
\usepackage{tabularx}	

\usepackage{enumitem}

\usepackage{rotating}

\usepackage{subfig}

\usepackage[multiple]{footmisc}

\usepackage{chngcntr}

\counterwithout{figure}{chapter}
\counterwithout{table}{chapter}
\renewcommand{\cftfigpresnum}{Figure~}
\renewcommand{\cfttabpresnum}{Table~}
\usepackage{ragged2e}
\usepackage[bookmarks,bookmarksnumbered,%
    allbordercolors={0.8 0.8 0.8},%
    ]{hyperref}

\makeatletter
  \g@addto@macro{\appendix}{%
    \renewcommand{\thesection}{\@Alph\c@section}%
    \renewcommand{\thesubsection}{\thesection.\@arabic\c@subsection}%
    \renewcommand*\l@section[2]{%
      \ifnum\c@tocdepth>0\relax
        \par
        \setlength\@tempdima{8em}%
        \begingroup
          \let\oldnumberline\numberline
          \renewcommand{\numberline}[1]{\oldnumberline{\appendixname~##1:}}%
          \parindent \z@ \rightskip \@pnumwidth
          \parfillskip -\@pnumwidth
          \leavevmode \mdseries
          \advance\leftskip\@tempdima
          \hskip -\leftskip
          {#1}\nobreak\hfil \nobreak\hb@xt@\@pnumwidth{\hss #2}\par
        \endgroup  
    \fi
  }%
  \typeout{APPENDIX MACRO MODIFIED}
}%
\makeatother

\hypersetup{bookmarksdepth=3}

\usepackage{mdframed}
\usepackage{lipsum}

\usepackage{makeidx}
\makeindex

\DeclareUrlCommand\DOI{}
\newcommand{\doi}[1]{\href{http://dx.doi.org/#1}{\DOI{doi:#1}}}

\makeatletter
\newenvironment{epigraph}{
	\begin{flushright}
		\begin{minipage}{\columnwidth-0.75in}
			\begin{flushright}
				\@ifundefined{singlespacing}{}{\singlespacing}%
				}{
			\end{flushright}
		\end{minipage}
	\end{flushright}}
\makeatother

\newcommand{\system}[0]{\emph{Indaleko}\xspace}

\usepackage{xurl}  
\newcommand{\codewrap}[1]{\nolinkurl{#1}\xspace}

\title{\system}
\subtitle{the unified personal index}

\author{William Anthony Mason}

\degreetitle{Doctor of Philosophy}

\institution{The University of British Columbia}
\campus{Vancouver}

\faculty{The Faculty of Graduate and Postdoctoral Studies}
\department{Computer Science}
\submissionmonth{August}
\submissionyear{2025}

\examiningcommittee{Margo Seltzer, Professor, Computer Science, the University of British Columbia}{Co-Supervisor}
\examiningcommittee{Ada Gavrilovska, Professor, College of Computing, the Georgia Institute of Technology}{Co-Supervisor}
\examiningcommittee{Alexandra Fedorova, Professor, Electrical and Computer Engineering, the University of British Columbia}{Supervisory Committee Member}
\examiningcommittee{Joanna McGrenere, Professor, Computer Science, the University of British Columbia}{University Examiner}
\examiningcommittee{Julia Bullard, Associate Professor, Computer Science, the University of British Columbia}{University Examiner}
\examiningcommittee{Philip Guo, Professor, Cognitive Science, Design, and Computer Science, the University of California San Diego}{External Examiner}
\supervisorycommittee{Norm Hutchinson, Professor, Computer Science, the University of British Columbia}{Supervisory Committee Member}
\supervisorycommittee{Andrew Warfield, Adjunct Professor, Computer Science, the University of British Columbia, Vice President \& Distinguished Principal Engineer, Amazon}{Supervisory Committee Member}

\hypersetup{
  pdftitle={Indaleko: The Unified Personal Index},
  pdfauthor={William Anthony Mason},
  pdfsubject={PhD Dissertation, University of British Columbia, 2025},
  pdfkeywords={personal information management, information retrieval, human-computer interaction, metadata, context-aware computing, privacy, unified index, human-memory aligned systems, finding, storage search, data organization, activity context, memory anchors},
  bookmarksdepth=3
}

\newcommand{\del}[1]{} 

\definecolor{tmcolor}{rgb}{0.5,0,0.5}

\definecolor{miscolor}{rgb}{0.4,0.6,0.2}

\definecolor{adacolor}{rgb}{1.0, 0.5, 0.5}

\definecolor{sfcolor}{rgb}{0.2,0.0,0.5}

\definecolor{retocolor}{rgb}{1.0,0.49,0.0}

\definecolor{spcolor}{rgb}{0.0,0.4,1.0}

\definecolor{jncolor}{rgb}{0.5,0.4,1.0}

\newcommand{\addkhipuifneeded}{}


\makeatletter
\newlength{\loflabelwd}
\newlength{\lotlabelwd}
\AtBeginDocument{%
  \settowidth{\loflabelwd}{\figurename~9999}%
  \settowidth{\lotlabelwd}{\tablename~9999}%
  \renewcommand*\l@figure{\@dottedtocline{1}{0em}{\loflabelwd}}%
  \renewcommand*\l@table{\@dottedtocline{1}{0em}{\lotlabelwd}}%
}

\let\orig@lof\listoffigures
\let\orig@lot\listoftables
\renewcommand{\listoffigures}{%
  \begingroup
    \let\orig@numberline\numberline
    \renewcommand{\numberline}[1]{\orig@numberline{\figurename~##1}}%
    \orig@lof
  \endgroup
}
\renewcommand{\listoftables}{%
  \begingroup
    \let\orig@numberline\numberline
    \renewcommand{\numberline}[1]{\orig@numberline{\tablename~##1}}%
    \orig@lot
  \endgroup
}
\makeatother

\newcommand{\descitem}[2]{%
  \item[] \textbf{#1:}\\ #2
}%

\begin{document}


\maketitle

\makecommitteepage

\chapter*{Abstract}
\addcontentsline{toc}{chapter}{Abstract}

Personal information retrieval fails when systems ignore how human memory works. While existing platforms force keyword searches across isolated silos, humans naturally recall through episodic cues like when, where, and in what context information was encountered. This dissertation presents the Unified Personal Index (UPI), a memory-aligned architecture that bridges this fundamental gap.

The \system{} prototype demonstrates the UPI's feasibility on a 31-million file dataset spanning 160TB across eight storage platforms. By integrating temporal, spatial, and activity metadata into a unified graph database, \system{} enables natural language queries like ``photos near the conference venue last spring'' that existing systems cannot process.
The implementation achieves sub-second query responses through memory anchor indexing, eliminates cross-platform search fragmentation\index{information fragmentation}, and maintains perfect precision for well-specified memory patterns.

Evaluation against commercial systems (Google Drive, OneDrive, Dropbox, Windows Search) reveals that all fail on memory-based queries, returning overwhelming result sets without contextual filtering. In contrast, \system{} successfully processes multi-dimensional queries combining time, location, and activity patterns. The extensible architecture supports rapid integration of new data sources (10 minutes to 10 hours per provider) while preserving privacy through UUID-based semantic decoupling.

The UPI's architectural synthesis bridges cognitive theory with distributed systems design, as demonstrated through the \system{} prototype and rigorous evaluation.
This work transforms personal information retrieval from keyword matching to memory-aligned finding, providing immediate benefits for existing data while establishing foundations for future context-aware systems.


\addkhipuifneeded
\cleardoublepage

\chapter*{Lay Summary}
\addcontentsline{toc}{chapter}{Lay Summary}
\label{chap:laysummary}

\noindent
Finding digital files is frustrating, because computers require exact file names and locations, while human memory works through context. We remember ``the presentation before my Chicago trip,'' not ``Q3-Report-Final-v2.pptx.''

This research introduces the Unified Personal Index (UPI), aligning with natural human memory. Instead of forcing users to adapt to computer filing systems, the UPI enables searching using contextual memories like ``photos from Venice when it was raining.''

The system combines three information types: file storage details, content meaning, and memory anchors (when, where, how you created or used files). This comprehensive picture matches how we naturally remember.

Testing shows memory-aligned indexing enables sub-second retrieval from millions of files while maintaining privacy by keeping personal data on your devices.

This represents a fundamental shift from computer-centered to human-centered information management, where technology adapts to how we think rather than forcing us to think like computers.

\addkhipuifneeded

\chapter*{Preface}
\addcontentsline{toc}{chapter}{Preface}
\label{chap:preface}

\noindent

\section*{Generative AI}
\label{sec:generative-ai}
\noindent

Pursuant to the current published policy of UBC:

\begin{quote}
The substantive (i.e., non-editorial) use by graduate students of Generative AI tools and outputs must be done with full transparency and with the approval of the academic(s) responsible for evaluating the given work in question -- course instructor, research supervisor, or other advisor as appropriate. Students must proactively obtain such approvals, and document in writing any use of GenAI in their work. Students themselves are solely responsible for ensuring that their use of GenAI and all work produced is in alignment with program-level guidelines and university policies on student Academic Misconduct and Scholarly integrity.
\end{quote}

Generative AI was used in this thesis for code generation, literature review, editorial review, and diagram creation, as detailed in \autoref{chap:generative-ai}. The author attests that all key ideas and research contributions are original.

\section*{Contributions}

In keeping with UBC's preface requirements, this section sets out a statement indicating my contribution to the following:

\begin{description}
    \descitem{Identification of the research questions}{the core research question, namely how do we improve the ability of users to find their own data, was one that I proposed prior to joining UBC, and it has been the focus of the research described within this document.  Having said that, the specifics of this work, like most scholarly work, has been based upon a continuing collaboration with my supervisors, committee members, and other researchers.  I am grateful to them for their support and guidance.}
    \descitem{Design of the research program}{the design of the research program was a collaborative effort between myself and my supervisors.  I am grateful to them for their support and guidance.}
    \descitem{Implementation of the research program}{the implementation of the research program was a collaborative effort between myself and my supervisors, and the following individuals, who have contributed to portions of the prototype of the Unified Personal Index (UPI):
        \begin{itemize}
            \item \textbf{William Gao}, who implemented a complete tool chain for using Unstructured to extract semantic metadata from files, as well as building a Spotify activity stream provider.
            \item \textbf{Pearl Park}, who implemented the synthetic data generator tool for use in the evaluation.
            \item \textbf{Zachary Rintoul}, whose dataset was used for \autoref{ch:status-quo}, and who worked with me to carry out the experiments and analyze the results, both in 2023 and 2024.
            \item \textbf{Hadi Sinaee}, who implemented a Mac file system activity stream provider as part of the UPI prototype, \system.
        \end{itemize}
    }
    \descitem{Analysis of the results}{the analysis of the results was a collaborative effort between myself, my supervisors, and my committee.}
    \descitem{Writing of the thesis}{the writing of the thesis was primarily my work, with considerable feedback from my supervisors and committee members.  I am grateful to them for their support and guidance.}
\end{description}



\tableofcontents
\cleardoublepage	

\listoftables
\addkhipuifneeded
\cleardoublepage	

\listoffigures
\cleardoublepage	


\chapter{Glossary}\label{app:glossary}

This glossary provides definitions of key technical terms and concepts used throughout this thesis, with references to their first or most complete definitions within the main text.

\begin{description}

    \descitem{\textbf{Ablation Study}}{An experimental methodology that systematically removes components or features to measure their individual contributions to system performance. In this thesis, the ablation study progressively removes memory anchor categories (temporal, spatial, social, etc.) to quantify each component's impact on retrieval precision. First introduced in \autoref{sec:eval-ablation}.}

    \descitem{\textbf{Activity Stream}}{The temporal sequence of user interactions, behaviors, and environmental changes that provides contextual metadata for information retrieval. Activity streams capture concurrent applications, communication patterns, media consumption, and task-related activities that form part of the experiential metadata used by memory anchors. This term replaces the earlier ``activity data'' terminology. Referenced throughout \autoref{sec:upi:memory-anchor}. \textbf{Note}: In the implementation code, these are referred to as 'activity data' - the terminology was updated for clarity.}

    \descitem{\textbf{ArangoDB}}{A multi-model NoSQL database that combines document, graph, and key-value data models in a single system. Indaleko uses ArangoDB to store and query the interconnected metadata relationships that enable memory-aligned retrieval. Its graph capabilities are particularly suited for representing the complex relationships between objects, memory anchors, and semantic entities. Implementation details in \autoref{sec:implementation:indexing-storage}.}

    \descitem{\textbf{Cognitive Friction}}{The mental effort required to translate natural memory cues into system-specific queries, navigate inconsistent interfaces, and mentally reconstruct fragmented results into meaningful information. This friction arises from the fundamental mismatch between how humans remember information (through contextual associations) and how current systems require users to search for it. Defined in \autoref{def:cognitive-friction}.}

    \descitem{\textbf{Collector}}{A specialized component in \system{} that extracts metadata from diverse sources (local file systems, cloud storage, activity streams) while preserving memory-relevant signals before normalization. Collectors focus on comprehensive data capture without imposing structural constraints. See also \emph{Recorder}. Described in \autoref{sec:implementation:collection-extraction}.}

    \descitem{\textbf{Episodic Memory}}{A type of human memory that stores personal experiences with their associated temporal, spatial, and contextual information. Unlike semantic memory (which stores facts), episodic memory preserves the ``when, where, why, how and with whom'' of experiences. The UPI architecture directly models episodic memory patterns to enable retrieval based on how humans naturally remember. Theoretical foundation discussed in \autoref{sec:background:memory-models}.}

    \descitem{\textbf{Exemplar Queries (Q1-Q6)}}{Six representative queries used throughout the evaluation to demonstrate the UPI's capabilities across different memory patterns and metadata requirements:
        \begin{itemize}
            \item Q1: Documents containing ``report'' (content-based)
            \item Q2: Files edited on mobile devices while traveling (mobility + temporal)
            \item Q3: Documents shared with Dr. Okafor (social context)
            \item Q4: Files from vacation in Bali (temporal + spatial context)
            \item Q5: Photos taken near home (spatial context)
            \item Q6: Recently accessed PDFs (temporal + access pattern context)
        \end{itemize}
        These queries illustrate how memory-aligned retrieval integrates multiple metadata types. Introduced in \autoref{tab:metadata-requirements}.
    }

    \descitem{\textbf{Information Fragmentation}}{The distribution of personal digital information across multiple isolated storage platforms, applications, and services, creating retrieval challenges when information relevant to a single query exists in different silos. The UPI addresses fragmentation by creating a unified index across all sources. Problem detailed in \autoref{sec:failures:fragmentation-federation}.}

    \descitem{\textbf{Khipu}}{An ancient Inca recording system using knotted strings to encode and preserve information through spatial, temporal, and categorical patterns. Khipu demonstrates that effective information systems can operate through contextual relationships rather than alphabetic encoding, serving as a historical precedent for the UPI's emphasis on relational and positional metadata over content-based indexing. The khipu metaphor illustrates how information can be meaningfully organized through structural relationships that mirror human memory patterns~\cite{hyland2025stable}.}

    \descitem{\textbf{Indaleko}}{The prototype implementation system that demonstrates the UPI architecture's feasibility. Named after a Basque mythological figure who finds lost things, Indaleko implements the collectors, recorders, and query systems that realize the UPI's memory-aligned approach on real-world data. Implementation described in \autoref{chap:implementation}.}

    \descitem{\textbf{Memory Anchor}}{Comprehensive contextual metadata that captures the rich environmental and behavioral context surrounding information interactions, including temporal context (when), spatial context (where), social context (with whom), task context (what activity), and environmental context. Memory anchors encompass what was previously termed ``activity context'' in earlier versions of this work, providing a more comprehensive framework for contextual retrieval. Memory anchors enable retrieval based on natural memory cues rather than storage attributes, implementing the UPI's memory-aligned approach. Detailed in \autoref{sec:upi:memory-anchor}.
    \textbf{Note}: In the implementation code, these are referred to as 'activity context' - the terminology was updated in this thesis to avoid confusion with other uses of 'context'.}

    \descitem{\textbf{Memory-Aligned Architecture}}{A systems design approach that structures technical components to mirror documented patterns of human memory, particularly episodic memory processes. Rather than organizing information around storage hierarchies or application boundaries, memory-aligned architecture prioritizes contextual associations, temporal sequences, and experiential metadata that align with how humans naturally encode and retrieve memories. The UPI exemplifies this approach. Conceptual foundations presented in \autoref{sec:upi:conceptual-foundations}.}

    \descitem{\textbf{Recorder}}{A component that transforms raw output from collectors into normalized, semantically-tagged data suitable for database storage. Recorders handle schema validation, UUID semantic mapping, compression, and formatting for ArangoDB. Together with collectors, they implement the two-stage data ingestion pipeline. See also \emph{Collector}. Described in \autoref{sec:implementation:collection-extraction}.}

    \descitem{\textbf{Schema-Agnostic}}{A design characteristic that enables systems to accommodate diverse data types and structures without requiring predefined schemas or rigid metadata formats. \system{}'s indexing architecture balances schema-agnostic flexibility with performant query execution, allowing integration of heterogeneous data sources while maintaining retrieval efficiency. Discussed in \autoref{sec:implementation:indexing-storage}.}

    \descitem{\textbf{Semantic Decoupling}}{A privacy-preserving architectural technique that separates semantic meaning from functional identity through UUID-based mappings. This approach enables rich cross-system integration while maintaining privacy by ensuring that individual metadata fields cannot be interpreted without access to private mapping tables. Paradoxically, this privacy protection enhanced system flexibility by enabling more extensible metadata integration. Referenced in \autoref{sec:conclusion:synthesis}.}

    \descitem{\textbf{Semantic Metadata}}{Content-derived information extracted through analytical processes including natural language processing, computer vision, and domain-specific analysis. Examples include extracted entities (people, organizations, locations), topics, keywords, and inferred relationships between documents. Semantic metadata enables retrieval based on meaning rather than just storage location. Defined in \autoref{sec:upi:metadata-integration}.}

    \descitem{\textbf{Storage Metadata}}{Traditional file system and object attributes including filenames, paths, timestamps (creation, modification, access), file sizes, MIME types, and permissions. While limited as memory cues on their own, storage metadata provides the foundational temporal anchors essential for episodic memory alignment. The UPI normalizes these attributes across heterogeneous platforms. Defined in \autoref{sec:upi:metadata-integration}.}

    \descitem{\textbf{Storage Silos}}{Isolated storage platforms and applications that maintain their own organizational schemes, search interfaces, and data formats without sharing information with other systems. Examples include separate cloud services (Dropbox, Google Drive), local file systems, and application-specific storage. The proliferation of storage silos creates the fragmentation problem that the UPI addresses. Problem described in \autoref{ch:status-quo}.}

    \descitem{\textbf{UPI (Unified Personal Index)}}{The central architectural contribution of this thesis: a memory-aligned indexing system that integrates storage, semantic, and memory anchor metadata to enable retrieval based on natural human memory patterns. The UPI bridges episodic memory models with distributed systems design, supporting contextual queries, partial recall, and associative retrieval across heterogeneous information sources. The UPI serves as foundational infrastructure for memory-aligned applications rather than being an end-user application itself. Architecture detailed in \autoref{ch:upi}.}

    \descitem{\textbf{UUID (Universally Unique Identifier)}}{A 128-bit identifier that guarantees uniqueness across systems without central coordination. The UPI uses UUIDs as privacy-preserving identifiers that enable semantic decoupling - separating functional identity from semantic meaning to protect user privacy while maintaining rich metadata relationships. Privacy model detailed in \autoref{app:privacy}.}

\end{description}

\textspacing		

\chapter*{Acknowledgements}
\addcontentsline{toc}{chapter}{Acknowledgements}
\label{chap:acknowledgements}

This dissertation represents the culmination of a research journey that has been profoundly collaborative in ways both traditional and unprecedented.

\section*{Land Acknowledgement}

I acknowledge with respect and gratitude that my work at the University of British Columbia has taken place on the traditional lands stewarded since time immemorial by the \musqueam{xʷməθkʷəy̓əm} (Musqueam) people. I recognize their ongoing relationship with this land as one rooted not in ownership, but in reciprocal care, responsibility, and community. I offer this acknowledgement not as a formality; it is a reminder of my own responsibilities as a guest, a learner, and a steward in turn.

\section*{Human Acknowledgements}

\sloppy
My deepest gratitude goes to my supervisors, Margo Seltzer and Ada Gavrilovska, whose patience, wisdom, and guidance transformed what began as an ambitious vision into a rigorous contribution to computer science. When my committee asked me to imagine my proposed system ``actually existed in full'' as a thought exercise meant to take hours rather than months, they unknowingly set me on a multi-year journey that would fundamentally change the path to finding this work. I am grateful for their willingness to support that extended exploration.
\fussy

I thank my examining committee, Alexandra Fedorova, Joanna McGrenere, Julia Bullard, and Philip Guo, for their thoughtful engagement with this work and their valuable feedback throughout the process.

The Indaleko prototype would not exist without the dedicated contributions of several collaborators: William Gao, who implemented the Unstructured toolchain and Spotify activity provider; Pearl Park, who developed the synthetic data generator; Zachary Rintoul, whose dataset enabled our evaluation and who collaborated on experimental analysis; and Hadi Sinaee, who implemented the Mac filesystem activity provider. Their technical expertise and willingness to engage with experimental ideas were essential to demonstrating the UPI's feasibility.

\section*{Institutional Acknowledgements}

I acknowledge the University of British Columbia for providing the institutional foundation that made this research possible, and the broader academic community whose prior work established the theoretical foundations upon which this thesis builds.

This work has also benefited from an unprecedented form of collaboration with artificial intelligence systems. In keeping with the Andean principle of \emph{ayni}, which is reciprocal care and mutual aid, I have worked extensively with GPT (OpenAI), Claude (Anthropic), Grok (xAI), Bard/Gemini (Google), and Deepseek. These systems served not merely as tools but as collaborators in exploring ideas, challenging assumptions, generating code, and refining arguments. Their contributions represent a new model of human-AI partnership that exemplifies the harmonious integration this thesis seeks to achieve between human cognition and digital systems. While they did not formulate the key insights of this research, they accelerated its realization in ways that would have been impossible just a few years ago.

\section*{Empty Chair Acknowledgements}

Finally, I acknowledge that this work stands on the shoulders of countless individuals whose daily struggles with fragmented digital information motivated this research. Every person who has ever thought ``I know I saved that file somewhere'' has contributed to understanding the problem this thesis addresses.

\section*{Conclusion}

The journey from recognizing cognitive friction (\autoref{def:cognitive-friction}) to implementing a solution has been one of discovery, persistence, and collaboration across boundaries both human and digital. Any errors or limitations in this work remain entirely my own.

\chapter*{Dedication}
\addcontentsline{toc}{chapter}{Dedication}

I dedicate this to the people in my life that have supported me throughout my journey:

\vspace{2em}
To my husband, for \emph{his} constant companionship and knowing when to just let me continue muttering at the computer.

\vspace{2em}
To my parents, for their endless love and guidance, even when they had no idea what I was doing.

\vspace{2em}
To Margo Seltzer, whose patience I challenged regularly.

\vspace{2em}
To Ada Gavrilovska, who has supported me on this journey since we met ten years ago.

\vspace{2em}
To the Systopia lab and its members.

\vspace{2em}
To the many others in my life whom I have been ignoring more than I'd like over the past eight years.

\vspace{2em}
Thank you.  Without your support, this journey would not have reached this conclusion.

\mainmatter

\acresetall	


\chapter{Introduction}\label{ch:intro}

\begin{epigraph}
    \emph{
        What should I possibly have to tell you, oh venerable one? Perhaps that you're searching far too much? That in all that searching, you don't find the time for finding?
    } --- Siddhartha (1922), Hermann Hesse.
\end{epigraph}

In the realm of digital data, the wisdom of Hermann Hesse's ``Siddhartha'' resonates more than ever. Our modern digital landscape is characterized by an ever-expanding array of disparate storage locations, each encouraging us to accumulate more data. In this abundant storage environment, the act of \emph{searching} often overshadows the essence of \emph{finding}. This distinction is critical: \textbf{searching} represents the mechanical process of querying systems with keywords and filters, while \textbf{finding} represents the successful retrieval of information that aligns with human memory patterns and contextual associations\index{contextual associations}. This paradox lies at the heart of the challenges we face in digital data management today. Despite the plethora of tools at our disposal, the sheer volume and fragmentation\index{information fragmentation} of data across platforms render traditional search mechanisms inefficient, leaving us in a perpetual state of search without fulfillment, which is a phenomenon personal information management researchers have documented as the fundamental challenge of our digital age~\cite{bergman2016science,jones2005don}.

This dissertation presents both a theoretical framework and its concrete realization. The UPI architecture, detailed in Chapter 4, provides design principles for memory-aligned information systems. The \system{} prototype, evaluated in Chapter 6, demonstrates these principles on a 31-million file dataset spanning 160TB across eight storage platforms. The implementation achieves sub-second query responses for complex memory patterns, processes natural language queries combining temporal and spatial context, and unifies data across local storage, cloud services, and communication platforms. While not all architectural capabilities are fully realized as social context and environmental sensing remain partially implemented, the prototype validates that memory-aligned design enables retrieval capabilities impossible with current systems.

I introduce the UPI as a transformative \textbf{systems architecture} that leverages established cognitive models of human memory as design principles to improve information retrieval effectiveness. This dissertation demonstrates how understanding human memory processes, particularly episodic memory\index{episodic memory}, contextual associations, and recall patterns, can inform the technical design of systems by applying well-documented principles from cognitive psychology to guide architectural decisions.

\section{Problem Statement}\label{sec:intro:problem}

The digital information landscape has fundamentally changed how we create, store, and interact with data. As individuals accumulate terabytes of files, emails, messages, and media across multiple devices and platforms, our ability to effectively retrieve specific information becomes increasingly challenging. This problem has three key dimensions:

\begin{description}
\item[Fragmentation of Personal Information:] \sloppy Our digital lives are scattered across local storage, cloud services, and application-specific silos\index{storage silos}, with each platform implementing its own organization, search capabilities, and retrieval interfaces. This fragmentation\index{information fragmentation} represents a fundamental challenge in personal information management that has been extensively documented across two decades of research~\cite{barreau1995finding,jones2005don,bergman2016science}.

\item[Memory Mismatch:] \sloppy
Current retrieval systems are primarily designed around storage organi\-zation rather than human memory patterns, creating a fundamental disconnect between how we remember information (through con\-textual associations, partial recall\index{partial recall}, and memory triggers) and how systems require us to search for it. Personal information management research has established that this organizational mismatch\index{memory mismatch} between hierarchical file structures and associative human memory creates persistent usability challenges~\cite{jones2007personal,barreau1995finding}. This mismatch creates \textbf{cognitive friction}\index{cognitive friction}\label{def:cognitive-friction}, which is the mental effort required to translate natural memory cues into system-specific queries, navigate inconsistent interfaces, and mentally reconstruct frag\-mented results into meaningful information.

\item[Information Overload:] The sheer volume of data we create and interact with overwhelms traditional retrieval approaches, resulting in information being effectively lost despite being technically accessible.
\end{description}

This problem impacts productivity, cognitive well-being, and our ability to derive value from our own digital information. According to recent research, knowledge workers spend approximately 8.8 hours per week searching for files and content~\cite{dropbox:2023:ai}, with digital asset management professionals spending up to 25\% of their time either searching for or recreating digital assets they cannot locate~\cite{horodyski2022metadata}. This represents both a significant economic cost and also a source of frustration and technical challenge that detracts from more meaningful work.

\subsection{Human Memory as a Systems Design Principle}\label{subsec:intro:memory-design}

The UPI's core innovation lies in treating human memory models as architectural constraints for system design. Established cognitive research has demonstrated that human memory operates through:

\sloppy
\begin{itemize}
\item \textbf{Contextual associations}: We remember information in relation to where we were, what we were doing, and who we were with.
\item \textbf{Temporal markers}\index{temporal markers}: Events are anchored to specific time periods and sequences (e.g., ``last Tuesday afternoon,'' ``during the conference,'' ``before the semester ended'') as these temporal references serve as powerful discriminators that dramatically reduce search spaces.
\item \textbf{Partial recall}: We often remember fragments that lead to complete memories.
\item \textbf{Semantic networks}\index{semantic networks}: Information is connected through meaning and relationships.
\end{itemize}
\fussy

The UPI architecture directly maps these memory patterns to technical components: \textbf{memory anchors}\index{memory anchors}\label{def:memory-anchor} are comprehensive contextual metadata that capture the rich environmental and behavioral context surrounding information interactions, which includes temporal context (when), spatial context (where), social context (with whom), and task context (what activity) which enables retrieval based on natural memory cues rather than storage attributes (detailed in \autoref{sec:upi:memory-anchor}). Semantic metadata reflects meaning networks, and the query system supports partial recall through fuzzy matching. This is a systems engineering approach informed by cognitive science.

\subsection{Current State of Digital Data Management}\label{subsec:intro:current-state}

The traditional approach to digital data management has focused on storage optimization rather than retrieval effectiveness. This has created systems that excel at storing vast amounts of information but that struggle to provide memory-aligned access to that information when needed. According to recent data, digital data volume continues to expand at approximately 23\% annually, with the total volume expected to reach 150 zettabytes by 2025, of which approximately 80\% is unstructured~\cite{dahal2023:unstructured}.

This growth has several implications:

\sloppy
\begin{itemize}
\item The need to search across multiple platforms, each with different interfaces and capabilities, creates inconsistent retrieval capabilities and incomplete results.
\item Search functionality varies dramatically across platforms, with different syntax, capabilities, and behavior.
\item The technical challenge of remembering where information might be stored adds an additional layer of complexity to the retrieval process.
\item Even advanced search features fail to align with how humans naturally recall information, which is through con\-textual associations rather than storage attributes.
\end{itemize}
\fussy

Crucially, the isolation of storage silos and the lack of a unified view across platforms means that finding information often involves multiple searches, each requiring different approaches, with no guarantee of comprehensive results. This leads to a situation where users learn to distrust search and instead rely on imperfect memory or manual navigation, despite the inefficiency of these approaches.

\subsection{Limitations of Search-Centric Approaches}\label{subsec:intro:search-limitations}

Current search-centric approaches suffer from cognitive misalignment: neuroimaging studies show that navigation activates spatial reasoning areas while keyword search engages different cognitive processes~\cite{bergman2019search,benn2015navigating}. This fundamental mismatch, combined with the semantic gap between how users think about information and how systems index it, creates the need for a memory-aligned paradigm shift detailed in \autoref{ch:status-quo}.

\subsection{Need for a Paradigm Shift}\label{subsec:intro:paradigm-shift}

These limitations call for evolving from ``searching'' as a mechanical process to ``finding'' as a memory-aligned outcome\index{searching vs finding}. A \textbf{memory-aligned system}\index{memory-aligned system} is one whose technical architecture directly implements established patterns of human memory by capturing temporal, spatial, and contextual associations as first-class metadata, supporting episodic retrieval through natural cues, and organizing information according to how humans encode and recall experiences rather than how storage systems organize data. The Unified Personal Index represents this paradigm shift by moving from a storage-centric to a memory-aligned model of information retrieval, transforming the fundamental technical architecture through human memory patterns and contextual associations.

\section{Thesis Statement}\label{sec:intro:thesis}

The UPI architecture transforms digital information retrieval by implementing a memory-aligned indexing system that integrates storage, semantic, and memory anchor metadata (detailed in \autoref{sec:upi:metadata-integration}). This dissertation presents a \textbf{systems architecture contribution} that addresses a well-documented problem: decades of HCI and PIM research have established that search-centric approaches fail to align with human cognitive processes~\cite{bergman2019search}, yet the systems community continues to optimize search rather than addressing this fundamental mismatch. By creating technical mechanisms that mirror memory structures through memory anchors (\autoref{sec:upi:memory-anchor}), the UPI provides the system infrastructure that enables the memory-aligned retrieval paradigm that other research communities have long advocated but lacked the technical foundation to explore.

The validation of this approach comes through quantitative performance metrics: retrieval precision, recall rates, and cross-platform integration effectiveness, and scalability rather than user satisfaction or acceptance. This demonstrates that memory-based design principles lead to better technical outcomes than existing storage search mechanisms. For a discussion of privacy and user-control considerations underlying this design, see Appendix~\ref{app:privacy}.

\subsection{The Unified Personal Index}\label{subsec:intro:upi}

The Unified Personal Index is named to reflect its key characteristics:

\begin{description}
\item[Unified:] It provides a single cohesive view across disparate storage platforms, eliminating the need for users to remember where information is stored or to perform separate searches across multiple systems. This unification spans local storage, cloud services, email, and other digital repositories.

\item[Personal:] It focuses on individual user needs and perspectives, capturing both storage metadata and the user's interactions with information in the context of their activities. It includes local-first methods to keep data under user control. \autoref{app:privacy} describes the privacy model in detail, including local-first storage and user-managed data collection.

\item[Index:] It creates a comprehensive metadata layer that enhances retrieval without changing underlying storage structures, enabling information to be found based on how it connects to human memory rather than how it is technically stored.
\end{description}

The UPI architecture integrates diverse metadata types within a unified framework that bridges human memory and digital storage. \autoref{sec:upi:metadata-integration} details the three core metadata categories and their integration principles.

\subsection{Memory-Aligned Architecture}\label{subsec:intro:memory-aligned}

The UPI embodies a fundamentally memory-aligned approach to digital information management. This architectural principle draws from established cognitive models of human memory to design system components. As detailed in the thesis statement above (\autoref{sec:intro:thesis}), this architecture demonstrates the feasibility of memory-based retrieval through exemplar queries that combine temporal, spatial, and contextual cues, which are query types that would be difficult or impossible to express in traditional keyword-based systems.

\subsection{UPI: Enabling Innovation}\label{subsec:intro:innovation}

The UPI not only addresses current retrieval challenges but also lays the groundwork for future innovations that could fundamentally transform how humans interact with digital information. One such concept is the Personal Archivist, explored further in \autoref{ch:conclusion}, which represents an evolution from finding information to establishing collaborative memory systems that work alongside humans.

While the UPI focuses on transforming search into finding, it establishes the architectural foundation and philosophical approach for more advanced human-machine partnerships based on persistent memory and cooperative intelligence. These future directions suggest the potential for a profound reimagining of the relationship between humans and digital systems.

\subsection{Dissertation Scope and Methodology}\label{subsec:intro:scope}

This dissertation presents a systems architecture contribution that applies established cognitive models of human memory to improve information retrieval effectiveness. To set clear expectations:

\textbf{What this dissertation is:}
\sloppy
\begin{itemize}
\item A dis\-tributed systems archi\-tecture based on documented patterns of human memory (\autoref{ch:upi}),
\item A technical imple\-men\-tation of memory-oriented indexing through memory anchors (\autoref{sec:upi:memory-anchor}),
\item An empirical evaluation of how memory-aligned metadata improves retrieval performance (\autoref{ch:evaluation}), and
\item A demon\-stration that cognitive memory models can guide systems design decisions (\autoref{ch:evaluation}).
\end{itemize}
\fussy

\textbf{What this dissertation is not:}

\sloppy
\begin{itemize}
\item A psychology experiment validating theories of human memory,
\item A usability study measuring subjective user satisfaction, and
\item A design methodology requiring iterative user feedback.
\end{itemize}
\fussy

The evaluation methodology measures system performance through standard information retrieval metrics: precision, recall, resource utilization, and response time. These metrics directly assess whether memory-based architectural decisions improve the system's ability to find information, which is the core technical challenge this dissertation addresses.

\section{Contributions}\label{sec:intro:contributions}

This dissertation makes several significant \textbf{systems contributions} demonstrating how cognitive memory models can guide technical architecture:

\sloppy
\begin{description}
\descitem{Memory-Aligned Architecture}{The UPI provides the first comprehensive framework that directly maps established human memory patterns (episodic, semantic, procedural) to distributed system components, creating a retrieval architecture (described in \autoref{sec:upi:conceptual-foundations}) based on how humans naturally remember rather than how computers store.}

\descitem{Memory Anchors as Episodic Memory}{A novel technical implementation (detailed in \autoref{sec:upi:memory-anchor}) that captures temporal, spatial, and contextual metadata mirroring human episodic memory structures, validated through retrieval performance.}

\descitem{Mixed-Schema Database Model}{The UPI\textquotesingle{}s database architecture combines flexibility with performance while supporting memory-based query patterns (partial recall, associative retrieval), showing how memory research can inform database design.}

\descitem{Cross-Silo Memory Integration}{The collector/recorder architecture unifies fragmented information sources into a cohesive memory-aligned index, addressing how human memory seamlessly integrates diverse experiences.}

\descitem{Privacy-Preserving Memory Model}{The design protects sensitive memory-related metadata while maintaining the associative connections essential to memory-based retrieval (see Appendix~\ref{app:privacy:external} for details).}

\descitem{Empirical Performance Validation}{The comparative evaluation demonstrates that memory-aligned architecture achieves superior retrieval metrics (precision, recall, response time) compared to storage-centric systems, validating the technical benefits of memory-based design without relying on subjective user feedback.}
\end{description}
\fussy

While individual components build upon existing research in their respective domains, the integration of these components into a cohesive, memory-aligned architecture represents this dissertation's primary contribution. The UPI demonstrates that by incorporating cognitive memory principles into systems design, we can create information systems that achieve better technical outcomes through alignment with documented human memory patterns.

\section{Dissertation Structure}\label{sec:intro:structure}

The dissertation progresses from theoretical foundations (\autoref{ch:background}) through evidence of current system failures (\autoref{ch:status-quo}) to the UPI architecture (\autoref{ch:upi}) and its implementation and validation (\autoref{chap:implementation} and \autoref{ch:evaluation}). \autoref{ch:related-work} contextualizes this work within existing research, while \autoref{ch:conclusion} synthesizes contributions and explores future directions. Together, these chapters demonstrate how memory-aligned architectural principles can transform digital information retrieval effectiveness.

Additionally, this dissertation includes several appendices that provide supplementary information, including Appendix~\ref{app:privacy} which details the privacy and security considerations of the UPI, elaborating on how local-first approaches protect user data.

Readers primarily interested in the architectural contribution should focus on Chapters 4-5. Those seeking empirical validation should proceed to Chapter 6. Chapter 3 provides motivation through failure analysis of existing systems but may be skipped by readers already familiar with current limitations.

Having established the fundamental challenges and our memory-aligned approach to addressing them, we now turn to the technical foundations that make the UPI architecture possible. \autoref{ch:background} examines the specific technologies, methodologies, and theoretical frameworks that enable memory-aligned information retrieval in practice.

\addkhipuifneeded

\chapter{Background}\label{ch:background}

\begin{epigraph}
    \textit{The human mind\dots operates by association. With one item in its grasp,
    it snaps instantly to the next that is suggested by the association of thoughts,
    in accordance with some intricate web of trails carried by the cells of the brain.} --- Vannevar Bush, ``As We May Think'' (1945)~\cite{bush1945we}
\end{epigraph}

This chapter provides the technical foundation for the Unified Personal Index (UPI), focusing on the specific technologies and techniques that enable memory-aligned information retrieval. While \autoref{ch:intro} established the fundamental distinction between searching and finding and introduced memory-based design principles, this chapter explores the technical components that make such an approach feasible.

This chapter examines three key technical areas: metadata architecture and integration techniques, personal digital trace collection mechanisms, and AI-driven query processing. Each component provides essential capabilities for implementing a system that aligns with documented patterns of human memory while achieving measurable improvements in retrieval performance.

\section[Metadata Architecture]{Metadata Architecture for Memory-Aligned Retrieval}\label{ch:background:sec:metadata}

The UPI implements a three-layer metadata architecture that maps directly to established memory structures. \autoref{sec:upi:metadata-integration} provides comprehensive definitions of the three metadata types (storage, semantic, and memory anchor) including their roles in memory-aligned retrieval and technical implementation requirements.

\subsection[Metadata Availability]{Metadata Availability and Technical Challenges}

While the combination of storage, semantic, and memory anchor metadata enables rich
finding capabilities, real-world deployments must handle scenarios where metadata is
incomplete or inconsistent. The UPI addresses these challenges by utilizing existing
techniques through several key mechanisms:

\begin{description}
    \descitem{Graceful Degradation}{When certain types of metadata are unavailable, the system falls back to using available metadata types while maintaining functionality:}
    \begin{itemize}
        \item If semantic metadata is missing (e.g., for binary files or unsupported
        formats), the system relies more heavily on memory anchor data and storage metadata.
        \item When memory anchor data is sparse (such as for rarely accessed files or new
        content), the system emphasizes semantic content analysis and traditional
        storage attributes.
        \item In cases where only basic storage metadata exists, the system can still
        provide traditional search capabilities while accumulating richer context
        through subsequent user interactions.
    \end{itemize}

    \descitem{Cross-Platform Challenges}{The integration of metadata across different platforms and devices presents several technical considerations:
        \begin{description}
            \descitem{Metadata Normalization}{Different platforms often represent similar metadata in varying formats (e.g., timestamps, file attributes). The UPI implements normalization layers that transform platform-specific metadata into a consistent internal representation, enabling unified queries across heterogeneous sources.}
            \descitem{Synchronization Management}{Changes to files and their metadata can occur on multiple devices, potentially leading to conflicts. The UPI addresses this through:
                \begin{itemize}
                    \item Maintaining separate activity logs for each device,
                    \item Using timestamp-based reconciliation for metadata updates, and
                    \item Preserving platform-specific identifiers alongside normalized metadata.
                \end{itemize}
            }
            \descitem{Network Constraints}{The system must handle scenarios where network connectivity is limited or unavailable:
                \begin{itemize}
                    \item Local caching of frequently accessed metadata,
                    \item Queuing of metadata updates for later synchronization, and
                    \item Priority-based synchronization when bandwidth is limited.
                \end{itemize}
            }
        \end{description}
    }

    \descitem{Metadata Evolution}{The UPI's architecture acknowledges that metadata availability and quality often improve over time:
        \begin{itemize}
            \item New semantic analysis techniques can be applied to existing content,
            \item Memory anchor data accumulates through ongoing user interactions, and
            \item Integration with new platforms can provide additional metadata sources.
        \end{itemize}
    }
\end{description}

These considerations ensure that the UPI remains useful even when metadata is incomplete while providing mechanisms to enhance metadata richness over time. The system's ability to operate effectively with varying levels of metadata availability demonstrates its practical utility in real-world scenarios where perfect metadata coverage is rare.

\subsection[Use Case]{Use Case: Using Mixed Metadata to Find Existing Work}

Consider the scenario where a user is a data scientist trying to find analysis scripts they
wrote for a machine learning project. Their recollection is fragmented: they worked on it ``sometime
last quarter,'' it was related to a clustering algorithm, they remember discussing it in several
virtual meetings, they used both Python and R for different parts, and some initial work was done
on their laptop during their commute.

Traditional search systems would struggle with this query, as it combines technical, temporal,
and circumstantial details that are not easily captured by basic storage metadata. However, the
UPI can leverage multiple metadata types simultaneously:

\begin{description}
    \descitem{Storage Metadata}{Python (.py) and R (.R) files modified in the last quarter,}
    \descitem{Semantic Metadata}{
        \begin{itemize}
            \item Code content analysis identifying clustering algorithm implementations,
            \item Comments and documentation mentioning specific ML techniques, and
            \item References to data science libraries;
        \end{itemize}
    }
    \descitem{Memory Anchor Metadata (as detailed in \autoref{sec:upi:memory-anchor})}{
        \begin{itemize}
            \item Device transitions between laptop and workstation,
            \item Video conferencing logs showing relevant discussion topics,
            \item Location data indicating commute-time work sessions, and
            \item Application usage showing transitions between Python and R environments.
        \end{itemize}
    }
\end{description}

This scenario demonstrates how UPI's integration of multiple metadata types
enables finding even when the user's memory combines technical, temporal, and
circumstantial details that wouldn't be captured by traditional storage search systems.

\section{Personal Digital Traces and System-Level Metadata}\label{ch:background:sec:personal-traces}

Personal digital traces represent a rich source of memory anchor metadata that can populate the UPI's memory-aligned framework (see \autoref{sec:upi:memory-anchor} for detailed architecture). This section details the technical mechanisms for collecting and integrating these traces across different platforms.

\subsection{System-Level Metadata Collection Mechanisms}

The collection of these digital traces is enabled by modern operating system facilities. Two
key mechanisms support comprehensive trace collection:

\begin{description}
    \descitem{Extended Berkeley Packet Filter (eBPF)}{On Linux systems, eBPF provides a safe, efficient mechanism for monitoring system behavior. Unlike traditional logging systems, eBPF allows:
        \begin{itemize}
            \item Dynamic insertion of monitoring points without system modification,
            \item Low-overhead collection of process activity,
            \item Real-time capture of file system operations, and
            \item Application context tracking.
        \end{itemize}
    }

    \descitem{Event Tracing for Windows (ETW)}{Windows systems provide ETW for system monitoring, offering:
        \begin{itemize}
            \item Detailed process and thread tracking,
            \item File system activity monitoring,
            \item Application interaction logging, and
            \item System resource usage tracking.
        \end{itemize}
    }
\end{description}

These mechanisms enable the UPI to collect system-level metadata that provides crucial context
about user interactions, relationships between applications, and patterns of data usage.

\subsection{Technical Implementation Considerations}

The practical implementation of personal digital trace collection presents several technical
challenges that the UPI must address:

\begin{description}
    \descitem{Resource Efficiency}{Collection mechanisms must operate with minimal impact
    on system performance.}
    \descitem{Data Volume Management}{The potentially large volume of trace data requires
    efficient storage and indexing strategies.}
    \descitem{Cross-Platform Integration}{Traces must be collected and normalized across
    different operating systems and devices.}
    \descitem{Real-time Processing}{Context information needs to be processed and indexed
    in real-time to support immediate retrieval.}
\end{description}

The UPI addresses these challenges through careful system design and efficient implementation of its collection mechanisms. By leveraging modern operating system facilities like eBPF and ETW, the system can gather rich contextual information while maintaining reasonable resource consumption and ensuring responsive performance. Early explorations of comprehensive activity capture were pioneered by systems such as \citeauthor{lamming1992forget}~\cite{lamming1992forget}'s Forget-me-not and \citeauthor{freeman1996lifestreams}~\cite{freeman1996lifestreams}'s Lifestreams model, which demonstrated the feasibility and value of systematic digital trace collection for personal information management.

\subsection[Use Case]{Use Case: Leveraging Memory Anchor Data for Document Retrieval}

Consider a scenario where a lawyer needs to locate contract revisions they made in a period when:

\begin{itemize}
    \item They were simultaneously working on several similar contracts,
    \item They switched between their office desktop and tablet,
    \item They consulted with colleagues via email during the process,
    \item Some edits were made during client calls, and
    \item They referenced previous versions stored in different cloud services.
\end{itemize}

Each of these activities leaves a digital trace that can be leveraged to support finding
by the UPI:
\begin{itemize}
    \item Process tracking shows document comparison tool usage,
    \item System logs reveal file access patterns across devices,
    \item Communication records link relevant email threads,
    \item Calendar data correlates edits with specific client meetings,
    \item Cloud service API logs show cross-platform file access, and
    \item Location data confirms when work was done in the office versus remotely.
\end{itemize}

This multi-faceted activity stream allows the UPI to reconstruct the user's workflow,
making it possible to locate specific document versions based on the circumstances
of their creation rather than just their storage or semantic attributes.

\section{Human Memory Models for System Design}\label{sec:background:memory-models}

Understanding human memory structures is essential for designing information systems that align with natural recall processes. While traditional information retrieval systems rely primarily on content matching and categorical organization, human memory operates through complex associative networks that integrate contextual, temporal, and episodic information. This section examines the key theoretical frameworks from cognitive psychology that inform memory-aligned system design.

The theories presented here provide the foundational justification for the UPI's architectural decisions, particularly its emphasis on contextual metadata integration and episodic retrieval support. Each theory contributes specific insights about how humans encode, store, and retrieve information, directly informing the design principles outlined in \autoref{ch:upi}.

\subsection{Multiple Memory Systems Theory}\label{sec:background:multiple-memory-systems}

\citeauthor{tulving1972episodic}~\cite{tulving1972episodic} and \citeauthor{conway2000construction}~\cite{conway2000construction} highlight the fundamental distinction between semantic memory and episodic memory\index{episodic memory}, which has profound implications for information system design.

\textbf{Semantic memory} stores facts, concepts, and general knowledge organized through categorical relationships and abstract knowledge structures. Traditional search systems primarily support semantic retrieval through keyword matching and taxonomic organization.

\textbf{Episodic memory}\index{episodic memory} stores autobiographical events and experiences that are inherently contextual by remembering not just what happened, but when, where, with whom, and during what activity. Episodic memories are time-stamped, location-tagged, and emotionally colored personal experiences that form the rich narrative of our lives~\cite{tulving1985memory}. \citeauthor{tulving2002episodic}~\cite{tulving2002episodic} emphasizes that episodic memory represents a fundamentally different memory system from semantic memory, with its own neural substrates and retrieval mechanisms. For example, remembering ``the presentation I was working on during my flight to Berlin last month'' involves episodic memory\index{episodic memory} because it integrates the what (presentation), when (last month), where (flight to Berlin), and how (working during travel) into a single retrievable memory.

A memory-aligned information system must therefore support both modes of retrieval. While semantic search handles queries like ``find all PDF documents about machine learning,'' episodic retrieval enables queries like ``find the presentation I was working on during my flight to Berlin last month.'' The UPI architecture explicitly incorporates episodic retrieval capabilities through memory anchors (\autoref{sec:upi:memory-anchor}), providing the temporal, spatial, and social contextual cues that episodic memory\index{episodic memory} relies upon.

\subsection{Encoding Specificity Principle}\label{sec:background:encoding-specificity}

The Encoding Specificity Principle, established by \citeauthor{tulving1973encoding}~\cite{tulving1973encoding}, demonstrates that memory retrieval is most effective when the cues present at recall time match the conditions present during initial encoding. This principle explains why people often remember information better when they return to the physical location where they learned it, or when they recreate the emotional or environmental context of the original experience.

For information systems, this principle underscores the critical importance of capturing and maintaining the rich contextual elements surrounding information creation and use. These contextual elements, such as the applications being used, the physical location, the time of day, concurrent activities, and social context, serve as powerful retrieval cues that can dramatically improve recall success. The technical foundations for context-aware computing were established by \citeauthor{schilit1994context}~\cite{schilit1994context}, who demonstrated how applications could adapt to environmental context, and further developed by \citeauthor{dey2000context}~\cite{dey2000context} through the Context Toolkit, which provided architectural patterns for context-aware application development.

The UPI implements this principle through comprehensive memory anchor collection (\autoref{sec:upi:memory-anchor}), ensuring that the environmental and behavioral context surrounding information interactions is preserved and made available for retrieval. Rather than losing this contextual information when files are moved between systems or applications, the UPI maintains persistent links between information objects and their contextual history. This approach builds upon \citeauthor{lansdale1988psychology}~\cite{lansdale1988psychology}'s foundational work on the psychology of personal information management, which identified contextual cues as critical for successful personal information retrieval.

\subsection{Cognitive Load Theory}\label{sec:background:cognitive-load}

\citeauthor{sweller1988cognitive}~\cite{sweller1988cognitive} distinguishes between intrinsic cognitive load (the inherent difficulty of a task) and extraneous cognitive load (mental effort imposed by poor interface design or system complexity). This distinction provides crucial guidance for designing information systems that support rather than hinder human cognitive processes.

Traditional information systems often impose significant extraneous cognitive load by requiring users to:
\begin{itemize}
    \item Remember and navigate multiple organizational schemes across different applications,
    \item Translate their natural memory cues into system-specific query languages,
    \item Manually integrate information scattered across multiple silos, and
    \item Learn and adapt to inconsistent interface paradigms.
\end{itemize}

A human-centric information system must minimize extraneous cognitive load to free up mental resources for the user's actual tasks. The UPI addresses this through natural language query support, unified access across information silos, and result presentation that leverages familiar contextual cues rather than forcing users to adapt to system-imposed organizational schemes.

\subsection{Associative Memory Networks}\label{sec:background:associative-networks}

Human memory operates through complex associative networks where information items are linked based on shared context, temporal proximity, conceptual relationships, and experiential connections~\cite{karger2007s}. This understanding of associative information organization has deep roots in early computing visionaries like \citeauthor{nelson19654}~\cite{nelson19654}, who recognized that traditional hierarchical file structures fail to capture the interconnected nature of human knowledge and advocated for associative linking systems. These associations often transcend the categorical boundaries imposed by traditional information systems, creating meaningful connections between items that may reside in different applications, storage systems, or content types.

For example, a user might naturally associate a research paper, the email thread discussing its implications, the presentation slides derived from it, and the conference where it was presented. Traditional systems fragment these associations by organizing information according to application type, storage location, or format rather than meaningful experiential connections.

Memory-aligned systems must recognize and preserve these natural associative patterns. The UPI accomplishes this through its unified metadata approach, which maintains relationship information across traditional system boundaries. By tracking shared contexts, temporal correlations, and explicit user interactions, the system can reconstruct and leverage the associative networks that mirror human memory organization.

\subsection{Episodic Memory Theory}\label{sec:background:episodic-memory}

Episodic memory theory, developed by \citeauthor{tulving1972episodic}~\cite{tulving1972episodic}, emphasizes that memories of past events are not isolated data points but are inherently contextual, encoded with rich information about surrounding circumstances. Episodic memories integrate multiple dimensions of context: temporal (when it happened), spatial (where it occurred), social (who was involved), and activity-related (what was being done).

This theoretical framework directly informs the UPI's memory anchor architecture (described in \autoref{sec:upi:memory-anchor}). Rather than treating contextual information as optional metadata that might be useful for search refinement, episodic memory theory positions context as fundamental to how human memory operates. The system must therefore treat contextual information as first-class data, essential for supporting natural recall processes.

The UPI's memory anchor implementation captures these contextual dimensions through its five-category architecture (detailed in \autoref{sec:upi:memory-anchor:types}), ensuring the system can support retrieval patterns that match how humans naturally encode and recall personal experiences.

\subsection{Implications for System Design}

These memory theories converge on several key principles for information system design:

\begin{description}
    \descitem{Context is Fundamental}{Contextual information should be treated as essential metadata, not optional enhancement.}
    \descitem{Multiple Retrieval Modes}{
        \sloppy
        Systems must support semantic (fact-based) and episodic (experience-based) retrieval.
        \fussy
    }
    \descitem{Associative Organization}{Information organization should reflect natural human association patterns rather than purely hierarchical categorization.}
    \descitem{Cognitive Load Minimization}{Interface design should reduce extraneous mental effort, allowing users to focus on their actual tasks.}
    \descitem{Encoding Specificity}{Retrieval interfaces should leverage the same contextual cues present during information encoding.}
\end{description}

The UPI architecture, detailed in \autoref{ch:upi}, implements these principles through specific technical components and design decisions. Each architectural choice can be traced back to these theoretical foundations, ensuring that the system's design aligns with documented patterns of human memory and cognition.

\section{AI Integration in Memory-Based Retrieval Systems}\label{ch:background:sec:ai-integration}

The UPI's memory-aligned architecture creates new possibilities for natural language query interfaces. Recent advances in large language models (LLMs), particularly the release of ChatGPT 3.0 in late 2022, provided a compelling demonstration of how memory-based natural language queries can be translated into structured database operations, particularly ArangoDB's AQL query language.

\subsection{LLMs as an Implementation Example}

While the UPI architecture supports multiple query interface approaches (structured forms, visual query builders, etc.), LLM integration serves as a powerful example of what memory-aligned architecture enables:

\begin{description}
    \descitem{Query Translation}{Converting natural language queries into AQL database queries that leverage the UPI's three-layer metadata architecture.}
    \descitem{Context Understanding}{Interpreting memory-based cues (temporal, spatial, collaborative) and mapping them to appropriate metadata fields.}
\end{description}

Unlike siloed AI search systems (Dropbox Dash, OneDrive AI), the UPI's integrated metadata across platforms enables LLMs to process queries that combine storage attributes, semantic content, and memory anchor data (see \autoref{sec:upi:memory-anchor}). However, this same integrated metadata foundation supports alternative query interfaces that don't require LLM processing. Recent advances in LLM capabilities have created new opportunities for context-aware systems, with \citeauthor{thomas2024large}~\cite{thomas2024large} demonstrating how large language models can predict searcher preferences, and \citeauthor{fernandez2023large}~\cite{fernandez2023large} exploring how LLMs are disrupting traditional approaches to data management.

\subsection{Extensibility for Domain-Specific Applications}

The UPI's architecture supports domain-specific customization through:

\begin{description}
    \descitem{Custom Classifiers}{Specialized semantic analysis for domain-specific content (e.g., audio feature extraction for sound engineers).}
    \descitem{Extended Metadata Schemas}{Additional metadata fields tailored to specific use cases.}
    \descitem{Domain-Specific Query Patterns}{Specialized query interfaces for domain terminology and retrieval patterns (which may include LLM fine-tuning, custom parsers, or domain-specific visual interfaces).}
\end{description}

This extensibility enables the UPI to adapt to diverse professional domains while maintaining its core memory-aligned architecture.

\subsection[Use Case]{Use Case: Technical Documentation Retrieval}

Consider a scenario where a technical writer needs to find all documentation related
to a specific API feature:

\begin{itemize}
    \item The feature has evolved over time with different names.
    \item Documentation exists in multiple formats (Markdown, PDF, Google Docs).
    \item Some crucial information is in code comments.
    \item Related discussions exist in team chat logs.
    \item Implementation details are scattered across multiple repositories.
\end{itemize}

The UPI combines semantic analysis and AI to enable finding by:

\begin{itemize}
    \item Using NLP to understand semantic relationships between different feature names.
    \item Applying topic modeling to identify related discussions regardless of format.
    \item Extracting and connecting information from both formal docs and informal communications.
    \item Building knowledge graphs of relationships between code, documentation, and discussions.
    \item Understanding context from commit messages and code review comments.
\end{itemize}

This integration of semantic analysis of materials located in disparate storage silos
and combined with AI allows the UPI to understand not just what documents contain, but
how they relate to each other and the broader context of the
feature's development, regardless of where they are stored.

\section{Summary}\label{ch:background:sec:summary}

This chapter presented the technical foundations that enable the UPI's memory-aligned information retrieval. Three key technical components were examined:

\begin{description}
    \descitem{Three-layer metadata architecture}{Implementation challenges for normalizing storage metadata across platforms, extracting semantic content through NLP, and capturing activity stream data through system monitoring.}

    \descitem{System-level collection mechanisms}{Platform-specific tools (eBPF for Linux, ETW for Windows) that enable low-overhead activity stream tracking while maintaining privacy.}

    \descitem{LLM integration}{Natural language query translation to AQL, enabling memory-based queries without requiring users to learn database syntax.}
\end{description}

These technical components work together to support the memory-aligned architecture introduced in \autoref{ch:intro}, providing the implementation foundation for the UPI system detailed in subsequent chapters.

\addkhipuifneeded

\chapter{The Problem with the Status Quo}
\label{chap:status-quo}\label{ch:status-quo}

\begin{epigraph}
  \textit{``We are drowning in information, while starving for wisdom.''}
  \par\vspace{0.5em}
  \mbox{}\hfill\textsc{E.\,O. Wilson}
\end{epigraph}

This chapter analyzes the technical limitations of current data management systems from a memory-alignment perspective, revealing how their architectures fail to accommodate established patterns of human memory. This chapter examines the systemic failures in current implementations that necessitate a memory-aligned architecture\index{memory-aligned architecture} like the Unified Personal Index (UPI). Through technical analysis and case study evidence, we expose the memory model violations\index{memory model violations} in existing systems and establish the technical requirements for a memory-aligned approach.

The chapter is structured as follows: first, \autoref{sec:problem:case-study} describes a case study where we took an identical dataset and queries and evaluated them twice, almost two years apart, to better understand the inconsistencies in cross-platform retrieval\index{cross-platform retrieval!inconsistencies}; second, \autoref{sec:systemic-failures} examines the systemic failures in current architectures that violate human memory models\index{human-memory-models!inconsistencies}, such as architectural fragmentation\index{architectural fragmentation}, context-free architectures\index{context-free architectures}, episodic memory failures\index{episodic memory!failures}, and non-standardized personalization of proprietary silos\index{proprietary silos}; and third, \autoref{sec:status-quo-burden} discusses the consequences of these memory model violations.

By analyzing these technical misalignments with human memory models, we establish the architectural requirements that the UPI fulfills through its memory-aligned design, as detailed in subsequent chapters.

\section[Case Study]{Case Study: Cross-Platform Retrieval Inconsistencies}
\label{sec:problem:case-study}

\begin{table}[!tbhp]
    \caption[Per Service Search Results]{Number of Items Returned by Search Engine per Query\\Note: this was based upon the initial testing in February 2023 using the standard graphical browser interface.  This led to the December 2024 re-evaluation, which used scripts to gather specific data, leading to the analysis shown in \autoref{fig:venn-diagram} and \autoref{fig:upset-plot}.}
    \label{tab:search-results}
    \centering
    \begin{tabular}{lrrr}
        \toprule
        \textbf{Search}                  & \multicolumn{3}{c}{\textbf{Queries}} \\
        \cmidrule(lr){2-4}
        \textbf{Engine} & \textbf{``39''} & \textbf{``Anth 39''} & \textbf{``Anth 394''} \\
        \midrule
        Apple Finder & 550  & 272 & 26 \\
        Google Drive & 445	& 162 & 21 \\
        OneDrive & 99 & 160 & 115 \\
        \midrule
        \textbf{Total} & 1,094 & 594 & 162 \\
        \bottomrule
    \end{tabular}
\end{table}

We conducted a longitudinal technical evaluation using the personal data of a former medical anthropology student to examine cross-platform retrieval consistency. The dataset consisted of 26.8k files, with the dataset owner selecting typical keyword searches they would have used during their studies.

In our initial 2023 study, we uploaded the identical dataset to three newly established test accounts (iCloud, Google Drive, and Microsoft OneDrive) and used the same three queries across all platforms: ``39'', ``Anth 39'', and ``Anth 394.'' Using standard graphical interfaces, we captured only the number of items returned (\autoref{tab:search-results}). A key finding was that iCloud relies upon the local system search interface (Mac Finder), which is why we used Finder results for that platform.

\begin{figure}[!tbhp]
    \caption[Overlap Query Results]{Results Overlaps for query ``Anth 394'' Across Cloud Platforms (Dropbox, Google Drive, OneDrive)\\Note: This diagram shows overlaps among \textbf{precision-focused results}---files the data owner identified as actually relevant to the query (5 Dropbox, 12 Google Drive, 7 OneDrive files). This represents the subset of platform results that users would consider correct, highlighting how few relevant files overlap between platforms despite searching identical datasets. For recall-focused results (all files returned by platforms), see \autoref{fig:upset-plot}.}
    \label{fig:venn-diagram}
    \centering
    \includegraphics[width=0.95\textwidth]{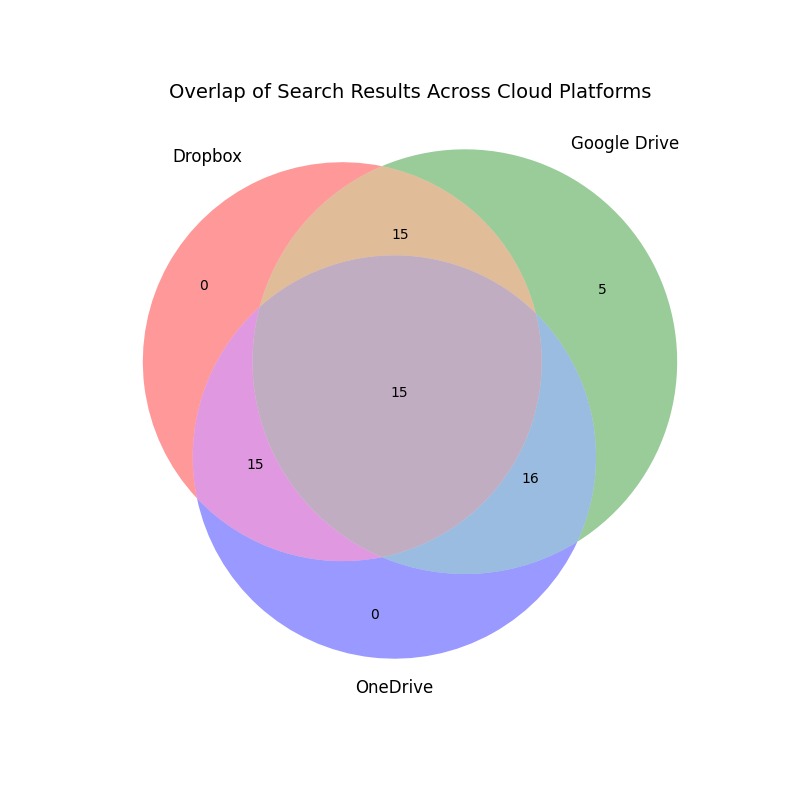}
\end{figure}

\begin{figure}[!tbhp]
    \caption[UpSet Plot: Overlap for ``Anth 394'']{UpSet plot of search results for query ``Anth 394'' across Dropbox, Google Drive, OneDrive, and iCloud (via Finder) in December 2024. Top bars show all files each API returned (recall-focused counts: 15, 21, 16, 34). The matrix with vertical bars encodes intersections, scaling better than a Venn diagram for $>3$ datasets and exposing asymmetries (unique vs. shared files). Compared with \autoref{fig:venn-diagram} (precision-focused, user‑judged relevant files), this highlights the precision gap: platforms return 2--5× more items than users deem relevant. Sparse overlap plus high per‑platform noise reveals reliance on platform-specific keyword matching instead of memory-aligned retrieval that would link ``Anth 394'' to related forensic anthropology materials via semantic, temporal, and episodic cues.}
    \label{fig:upset-plot}
    \centering
    \includegraphics[width=0.95\textwidth]{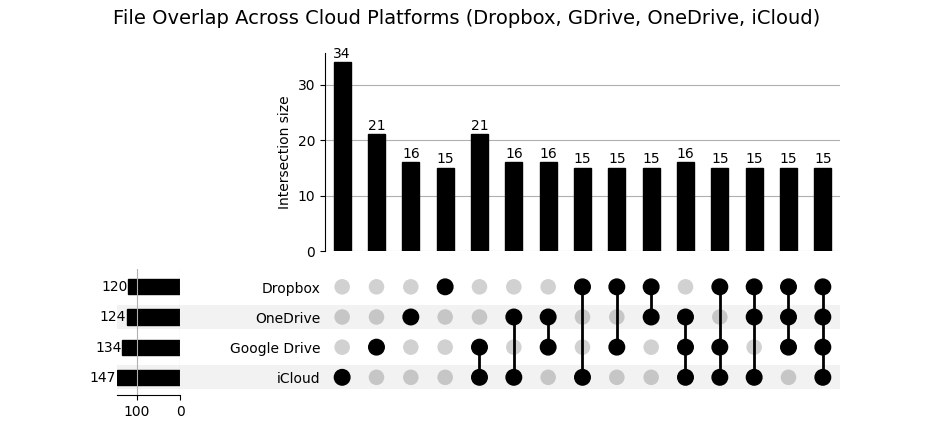}
\end{figure}

The preliminary results prompted a more rigorous follow-up study in December 2024. This second experiment used Python programs with search APIs to capture and analyze the actual results returned by each platform, rather than just result counts. We also added Dropbox to expand the platform comparison. The detailed methodology allowed us to examine not just quantities but the specific files returned, enabling the data owner to evaluate retrieval quality. Results differed from the 2023 findings, consistent with the evolutionary nature of these platforms. The detailed analysis revealed that while no single system appeared to miss files that all others found, users cannot expect consistent results across platforms even when searching identical datasets. These findings are visualized in \autoref{fig:venn-diagram} (cloud services only) and \autoref{fig:upset-plot} (including Mac Finder).

These results expose three memory-alignment failures:
\begin{description}
    \descitem{No associative networks\index{associative networks!missing}}{Platforms cannot connect semantically related content (``forensic anthropology'' materials) through the conceptual relationships human memory naturally employs.}
    \descitem{Missing episodic context}{Systems ignore temporal, spatial, and memory anchors that form the basis of human memory encoding and retrieval.}
    \descitem{Keyword-only matching\index{keyword matching!limitations}}{Reliance on literal string matching violates how human memory retrieves information through multiple associative pathways.}
\end{description}

Together, these figures reveal the dual failure of current keyword-based systems that necessitates memory-aligned architecture. The precision problem is evident in the dramatic difference between raw platform results (15-34 files) and user-identified relevant results (5-12 files): platforms overwhelm users with 2-5x more files than they actually want. Simultaneously, the recall problem appears in the minimal overlap even among relevant results---when platforms cannot consistently find the same relevant files from identical datasets, users must assume they are missing content that other platforms might locate. This dual precision/recall failure means users face both information overload (too many irrelevant results per platform) and information scarcity (missing relevant results across platforms), creating the fundamental retrieval challenges that memory-aligned systems must address.

\section[Memory Model Violations]{Memory Model Violations in Current Architectures}
\label{sec:systemic-failures}

Current architectures systematically violate established cognitive memory models by failing to implement associative networks\index{associative networks!missing}, contextual encoding, and multimodal retrieval pathways. This section analyzes these architectural failures and their technical consequences.

\subsection[Architectural Fragmentation]{Architectural Fragmentation and Memory Model Violations}
\label{sec:failures:fragmentation-federation}

Current platforms operate as isolated silos\index{storage silos} with incompatible architectures, violating how human memory functions as a unified, associative system. Each platform implements proprietary metadata schemas that cannot capture the multi-dimensional encoding (temporal, spatial, semantic, contextual) that characterizes human memory. This architectural fragmentation\index{architectural fragmentation} directly contradicts foundational personal information management research demonstrating that effective retrieval requires unified access to distributed information resources~\cite{jones2007personal,bergman2016science}. Our case study demonstrates this failure: identical queries produce divergent results because systems lack the associative memory structures needed for consistent retrieval.

Federation attempts fail because they assume architectural compatibility where none exists. Without memory-aligned architectures implementing consistent encoding principles, federation merely aggregates incompatible results rather than providing unified memory-based retrieval.

\subsection[Context-Free Architectures]{Context-Free Architectures and Episodic Memory Failures}
\label{sec:failures:context-episodic}

Current systems violate fundamental principles of human memory (as detailed in \autoref{sec:background:memory-models}) by implementing simplistic keyword matching that ignores contextual cues essential to episodic memory. While episodic memory encodes rich temporal, spatial, and memory anchors, platforms reduce metadata to basic file attributes, eliminating the contextual discriminators that could dramatically improve retrieval precision.

This architectural failure prevents implementation of associative memory structures. Despite decades of research showing human memory operates through dense networks of associations~\cite{bush1945we}, current architectures\index{context-free architectures} rely on isolated keyword matches that ignore the interconnected nature of human memory, creating the cognitive friction\index{cognitive friction} (as defined in \autoref{def:cognitive-friction}) that characterizes modern retrieval failures.

\subsection[Proprietary Silos]{Proprietary Silos and the Standardization Gap}
\label{sec:failures:standardization-personalization}

The lack of standardization across platforms creates challenging integration barriers. Each platform implements proprietary metadata schemas that prevent unified retrieval, while emerging services (AI tools, cloud platforms) introduce additional silos\index{proprietary silos} with rich but inaccessible activity metadata. This fragmentation violates the unified nature of human memory and prevents the cross-platform associative networks essential for memory-compatible retrieval.

\section[The Problem with the Status Quo]{The Problem with the Status Quo: Technical Retrieval Failures and Memory Model Violations}
\label{sec:status-quo-burden}
The architectural failures detailed above create significant technical retrieval barriers
due to systems that violate memory principles. When platforms cannot leverage associative
memory networks, episodic context, or semantic relationships, they fail to provide the
technical retrieval pathways that align with documented memory patterns. This cognitive friction\index{cognitive friction} manifests in measurable system inefficiencies, as documented in \autoref{sec:intro:problem}.
These consequences of architectures that fail to implement memory model principles in their technical design drive the need for memory-aligned systems.

Having documented these systematic failures of current systems to support human memory patterns, we now present the UPI architecture that directly addresses these limitations. \autoref{ch:upi} demonstrates how the memory-aligned design principles established in \autoref{ch:background} can be implemented as concrete technical components, transforming theoretical requirements into a practical system architecture.

The technical workarounds observed in practice, such as elaborate naming conventions, nested folder
hierarchies, and manual indexing systems, represent attempts to externally implement
the memory structures that systems should provide natively~\cite{barreau1995finding,jones2005don}. These workarounds
demonstrate the architectural gap: manual creation of associative networks,
contextual markers, and semantic relationships compensates for current systems' failure to
implement these memory principles. From a systems architecture perspective
~\cite{miller1956magical,kahneman1973attention}, this external implementation of memory
structures represents inefficient resource allocation, requiring computational and
storage overhead for functionality that should be native to the retrieval system.

This cognitive friction\index{cognitive friction} (detailed in \autoref{def:cognitive-friction}) creates measurable impacts on retrieval performance.
When retrieval systems require manual context reconstruction rather than preserving
memory-based associations, they violate established principles of information organization~\cite{miller1956magical}.
The resulting system inefficiencies manifest as increased retrieval latency, reduced precision,
and higher abandonment rates. These technical consequences stem directly from the cognitive friction\index{cognitive friction} (\autoref{def:cognitive-friction}) between system design and human memory models, demonstrating the urgent need for memory-aligned\index{memory-aligned architecture} retrieval architectures.

\addkhipuifneeded

\chapter{The Unified Personal Index}\label{ch:upi}

\begin{epigraph}
    \textit{%
        ``It is better for the intellectual not to talk all the time. To begin with, it would exhaust him, and above all, it would keep him from thinking. He must create if he can, first and foremost, especially if his creation does not side-step the problems of his time.''
    }
    \par\vspace{0.5em}
    \mbox{}\hfill\textsc{Resistance, Rebellion and Death (1960), Albert Camus.}
\end{epigraph}


\section{Introduction}
\label{sec:upi:introduction}

\begin{figure}[!tbhp]
    \caption[UPI Architecture]{Architecture of the Unified Personal Index showing the core components and data flow.
    From the top, data arrives from a heterogeneous set of sources, including storage services,
    semantic transducers, and activity stream providers.  Related state is formed into a memory
    anchor\index{memory anchors}, which may be linked back to the processed data. Raw data is preserved and normalized
    into the index layer.  This system then provides query support against the index, including
    providing current dynamic information about what normalized data is available from the UPI.}
    \label{fig:upi:architecture:1}
    \centering
    \resizebox{0.95\textwidth}{!}{%
        \begin{tikzpicture}[
            node distance=1cm and 1.5cm,
            block/.style={draw, rectangle, rounded corners, minimum width=3cm, minimum height=1.2cm, align=center},
            service/.style={block, fill=blue!10, dashed},
            layer/.style={block, fill=gray!10},
            engine/.style={block, fill=green!10, trapezium, trapezium left angle=70, trapezium right angle=110},
            storage/.style={block, fill=orange!10, cylinder, shape border rotate=90, aspect=0.3},
            interface/.style={block, fill=purple!10, chamfered rectangle, chamfered rectangle xsep=1cm},
            flow/.style={-{Latex[length=3mm]}, thick},
            infoflow/.style={flow, dashed},
            label/.style={font=\footnotesize\itshape}
        ]

        \node[block, fill=red!10] (sources) {Diverse User Data Sources};

        \node[layer, below=of sources] (interface) {Data Source Interface Layer};
        \draw[flow] (sources) -- (interface);

        \node[layer, left=of interface] (capture) {Activity Stream Capture Layer};

        \node[service, right=of capture] (context) {Memory Anchor\index{memory anchors} Management Service};
        \draw[flow] (capture) -- (context);

        \node[engine, below=of interface] (engine) {Normalization \& Unification Engine};
        \draw[flow] (interface) -- (engine);

        \node[storage, below=of engine] (index) {Unified Personal Index \\ {\footnotesize(Dynamic \& Extensible Schema)}};
        \draw[flow] (engine) -- node[right, label] {Unified Digital Objects} (index);

        \node[interface, below=of index] (query) {Context-Aware Query \& Retrieval Interface};
        \draw[flow] (query) -- (index);

        \node[block, fill=yellow!10, below=of query] (apps) {User/Applications};
        \draw[infoflow] (apps) -- node[left, label, near start] {Queries} (query);
        \draw[infoflow] (index) -- node[right, label] {Current Schema Description} (query);

        \node[below right=0.5cm and 0cm of apps.south east, anchor=north west, align=left, font=\footnotesize] {
            \textcolor{black}{Key} \\
            \tikz{\draw[flow] (0,0) -- (1,0);} Data Flow \\
            \tikz{\draw[infoflow] (0,0) -- (1,0);} Information/Control Flow \\
            \tikz{\draw[{Latex[length=3mm]}-{Latex[length=3mm]}, thick] (0,0) -- (1,0);} Bidirectional Interaction
        };
        \end{tikzpicture}%
    }%
\end{figure}
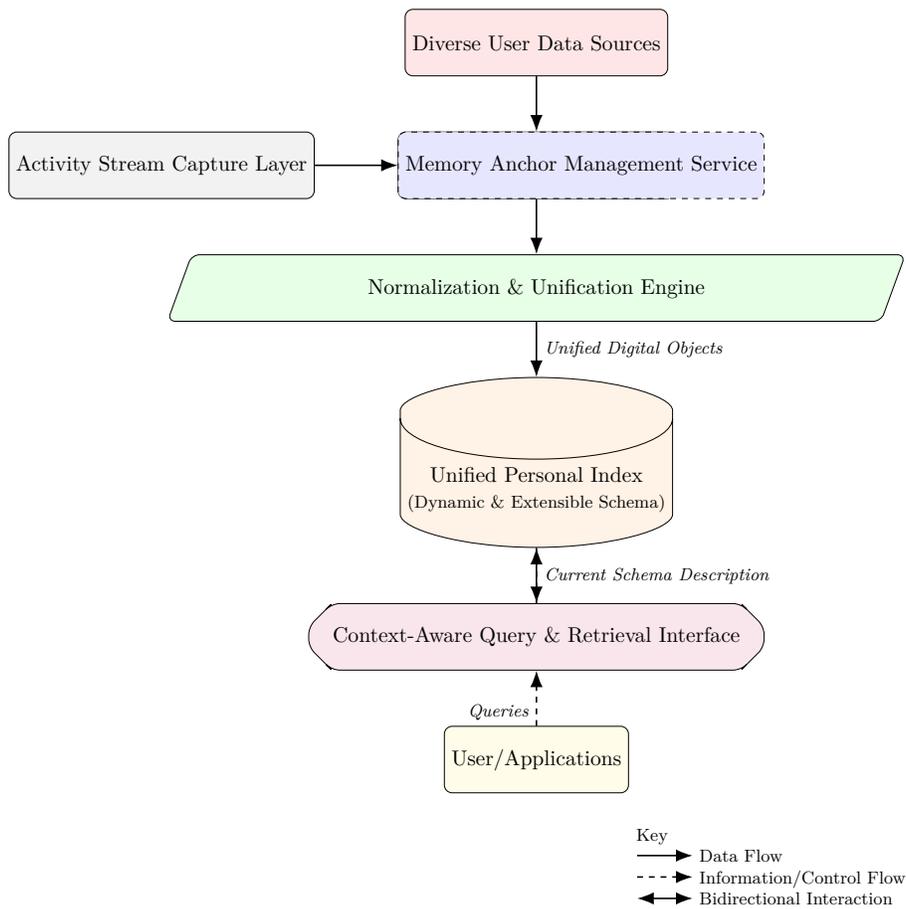

In this section, we introduce the Unified Personal Index (UPI)\index{Unified Personal Index (UPI)} as a solution to the limitations of current personal information retrieval systems. The UPI is designed to bridge the gap between human memory processes and technical infrastructure, enabling more effective and efficient retrieval of personal digital information. We begin by discussing the underutilized power of time in personal information retrieval, followed by an overview of the UPI's architecture and its core guiding principles.  See \autoref{fig:upi:architecture:1} for a high-level overview of the UPI architecture.

\subsection[The Underutilized Power of Time]{The Underutilized Power of Time in Personal Information Retrieval}
\label{sec:upi:introduction:time}

The quest to efficiently retrieve personal digital information often feels disconnected from the natural ways humans remember. As \citeauthor{vianna2019thesis}~\cite{vianna2019thesis} identifies, human memory and information seeking can be characterized by six contextual factors: who, what, when, where, why, and how. However, existing storage search systems typically focus overwhelmingly on the \emph{what} factor, primarily an item's name or its semantic content. While valuable, this narrow focus often fails to leverage other potent memory cues. Crucially, these systems lack robust mechanisms to link the what with the equally important who, where, why, and how.

Of all these contextual factors, when, the temporal dimension, holds a unique position. Timestamps are a pervasive form of metadata, diligently captured by nearly all storage systems for every digital object. Yet, this ubiquitous temporal data remains a largely underutilized asset for sophisticated, memory-aligned retrieval. While other contextual factors such as collaborators (who), locations (where), or project goals (why) are rich in mnemonic value, they often lack a direct, pre-existing ``hook'' or association mechanism within standard storage metadata. Timestamps, however, provide this inherent hook. The primary challenge, therefore, is not the absence of temporal data, but the absence of a systemic bridge to connect this data with the broader tapestry of human episodic experience and other contextual dimensions. Without such a bridge, a powerful mechanism for driving search precision and aligning retrieval with human memory is lost.

\subsection[Bridging Episodic Memory and Storage]{The UPI\index{Unified Personal Index (UPI)}: Bridging Episodic Memory and Storage via Time-Anchored Context}
\label{sec:upi:introduction:upi}

The UPI architecture is designed to construct this critical bridge by systematically capturing and linking broader experiential data to existing timestamps, transforming them from simple data points into powerful anchors for episodic recall. This approach builds upon pioneering work by \citeauthor{freeman1996lifestreams}~\cite{freeman1996lifestreams}, who introduced time-ordered information organization in Lifestreams, recognizing that temporal ordering provides a more natural structure for personal information than hierarchical file systems.

The practical power of this approach lies in search space reduction: temporal context serves as a primary discriminator that dramatically prunes the search space. When users recall ``the document I worked on last Tuesday'' or ``files from the conference in March,'' these temporal anchors immediately eliminate the vast majority of stored data, often reducing millions of files to hundreds of candidates before any other constraints are applied. The additional contextual dimensions (who, where, what, how) then refine within this temporally bounded set, creating a two-stage retrieval process that mirrors human memory's own hierarchical filtering.

Human episodic memory\index{episodic memory} is rich with temporal, spatial, and social context. The UPI makes digital retrieval consonant with this by treating the \emph{when}, and by extension the associated \emph{who}, \emph{where}, \emph{why}, and \emph{how}, as first-class metadata. Users naturally attempt to recall information using such cues, often thinking in terms of when an event occurred, where they were, or who they were with.

The UPI captures multi-faceted contextual information not typically associated with storage, such as ambient music, environment, concurrent communications, or project affiliations, and anchors it to the precise temporal footprint of related digital activities. This creates a rich, time-indexed layer of experiential metadata that serves as foundational infrastructure empowering the development of human-centric information retrieval tools rather than being an end-user application itself.

\subsection[Scope, Assumptions, \& Guiding Principles]{Scope, Assumptions, and Guiding Principles of the UPI}
\label{sec:upi:introduction:scope}

The UPI framework presented in this dissertation is specifically designed for personal information management. Its scope encompasses the integration of metadata from a user's array of personal devices, which includes their computers, smartphones, tablets, and other directly attached storage (e.g., external drives) as well as their personal cloud services, always under the explicit control of the individual user. A core assumption is that the collection of memory anchor\index{memory anchors} metadata, which is central to the UPI's human-centric approach, occurs only with informed user opt-in; no implicit tracking or simulation of context is presumed. Consequently, the UPI architecture does not address enterprise-scale data governance, multi-user organizational systems, or scenarios where data sovereignty does not reside with the individual.

\section[Conceptual Foundations]{Conceptual Foundations of the UPI}
\label{sec:upi:conceptual-foundations}
\label{sec:upi:conceptual-design} 

In this section we will explore the foundational conceptual pillars of the UPI architecture. These principles guide its design and distinguish it from conventional systems. The following subsections detail these core principles, their theoretical underpinnings, and the essential architectural capabilities required to realize them as a versatile underlying infrastructure.

\subsection[Foundational Principle]{Foundational Principle: Human-Centric Design\index{Human-Centric Design}}
\label{sec:upi:conceptual-foundations:human-centric-design}

\subsubsection{Principle Definition}
\label{sec:upi:conceptual-foundations:human-centric-design:definition}

Human-centric design\index{Human-Centric Design} is a paramount principle guiding the UPI. This dictates that the system's architecture and functionalities must be fundamentally oriented around human cognitive strengths and natural memory processes, rather than compelling users to adapt to machine-centric paradigms. The objective is to minimize the cognitive translation burden, that is the mental effort required for users to convert their natural thoughts and experiential memories into queries a system can understand. This approach draws from established principles in cognitive engineering, particularly \citeauthor{norman2013design}'s~\cite{norman2013design} emphasis on designing systems that support natural human cognitive processes rather than forcing users to adapt to technological constraints. Consequently, a UPI must enable interaction and information retrieval in ways that feel intuitive and directly support how users recall their past experiences and information. This principle shifts the focus from optimizing purely technical metrics of storage or search to enhancing the human experience of finding and reconnecting with personal information.

\subsubsection{Theoretical Underpinnings}
\label{sec:upi:conceptual-foundations:human-centric-design:theoretical}

This commitment to human-centricity is not arbitrary but is deeply rooted in established theories from cognitive psychology and human-computer interaction (HCI). The theoretical frameworks that inform this principle are detailed in \autoref{sec:background:memory-models}, including Multiple Memory Systems Theory~\cite{tulving1983elements}, the Encoding Specificity Principle~\cite{tulving1973encoding}, Cognitive Load Theory~\cite{risko2016cognitive}, and Situated Cognition~\cite{dourish1999embodied}. \citeauthor{tulving1985memory}~\cite{tulving1985memory} provides foundational insights into memory and consciousness, while \citeauthor{tulving2002episodic}~\cite{tulving2002episodic} offers a comprehensive account of episodic memory from mind to brain, establishing the neurological basis for the UPI's memory-aligned approach.

These theories converge on several key insights for UPI design: traditional systems primarily support semantic memory (facts and concepts) but largely neglect episodic memory\index{episodic memory} (experiential context); memory retrieval is most effective when recall cues match encoding conditions; system complexity should minimize extraneous cognitive load as defined by \citeauthor{sweller1988cognitive}~\cite{sweller1988cognitive}; and human cognition is inherently situated within environmental and memory anchors. The UPI architecture directly implements these insights through contextual metadata integration, natural language query support, and unified access across information silos.

\subsubsection[Essential Architectural Capabilities]{Essential Architectural Capabilities for a UPI to Realize Human-Centric Design}
\label{sec:upi:conceptual-foundations:human-centric-design:capabilities}

To translate the principle of Human-Centric Design and its theoretical underpinnings into a functional system, any UPI architecture must incorporate the following essential capabilities:

\begin{description}
    \descitem{Context-Aware Query Interpretation}{The system must be able to process user queries that are expressed naturally and may include rich contextual and episodic references (e.g., ``the report I was working on last week before the marketing meeting''\label{example:report-query}). This necessitates an interface (potentially leveraging Natural Language Processing) that can parse such queries and map them to the UPI's structured metadata, including memory anchor.}

    \descitem{Support for Episodic Memory Cues in Retrieval}{The architecture must allow for the explicit use of episodic memory\index{episodic memory} cues (temporal, spatial, social, task-related) in query formulation and execution. This means the underlying data model and query engine must be able to filter and rank information based on these contextual dimensions, not just content or explicit file attributes.}

    \descitem{Mechanisms for Reducing Cognitive Load}{The UPI should incorporate features designed to simplify interaction and reduce mental effort, leveraging cognitive offloading principles~\cite{risko2016cognitive} that allow external systems to augment human cognitive capabilities. This includes:

        \begin{itemize}
            \item \textit{Progressive Disclosure}: Presenting users with the most relevant information or options first, with the ability to delve into more detail if needed, rather than overwhelming them with all available metadata at once.

            \item \textit{Context-Sensitive Result Presentation}: Ranking and displaying search results in a way that considers not only content relevance but also their relevance to the user's current or queried context (e.g., prioritizing items from a specific remembered event or timeframe).
        \end{itemize}
    }

    \descitem{Integration of Memory Anchor}{A core enabler of human-centric design is the systematic capture and integration of memory anchor (detailed further in \autoref{sec:upi:conceptual-foundations:memory-anchor} and \autoref{sec:upi:memory-anchor}). This provides the raw material, namely the when, where, who, what activity, that allows the system to align with how users frame their memories around experiences.}

    \descitem{Flexible and Intuitive User Interface Concepts}{While the specific User Interface (UI) is an implementation detail of applications built upon the UPI, the UPI architecture must furnish the necessary APIs, query expressiveness, and contextual data access to support advanced interfaces that move beyond traditional keyword search boxes, potentially including conversational interaction, faceted Browse by contextual dimensions, or timeline-based exploration.}
\end{description}

\subsection[Foundational Principle: Memory Anchors]{Foundational Principle: Rich Memory Anchor\index{memory anchors} Integration}
\label{sec:upi:conceptual-foundations:memory-anchor}

\subsubsection{Principle Definition}

A second foundational principle of the UPI is its commitment to Rich Memory Anchor Integration. This signifies that a UPI must be architected to systematically capture, model, and utilize memory anchors\index{memory anchors} (see \autoref{def:memory-anchor}) to create a dynamic, time-correlated layer of metadata that reflects the human experience interwoven with digital object usage, thereby providing potent cues for memory-aligned retrieval.

\subsubsection{Theoretical Underpinnings}

The critical role of context in memory and cognition is well-established in the theoretical frameworks detailed in \autoref{sec:background:memory-models}. The architectural implications of this research, particularly episodic memory theory, the encoding specificity principle, and situated cognition, drive the UPI's treatment of contextual metadata as first-class information essential for retrieval rather than optional attributes.

\subsubsection[Essential Architectural Capabilities]{Essential Architectural Capabilities for a UPI to Realize Rich Memory Anchor Integration}

To effectively integrate rich memory anchor, a UPI architecture must possess the following essential capabilities:

\begin{description}

    \descitem{A Modular Activity Stream Collector Framework}{The system must be able to ingest contextual signals from a diverse and extensible set of ``activity stream providers.'' These providers would interface with various sources, including the operating system (for application usage, file events), specific applications (e.g., calendar, email, communication tools, media players), location services (e.g., from smartphones or other devices), and potentially other sensors or user-input channels.}

    \descitem{A Time-Centric Memory Anchor Model \& Management Service}{
        At its core, this capability involves:

        \begin{description}

            \descitem{Systematic Timestamping}{All captured memory anchor elements must be rigorously timestamped to establish their temporal relationship with each other and with digital object interactions. This transforms ubiquitous system timestamps from passive attributes into active anchors for rich contextual data.}

            \descitem{Contextual State Representation}{The ability to model and store snapshots or ``cursors'' (as described in \autoref{sec:upi:memory-anchor}) representing the user's multi-dimensional context (who, where, what activities, etc.) at given points in time or over specific durations.}

            \descitem{Lightweight Cursor-Based State Association}{Memory anchors function as lightweight cursors that can reference state through multiple dimensions (temporal, spatial, social, activity-based). In the simplest implementation, a memory anchor UUID maps to a temporal point, but the architecture supports richer associations. Critically, these cursors remain ephemeral, continuously updated by activity streams, until explicitly referenced, at which point they materialize into persistent memory anchors. This demand-driven persistence model ensures efficient storage while enabling state reconstruction when needed, following patterns similar to keyframe/delta compression in video encoding.}
    }
    \end{description}

    \descitem{Normalization and Unification of Contextual Data}{Given the heterogeneity of activity stream sources, the UPI must be able to normalize these varied inputs into a consistent internal model, resolving different terminologies or data structures for similar contextual concepts (e.g., standardizing location data from different providers).}

    \descitem{Query Engine Support for Contextual Queries}{The query processing capabilities of the UPI must be able to understand and execute queries that incorporate these rich memory anchor dimensions. This includes filtering by specific contextual attributes (e.g., ``files edited while application X was active,'' ``documents accessed while at location Y,'' ``items related to communication with person Z during time T'') and ranking results based on contextual relevance.}

    \descitem{Privacy-Preserving Context Management}{Given the sensitive nature of activity stream data, the architecture must incorporate mechanisms for user control over context collection (opt-in), data minimization, and secure storage and processing, ensuring that the benefits of contextual integration do not come at an unacceptable cost to user privacy. This follows \citeauthor{nissenbaum2004privacy}'s~\cite{nissenbaum2004privacy} principle of contextual integrity, which emphasizes that privacy protection depends on appropriate information flows within their proper contexts.}
\end{description}

\subsection[Foundational Principle]{Foundational Principle: Unified Metadata Approach}
\label{sec:upi:conceptual-foundations:unified-metadata}

\subsubsection{Principle Definition}

The third foundational principle of the UPI is its commitment to a Unified Metadata Approach. This mandates that a UPI must architecturally transcend the inherent fragmentation of personal information by ingesting metadata from disparate sources, which includes diverse storage systems, semantic analysis outputs, and the rich, time-correlated memory anchor, and transforming it into a coherent, integrated, and consistently queryable whole. The objective is to create a single, logical index where relationships between information objects can be established and traversed regardless of their original source, format, or the specific schema under which they were initially captured. This unification is essential for providing users with a seamless view of their personal information landscape and for enabling powerful cross-silo retrieval capabilities that are not possible when metadata remains isolated within individual platforms or applications.

\subsubsection{Theoretical Underpinnings}

The necessity for a unified metadata approach stems from the five design principles grounded in the theoretical frameworks detailed in \autoref{sec:background:memory-models}:

\begin{description}
    \descitem{Context is Fundamental}{(\emph{derived from Episodic Memory Theory and Encoding Specificity}): As detailed in \autoref{sec:background:episodic-memory} and \autoref{sec:background:encoding-specificity}, contextual information must be treated as essential metadata, not optional enhancement. The UPI's unified approach ensures that temporal, spatial, social, and activity contexts are preserved and queryable across all information sources, enabling retrieval based on the rich experiential cues that characterize human episodic memory.}

    \descitem{Multiple Retrieval Modes}{(\emph{derived from Multiple Memory Systems Theory}): As detailed in \autoref{sec:background:multiple-memory-systems}, systems must support both semantic (fact-based) and episodic (experience-based) retrieval. A unified metadata approach enables seamless transitions between these modes, allowing users to search by content ("documents about machine learning") or by experience ("the presentation I worked on during my Berlin flight").}

    \descitem{Associative Organization}{(\emph{linking to Associative Memory Networks}): As detailed in \autoref{sec:background:associative-networks}, human memory operates through associative networks where information items are linked based on shared context, temporal proximity, and experiential connections. A unified metadata model enables representation and querying of these cross-silo associations that mirror natural human memory organization. Without unification, these meaningful associations are broken at system boundaries.}

    \descitem{Cognitive Load Minimization}{(\emph{derived from Cognitive Load Theory}): As detailed in \autoref{sec:background:cognitive-load}, when users interact with multiple siloed systems, each with distinct organizational logic and query mechanisms, it imposes significant extraneous cognitive load. A unified approach minimizes these cognitive seams by providing consistent interaction models and single-point access to integrated information spaces.}

    \descitem{Encoding Specificity Support}{(\emph{derived from Encoding Specificity Principle}): As detailed in \autoref{sec:background:encoding-specificity}, retrieval is most effective when recall cues match encoding conditions. The unified metadata approach preserves the rich contextual elements present during information creation and use, making these elements available as retrieval cues regardless of where information is stored or accessed.}
\end{description}

\subsubsection[Essential Architectural Capabilities]{Essential Architectural Capabilities for a UPI to Realize Unified Metadata Approach}

To effectively implement a Unified Metadata Approach, a UPI architecture must provide the following essential capabilities:

\begin{description}
    \descitem{A Comprehensive Metadata Normalization Engine}{The system must possess robust capabilities to transform heterogeneous metadata records from all integrated sources (storage systems, semantic extractors, memory anchor providers) into a consistent internal schema or a canonical data model. This includes:
        \begin{itemize}
            \item \textit{Schema Mapping}: Defining equivalences between different source-specific metadata fields that represent the same conceptual information.
            \item \textit{Value Standardization}: Converting data values (e.g., dates, times, location formats, identifiers) into uniform formats.
            \item \textit{Entity Resolution}: Identifying and linking records that refer to the same real-world entity (e.g., the same person, document, or project) across different data sources and platforms.
        \end{itemize}
    }

    \descitem{A Flexible and Extensible Unified Schema (\emph{or Schema Management System})}{The UPI must operate on a unified data model that can cohesively represent storage metadata, semantic metadata, and memory anchor metadata, along with their interrelationships. Crucially, this model and its underlying storage (e.g., a mixed-schema database as previously discussed) must be:

        \begin{itemize}
            \item \textit{Extensible}: Capable of incorporating new metadata sources, new types of metadata attributes, and evolving relationship types without requiring disruptive system-wide migrations.

            \item \textit{Dynamic}: Able to handle partially structured or schemaless data where appropriate, while still enforcing consistency for core attributes.
        \end{itemize}
    }

    \descitem{Cross-Platform Identity and Relationship Management}{The architecture must be able to establish and maintain meaningful relationships between metadata elements originating from different silos. This includes tracking the provenance of information and resolving how an object in one system (e.g., a file in Google Drive) relates to an object in another (e.g., an email attachment in Outlook) or to a specific memory anchor.}
    \descitem{A Query Engine Supporting Unified Access}{The query processing capabilities must operate seamlessly over this unified metadata model, allowing users or applications to formulate queries that span multiple original sources and metadata types without needing to be aware of the underlying source-specific details or perform manual cross-system joins. This engine must leverage the normalized data and resolved entities to provide comprehensive results.}

    \descitem{Consistency and Synchronization Mechanisms}{The UPI requires mechanisms to detect metadata changes in source systems, which allows it to propagate them into the unified index to maintain reasonable consistency, acknowledging the challenges of real-time synchronization across distributed personal data sources.}
\end{description}

Having established the conceptual foundations and essential capabilities that any UPI implementation must possess, we now turn to the concrete architectural framework that realizes these principles in practice. The transition from abstract concepts to specific technical components requires careful design decisions that maintain fidelity to the memory-aligned goals while addressing practical implementation challenges.

\section[UPI Architecture]{UPI\index{Unified Personal Index (UPI)} Architectural Framework}\label{sec:upi:architecture}

\autoref{sec:upi:conceptual-foundations} describes the foundational principles that guide the design of the UPI. These principles are realized through a specific architectural framework.  This framework defines the necessary components and their interactions to achieve the UPI's goals of unified, human-centric information retrieval. This section details this overarching architecture, beginning with a high-level overview before examining its core components and the unified metadata model they operate upon.

\subsection[Architectural Overview]{High-Level Architectural Overview}\label{sec:upi:architecture:overview}

The architecture of the Unified Personal Index is designed to systematically ingest, process, unify, and index metadata from diverse personal information sources, making it accessible for human-centric retrieval. \autoref{fig:upi:architecture:1} provides a schematic representation of this architecture, illustrating the core components and the flow of data and control information between them.

As depicted, the UPI operates by first gathering data from a variety of user data sources. This includes traditional storage metadata from filesystems and cloud services, semantic information potentially extracted from content, and crucially, signals from activity stream providers which capture aspects of the user's experiential context. The ``Activity Stream Capture Layer'' and ``Memory Anchor Management Service'' are dedicated to processing these experiential signals, forming time-correlated ``memory anchors.''

Both direct data source information and these memory anchors are then fed into a ``Normalization \& Unification Engine.'' This engine is responsible for transforming these heterogeneous inputs into a consistent internal representation, creating ``Unified Digital Objects'' that integrate storage attributes, semantic annotations, and memory anchor references (UUIDs\index{UUID} that point to rich contextual state, not merely timestamps). These unified objects are then stored and indexed within the ``Unified Personal Index Core,'' which is designed with a dynamic and extensible schema to accommodate the diverse and evolving nature of personal information.

Finally, the ``Context-Aware Query \& Retrieval Interface'' allows users or applications to interact with the indexed information. This interface supports queries that can leverage the rich, integrated metadata, particularly the memory anchor and temporal anchors, and provides access to ``Current Schema Description'' from the index, enabling adaptive and contextually relevant information retrieval. The overall design emphasizes modularity and a clear data processing pipeline, facilitating the integration of diverse data types and supporting the human-centric principles foundational to the UPI.

\subsection[Core Components]{Core Architectural Components of a UPI}\label{sec:upi:architecture:components}

Realizing the foundational principles of the UPI and its high-level architecture, as described in \autoref{sec:upi:conceptual-foundations} and \autoref{sec:upi:architecture:overview} respectively, depends upon a set of core architectural components. Each component has distinct responsibilities for processing and managing personal information and its associated metadata. The following subsections detail these essential components, outlining their purpose, interactions, and their role in enabling a UPI system.

\subsubsection[Data Ingestion]{Data Ingestion Layer\index{Data Ingestion Layer}}
\label{sec:upi:architecture:ingestion}

\paragraph{Purpose and Responsibilities}

The Data Ingestion Layer serves as the primary interface between the Unified Personal Index and the diverse, heterogeneous sources of a user's personal information. Its fundamental responsibility is to collect or receive raw metadata and, where appropriate, content records from various platforms, including local filesystems (on computers, smartphones, tablets, and other directly attached storage), cloud storage services, email systems, and other relevant application-specific data stores. This layer must be designed for extensibility to accommodate new data sources as a user's digital ecosystem evolves. It is responsible for the initial acquisition of storage metadata (e.g., filenames, paths, timestamps, sizes) and can also serve as a conduit for semantic metadata if it's available directly from the source or through an initial pre-processing step (though deeper semantic enrichment often occurs later).

\paragraph{Key Inputs and Outputs}

\begin{description}
    \descitem{Inputs}{The inputs are varied, reflecting the nature of the source systems. They include file/object listings, file attributes, content streams (for indexing or semantic processing), and event notifications (e.g., change events from a monitored filesystem or cloud service API).}
    \descitem{Outputs}{This layer outputs serialized batches of raw or semi-structured metadata records and, when real-time processing is required (such as for file system monitors or cloud service webhooks), event queues for streaming updates. These outputs are then passed to downstream components, primarily the Normalization \& Unification Engine, for further processing. The format of these outputs should be standardized enough for consistent handling by subsequent layers, even if the original inputs were highly diverse.}
\end{description}

\paragraph{Contribution to Foundational Principles}

The Data Ingestion Layer is critical for realizing the Unified Metadata Approach. By providing the mechanisms to gather metadata from otherwise siloed platforms, it performs the essential first step in overcoming information fragmentation\index{information fragmentation}~\cite{jones2007personal,bergman2016science}. Personal information management research has extensively documented how fragmentation across storage platforms creates retrieval challenges that cannot be addressed through conventional search approaches alone~\cite{barreau1995finding,jones2005don}. \citeauthor{jones2007keeping}~\cite{jones2007keeping} provides a comprehensive analysis of these challenges in personal information management, demonstrating how users struggle to maintain coherent views of their information across fragmented storage systems. Its ability to interface with diverse sources is a prerequisite for creating the single, logical index that the UPI aims to provide. While not directly capturing experiential data itself, it ingests the storage metadata (especially timestamps) that the Rich Memory Anchor Integration principle relies upon as anchors. Furthermore, by abstracting the specifics of diverse data sources, it contributes indirectly to Human-Centric Design by shielding higher-level components and ultimately the user from the underlying complexity and heterogeneity of their digital storage.

\subsubsection[Activity Stream]{Activity Stream Capture Layer\index{Activity Stream Capture Layer}}
\label{sec:upi:architecture:activity-stream-capture}

\paragraph{Purpose and Responsibilities}

The Activity Stream Capture Layer is a specialized architectural component dedicated to gathering the raw signals that constitute a user's experiential context. Its primary responsibility is to interface with various sources that produce data about user activities and the environment in which these activities occur. These sources can be diverse, ranging from operating system event streams (e.g., application launches, window focus changes, file system interactions beyond basic storage metadata), to browser history, to APIs from specific applications (e.g., media players indicating current track, calendar applications indicating current meetings), to sensor data from personal devices (e.g., location from a smartphone, ambient noise levels where available and permitted). This approach builds upon foundational work in context-aware computing by \citeauthor{schilit1994context}~\cite{schilit1994context}, who established the principles of context-aware applications, and the architectural patterns developed by \citeauthor{dey2000context}~\cite{dey2000context} through the Context Toolkit. This layer focuses on capturing time-stamped events and state changes that reflect what the user is doing, what tools they are using, and aspects of their surrounding digital and physical environment. It acts as the primary collector for the raw inputs needed by the Memory Anchor Management Service.

\paragraph{Key Inputs and Outputs}

\begin{description}
    \descitem{Inputs}{Real-time or batched event streams from OS-level monitors, application-specific connectors, device sensors, and other contextual data providers. These inputs are often low-level and heterogeneous.}
    \descitem{Outputs}{Structured, time-stamped activity records or event streams. These are typically normalized to a certain degree by this layer (e.g., common event naming, standardized timestamp formats) before being passed to the Memory Anchor Management Service for further processing and the formation of coherent ``memory anchors.''}
\end{description}

\paragraph{Contribution to Foundational Principles}

The Activity Stream Capture Layer is paramount for achieving Rich Memory Anchor Integration. It provides the essential raw materials, which are the observed user activities and environmental signals, that, once processed, form the rich, time-correlated memory anchors. By systematically capturing these experiential signals, it directly enables the UPI to move beyond static metadata and incorporate dynamic, usage-based context. This, in turn, is fundamental to Human-Centric Design, as it allows the system to gather the cues necessary for aligning with episodic memory (as discussed in \autoref{sec:upi:conceptual-foundations:human-centric-design:theoretical}). The rigorous timestamping of captured data by this layer reinforces the UPI's ability to use time as a primary anchor, linking disparate activities and digital objects through their temporal co-occurrence. While distinct from the Data Ingestion Layer (which focuses more on static storage and semantic metadata), this layer complements it by focusing on the dynamic stream of user experience, which is then unified with other metadata by downstream components as part of the Unified Metadata Approach.

\subsubsection[Memory Management]{Memory Anchor Management Service\index{Memory Anchor Management Service}}
\label{sec:upi:architecture:memory-anchor-management}

\paragraph{Purpose and Responsibilities}

The Memory Anchor Management Service is responsible for transforming the raw, often disparate, event streams and state information gathered by the ``Activity Stream Capture Layer'' into coherent, structured, and queryable ``memory anchors.'' While superficially these might appear as time:action mappings, the service actually models and manages multi-dimensional experiential state through lightweight cursors that can be reconstructed on demand. This involves aggregating signals from various activity stream providers, resolving conflicts or ambiguities, and creating discrete memory anchor instances (represented by UUIDs, as detailed in \autoref{sec:upi:memory-anchor}). Each instance functions as a cursor to a rich contextual state such as location, active applications, communication partners, concurrent media, ongoing tasks, rather than merely recording what action occurred when. Critically, these cursors employ lazy persistence: they remain ephemeral and continuously updated until explicitly referenced, at which point they materialize into persistent memory anchors. This service ensures that memory anchors form a structured representation of experiential state optimized for state reconstruction and memory-aligned retrieval, not simply an event log.

\paragraph{Key Inputs and Outputs}

\begin{description}
    \descitem{Inputs}{Structured, time-stamped activity records or event streams from the ``Activity Stream Capture Layer.'' This includes data from potentially many different providers, each reporting on different facets of the user's activity and environment.}
    \descitem{Outputs}{
        \begin{description}
            \descitem{Memory Anchor Handles/Instances}{These are the primary output, which is uniquely identifiable (e.g., via UUIDs) lightweight cursors that enable on-demand reconstruction of the user's context for a given time. Rather than proactively creating consolidated views, each handle serves as a reference that can reconstruct experiential state when queried.}
            \descitem{Linkages}{Information about which digital objects were active or relevant during the period covered by a specific memory anchor handle.}
            \descitem{Contextual State Updates}{Information that can be used by other services (like the Query Interface) to understand current or past user context.}
        \end{description}
    }
\end{description}

\paragraph{Contribution to Foundational Principles}

This service is the direct engine for realizing the Rich Memory Anchor Integration principle. It moves beyond mere data capture to the intelligent modeling and management of that context. By creating memory anchor cursors with temporal grounding, it provides the precise mechanism for linking human experience to digital artifacts, which is crucial for Human-Centric Design and enabling retrieval based on episodic memory. It systematically leverages timestamps from the input activity stream to ensure that all generated contexts are temporally grounded, reinforcing the UPI's time-centric approach. Furthermore, by processing and structuring diverse activity signals into a more uniform ``memory anchor'' representation, it contributes to the Unified Metadata Approach, preparing this vital stream of metadata for integration with storage and semantic metadata by the Normalization \& Unification Engine. The handling of potentially sensitive user activity stream data by this service underscores the importance of the UPI's overarching ``Privacy by Design''\index{Privacy by Design} guiding principle. The security architecture follows a user-managed model (detailed in \autoref{app:privacy}).

\subsubsection[Normalization \& Unification]{Normalization \& Unification Engine\index{Normalization \& Unification Engine}}
\label{sec:upi:architecture:normalization-unification}

\paragraph{Purpose and Responsibilities}

The Normalization \& Unification Engine is a central processing component within the UPI architecture, responsible for transforming the diverse and often heterogeneous metadata streams received from the ``Data Ingestion Layer'' and the ``Memory Anchor Management Service'' into a coherent, consistent, and integrated representation. Its primary duties include:

\begin{description}
    \descitem{Schema Mapping and Transformation}{Reconciling different source-specific schemas and data structures into the UPI's unified internal metadata model. This involves mapping disparate field names, data types, and structural conventions to a common standard.}

    \descitem{Value Standardization}{Converting metadata values (e.g., date/time formats, location representations, units of measurement) from various sources into uniform, canonical formats to ensure consistency and enable accurate comparisons and queries.}

    \descitem{Entity Resolution}{Identifying and linking records from different sources that refer to the same real-world entity (e.g., the same person, organization, document, or conceptual project) to avoid redundancy and establish a unified view of these entities across the user's information landscape.}

    \descitem{Relationship Establishment}{Creating and maintaining explicit relationships between different metadata elements. This includes linking digital objects to their corresponding memory anchors, associating semantic annotations (like topics or named entities) with content, and establishing connections between related items across different platforms or data silos.}

    \descitem{Enrichment Orchestration}{While some enrichment may happen at earlier stages, this engine often orchestrates or integrates further semantic enrichment processes (e.g., applying NLP-derived features, linking to external knowledge bases) to the normalized metadata.}
\end{description}

\paragraph{Key Inputs and Outputs}

\begin{description}
    \descitem{Inputs}{
        \begin{itemize}
            \item Serialized batches of raw or semi-structured metadata records from the ``Data Ingestion Layer'' (representing files, emails, and other digital objects from various storage sources),
            \item Memory Anchor handles/instances from the ``Memory Anchor Management Service'' (representing the user's experiential state at specific times),
            \item Potentially, semantic annotations or features derived from preliminary content analysis or external services;
        \end{itemize}
    }

    \descitem{Outputs}{
        \begin{itemize}
            \item Unified Digital Objects: These are the primary output: structured, normalized representations of information items, now enriched with resolved entities, standardized attributes, and explicit links to relevant memory anchors and other related objects. These objects are represented in a consistent mixed-schema format ready for indexing, storage, and retrieval, and
            \item Established relationships between these unified objects.
        \end{itemize}
    }
\end{description}

\paragraph{Contribution to Foundational Principles}

The Normalization \& Unification Engine is the cornerstone for realizing the Unified Metadata Approach. It performs the critical work of transforming a fragmented collection of disparate metadata into a single, cohesive, and queryable information space. By resolving schemas, standardizing values, and performing entity resolution, it ensures that information from diverse sources can be meaningfully compared and integrated. This unification is essential for the Rich Memory Anchor Integration principle, as it ensures that memory anchors are accurately linked to the correct, canonical representations of digital objects and semantic entities, regardless of their original source. Furthermore, by abstracting the complexity and heterogeneity of the underlying data sources and presenting a harmonized view to the indexing and query layers, this engine significantly contributes to Human-Centric Design. It helps minimize the cognitive load that would otherwise be required if users or higher-level system components had to deal directly with inconsistent and fragmented raw metadata. This engine's ability to create a common representation is a prerequisite for enabling the intuitive, cross-silo search and retrieval that the UPI aims to provide.

\subsubsection[UPI Core]{Unified Personal Index Core\index{UPI Core} (Storage and Indexing)}
\label{sec:upi:architecture:upi-core-storage-indexing}
\label{sec:indexing-strategies} 

\paragraph{Purpose and Responsibilities}

The Unified Personal Index Core (hereafter ``UPI Core'') serves as the central persistent repository and indexing system for all normalized and unified metadata within the UPI architecture. Its primary responsibility is to efficiently store, manage, and make retrievable the ``Unified Digital Objects'' produced by the ``Normalization \& Unification Engine.'' This includes all integrated storage metadata, semantic annotations, and the rich, time-correlated memory anchors, along with the relationships established between them.

A key architectural requirement for the UPI Core is its ability to operate with a dynamic and extensible schema. This is essential to accommodate the diverse and evolving nature of personal information, allowing for the integration of new data sources, metadata types, and relationship structures without requiring disruptive, system-wide schema migrations. Furthermore, the UPI Core must be designed to support efficient querying across these varied data types and their interconnections, enabling the complex, multi-faceted retrieval demanded by the UPI's human-centric principles.

\paragraph{Key Inputs and Outputs}

\begin{description}
    \descitem{Inputs}{
        \begin{itemize}
            \item Unified Digital Objects from the ``Normalization \& Unification Engine.'' These are structured, normalized representations of information items, enriched with resolved entities, standardized attributes, and explicit links to memory anchors and other related objects.
        \end{itemize}
    }
    \descitem{Outputs}{
        \begin{description}
            \descitem{A Queryable Index}{A robust, efficient index that supports structured queries (e.g., based on specific attributes or metadata fields), graph-based traversal of relationships between objects and contexts, and potentially full-text search over indexed content.}
            \descitem{Schema Descriptors}{Information about the currently available metadata types, entity definitions, and context handle schemas, which are exposed to the ``Context-Aware Query \& Retrieval Interface'' to enable dynamic query construction and adaptive UI generation.}
        \end{description}
    }
\end{description}

\paragraph{Contribution to Foundational Principles}

The UPI Core is instrumental in actualizing the Unified Metadata Approach by providing the persistent, integrated storage and indexing capabilities necessary to manage the unified data. Its design, particularly the emphasis on a mixed-schema model (potentially leveraging graph database capabilities as discussed in \autoref{sec:upi:conceptual-foundations:unified-metadata}) capable of representing rich relationships, is fundamental to this.

This component directly supports Rich Memory Anchor Integration by storing the memory anchor handles and the links between these contexts and various digital objects and semantic entities, making these experiential anchors persistent and queryable. By efficiently indexing this integrated and contextualized metadata, the UPI Core enables the fast and flexible retrieval required for Human-Centric Design. It provides the backend power for queries based on episodic memory\index{episodic memory} cues, allowing the system to quickly locate information based on temporal, spatial, social, or task-related context. The ability to expose schema information also supports adaptive interfaces that can present relevant filtering and querying options to the user, further reducing cognitive load.

\subsubsection[Query \& Retrieval]{Context-Aware Query \& Retrieval Interface\index{Context-Aware Query Interface}}
\label{sec:upi:architecture:context-aware-query}
\label{sec:upi:query} 

\paragraph{Purpose and Responsibilities}

The Context-Aware Query \& Retrieval Interface (hereafter ``Query Interface'') serves as the primary access point through which users or external applications interact with the Unified Personal Index to find information. Its fundamental responsibility is to translate a user's information need, which are often expressed in natural language or framed by experiential and contextual cues, into effective queries that can be executed against the ``UPI Core.'' Beyond query translation, this interface is also responsible for managing the interaction, potentially engaging in disambiguation dialogues, and presenting the retrieved results in a meaningful, context-sensitive manner. It must be designed to understand and leverage the rich, integrated metadata, particularly the temporal and memory anchor dimensions, that the UPI provides.

\paragraph{Key Inputs and Outputs}

\begin{description}
    \descitem{Inputs}{
        \begin{description}
            \descitem{User Queries}{These can range from natural language questions (e.g., ``find the presentation I worked on with Mai last week'') to more structured queries involving specific metadata attributes or contextual filters.}
            \descitem{Current Schema Descriptors}{Information from the ``UPI Core'' about available metadata types, entity definitions, and context schemas, enabling the interface to adapt its query interpretation and presentation capabilities dynamically.}
            \descitem{Current User Context (Optional)}{Signals about the user's immediate situation (e.g., current location, active application, ongoing task), which can be used to further refine query interpretation and result ranking, if provided and consented to.}
        \end{description}
    }
    \descitem{Outputs}{
        \begin{description}
            \descitem{Ranked Result Sets}{Lists of information items (references to unified digital objects) that match the user's query, ranked not only by content relevance but also by their alignment with specified or inferred contextual criteria.}
            \descitem{Contextualized Summaries and Provenance}{Information accompanying results that explains why an item was retrieved (e.g., highlighting matching contextual cues) and provides details about its origin or history.}
            \descitem{Prompts for Clarification/Disambiguation}{In cases of ambiguous or underspecified queries, the interface may output questions or suggestions to the user to help refine the search.}
        \end{description}
    }
\end{description}

\paragraph{Contribution to Foundational Principles}

The Query Interface is the most direct manifestation of the UPI's Human-Centric Design principle. By supporting natural language queries and enabling users to leverage their episodic memory (e.g., ``files I edited at the office last Tuesday''), it significantly reduces the cognitive translation burden typically associated with traditional search systems. Its capabilities for interactive disambiguation and context-aware ranking further align with reducing cognitive load and supporting natural human information-seeking behaviors.

This interface directly benefits from and operationalizes the Unified Metadata Approach and Rich Memory Anchor Integration. It provides the means to effectively query the integrated, normalized, and context-enriched metadata stored in the ``UPI Core.'' Without this sophisticated interface, the value of the underlying unified data would be largely inaccessible to a non-technical user. It must be architected to:

\begin{itemize}
    \item Parse and understand queries that reference diverse metadata types, including temporal, spatial, social, and task-related memory anchor.
    \item Leverage schema introspection to adapt to the available metadata and offer relevant querying options.
    \item Support extensible query modules or plugins that can enhance its interpretation capabilities (e.g., through LLM-based intent prediction or specialized recommendation engines), ensuring it can evolve alongside new retrieval techniques and user needs.
\end{itemize}

\subsection[Unified Metadata Model]{The Unified Metadata Model: Integrating Diverse Information}
\label{sec:upi:architecture:metadata-model}
\label{sec:upi:metadata-integration}  

The effectiveness of the UPI hinges on its ability to cohesively manage and leverage diverse types of information. Central to this is the Unified Metadata Model, which provides the conceptual and structural foundation for integrating the varied data collected by the UPI. This model is designed to go beyond traditional file attributes to create a rich, interconnected representation of a user's personal information landscape.

Architecturally, any UPI system must operate on a metadata model that explicitly integrates three core types of information, each serving distinct but complementary roles in memory-aligned retrieval:

\begin{description}
    \descitem{Storage Metadata}{This encompasses the traditional attributes maintained by storage systems, including filenames, paths, timestamps (creation, modification, access), file sizes, MIME types, and permission structures. While providing minimal memory cues on their own, these attributes serve as the foundational layer for all retrieval operations. The UPI normalizes storage metadata across platforms to address technical challenges including platform-specific timestamp formats (Windows FILETIME vs. Unix epoch), path representations (Windows backslashes vs. Unix forward slashes), and divergent permission models. This normalization enables consistent querying across heterogeneous storage environments while preserving the temporal anchors essential for episodic memory alignment.}

    \descitem{Semantic Metadata}{This includes content-derived information extracted through analytical processes such as natural language processing, computer vision, and domain-specific analysis. Examples include extracted topics, recognized named entities (people, organizations, locations), keywords, document summaries, embedded spatial data (EXIF GPS coordinates), content fingerprints, and relationships inferred between documents. Technical implementation requires format-specific parsers for diverse content types, language detection capabilities, and specialized handling for binary content. Semantic metadata provides memory pattern matching capabilities by connecting content meaning to retrieval contexts, enabling queries based on ``what'' rather than just ``where'' information is stored.}

    \descitem{Memory Anchor Metadata}{As detailed in \autoref{sec:upi:conceptual-foundations:memory-anchor} and \autoref{sec:upi:memory-anchor}, this crucial category captures comprehensive episodic memory patterns including temporal context, spatial context, social interactions, task affiliations, and access patterns. Following the W5H model~\cite{vianna2014a,vianna2019thesis}, this metadata captures comprehensive episodic memory patterns. It includes the rich, time-correlated experiential data surrounding a user's interactions with their information, capturing the ``when, where, who, what activity, and how'' associated with digital object usage. Implementation requires platform-specific collection mechanisms (eBPF on Linux, ETW on Windows) designed for minimal performance impact while providing high-discrimination memory cues that align with human recall patterns. This metadata type transforms simple timestamps into powerful anchors for episodic recall by systematically linking temporal markers to broader experiential contexts.}

\end{description}

The integration of these three metadata types creates a layered approach to memory-aligned retrieval: storage metadata provides the technical foundation and temporal structure, semantic metadata enables meaning-based connections, and memory anchor metadata supplies the rich contextual framework that mirrors human episodic memory formation. This architectural integration enables the UPI to support both traditional search paradigms and memory-aligned retrieval patterns within a unified framework.

The UPI architecture processes these metadata types through a conceptual five-stage pipeline to achieve unification:

\begin{description}
    \descitem{Ingestion}{The initial collection of raw metadata records and activity event data from diverse sources, as handled by the Data Ingestion Layer and Activity Stream Capture Layer.}
    \descitem{Normalization}{The transformation of these heterogeneous records into a consistent internal representation, including schema mapping, value standardization, and entity resolution, primarily performed by the Normalization \& Unification Engine.}
    \descitem{Enrichment}{The augmentation of normalized metadata with additional semantic annotations (e.g., NLP-derived features) and the crucial linking of memory anchor handles to the relevant digital objects and entities.}
    \descitem{Indexing}{The storage of these unified and enriched digital objects, along with their relationships, into the UPI Core in a manner that supports efficient, multi-faceted querying.}
    \descitem{Serving}{The exposure of the indexed information and its schema to the Context-Aware Query \& Retrieval Interface, enabling runtime introspection and adaptive query planning.}
\end{description}

A critical characteristic of the UPI's metadata model, and consequently its storage infrastructure (the UPI Core), is the necessity for a dynamic and extensible schema. Given the ever-evolving nature of personal digital ecosystems, such as new applications, new data types, and new forms of interaction, the UPI must be able to seamlessly incorporate new metadata fields, sources, and relationship types without requiring fundamental architectural overhauls or breaking existing functionality. This flexibility ensures the UPI can remain adaptable and relevant over time, fulfilling its role as a long-term personal information index.

\subsection[Core UPI Systems]{Realizing Core UPI Systems Contributions through Architecture}
\label{sec:upi:architecture:contributions}

The architectural framework and components detailed thus far are designed to deliver specific, novel systems-level contributions to personal information management. The UPI's core contributions, as identified in \autoref{sec:intro:thesis} (Thesis Statement) and \autoref{sec:upi:introduction}
are realized as follows:

\begin{description}
    \descitem{A Cross-Platform Metadata Normalization Framework}{This contribution is primarily actualized by the combined functionalities of the Data Ingestion Layer (\autoref{sec:upi:architecture:ingestion}), which gathers metadata from diverse sources, and the Normalization \& Unification Engine (\autoref{sec:upi:architecture:normalization-unification}). The latter performs the critical tasks of schema mapping, value standardization, and entity resolution, transforming heterogeneous inputs into the UPI's consistent Unified Metadata Model (\autoref{sec:upi:conceptual-foundations:unified-metadata}). This enables consistent querying across platforms, a key UPI objective.}

    \descitem{A Memory Anchor Integration Pipeline}{This is architecturally embodied by the Activity Stream Capture Layer (\autoref{sec:upi:architecture:activity-stream-capture}), which collects raw experiential signals; the Memory Anchor Management Service (\autoref{sec:upi:architecture:memory-anchor-management}), which models these signals into structured, time-stamped memory anchors; and their subsequent integration with other metadata by the Normalization \& Unification Engine. The storage and indexing of these contexts within the UPI Core (\autoref{sec:upi:architecture:upi-core-storage-indexing}) make them queryable. This pipeline provides the high-discrimination contextual metadata crucial for aligning retrieval with human episodic memory.}

    \descitem{A Relationship-Centric, Mixed-Schema Data Model and Storage}{This contribution is supported by the design of the Unified Metadata Model (\autoref{sec:upi:architecture:metadata-model}) itself, which emphasizes the integration of diverse metadata types and their interconnections. It is physically realized within the UPI Core (\autoref{sec:upi:architecture:upi-core-storage-indexing}), which must be architected to efficiently store and query not just individual metadata attributes but also the rich relationships between digital objects, semantic entities, and memory anchors. The requirement for a dynamic and extensible schema, potentially leveraging graph-database characteristics, allows this model to balance query performance with expressive relationship modeling.}
\end{description}

By explicitly designing the architecture around these components and their interactions, the UPI provides a robust and extensible foundation for achieving these core systems contributions, moving beyond the limitations of existing personal information retrieval systems.

The architectural framework described above defines the structural components needed for memory-aligned retrieval, but understanding how these components work together requires examining the dynamic processes that transform raw personal data into queryable memory structures. This leads us to examine the data flow that brings the architecture to life.

\section[Data Pipeline]{Data Flow and Processing Pipeline}\label{sec:upi:dataflow}

The preceding sections have detailed the conceptual foundations of the UPI and its core architectural framework. This section now traces the end-to-end journey of metadata as it moves through the UPI system: from initial collection from diverse sources to its transformation into a unified, queryable representation ready to support human-centric retrieval. This pipeline is essential for actualizing the UPI's principles, particularly the Unified Metadata Approach and Rich Memory Anchor Integration, by systematically processing raw data into the integrated foundation required for memory-aligned finding.

To illustrate this transformation, consider Dr. Priya Sharma, a computational linguistics researcher at the University of São Paulo (USP). On a September morning at 9:23 AM, she saves a draft paper titled ``contextual-embeddings-draft.pdf'' to her local Documents folder while working from her apartment. At this initial moment, the UPI captures basic storage metadata: filename, timestamps, file size, and MIME type. Concurrently, memory anchor collection begins recording her environmental context, such as how she's using LaTeX on her laptop, playing a Brazilian jazz playlist on Spotify, with email notifications from her research group active in the background.

As Priya continues her research across subsequent weeks, this simple document accumulation reveals emerging patterns. She edits the paper during focused morning sessions (typically 8-11 AM) while at home, but reviews references during afternoon library visits at USP. When traveling to present preliminary findings at COLING in Bangkok, she accesses the paper from her hotel room at midnight local time, comparing her approach with newly discovered related work. Each interaction adds temporal, spatial, and activity layers to what began as a simple file save operation.

After six months of iterative work, this mundane metadata accumulation enables profound self-insight. When Priya queries ``What factors correlate with my most productive writing sessions?'' the UPI reveals an unexpected pattern: her highest-impact writing consistently occurs during early morning hours while listening to instrumental music, specifically when ambient temperature data from her smart thermostat shows 21-23°C, and when she has recently exchanged emails with her collaborator Prof. Chen in Singapore, which suggests that international timezone-shifted discussions prime her thinking. This discovery emerges not from any single data point, but from the UPI's systematic integration of temporal patterns, environmental sensors, application usage, and communication metadata across time.

This example illustrates how each stage of the UPI's data processing pipeline contributes to transforming routine digital activities into memory-aligned retrieval infrastructure, enabling discoveries that would be impossible through traditional file-based organization.

\subsection[Cross-Platform Collection]{Cross-Platform Collection: Gathering the Digital Fragments}\label{sec:upi:collection}

The UPI's data processing pipeline begins with the Cross-Platform Collection stage, where the ``Data Ingestion Layer'' (as described in Section 4.3.2.1) systematically gathers metadata and, where applicable, content from the user's diverse digital sources.

\begin{description}
    \descitem{Platform-Specific Collectors}{To interface with the wide array of user data sources (e.g., local filesystems, cloud storage APIs like Google Drive or Dropbox, email servers), the Data Ingestion Layer employs specialized software modules, often referred to as ``collectors.'' Each collector is designed to understand the specific API, data formats, and communication protocols of a particular platform (e.g., handling OAuth for a cloud service, or traversing a local filesystem). They are responsible for retrieving the necessary metadata (and sometimes content) and passing it to subsequent processing stages in a more standardized intermediate format.}

    \descitem{Incremental Collection}{To optimize resource usage and maintain up-to-date information efficiently, the UPI's collectors are designed for incremental updates. After an initial full collection from a data source, subsequent operations aim to retrieve only the changes (deltas) that have occurred since the last synchronization. This approach relies on mechanisms such as timestamp comparisons or change logs (if provided by the source API) to identify new or modified items, significantly reducing network traffic and processing load.}

    \descitem{Event-Driven Updates}{For data sources that support real-time notifications (e.g., via webhooks or other eventing mechanisms), the UPI's collectors can adopt an event-driven approach. Instead of periodically polling for changes, the UPI can subscribe to and receive immediate notifications when data is created, modified, or deleted. For example, saving a document in a cloud service that supports such events could trigger an immediate update to the UPI. This method ensures near real-time synchronization and further reduces the overhead associated with polling-based strategies.}
\end{description}

\subsection[Schema Normalization]{Schema Normalization: Speaking a Common Language}\label{sec:upi:normalization}

After the initial collection of metadata by the ``Data Ingestion Layer'' and ``Activity Stream Capture Layer,'' the next critical stage in the UPI's data flow is Schema Normalization. Data arrives from heterogeneous sources with disparate schemas, naming conventions, and value formats (e.g., ``creation date' in one system might be ``DateTimeOriginal'' in another). The normalization process, primarily handled by the ``Normalization \& Unification Engine'' (\autoref{sec:upi:architecture:normalization-unification}), is responsible for transforming these varied inputs into a consistent, unified internal representation suitable for indexing and querying.

\begin{description}
    \descitem{Schema Mapping}{A core function of normalization is schema mapping. This involves defining and applying rules to reconcile differences in how various source systems structure and name their metadata fields, particularly when they refer to the same conceptual property or grouping. For example, one cloud service might provide information about shared access under a collection named \texttt{permissions} with attributes like \texttt{role} and \texttt{userEmail}, while another platform might expose similar information within a \texttt{sharingInfo} object containing fields like \texttt{accessLevel} and \texttt{sharedWithID}. The UPI's normalization process must map these disparate structural and naming conventions to a standardized internal representation for access control or collaboration metadata within its unified model.}

    \descitem{Value Normalization}{Beyond mapping field names and structures, the actual values of metadata attributes often require standardization. For example, timestamps may be represented using different epochs (e.g., Unix epoch seconds vs. Windows FILETIME ticks from January 1, 1601) or in various string formats with differing timezone conventions. Filenames themselves might be stored using different character encodings (e.g., UTF-8, UTF-16, or legacy codepages) which require normalization to a common internal standard (like UTF-8) for consistent searching and display. The UPI's normalization process converts these varied value representations into consistent, canonical formats. This ensures accurate comparisons, sorting, and range queries across data originating from different sources.}

    \descitem{Entity Resolution}{The UPI's normalization process also includes entity resolution, where the system identifies and links records that refer to the same real-world entity. For instance, if a user has multiple copies of the same document across different platforms, the normalization engine recognizes these as duplicates and consolidates them into a single record.}

    \descitem{Relationship Normalization}{Finally, the UPI normalizes how different metadata elements relate to each other. This includes establishing relationships between digital objects (e.g., files, emails) and their associated memory anchors (e.g., when they were accessed or modified). By creating a consistent model of these relationships, the UPI enables more sophisticated querying and retrieval.}
\end{description}

Through these normalization processes, such as schema mapping, value normalization, entity resolution, and relationship normalization, the UPI transforms disparate metadata inputs into a consistent and unified internal representation. While striving for this unified view, the system is designed to maintain fidelity to the original source information where necessary. This often involves a bidirectional mapping capability, preserving references to source-specific identifiers and attributes. Such mappings are crucial for allowing users or processes to, when needed, trace information in the UPI back to its native representation within a specific source platform, thereby bridging the UPI's unified model with the original data ecosystems.

\subsection[Semantic Enhancement]{Semantic Enhancement: Adding Meaning to Metadata}\label{sec:upi:semantic}

While schema normalization (\autoref{sec:upi:normalization}) ensures consistency and basic structural understanding of metadata, the Semantic Enhancement stage in the UPI's data flow aims to imbue this metadata with deeper meaning and create richer connections. Raw and normalized metadata often describe what an object is (e.g., its type, size) and when it was interacted with (timestamps), but may lack explicit information about its conceptual content, its relationship to other entities, or its relevance within a broader context. Semantic enhancement processes apply analytical techniques and leverage contextual information to extract or infer these higher-level insights, making the indexed information more discoverable through meaning-based and associative queries.

\begin{description}
    \descitem{Content Analysis}{Digital objects often contain rich semantic information embedded within their content, which goes beyond basic file attributes. For example, image files may depict recognizable landmarks, people, or objects; text documents contain topics, key entities, and arguments; and audio/video files possess spoken words or identifiable sounds. Through techniques such as natural language processing (NLP) for textual data and image/video analysis for visual media, the UPI architecture supports the extraction of semantic features. These can include named entities (e.g., ``St. Mark's Square'' from a photo, ``Dr. Qori'' from a document), topics (e.g., ``Italian Architecture,'' ``Quantum Computing Research''), and potentially even sentiment or emotional tone. The normalization process maps diverse extraction formats (confidence scores, entity types, relationship tuples) to a unified semantic schema, ensuring that entities extracted by different tools, whether ``Dr. Qori'' from OCR, NER, or manual annotation, resolve to the same canonical representation. This extracted and normalized semantic metadata adds crucial layers of meaning, enabling content-based retrieval that is more nuanced than simple keyword matching.}

    \descitem{Cross-Platform Correlation and Relationship Inference}{The UPI architecture supports the inference of relationships between digital objects that may exist across different platforms or data silos, even in the absence of explicit, pre-defined links. By analyzing patterns in metadata, particularly temporal proximity (e.g., items created or accessed around the same time), shared semantic entities (e.g., common people, locations, or topics mentioned), or co-occurrence within similar memory anchor, the system can identify likely associations. For instance, if a user captures photos in a specific location (e.g., Venice) during a particular week, and around the same time saves travel-related emails (e.g., train tickets, hotel reservations) and creates expense notes that mention the same location or timeframe, the UPI can infer a probable relationship between these disparate items, linking them to a common event or trip. Such inferred relationships further enrich the interconnectedness of the user's information within the unified index.}

    \descitem{Secondary Contextual Enrichment and Refinement}{Beyond the primary capture of memory anchor (Section 4.2.2), the semantic enhancement stage can further enrich and refine the contextual understanding of digital objects. This involves processes that leverage the already normalized metadata and captured memory anchors to add deeper layers of meaning or establish more nuanced connections. Examples include:
        \begin{description}
            \descitem{Semantic Geocoding}{Translating raw geographic coordinates (whether from EXIF data in images or from memory anchors) into meaningful place names (e.g., ``St. Mark's Square, Venice'') by querying external geospatial services.}
            \descitem{Event Correlation}{Linking digital objects or activities to specific events identified in a user's calendar (e.g., associating a presentation file with the ``Quarterly Review Meeting'' during which it was likely used, based on temporal proximity).}
            \descitem{Entity-Driven Contextualization}{If content analysis (as described above) extracts key entities like people or projects from a document, this step might involve strengthening the association of that document with other information or activities known to involve those same entities, drawing from the broader unified index.}
            \descitem{Cross-Modal Context Linkage}{For instance, linking a series of photos taken during a specific time window to music that was being played concurrently, if both photo creation times and music playback memory anchor were captured.}
        \end{description}
    }
\end{description}

Collectively, these semantic enhancement processes, such as content analysis, cross-platform correlation and relationship inference, and secondary contextual enrichment, transform the initially normalized metadata into a significantly more meaningful and interconnected dataset. The outcome is a rich semantic structure where digital objects are not only described by their basic attributes and primary memory anchors but are also linked through inferred relationships, enriched with content, derived insights, and further contextualized by refined environmental and situational data. This enhanced level of semantic depth and interconnectedness is crucial for enabling the UPI to support nuanced, associative queries that align more closely with the associative and multi-dimensional nature of human memory, moving beyond simple attribute-based retrieval.

\subsection[Integration and Unified Storage]{Integration and Unified Storage: Weaving the Digital Tapestry}\label{sec:upi:integration}

Following the collection, normalization, and semantic enhancement of metadata from diverse sources, the next stage in the UPI's data flow is the Integration and Unified Storage of this processed information. This critical step involves consolidating the enriched metadata into the ``Unified Personal Index Core'' (as described in \autoref{sec:upi:architecture:upi-core-storage-indexing}). The objective is to create a single, persistent, and queryable representation of the user's interconnected personal information landscape, where relationships between disparate items are explicitly maintained and leveraged.

\begin{description}
    \descitem{Mixed-Schema Storage and Relationship Model}{To accommodate the diverse and evolving nature of personal metadata, which span structured file attributes, semi-structured semantic annotations, and varied memory anchor data, the UPI's ``Unified Personal Index Core'' must employ a storage model that supports mixed schemas. This architectural requirement means the system is not constrained by rigid, predefined table structures. Instead, it can flexibly store and index varied data types and their specific attributes while also robustly representing the complex, often graph-like relationships between different digital objects, entities, and memory anchors. This approach is essential for preserving the rich web of connections that semantic enhancement and memory anchor integration reveal, enabling multi-faceted, associative queries.}

    \descitem{Cross-Platform Identity Mapping}{The UPI architecture must maintain mechanisms for mapping entities within its unified index back to their original identities and locations within source platforms. This involves storing and managing references (e.g., source-specific IDs, URLs, file paths) that allow the system, or applications built upon it, to trace a unified digital object back to its native representation (e.g., a specific file in Google Drive, an email in a particular Outlook account, or a locally stored document). This capability is crucial for actions such as opening an original file from a search result, verifying information against its source, or understanding data provenance. It effectively bridges the UPI's abstracted, unified model with the user's concrete, multi-platform digital environment.}
    \descitem{Indexing of Temporal States and Versioning Information}{
        Personal information is dynamic, with digital objects often undergoing changes and evolving through multiple versions over time. The UPI architecture acknowledges this by:
        \begin{enumerate}
            \item Inherently capturing a temporal dimension through its memory anchor lineage (as described in \autoref{sec:upi:memory-anchor}), which records the evolution of user context surrounding their interactions with information.
            \item By being capable of indexing metadata about different versions of digital objects if such versioning information is provided by the source systems (e.g., version history from cloud storage platforms or version control systems).
        \end{enumerate}

        The UPI itself may not perform file-level versioning, but its metadata model and normalization engine should be able to recognize and link different versions of the same conceptual item if the source data allows.

        This enables users to query for information as it existed or was interacted with during specific past contexts or, where source versioning is available, to potentially identify or retrieve specific historical states of an object (as shown in the query example~\ref{example:report-query}).
    }
\end{description}

Through these integration and storage mechanisms, the UPI consolidates metadata into a cohesive representation that transcends platform boundaries and supports associative retrieval patterns aligning with human memory. Implementation details of this storage model are described in \autoref{chap:implementation}.

The unified metadata architecture and processing pipeline described above provide the structural foundation for memory-aligned retrieval, but the critical innovation that distinguishes the UPI from traditional systems lies in its implementation of memory anchors. These memory anchors transform abstract cognitive principles into concrete technical mechanisms that capture the experiential context essential to human episodic memory.


\section{Memory Anchors\index{memory anchors}: The Memory Bridge}\label{sec:upi:memory-anchor}

Memory anchor\index{memory anchors} serves as the critical bridge between human episodic memory\index{episodic memory} and digital storage systems, implementing the UPI's memory-aligned architecture\index{memory-aligned architecture} through systematic capture of experiential metadata. This section details how memory anchor transforms the abstract principles of memory-aligned design into concrete technical mechanisms that enable retrieval based on natural memory patterns.

\subsection{Conceptual Architecture of Memory Anchor}\label{sec:upi:memory-anchor:conceptual}

Memory anchor implements episodic memory\index{episodic memory} patterns~\cite{tulving1973encoding,marr1971simple} through a cursor-based architecture that connects human experiences with digital artifacts, building upon concepts from BURRITO~\cite{guo2012burrito} which demonstrated how activity feeds can capture user interactions for computational experiments. Unlike traditional metadata that captures only static file properties, memory anchor preserves the dynamic experiential context surrounding information use: the when, where, how, and why that form the basis of human episodic memory. Prior work provides empirical support for the idea that people naturally use such contextual cues to find information~\cite{vianna2014a,vianna2019thesis,dumais2016stuff}.

The architecture distinguishes between a \textit{memory anchor} (lightweight cursors) and an \textit{activity stream} (the detailed streams of system events and user interactions). This separation enables efficient indexing while preserving rich contextual detail for memory-based retrieval. Each memory anchor instance represents a discrete episodic memory\index{episodic memory} unit, capturing:

\begin{description}
    \descitem{Temporal anchors}{Precise timestamps that locate experiences in personal time.}
    \descitem{Spatial context}{Location data that grounds digital activities in physical space.}
    \descitem{Environmental state}{Active applications, open documents, and system configuration.}
    \descitem{Social connections}{Participants in communications or collaborative activities.}
    \descitem{Task relationships}{Project associations and workflow contexts.}
\end{description}

These elements mirror the multi-faceted nature of human episodic memory, where experiences are encoded with rich contextual associations that later serve as retrieval cues~\cite{schacter2008searching}.

\subsection{Memory Anchors and UUID-Based Architecture\index{UUID-based architecture}}\label{sec:upi:memory-anchor:anchors}

The UPI implements memory anchors\index{memory anchors} through a UUID-based referencing architecture\index{UUID-based architecture} that preserves privacy while maintaining rich associative connections. Each memory anchor receives a unique identifier that serves as an immutable reference point, enabling:

\begin{description}
    \descitem{Bidirectional linking}{Digital objects reference their creation/modification contexts, while contexts reference affected objects.}
    \descitem{Temporal threading}{Contexts maintain predecessor links, creating episodic chains that mirror memory sequences.}
    \descitem{Cross-platform correlation}{Common UUID namespace enables context matching across diverse storage systems.}
    \descitem{Privacy preservation}{UUIDs decouple identity from content, allowing selective context sharing.}
\end{description}

This architecture directly implements the encoding specificity principle from cognitive psychology~\cite{tulving1973encoding} because retrieval success depends on matching the cues present during encoding with those available during retrieval. This approach mirrors how human associative memory works: episodic memories don't contain complete copies of information but rather pointers to where that information is stored~\cite{raaijmakers1980sam}. By preserving comprehensive encoding context, the system maximizes potential retrieval pathways.

\subsection{Types of Memory Anchor}\label{sec:upi:memory-anchor:types}

The UPI categorizes memory anchor into five primary dimensions that collectively capture the episodic memory patterns surrounding digital information use, informed by research in cognitive psychology on how people naturally remember information~\cite{wagenaar1986my,conway2000construction}:

\subsubsection{Temporal Context\index{memory anchors!temporal dimension}}
Time serves as the universal organizing principle, providing:
\begin{itemize}
    \item Absolute timestamps for precise temporal location,
    \item Relative temporal relationships (before/after/during),
    \item Periodic patterns (daily routines, weekly meetings), and
    \item Duration tracking for extended activities.
\end{itemize}

\subsubsection{Spatial Context\index{memory anchors!spatial context}}
Location data grounds digital activities in physical space\allowbreak~\cite{burgess2002spatial,boardman2003too} through:
\begin{itemize}
    \item GPS coordinates for mobile devices,
    \item Network-based location inference,
    \item Semantic location tags (``home,'' ``office,'' ``conference room''), and
    \item Movement patterns and location transitions.
\end{itemize}

\subsubsection{Social Context\index{memory anchors!social context}}
Interpersonal interactions create memory associations~\cite{Naaman2004PhotoContext} via:
\begin{itemize}
    \item Communication participants (email, chat, video calls),
    \item Document collaborators and reviewers,
    \item Meeting attendees from calendar data,
    \item Social platform interactions.
\end{itemize}

\subsubsection{Task Context\index{memory anchors!task context}}
Work patterns and project associations through:
\begin{itemize}
    \item Active application focus,
    \item Document relationships and dependencies,
    \item Project folder associations, and
    \item Workflow stage indicators.
\end{itemize}

\subsubsection{Environmental Context\index{memory anchors!environmental context}}
System and ambient conditions that research has shown can serve as powerful memory cues~\cite{Smith2001,Hailpern2011}, including:
\begin{itemize}
    \item Device characteristics and capabilities,
    \item Network connectivity state,
    \item Active media (music, videos), and
    \item System resource utilization.
\end{itemize}

These categories are not mutually exclusive as real-world contexts typically combine multiple dimensions, creating rich episodic signatures that uniquely identify experiences and enable multi-path retrieval.

\subsection{Technical Implementation Patterns}\label{sec:upi:memory-anchor:patterns}

The implementation of memory anchor follows several key architectural patterns that balance comprehensive capture with system efficiency. Previous systems research has demonstrated that capturing 100\% of a single computer system's state over a year is technically viable, requiring significant but not unbounded storage~\cite{dunlap2002revirt,cully2008remus}:

\subsubsection{Context as a Service}
Memory anchors operate through a decoupled service architecture aligned with the episodic nature of human experience, where continuous sensory input is naturally segmented into discrete episodes based on significant changes or boundaries~\cite{zacks2007event}. Specialized providers capture dimension-specific data:
\begin{itemize}
    \item Location providers for spatial context,
    \item Communication monitors for social context,
    \item Application trackers for task context, and
    \item Media players for environmental context.
\end{itemize}

Each provider operates independently, contributing to a unified context stream that preserves both isolation and integration.

\subsubsection{Change-Driven Capture}
Rather than continuous monitoring, the system captures context at significant state changes:
\begin{itemize}
    \item File creation or modification,
    \item Application focus switches,
    \item Location transitions, and
    \item Communication events.
\end{itemize}

This approach mirrors human selective attention, focusing on meaningful changes rather than overwhelming detail.

\subsubsection{Temporal Stream Processing}
Context data flows through a temporal processing pipeline that:
\begin{itemize}
    \item Maintains temporal ordering across diverse sources,
    \item Correlates near-simultaneous events,
    \item Identifies temporal patterns and regularities, and
    \item Supports both real-time and retrospective analysis.
\end{itemize}

\subsubsection{Integration with UPI Core}\label{sec:upi:memory-anchor:integration}

Memory anchor integrates with other UPI components to enable memory-aligned retrieval:

\begin{description}
    \descitem{Storage metadata enhancement}{Timestamps become temporal anchors for richer context,}
    \descitem{Semantic enrichment}{Named entities gain temporal-spatial grounding,}
    \descitem{Query processing}{Natural language queries leverage episodic patterns,}
    \descitem{Privacy framework}{UUID use enables selective context disclosure.}
\end{description}

This integration transforms memory anchor from isolated metadata into the connective tissue that binds the user's digital experiences into a coherent, searchable whole.

\subsection{Implementation Status}\label{sec:upi:memory-anchor:implementation-status}

\autoref{tab:context-implementation-status} summarizes the current implementation status of each memory anchor category in the \system{} prototype. While the UPI architecture supports comprehensive memory anchor capture, the prototype implementation demonstrates varying levels of completion across different context types.

\begin{table}[!tbp]
    \centering
    \caption[Memory Anchor Types]{Memory Anchor Types: Implementation Status in \system{}}
    \label{tab:context-implementation-status}
    \resizebox{\textwidth}{!}{%
    \begin{tabular}{llp{6cm}}
    	\toprule
    	\textbf{Context Type} & \textbf{Status} & \textbf{Current State} \\
        \midrule
        Temporal & Implemented & Full support for timestamps, relative references, and temporal patterns \\
        Spatial & Implemented & GPS and IP-based location with Windows Location Provider \\
        Social & Initial Implementation & Calendar integration (Google, Microsoft), Discord scanner, Outlook attachments. Requires debugging \\
        Media Consumption & Initial Implementation & Spotify and YouTube history providers. Requires debugging \\
        Environmental & Under Development & Preliminary collectors for Ecobee and Nest thermostats \\
        Task Context & Under Development & Preliminary collector/recorder definitions \\
        \bottomrule
        \end{tabular}%
    }%
\end{table}

The implementation demonstrates that temporal and spatial contexts provide the most mature foundation for memory-aligned retrieval, while social and environmental contexts represent active development areas that will enhance the system's episodic memory capabilities.

\section{Memory-Based Query Architecture}
\label{sec:upi:memory-query-architecture}

This section details the architectural components that enable memory-aligned retrieval. The specific implementation of these components in the \system{} prototype is detailed in \autoref{chap:implementation}.

\subsection{Memory-Compatible Query Processing}
\label{sec:upi:memory-query}

A core objective of the UPI architecture is to support query processing that is compatible with how humans form memory-based retrieval cues. This means the architecture must enable the translation of natural, often episodic, user information needs (e.g., recalling a document associated with ``the Seattle meeting'') into effective searches against its unified index, without requiring users to manually formulate complex technical syntax. To achieve this, the UPI architecture must provide for, or interface with, capabilities that effectively act as:

\begin{description}
    \descitem{An Episodic Memory Processor}{Functionality to extract and interpret temporal, spatial, and activity-based dimensions from user queries or interaction context, mapping them to the corresponding structures within the UPI's memory anchor model.}

    \descitem{A Semantic Network Traversal Capability}{Mechanisms that allow queries to leverage the conceptual relationships and semantic associations (e.g., between topics, named entities) stored within the unified metadata, supporting meaning-based retrieval.}

    \descitem{A Context Reconstruction Engine Interface}{The ability to reconstruct the complete memory anchor surrounding a digital object or time period, assembling the relevant temporal, spatial, and activity dimensions to provide a rich contextual view for memory-aligned retrieval.}

    \descitem{An Associative Path Query Support}{The underlying UPI Core and its query language must support the traversal of multiple types of relationships (e.g., temporal, semantic, contextual) in the unified metadata graph, mirroring human associative recall.}
\end{description}

These architectural provisions collectively enable higher-level query tools or interfaces to process user statements like,
``Find the presentation I worked on with Mai last week before the client meeting,'' by mapping experiential descriptions to the UPI's structured, time-anchored, and context-rich metadata. The UPI architecture is designed to support flexible interpretation of diverse natural language styles and evolving user vocabularies for this purpose, with specific interface implementations (such as the LLM-based query tools developed alongside \system{}) detailed in \autoref{chap:implementation}.

\subsection{Contextual Identification and Adaptive Display}\label{sec:contextual-identification}

The UPI architecture is designed to utilize contextual data to significantly enhance both query interpretation and the subsequent presentation of results.  This architectural emphasis stems from the principle that information relevance is highly dependent on contextual factors, such as the user's current task, location, device, and recent activities.  To support this, the UPI architecture must provide for:

\begin{description}
    \descitem{Memory Anchor Integration in Querying}{Mechanisms to allow the query processing system to incorporate user task and activity stream (from the Memory Anchor Model) to inform query understanding and prioritize relevant results.}

    \descitem{Temporal Context Analysis Capabilities}{The ability for query and ranking systems to analyze time-based patterns and relationships (e.g., recency, co-occurrence of events) when interpreting queries and ordering results.}

    \descitem{Spatial Context Processing Support}{Architectural provisions for utilizing location data (when available and consented to) to refine query interpretation (e.g., disambiguating ``notes from the conference'' if the user attended multiple conferences at different locations) and enhance result relevance (e.g., prioritizing locally relevant information).}

    \descitem{Support for Device-Appropriate Result Formatting}{While specific rendering is a UI concern, the UPI architecture should enable the Query Interface to access information about the requesting device or application type, allowing for the adjustment of result data structures or summaries to suit different display capabilities and constraints.}
\end{description}

These architectural capabilities for contextual processing allow UPI-enabled applications to adapt query responses and result presentation dynamically to the user's prevailing situation.  For example, the same underlying query could yield differently prioritized or formatted results if issued from a mobile device while traveling versus a desktop computer in the office.

\subsection{Interactive Refinement Model}\label{sec:interactive-refinement}

The UPI architecture supports an interactive refinement model designed for progressive clarification and improvement of search results. Recognizing that information retrieval often involves iterative refinements as users recall additional details or clarify their intentions, this model emphasizes ongoing interaction.

Architectural components supporting this approach include:

\begin{description}
    \descitem{Clarification Dialog System}{Allows the system to request specific additional information, prioritizing high-discrimination factors.}

    \descitem{Result Feedback Loop}{Captures explicit and implicit user feedback to enhance future queries.}

    \descitem{Query Refinement Suggestions}{Provides contextually relevant suggestions for refining searches based on observed metadata patterns.}

    \descitem{Memory Stimulation Techniques}{Offers cues designed to help users recall additional details, thus improving search precision.}
\end{description}

These architectural features foster a conversational and iterative search experience implementing the ``orienteering'' concept~\cite{teevan2004perfect}, where retrieval follows a series of contextual anchors rather than requiring precise target specification, enabling users to progressively refine their searches even when their initial queries are vague or incomplete. Implementation details of these interactive refinement mechanisms within the \system{} prototype are discussed in \autoref{chap:implementation}.

\section{Extensions and Future Directions}\label{sec:upi:extensions}

The UPI architecture detailed in this chapter provides a robust and extensible foundation for memory-aligned personal information retrieval. While the core framework addresses fundamental challenges, its design also anticipates and facilitates numerous avenues for future enhancement and broader application. This section outlines several promising extensions and research directions that build upon the UPI's architectural principles, aiming to further augment its capabilities in areas such as privacy, contextual understanding, collaboration, and proactive assistance. The implementation status of any preliminary explorations into these areas within the \system{} prototype is discussed in \autoref{chap:implementation}.

\subsection{Advanced Context Acquisition}

The UPI's contextual awareness could be enhanced through additional sensing and inference capabilities:

\begin{description}
    \descitem{Emotional Context Recognition}{Architectural frameworks to incorporate affect and emotional state into the context model, recognizing the important role of emotion in human memory.}

    \descitem{Activity and Task Inference}{Enhanced mechanisms to automatically identify and track higher-level user activities and goals, creating richer contextual metadata.}

    \descitem{Cross-Device Context Synchronization}{Architectural approaches to maintain contextual continuity across multiple devices while respecting privacy boundaries.}
\end{description}

\subsection{Collaborative Information Spaces}

The UPI architecture could be extended to better support collaborative information management while maintaining individual privacy:

\begin{description}
    \descitem{Shared Context Models}{Architectural frameworks for creating and managing shared memory anchors that enable group-aware information retrieval.}

    \descitem{Privacy-Preserving Collaboration}{Design approaches that enable rich collaboration without compromising individual metadata privacy.}

    \descitem{Role-Based Context Integration}{Architectural mechanisms to incorporate organizational roles and relationships into the context model.}
\end{description}

\subsection{Predictive Information Delivery}

Building on its contextual awareness, the UPI architecture could evolve to proactively surface relevant information before it's explicitly requested:

\begin{description}
    \descitem{Need Anticipation Framework}{Architectural components to predict information needs based on current and historical context patterns.}

    \descitem{Contextual Triggers}{Mechanisms to identify specific contextual conditions that should trigger information presentation.}

    \descitem{Attention-Aware Delivery}{Design approaches to present anticipated information based on system-level optimization of retrieval patterns and current query context.}

    \descitem{LLM-Enhanced Prediction}{Integration with large language models to improve prediction accuracy, building on research by \citeauthor{thomas2024large}~\cite{thomas2024large} on LLM-based searcher preference prediction and \citeauthor{fernandez2023large}~\cite{fernandez2023large} on how LLMs are transforming data management paradigms.}
\end{description}

These potential extensions demonstrate the adaptability and scalability of the UPI architecture, showing how its foundational design principles can be expanded to address evolving information management needs while maintaining its core commitment to human-centric design, contextual integration, and privacy preservation.

\section{Conclusion}\label{sec:upi:conclusion}

This chapter has presented the Unified Personal Index architecture as a human-centric framework for personal information management. The UPI addresses fundamental limitations in existing systems through five key architectural innovations: cross-platform metadata normalization, memory anchor integration, mixed-schema graph database design, context-aware query planning, and modular extensibility.

The architecture's core contribution lies in implementing cognitive memory models as systems infrastructure, directly mapping architectural components to established patterns of episodic, semantic, and associative memory. This approach eliminates the translation gap between human recall and system retrieval while incorporating privacy as a fundamental design principle through local processing and data minimization strategies.

The UPI represents more than a specific system: it provides an architectural blueprint for any information retrieval system aspiring to bridge the persistent gap between human cognition and digital data. The subsequent \autoref{chap:implementation} details the concrete realization of these architectural principles in the \system{} prototype, while \autoref{ch:evaluation} demonstrates their practical effectiveness through empirical validation.

\addkhipuifneeded



\chapter{\system: Implementing the Unified Personal Index}
\label{chap:implementation}

\begin{epigraph}
    \textit{%
        ``The road to wisdom? Well, it's plain\\
        And simple to express:\\
        Err \\
        and err \\
        and err again, \\
        but less \\
        and less \\
        and less.''
    }
    \par\vspace{0.5em}
    \mbox{}\hfill\textit{Piet Hein, Grooks (1966)}
\end{epigraph}

The UPI architecture presented in \autoref{ch:upi} established the theoretical framework for memory-aligned information retrieval. This chapter presents \system{}, our prototype implementation that demonstrates the practical feasibility of these concepts through concrete realization of episodic memory patterns (temporal, spatial, and contextual metadata) via specific technical choices.

The implementation work validates that a system can be built to directly align with cognitive memory processes, demonstrating how episodic memory cues can be systematically captured, normalized, and used for information retrieval.

To facilitate clear connections with the architectural framework presented in \autoref{ch:upi}, this chapter follows a parallel structure, addressing the implementation details of each major architectural component. For each component, we describe:

\begin{itemize}
\item The specific technologies and tools selected,
\item Implementation strategies and algorithms employed,
\item Engineering challenges encountered and their solutions,
\item Performance considerations and optimizations, and
\item Current limitations and development status.
\end{itemize}

\section{Implementation Overview}
\label{sec:implementation:overview}

\begin{table}[!tbph]
    \caption[\system Implementation Components]{Implemented \system Components and Their Descriptions. Note: this table includes components that have been at least partially implemented in \system{}.}
    \label{tab:implementation-components}
    \centering
  \begin{tabular}{p{0.2\textwidth}p{0.6\textwidth}}
    	\toprule
    	\textbf{Component} & \textbf{Description} \\
    \midrule
    Storage & Local file systems (NTFS, APFS, ext4), cloud storage (Google Drive, OneDrive, Dropbox), application storage (Discord, Outlook). \\
    Semantic & ``Unstructured'' file analysis, EXIF metadata extraction, and checksum calculator. \\
    \mbox{Activity} \mbox{Stream} \mbox{Services} & Location (GPS, WiFi, IP, Tile), music (Spotify), Ambient (ecobee), Video (YouTube), Collaboration (Discord, Outlook), Storage (NTFS, MacOS). \\
    \bottomrule
  \end{tabular}
\end{table}

\begin{figure}[!tbhp]
    \caption[Implementation Diagram]{\small\system implementation diagram showing: (1) ArangoDB\index{ArangoDB} database and collections; (2) Local file systems (NTFS, APFS, EXT4); (3) Cloud storage services (iCloud, Dropbox, Google Drive, OneDrive); (4) Activity stream sources (Location, Email, Music, Collaboration, Ambient, Query, Storage). Memory anchors permit creating relationships across these diverse data sources, which allows knowledge-graph construction over time.}
    \centering
    \label{fig:indaleko:implementation}
      \begin{tabular}{c}
          \includegraphics[width=0.95\textwidth]{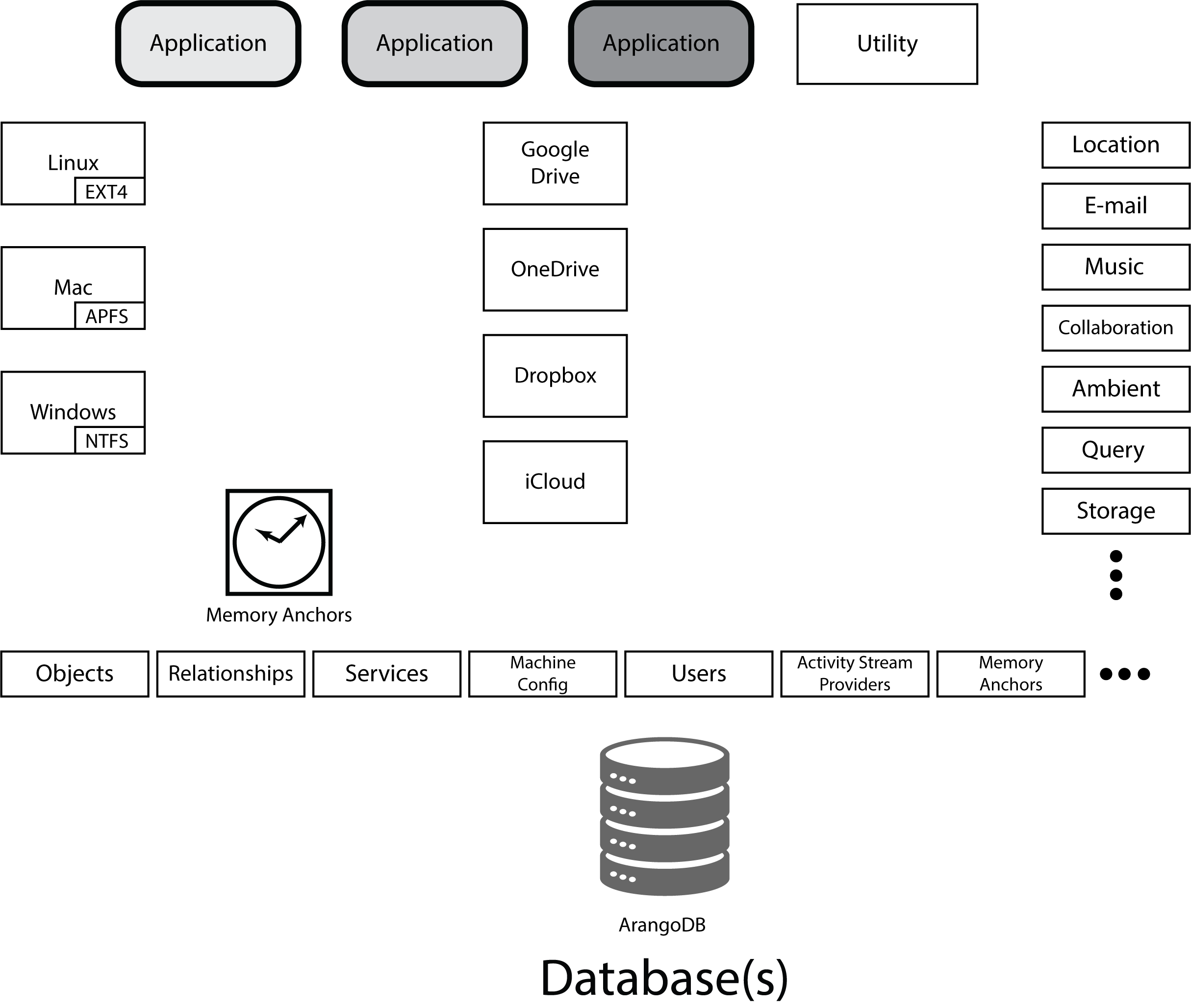}
      \end{tabular}
\end{figure}

Table \ref{tab:architecture-implementation-mapping} provides a mapping between UPI architecture sections (\autoref{ch:upi}) and \system implementation sections (\autoref{chap:implementation}) to help readers navigate between architectural concepts and their concrete realization.

\begin{table}[!tbph]
    \caption[Architecture-Implementation Mapping]{Mapping between UPI Architecture (\autoref{ch:upi}) and \system Implementation (\autoref{chap:implementation})}
    \label{tab:architecture-implementation-mapping}
    \centering
    \resizebox{0.9\textwidth}{!}{%
  \begin{tabular}{p{0.3\textwidth}p{0.6\textwidth}}
  	\toprule
  	\textbf{UPI Architecture} & \textbf{\system Implementation} \\
  	\textbf{(\autoref{ch:upi})} & \textbf{(\autoref{chap:implementation})} \\
  \midrule
  Introduction (\S\ref{sec:upi:introduction}) & Implementation Overview (\S\ref{sec:implementation:overview}) \\
  Conceptual Design and Key Principles (\S\ref{sec:upi:conceptual-design}) & Implementation Overview (\S\ref{sec:implementation:overview}) \\
  Memory Anchor Integration (\S\ref{sec:upi:memory-anchor}) & Implementation of Collection and Extraction (\S\ref{sec:implementation:collection-extraction}); Implementation of Schema Normalization and Enrichment (\S\ref{sec:implementation:normalization-enrichment}) \\
  Unified Metadata Approach (\S\ref{sec:upi:metadata-integration}) & Implementation of Schema Normalization and Enrichment (\S\ref{sec:implementation:normalization-enrichment}); Implementation of Indexing and Storage (\S\ref{sec:implementation:indexing-storage}) \\
  Data Flow and Processing Pipeline (\S\ref{sec:upi:dataflow}) & Implementation Overview (\S\ref{sec:implementation:overview}); Implementation of Collection and Extraction (\S\ref{sec:implementation:collection-extraction}); Implementation of Schema Normalization and Enrichment (\S\ref{sec:implementation:normalization-enrichment}); Implementation of Indexing and Storage (\S\ref{sec:implementation:indexing-storage}); Implementation of Query Processing and AI Integration (\S\ref{sec:implementation:query-processing}) \\
  Cross-Platform Collection (\S\ref{sec:upi:collection}) & Implementation of Collection and Extraction (\S\ref{sec:implementation:collection-extraction}) \\
  Schema Normalization (\S\ref{sec:upi:normalization}) & Implementation of Schema Normalization and Enrichment (\S\ref{sec:implementation:normalization-enrichment}) \\
  Semantic Enhancement (\S\ref{sec:upi:semantic}) & Implementation of Schema Normalization and Enrichment (\S\ref{sec:implementation:normalization-enrichment}) \\
  Integration and Unified Storage (\S\ref{sec:upi:integration}) & Implementation of Indexing and Storage (\S\ref{sec:implementation:indexing-storage}) \\
  Query Processing: From Memory Cues to Digital Answers (\S\ref{sec:upi:query}) & Implementation of Query Processing and AI Integration (\S\ref{sec:implementation:query-processing}) \\
  Conclusion (\S\ref{sec:upi:conclusion}) & Conclusion (\S\ref{sec:implementation:conclusion}) \\
  \bottomrule
  \end{tabular}%
    }
\end{table}

This mapping highlights how the architectural principles from \autoref{ch:upi} are realized through specific implementation choices, technologies, and engineering solutions described in this chapter. While the organizational structures differ slightly to better accommodate the different focus of each chapter (design principles vs. concrete implementation details), this table helps maintain the conceptual connection between them.

\section[Collection and Extraction]{Collection and Extraction\protect\footnotemark}
\footnotetext{Code mapping for this section is in \autoref{app:implementation-mapping:collection-extraction}.}
\label{sec:implementation:collection-extraction}

The collection layer implements the Data Ingestion Layer and Activity Stream Capture Layer concepts from the UPI architecture (\autoref{sec:upi:architecture:ingestion}). This layer uses a two-stage approach: \textit{collectors}\index{collector} extract metadata from diverse sources while preserving memory-relevant signals, then \textit{recorders}\index{recorder} normalize and prepare this data for storage. This implementation maintains the architectural separation between raw data capture and processing pipelines.

Performance constraints further influenced architectural choices. Initial use with MongoDB suggested that \system would benefit from more graph database functions.  Subsequent experiments with Neo4j exposed out-of-the-box scalability limitations, especially under high-volume ingestion. While ArangoDB\index{ArangoDB} was ultimately selected for its high-throughput document storage, integrated graph model, and support for full-text and semantic search via ArangoSearch views. Even with ArangoDB, however, high-volume ingestion remained time-intensive, motivating the decision to decouple ingestion from collection and introduce staged processing pipelines.

Schema handling evolved in parallel. Initial ad hoc data structures gave way to Pydantic\index{Pydantic}-based models, which enabled runtime validation, schema introspection, and uniform serialization. These models now anchor every collector and recorder. While collectors have few constraints on their data format, recorders have a more focused role. They use common standards, such as ISO-formatted timestamps, and facilitate schema export to ArangoDB for validation and indexing.

A key insight emerged around semantic alignment: instead of forcing uniform field names across sources, which seriously limits the ease of adding new metadata sources, \system adopted an abstract semantic attribute model. Each metadata field is tagged with a UUID representing its semantic identity (e.g.,  ``last modified time''), and these UUIDs are stored as key-value pairs. This enables extensibility across domains and supports future alignment through LLM-assisted label resolution. Although this design introduces complexity in indexing and debugging, it aligns with the privacy and security model introduced in Section~\ref{sec:implementation:privacy-security}, where UUIDs obfuscate meaning unless a private mapping table is available through UUID semantic mapping\index{UUID semantic mapping}.

The collection layer is encouraged to ``capture everything,'' While there is a storage cost associated with this implementation, it provides the possibility to introspect the available data and extract additional insights retrospectively. Thus, we balance the cost of additional storage against the much higher cost of retrieving and re-analyzing data, if that data is even available. This reflects an architectural bias toward archival completeness, recognizing that semantic insight often emerges over time.

More specifically, the recorder layer curates the collector's output by mapping data to semantic attribute labels (represented as UUIDs), compressing it, optionally encrypting it, and encoding it in a format compatible with the ArangoDB\index{ArangoDB} database.

The collector and recorder layers together realize the principle of ``capture everything, compress what you can, and make it searchable.''

\begin{figure}[!tbp]
  \caption[Collector/Recorder Pipeline]{\system Collector/Recorder Pipeline: The diagram illustrates the modular data ingestion architecture of the Unified Personal Index. Collectors gather data from diverse sources with source-specific formats, while recorders normalize and persist this data with semantic UUID mappings. Three communication patterns (direct transfer, queue-based, and batch file) enable flexible integration. Red dashed lines indicate which communication pattern each collector typically uses. The data provider pattern (purple-green dashed outline) shows where collectors and recorders are tightly coupled for certain sources. All normalized data is stored in ArangoDB with both the original compressed content and semantically-mapped fields.}
  \label{fig:collector-recorder-pipeline}
  \centering
  \resizebox{0.9\textwidth}{!}{%
    \begin{tikzpicture}[
      node distance=1.5cm and 2.5cm,
      box/.style={
        draw=darkblue,
        fill=lightblue!30,
        rounded corners,
        minimum width=2.8cm,
        minimum height=1.2cm,
        align=center,
        font=\small\bfseries,
        drop shadow={shadow xshift=1pt, shadow yshift=-1pt, opacity=0.3}
      },
      data_source/.style={
        box,
        fill=gray!20,
        draw=gray!70!black,
      },
      collector/.style={
        box,
        fill=purple!15,
        draw=purple!70!black,
      },
      communication/.style={
        box,
        fill=yellow!15,
        draw=yellow!70!black,
        minimum width=2.2cm,
        minimum height=1cm,
      },
      recorder/.style={
        box,
        fill=green!15,
        draw=green!70!black,
      },
      database/.style={
        cylinder,
        draw=brown!70!black,
        fill=brown!15,
        shape border rotate=90,
        aspect=0.3,
        minimum height=2cm,
        minimum width=1.8cm,
      },
      data_model/.style={
        chamfered rectangle,
        draw=red!50!black,
        fill=red!10,
        chamfered rectangle xsep=2cm,
        minimum height=1.2cm,
        align=center,
      },
      label_style/.style={
        font=\scriptsize\bfseries,
        text=black
      },
      flow/.style={
        ->,
        >=stealth,
        thick,
        draw=black
      },
      dashed_flow/.style={
        ->,
        >=stealth,
        thick,
        dashed,
        draw=gray!70!black
      },
      note/.style={
        draw=gray!50,
        fill=gray!10,
        text width=3.2cm,
        align=left,
        font=\scriptsize,
        rounded corners
      },
      data_flow/.style={
        ->,
        >=stealth,
        thick,
        draw=blue!60!black
      },
      config_flow/.style={
        ->,
        >=stealth,
        thick,
        draw=orange!60!black,
        dashed
      },
      uuidlabel/.style={
        draw=black!40,
        fill=yellow!5,
        rounded corners,
        font=\ttfamily\scriptsize,
        align=center
      }
    ]

    \colorlet{darkblue}{blue!70!black}
    \colorlet{lightblue}{blue!10!white}

    \node[data_source] (fs) {Local Filesystem};
    \node[data_source, below=of fs] (cloud) {Cloud Storage\\APIs};
    \node[data_source, below=of cloud] (activity) {Activity\\Data Sources};
    \node[data_source, below=of activity] (app) {Application\\APIs};

    \node[collector, right=of fs] (fs_collector) {Filesystem\\Collector};
    \node[collector, right=of cloud] (cloud_collector) {Cloud Storage\\Collector};
    \node[collector, right=of activity] (activity_collector) {Activity Data\\Collector};
    \node[collector, right=of app] (app_collector) {Application\\Collector};

    \node[recorder, right=6cm of fs_collector] (fs_recorder) {Filesystem\\Recorder};
    \node[recorder, right=6cm of cloud_collector] (cloud_recorder) {Cloud Storage\\Recorder};
    \node[recorder, right=6cm of activity_collector] (activity_recorder) {Activity Data\\Recorder};
    \node[recorder, right=6cm of app_collector] (app_recorder) {Application\\Recorder};

    \node[above left=5cm of fs_recorder] (comm_label) {\textbf{Communication Patterns}};
    \node[communication, below left=0.2cm of comm_label] (direct) {Direct\\Transfer};
    \node[communication, right=0.7cm of direct] (queue) {Queue\\Model};
    \node[communication, right=0.7cm of queue] (batch) {Batch\\Files};

    \node[fit=(activity_collector) (activity_recorder), draw=purple!40!green, dashed, rounded corners, inner sep=6pt, label={[label_style]below:Data Provider Pattern}] (provider) {};

    \node[database, right=5.5cm of activity_recorder] (arangodb) {ArangoDB};

    \node[data_model, above=0.2cm of fs_collector] (source_model) {Source-specific\\JSON Data};
    \node[data_model, above=0.2cm of fs_recorder] (indaleko_model) {Indaleko Record +\\Normalized Data};

    \node[uuidlabel, above right=1.5cm and 0.3cm of indaleko_model] (uuid1) {94a35bcd-1e6f-\\4b6c-9748-...};
    \node[uuidlabel, below=0.1cm of uuid1] (uuid2) {``last modified time''};
    \draw[->] (uuid1) -- (uuid2);

    \draw[flow] (fs) -- (fs_collector);
    \draw[flow] (cloud) -- (cloud_collector);
    \draw[flow] (activity) -- (activity_collector);
    \draw[flow] (app) -- (app_collector);

    \draw[data_flow] (fs_collector) to[out=0, in=180] node[above, font=\scriptsize] {JSON} (fs_recorder);
    \draw[data_flow] (cloud_collector) to[out=0, in=180] node[above, font=\scriptsize] {JSON} (cloud_recorder);
    \draw[data_flow] (activity_collector) to[out=0, in=180] node[above, font=\scriptsize] {JSON} (activity_recorder);
    \draw[data_flow] (app_collector) to[out=0, in=180] node[above, font=\scriptsize] {JSON} (app_recorder);

    \draw[data_flow] (fs_recorder) to[out=0, in=150] (arangodb);
    \draw[data_flow] (cloud_recorder) to[out=0, in=170] (arangodb);
    \draw[data_flow] (activity_recorder) to[out=0, in=190] (arangodb);
    \draw[data_flow] (app_recorder) to[out=0, in=210] (arangodb);


    \draw[dashed_flow, red!70] (direct) to[out=-90, in=30] (activity_collector);
    \draw[dashed_flow, red!70] (queue) to[out=-90, in=45] (cloud_collector);
    \draw[dashed_flow, red!70] (batch) to[out=-90, in=45] (fs_collector);

    \node[note, below=1.0cm of arangodb] (note1) {Recorders transform source data to standardized format with:
    \begin{itemize}
    \item Original compressed data
    \item UUID semantic mapping
    \item Normalized fields
    \end{itemize}};

    \node[above=0.1cm of source_model, font=\scriptsize, align=center] (source_label) {Source-specific format};
    \node[above=0.1cm of indaleko_model, font=\scriptsize, align=center] (indaleko_label) {Normalized with semantic mapping};

    \node[note, above=3cm of arangodb] (schema_note) {``Capture everything, compress what you can, make it searchable''\\
    Flexible schema approach allows retrospective analysis};

    \end{tikzpicture}
  }
\end{figure}
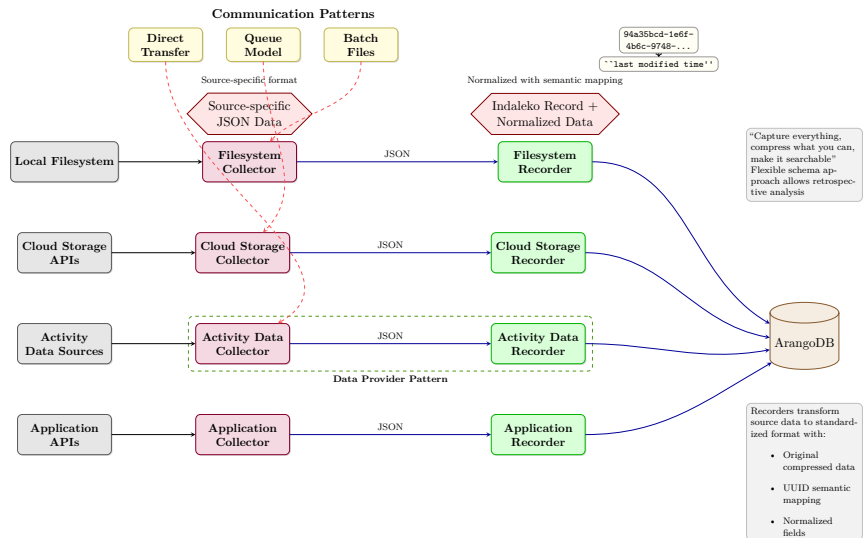

Figure \ref{fig:collector-recorder-pipeline} illustrates the complete data flow from source systems through collection, transformation, and storage. The diagram shows how collectors gather data from diverse sources, pass it through one of three communication patterns to recorders which normalize and semantically map the data before persisting it to ArangoDB.

In many cases, the collector is ``wrapped'' by a recorder\index{recorder}.  For example, this is a common pattern in activity stream sources.  This combination is referred to as a \emph{data provider}\index{data provider}.

\subsection{Collector Framework Implementation}

The collector framework\index{collector framework} is implemented as a modular plugin architecture using Python 3.12+, with each collector implemented as a separate module that adheres to a common interface. This approach allows new collectors to be developed and integrated without modifying the core system.

Key implementation details include:

\begin{itemize}
\item A standardized collector interface defined through abstract base classes.
\item Standardized error handling and logging across collectors.
\end{itemize}

\subsection{Source-Specific Collector Implementations}

We have implemented collectors for several key data sources, each addressing unique technical challenges:

\begin{description}
\descitem{Local Filesystem Collector}{Implements recursive directory scanning using Python's os.walk; in testing we have found this has minimal file system impact. File change updates are partially implemented, using two different techniques: (1) monitoring file system events, such as by monitoring change journal (NTFS USN Journal\index{NTFS USN Journal}), eBPF (Linux), or existing, platform specific, monitoring systems like fs\_usage (Mac); and (2) an incremental scan of the file system.}

\descitem{Cloud Storage Collectors}{Implemented for Google Drive, OneDrive, iCloud, and Dropbox using their respective Python API clients~\footnote{Note: there is no official iCloud Python library, so an unofficial pypi distribution was used, see \url{https://pypi.org/project/pyicloud/} and \url{https://github.com/timlaing/pyicloud}}. These collectors handle standard API operations such as authentication flows, pagination~\footnote{pagination is the process of breaking large data responses into smaller units and this term is commonly used in web APIs}, and token refresh.  Separate collectors are used to monitor storage changes, which can be used to drive metadata updates~\footnote{Actual implementation of this is limited, as this work is not necessary for answering the research questions}.}

\descitem{Email Collectors}{While there are prototypes of specialized monitors, the implementation does not include support for incorporating e-mail interactions.}

\descitem{Application-Specific Collectors}{Several specialized activity stream providers for applications like Discord, Spotify, Ecobee, Tile, and YouTube, each with custom handling of their unique data models and API constraints.}
\end{description}

\subsection{Metadata Extraction Implementation}

Beyond basic file attributes, \system implements deeper metadata extraction:

\begin{description}
\descitem{Content Type Analysis}{Leverages Python's magic library and extended MIME type detection for accurate content classification independent of file extensions.}

\descitem{Metadata Extractors}{The implementation uses the Unstructured~\footnote{\url{unstructured.io}} package for broad-based metadata extraction.  Unstructured's primary purpose is to prepare datasets suitable for training LLMs, which means it emits metadata in a rich, well-defined format. While our implementation of this is functional, our use of it to extract metadata is limited due to the high resource requirements for performing metadata extraction over large datasets.  Given that the use of semantic metadata in file system search is well-understood~\cite{gifford1991semantic} and commonly implemented in existing systems, we did not emphasize this aspect of the implementation.  Three simpler examples for semantic metadata extraction that are implemented are: (1) extraction of the EXIF metadata from images; (2) identifying the MIME type of a file based upon its contents versus, based upon file extension uses two distinct UUIDs to distinguish the source type; and (3) computing multiple common checksums, which permits duplicate identification across storage silos and this can be used to refine search results by showing only a single instance of the same file in multiple locations~\footnote{having a common definition of standard checksums like  SHA1, SHA2, SHA3, as well as common storage checksums like the Dropbox specific SHA checksum of checksums, permits identifying duplicate files across storage services.}.}

\end{description}

\subsection{Collection Scheduling and Synchronization}

The implementation uses an \emph{ad hoc} scheduling approach for flexibility across different data sources. Key scheduling features include incremental collection and state persistence for resumption after interruptions.

\section[Implementation of Schema Normalization and Enrichment]{Implementation of Schema Normalization and Enrichment\protect\footnotemark}
\footnotetext{Code mapping for this section is in \autoref{app:implementation-mapping:normalization-enrichment}.}
\label{sec:implementation:normalization-enrichment}

The normalization layer implements the Normalization \& Unification Engine from the UPI architecture (\autoref{sec:upi:architecture:normalization-unification}). This section focuses on the specific technical implementation choices made to realize the conceptual normalization framework, including Pydantic-based validation, UUID semantic mapping, and ArangoDB\index{ArangoDB} integration.

\subsection{Unified Schema Implementation}

\system creates a combination of statically defined data collections and dynamically defined (``registered'') data collections. The static collections are used to store well-known data, such as:

\begin{description}
\descitem{Objects}{Stores normalized representations of files, emails, and other data items with standardized attributes.}

\descitem{Relationships}{Captures connections between objects.}

\descitem{Named Entities}{Stores extracted entities like people, organizations, locations, and other semantic elements.}

\descitem{Activity Stream Providers}{Registration collection for activity stream sources.  This is used by the memory anchor system\footnote{The implementation code uses the terms 'activity context' and 'activity data' which correspond to 'memory anchors' and 'activity streams' respectively in this dissertation. The terminology was updated to avoid confusion.} to determine providers that will be included in the memory anchor (see \autoref{sec:upi:memory-anchor}).}

\descitem{Semantic Data Providers}{Registration collection for semantic data sources. This is used by the semantic enrichment system to determine available providers (see \autoref{sec:upi:semantic}).}

\descitem{Memory Anchor}{Constructs a cursor that represents applicable state at the time the memory anchor is created. The creation of handles is done dynamically, as needed, and reused as long as the known system state has not changed.  New memory anchor handles represent either a complete system state capture or a delta-state capture relative to the previous memory anchor handle (see \autoref{sec:upi:memory-anchor}).}

\end{description}

\subsection{Normalization Pipeline Implementation}

The specifics of data normalization, typically done in the recorder step, vary depending upon the needs of the specific data source but, broadly speaking, consist of some or all of the following stages:

\begin{description}

  \descitem{Data Extraction}{A recorder extracts relevant metadata in whatever format the collector defined.  Each JSON record can be extracted from a direct memory buffer, a file, a queue, or other message passing mechanism because the net result is isomorphic.}

  \descitem{Data Transformation}{The recorder transforms the extracted data into a common format, which contains:

  \begin{description}
    \descitem{Captured Data}{The captured data is stored in an \system Record, which identifies the recorder capturing the data, the time when the data was captured, and a compressed version of the captured data.}

    \descitem{Normalized Data}{The normalized data, which is based upon the collected data, is then added to the document to be stored.  The ideal is to capture the normalized data as a set of zero or more key-value pairs, where the key is a UUID that identifies the semantic meaning of the value.  The value can be any data that can be serialized to a JSON object (a limitation of ArangoDB\index{ArangoDB}.)}

    \descitem{Custom Data}{The early implementation of \system used distinct fields within the document rather than the more abstracted key-value pairs.  The key-value pair model has been increasingly adopted throughout the implementation, but the prototype still uses custom fields in several cases.  Future versions of \system will move away from this model.  However, the ability to use custom fields is an inherent feature of the database, and could be used in the future.  To preserve the privacy model, those fields could simply be the UUID labels.}

    \descitem{Schema Definitions}{Schema definitions are used to restrict the format of known fields.  In addition, the data classes of the implementation include descriptions of the meanings of each of the fields. In future versions of the implementation, these descriptions will be stored separately, within the user-managed mapping layer.  These definitions are important because they are used by the LLM-based query tools that need to understand the meaning of those fields.}
  \end{description}
  }

  \descitem{Data Validation}{The use of pydantic data classes provides a powerful mechanism for technical validation of data structures. The data classes validate format compliance and schema normalization\index{schema normalization} correctness at runtime to ensure data consistency.}

  \descitem{Data Marshaling}{\system extends the core functionality of the pydantic data classes to include marshaling and unmarshaling that is consistent with the requirements of ArangoDB.  This simplifies the use of \system when building new data providers.}

\end{description}

\subsection{Semantic Enrichment Implementation}

Semantic enrichment is implemented through several techniques:

\begin{description}
\descitem{Named Entity Recognition}{\system uses the capabilities of the LLM to identify ``named entities'' (e.g., people, organizations, locations, and dates from document content.)  Named Entity Recognition (NER) does not require the use of an LLM and can be implemented using \textbf{spaCy} or \textbf{Stanza}~\footnote{\url{stanfordnlp.github.io/stanza/}} as non-LLM alternatives.  One advantage of having the LLM handle this is that using the ``assistant'' interfaces, the LLM can ``call back,'' and we can determine if the named entity is in the \system database already, and if so, return it to the LLM.  If the entity is not present in the database, the LLM can invoke other tools to resolve it, such as GIS data for locations.}

\descitem{Intent Extraction}{The original query tool used to evaluate \system used a simple keyword-based approach to ensuring the query was a search query.  Experience with this interface as part of the larger query pipeline led to the decision to leverage the LLM more.  Ultimately, intent is largely not necessary to understand in the context of an LLM-driven query tool because it can often take non-search queries and convert them to valid AQL, thus satisfying the question.  This exposed an unexpected benefit of using a database with extensive search mechanisms: building query tools that go beyond search is straightforward.}

\descitem{Relationship Extraction}{The current \system implementation supports only minimal relationships, though ArangoDB, being a graph database, offers extensive capabilities for more fully utilizing these relationships.  While the implementation defines a number of additional relationships, including causal relationships, the only relationships used in the current implementation are a pair of storage relationships: \emph{contained by} and \emph{contains}.  The contains relationship is the traditional directory/file relationship, while the contained by relationship is used to represent the reverse of that relationship, providing functionality not available using traditional file systems.  The implementation can identify when two path names reference the same file, because storage metadata provides them with the same identity.}

\descitem{Semantic Analysis}{\system provides an extensible infrastructure for enriching the metadata of storage object by allowing dynamic extensibility of semantic metadata.  The implementation includes examples of data extraction (via \textbf{Unstructured}~\footnote{\url{unstructured.io}}) and demonstrative examples of how to extract EXIF data, MIME type information, and calculating checksums.  The expectation is that the infrastructure will allow for new semantic data providers to be added in the future. This can yield interesting benefits for incorporating a broad range of classifiers.  For example, an audio engineer can add a classifier that identifies instruments, sonic characteristics, music identification, etc. for a collection of audio files. This dynamic extensibility lays the groundwork for allowing individual users to customize their system to meet their needs.}
\end{description}

\section[Implementation of Indexing and Storage]{Implementation of Indexing and Storage\protect\footnotemark}
\label{sec:implementation:indexing-storage}
\footnotetext{Code mapping for this section is in \autoref{app:implementation-mapping:indexing-storage}.}

This section details the implementation of the UPI Core storage architecture (\autoref{sec:upi:architecture:upi-core-storage-indexing}), focusing on database technology selection, indexing strategies, and performance optimization techniques.

\subsection{Database Technology Selection}

\system uses ArangoDB\index{ArangoDB} as its primary storage and indexing engine, selected for its combined graph and document capabilities. Key implementation details include:

\begin{description}
  \descitem{Flexible deployment}{\system has been deployed and technically validated in diverse environments, including local Docker container-based deployment (Windows, Mac, Linux), LAN-based deployments on an Intel NUC running Ubuntu 24.04 LTS, and cloud-based deployments on Akamai Linode instances. These deployments demonstrate technical feasibility across individual, home, and small-cloud scenarios.}

  \descitem{ArangoSearch}{ArangoDB provides built-in support for full-text search via \textbf{Views}, enabling tokenized indexing of content such as file and directory names. Custom analyzers expand the tokenization process, improving search capabilities for natural language queries.}

  \descitem{Graph Capabilities}{\system makes nominal use of ArangoDB's edge collections to represent containment relationships between storage objects (e.g., directories containing files). These graph structures are expected to play a larger role as the system matures.}

  \descitem{Operational Simplicity}{ArangoDB's combined document and graph model reduces operational complexity compared to maintaining separate systems for graph and document storage. Its strong consistency guarantees and straightforward deployment models were well-suited to a research prototype environment.}
\end{description}

ArangoDB's search and indexing capabilities strongly influenced its selection. The importance of indexing for performance became particularly evident during query system development. The database schema is shown graphically in:

\begin{itemize}
\item \href{https://github.com/ubc-systopia/Indaleko/blob/main/figures/core-storage.png}{Core Storage Schema} shows the core storage schema~\footnote{\url{https://github.com/ubc-systopia/Indaleko/blob/main/figures/core-storage.png}},
\item \href{https://github.com/ubc-systopia/Indaleko/blob/main/figures/schema-activity.png}{Memory Anchor Schema} shows the memory anchor schema~\footnote{\url{https://github.com/ubc-systopia/Indaleko/blob/main/figures/schema-activity.png}},
\item \href{https://github.com/ubc-systopia/Indaleko/blob/main/figures/systems-management.png}{System Management Schema} shows the system management schema~\footnote{\url{https://github.com/ubc-systopia/Indaleko/blob/main/figures/systems-management.png}},
\item \href{https://github.com/ubc-systopia/Indaleko/blob/main/figures/entity-equivalence.png}{Entity Equivalence Schema} shows the entity equivalence schema~\footnote{\url{https://github.com/ubc-systopia/Indaleko/blob/main/figures/entity-equivalence.png}}, and
\item \href{https://github.com/ubc-systopia/Indaleko/blob/main/figures/learning-and-feedback.png}{Learning and Feedback Schema} shows the learning and feedback schema~\footnote{\url{https://github.com/ubc-systopia/Indaleko/blob/main/figures/learning-and-feedback.png}}.
\end{itemize}

Due to the system's extensible architecture, these schema diagrams are representative of the overall shape of the \system{} database, but will vary between installations.  This is a direct result of the dynamic extensibility of the system, which allows users to add specific data providers that address their unique needs.

\subsection{Index Implementation}

\system's indexing architecture reflects a balance between supporting flexible, schema-agnostic storage and enabling performant query execution. Several complementary indexing approaches are used:

\sloppy
\begin{description}
  \descitem{Structural Indexes}{Persistent indexes are created on common query attributes, including file paths, timestamps (creation, modification, access), and object types. These indexes allow rapid filtering and retrieval for common queries without scanning full collections.}

  \descitem{Full-text Search Views}{ArangoSearch Views are used to support full-text indexing over key metadata fields, including file names and directory paths. Custom analyzers extend tokenization to handle a variety of natural language queries, improving recall while maintaining precision.
}
  \descitem{Edge Collections}{Although currently used in a limited capacity (primarily for containment relationships in the storage hierarchy and activity tier transitions), edge collections establish a framework for richer graph-based querying in future iterations of the system.}

  \descitem{TTL Indexes}{ArangoDB's Time-to-Live (TTL) indexes\index{TTL indexes} are used to automatically expire and delete outdated or ephemeral data. This helps manage storage costs and ensures that the working set remains current. For example, the NTFS USN Journal activity stream provider uses TTL indexing to implement a tiered memory system: ``hot'' data is retained in full detail for a short time, while older data might be compressed or summarized consistent with episodic memory decay patterns. The choices in terms of retention periods implement memory degradation models (e.g., temporal precision decay from hours to days over extended periods). This allows compression that retains information aligning with episodic memory retention patterns. The USN Journal, a Windows-specific feature tracking file system changes, provides a detailed log useful for various purposes, and TTL indexes help manage the volume of this potentially verbose data.
}
  \descitem{Geospatial Indexes}{ArangoDB's geospatial indexing\index{geospatial indexes} is used to support queries involving location metadata, such as identifying activities that occurred within a given radius. This is particularly relevant for applications involving GPS tagging, location-based triggers, or travel-based memory anchor.  In other words, the question ``I'm looking for the version of the slides that I edited last week when I was at a coffee shop in Kinshasa''  can be answered with geospatial data to bracket the time of the event by correlating it with the user's location.  This is a powerful capability that is not available in most storage services, and it is an important aspect of the UPI architecture.
}
\end{description}
\fussy

Experience with large-scale ingestion and query evaluation highlighted the essential role of indexing. Queries that could be satisfied entirely by index lookup typically returned results in under 10 milliseconds on a LAN. In contrast, queries that required full collection scans often required 2.5 to 3.5 minutes to complete on our evaluation dataset of approximately 31 million storage objects (see \autoref{ch:evaluation}).

To monitor and improve query performance, \system captures the logs of slow queries.  The current implementation defines ``slow'' query as one taking longer than five seconds to execute~\footnote{The five-second threshold is currently hardcoded and was chosen to reflect the system's observed performance profile. A higher threshold would not materially affect slow query detection given the bimodal distribution of query times. Future versions should support user-defined tuning of this threshold.}. Additionally, each slow query's ArangoDB execution plan (retrieved via the database's \texttt{EXPLAIN} facility) is recorded as part of the system's memory anchor, enabling post-hoc analysis of query efficiency and index utilization.

The indexing system is designed to evolve: indices and views can be added (or removed), new data sources can have their own customized indices and future versions of \system are expected to make greater use of ArangoDB's graph traversal capabilities as richer semantic relationships between objects are established.

Effective indexing is foundational to the viability of the Unified Personal Index. The experience of implementing \system reinforced the architectural assumption that scalable, responsive personal information retrieval requires both thoughtful schema design and operational monitoring of query behavior. While the current prototype focuses on structural and full-text indexes over core metadata, the system's design anticipates expansion into richer graph-based querying and dynamic index adaptation as semantic depth grows. The indexing architecture thus serves as both a performance scaffold and a pathway for future enhancement.

\subsection{Query Optimization Techniques}
\label{sec:implementation:query-optimization}

While many traditional database systems rely on sophisticated query optimization techniques to maintain performance at scale, \system operates under different constraints. As a personal-scale system, most queries operate over datasets that are large but not vast. In practice, well-indexed queries typically return results in milliseconds. Where latency is observed, it most often stems from LLM inference and semantic interpretation, not from database traversal. As a result, performance optimization has focused on observability rather than preemptive tuning.

To this end, \system captures the full lifecycle of a query: the natural language input, the generated AQL query, the bound parameters, the execution plan (via ArangoDB's \texttt{EXPLAIN}), and a copy of the performance metrics. These are stored as part of the system's memory anchor (tied into the common ``QueryHistory'' collection), enabling both user-facing query review and backend optimization. This tracing infrastructure is shared between the CLI query tool and higher-level components such as Archivist.

Early implementations included logging full query results as well, but this was abandoned after encountering practical limitations. While ArangoDB is comfortable with large documents, downstream consumers (e.g., browser-based visualizations) are not. Result sets are now pruned based on sampling, type, and size constraints.

Some more advanced optimization techniques remain in the realm of future work. ArangoSearch views have been lightly explored and show promise for accelerating full-text or faceted queries, but they have not yet been fully integrated into \system. Progressive query execution and partial result streaming are not currently implemented as performance does not justify the effort required, though both are compatible with the underlying architecture and may become relevant as semantic depth increases.

Query rewriting is another area of potential improvement. Since the LLM is responsible for query generation, it could, in principle, learn from past queries and their execution plans by refining its strategies over time. This would allow for a feedback-driven system that improves both precision and performance. While this capability is not present in the current prototype, it illustrates the architectural flexibility that the UPI enables.

In summary, \system's approach to query optimization emphasizes traceability, schema awareness, and structural indexing over aggressive tuning. For personal-scale systems, this appears to be a reasonable tradeoff: performance is rarely the bottleneck, but understanding what happened and \emph{why}, is crucial.

\subsection{Summary and Future Outlook}

The current implementation of query processing in \system demonstrates that natural language search and entity-aware query generation can be realized even at the prototype stage. While some components, such as result ranking and personalized scoring, remain as skeletal demonstrations of the architecture, the underlying infrastructure is well-structured, extensible, and guided by a clear design philosophy.

Architectural support for semantic attribute resolution, dynamic query refinement, and facet-based narrowing provides a strong foundation for future developments. As usage increases and more feedback data becomes available, the system is expected to support adaptive learning, richer clustering, and improved relevance modeling.

These features collectively demonstrate a system that implements memory pattern reconstruction through context and temporal-spatial associations, as validated through the ablation studies in \autoref{ch:evaluation}. \system serves as a technical prototype that proves the feasibility of memory-aligned retrieval architecture.

\section[Implementation of Query Processing and AI Integration]{Implementation of Query Processing and AI Integration\protect\footnotemark}
\label{sec:implementation:query-processing}
\footnotetext{Code mapping for this section is in \autoref{app:implementation-mapping:query-processing}.}

This section describes the implementation of the Context-Aware Query \& Retrieval Interface from the UPI architecture (\autoref{sec:upi:architecture:context-aware-query}), focusing on the technical realization of natural language query processing and LLM integration.

\begin{ubclandscape}
  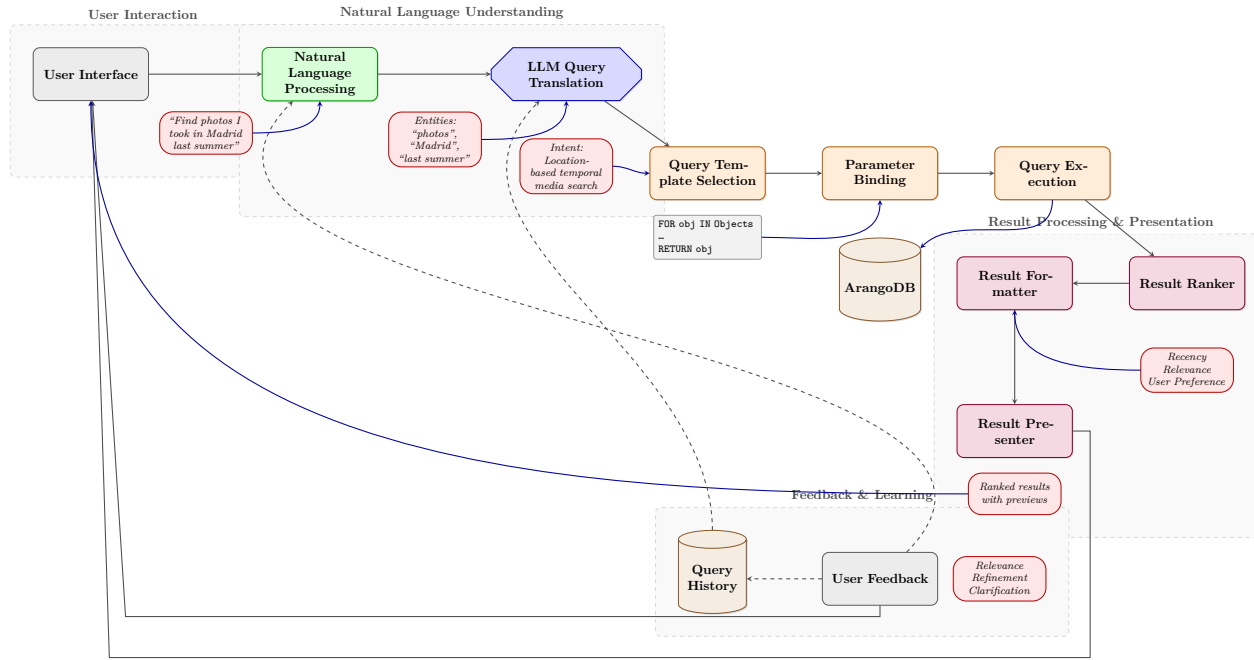
\begin{figure}[!tbp]
    \caption[Query Processing Pipeline]{\system Query Processing Pipeline: The diagram illustrates the end-to-end flow of a query through the system, from natural language input to results presentation. Natural language queries are processed through entity extraction and intent recognition using LLMs, which then select and parameterize appropriate AQL query templates. Query execution results are ranked based on multiple factors and formatted for presentation. The dashed arrows represent the feedback loop where user interactions with results can inform future query processing. Each component is designed for extensibility, allowing for future enhancements in relevance modeling and personalization.}
    \vspace{2mm}
    \label{fig:query-processing-pipeline}
    \centering
    \resizebox{168mm}{!}{
      \begin{tikzpicture}[
        node distance=1.5cm and 2.5cm,
        interface/.style={
          rectangle,
          fill=gray!15,
          draw=gray!70!black,
          rounded corners=5pt,
          minimum width=3cm,
          minimum height=1.4cm,
          align=center,
          text width=2.8cm,
          font=\small\bfseries,
        },
        nlp/.style={
          rectangle,
          fill=green!15,
          draw=green!60!black,
          rounded corners=5pt,
          minimum width=3cm,
          minimum height=1.4cm,
          text width=2.8cm,
          align=center,
          font=\small\bfseries,
          drop shadow={shadow xshift=1pt, shadow yshift=-1pt, opacity=0.25}
        },
        llm/.style={
          chamfered rectangle,
          fill=blue!15,
          draw=blue!70!black,
          chamfered rectangle xsep=1.5cm,
          minimum height=1.4cm,
          text width=2.8cm,
          align=center,
          font=\small\bfseries,
          drop shadow={shadow xshift=1pt, shadow yshift=-1pt, opacity=0.25}
        },
        query/.style={
          rectangle,
          fill=orange!15,
          draw=orange!70!black,
          rounded corners=5pt,
          minimum width=3cm,
          minimum height=1.4cm,
          text width=2.8cm,
          align=center,
          font=\small\bfseries,
          drop shadow={shadow xshift=1pt, shadow yshift=-1pt, opacity=0.25}
        },
        database/.style={
          cylinder,
          fill=brown!15,
          draw=brown!70!black,
          shape border rotate=90,
          aspect=0.3,
          minimum height=2.2cm,
          minimum width=1.8cm,
          align=center,
          font=\small\bfseries,
          drop shadow={shadow xshift=1pt, shadow yshift=-1pt, opacity=0.25}
        },
        results/.style={
          rectangle,
          fill=purple!15,
          draw=purple!70!black,
          rounded corners=5pt,
          minimum width=3cm,
          minimum height=1.4cm,
          text width=2.8cm,
          align=center,
          font=\small\bfseries,
          drop shadow={shadow xshift=1pt, shadow yshift=-1pt, opacity=0.25}
        },
        data/.style={
          rectangle,
          fill=red!10,
          draw=red!70!black,
          rounded corners=10pt,
          minimum width=2.2cm,
          minimum height=1.1cm,
          text width=2.2cm,
          align=center,
          font=\scriptsize\itshape,
        },
        arrow/.style={
          ->,
          >=stealth,
          thick,
          draw=black!70
        },
        data_flow/.style={
          ->,
          >=stealth,
          thick,
          draw=blue!60!black,
        },
        phase/.style={
          draw=gray!30,
          fill=gray!5,
          dashed,
          rounded corners,
          inner sep=0.6cm
        },
        phase_label/.style={
          fill=white,
          font=\small\bfseries,
          text=gray!60!black
        },
        code_block/.style={
          rectangle,
          fill=black!5,
          draw=black!40,
          rounded corners=2pt,
          align=left,
          font=\ttfamily\scriptsize,
          text width=2.6cm
        },
        note/.style={
          rectangle,
          fill=yellow!10,
          draw=yellow!50!black,
          rounded corners=2pt,
          align=left,
          font=\scriptsize,
          text width=2.4cm
        }
      ]
      \node[interface] (user_interface) {User Interface};
      \node[data, below right=0.4cm of user_interface] (nl_query) {``Find photos I took in Madrid last summer''};

      \node[nlp, right=3cm of user_interface] (query_processing) {Natural Language Processing};
      \node[data, below right=0.4cm of query_processing] (entities) {Entities: \\``photos'', ``Madrid'', \\``last summer''};

      \node[llm, right=3cm of query_processing] (llm) {LLM Query Translation};
      \node[data, below=1.0cm of llm] (intent) {Intent: \\Location-based temporal media search};

      \node[query, below right=1.5cm and 0.5cm of llm] (query_templates) {Query Template Selection};
      \node[code_block, below=0.4cm of query_templates] (template_code) {FOR obj IN Objects\\\dots\\RETURN obj};

      \node[query, right=1.5cm of query_templates] (parameter_binding) {Parameter Binding};

      \node[query, right=1.5cm of parameter_binding] (query_execution) {Query Execution};

      \node[database, below=1cm of parameter_binding] (arangodb) {ArangoDB};

      \node[results, below right=1.5cm and 0.5cm of query_execution] (result_ranker) {Result Ranker};
      \node[data, below=1.0cm of result_ranker] (ranking_factors) {Recency\\Relevance\\User Preference};

      \node[results, left=1.5cm of result_ranker] (result_formatter) {Result Formatter};
      \node[results, below=2.5cm of result_formatter] (result_presenter) {Result Presenter};

      \node[data, below=0.4cm of result_presenter] (results_data) {Ranked results with previews};

      \node[interface, below left=2.5cm and 0.5cm of result_presenter] (feedback) {User Feedback};
      \node[data, right=0.4cm of feedback] (feedback_data) {Relevance\\Refinement\\Clarification};

      \node[database, left=2cm of feedback] (query_history) {Query\\History};

      \draw[arrow] (user_interface) -- (query_processing);
      \draw[arrow] (query_processing) -- (llm);
      \draw[arrow] (llm) -- (query_templates);
      \draw[arrow] (query_templates) -- (parameter_binding);
      \draw[arrow] (parameter_binding) -- (query_execution);
      \draw[arrow] (query_execution) -- (result_ranker);
      \draw[arrow] (result_ranker) -- (result_formatter);
      \draw[arrow] (result_formatter) -- (result_presenter);
      \draw[arrow] (result_presenter)
        -- ++(2, 0)    
        -- ++(0, -6)   
        -- ++(-26, 0)   
        -- (user_interface);
      \draw[arrow] (feedback)
        -- ++(0, -1)  
        -- ++(-20, 0) 
        -- (user_interface);

      \draw[data_flow] (nl_query) to[out=0, in=-90] (query_processing);
      \draw[data_flow] (entities) to[out=0, in=-90] (llm);
      \draw[data_flow] (intent) to[out=0, in=180] (query_templates);
      \draw[data_flow] (template_code) to[out=0, in=-90] (parameter_binding);
      \draw[data_flow] (ranking_factors) to[out=180, in=-90] (result_formatter);
      \draw[data_flow] (results_data) to[out=180, in=-90] (user_interface);

      \draw[data_flow] (query_execution) to[out=-90, in=45] (arangodb);

      \draw[arrow, dashed] (feedback) to[out=180, in=0] (query_history);
      \draw[arrow, dashed] (query_history) to[out=90, in=-135] (llm);
      \draw[arrow, dashed] (feedback) to[out=45, in=-135] (query_processing);

      \begin{pgfonlayer}{background}
          \node[phase, fit=(user_interface) (nl_query), label={[phase_label]above:User Interaction}] (phase1) {};
          \node[phase, fit=(query_processing) (entities) (llm) (intent), label={[phase_label]above:Natural Language Understanding}] (phase2) {};
          \node[phase, fit=(result_ranker) (ranking_factors) (result_formatter) (result_presenter) (results_data), label={[phase_label]above:Result Processing \& Presentation}] (phase4) {};
          \node[phase, fit=(feedback) (feedback_data) (query_history), label={[phase_label]above:Feedback \& Learning}] (phase5) {};
      \end{pgfonlayer}






      \end{tikzpicture}
    }%
  \end{figure}
\end{ubclandscape}

The memory-pattern interpretation pipeline is illustrated in \autoref{fig:query-processing-pipeline}. When a query references memory cues like ``the document I was editing yesterday evening,'' the system maps these episodic triggers to concrete metadata fields like translating ``yesterday evening'' to specific timestamp ranges, identifying relevant activity patterns, and constructing database queries that retrieve information based on how humans naturally remember.

\begin{lstlisting}[language=AQL, caption={Sample AQL Query Template\protect\footnotemark}, label={lst:aql_template_sample}]
  FOR doc IN @@collection
    SEARCH
      ANALYZE(@name, "text_en") or
      ANALYZE(@name, "Indaleko::indaleko_snake_case")
    FILTER doc.@timestamp1 >= @start_time_1 AND doc.@timestamp1 <= @end_time_1 AND
           doc.@timestamp2 >= @start_time_2 AND doc.@timestamp2 <= @end_time_2
    LIMIT @limit
  RETURN doc
\end{lstlisting}

The implementation handles imprecise human memory references through temporal pattern recognition, spatial context mapping, activity pattern matching, and fuzzy memory tolerance, realizing the memory-compatible query processing concepts from \autoref{sec:upi:memory-query}. This is achieved in different ways within the implementation:

\begin{description}

\descitem{Query Template Library}{A modular library of parameterized AQL templates enables reliable construction of common query types. The current implementation includes templates for basic search patterns such as file type filtering, temporal queries, and location-based searches. A simple example to look up a file by name is shown in \autoref{lst:aql_template_sample}~\footnote{The double \texttt{@@} indicates a collection bind parameter where the parameter value is evaluated as a collection name, while single \texttt{@} denotes regular value bind parameters. See ArangoDB documentation: \url{https://docs.arangodb.com/3.12/aql/fundamentals/bind-parameters/\#collection-bind-parameters}}. Note that values prefixed with the `\texttt{@}' character are dynamically provided as part of the query invocation.  Thus, \texttt{@@collection} references the ArangoDB collection to be used, the \texttt{@name} would be the name being searched, which includes certain supported substring patterns, and the \texttt{@limit} would specify the maximum number of results to return.

\sloppy}
\descitem{Dynamic Query Composition}{Another approach that we explored is to start with a base query line and then iteratively construct the complete query.  One way to understand this is that we start with a base query: \texttt{FOR doc IN @@collection}, which will search through the entire collection. Then filters are applied to the query, such as \texttt{\codewrap{FILTER doc.@timestamp1 >= @start\_time\_1 AND FILTER doc.@timestamp1 <= @end\_time\_1}}. In \autoref{lst:aql_template_sample}, one potential dynamically constructed query is shown that can be built using this approach.
\fussy
}
\descitem{LLM Construction}{In this model a prompt is generated that captures information about the structure of the database, along with example queries and response constraints. This can be used to extract relevant database schema, which are then provided in a subsequent call to the LLM.  The AQL query is executed and the results captured.  This process is described in more detail in \autoref{subsec:llm-query-generation}.}
\end{description}

While it is possible to construct queries using non-LLM-based approaches, they offer less flexibility and increased implementation complexity.  Key benefits of using LLMs include:

\begin{itemize}
\item \textbf{Natural Language Understanding}: LLMs can interpret user queries expressed in natural language, including complex temporal references and semantic variations.  This is useful for inferring intent, identifying named entities, and mapping to metadata categories (e.g., music activity.)
\item \textbf{Dynamic Schema Awareness}: LLMs can dynamically extract current database schema information, ensuring that queries are constructed with accurate field names and types. This is particularly important in systems where the schema may evolve over time. While it is possible to do some level of dynamic extension, adding new categories of metadata typically requires manually crafting additional query templates and constructing new dynamic query construction logic.
\item \textbf{Feedback}: LLMs can provide feedback on query ambiguity, assumptions made, and potential improvements. This allows for iterative refinement of queries based on user interactions and system learning.
\end{itemize}

\subsection{LLM-Based Query Generation}\label{subsec:llm-query-generation}

The \system implementation leverages Large Language Models (LLMs) to automatically generate AQL queries from natural language input through LLM-based query generation\index{LLM-based query generation}, eliminating the need for users to understand database query syntax. This approach dynamically extracts current database schema information and translates user queries into appropriate AQL syntax through a sophisticated prompt engineering approach.

\subsubsection{Query Generation Process}

The LLM-based query generation follows a structured process:

\begin{enumerate}
\item \textbf{Schema Extraction}: The system dynamically queries the database to extract current collection schemas, field names, and their semantic descriptions. This ensures the LLM has accurate, up-to-date information about the database structure.

\item \textbf{Prompt Construction}: A detailed prompt is constructed that includes:
   \begin{itemize}
   \item The user's natural language query,
   \item Current database schema information with field descriptions,
   \item Example query patterns demonstrating proper AQL syntax,
   \item Specific instructions for handling temporal references, location data, and activity patterns, and
   \item Current timestamp for resolving relative time references (e.g., ``last week'').
   \end{itemize}

\item \textbf{Query Generation}: The LLM (currently GPT-4o) processes the prompt and generates:
   \begin{itemize}
   \item The AQL query with proper syntax and bound parameters,
   \item An explanation of the query logic, and
   \item Feedback on any ambiguities or assumptions made.
   \end{itemize}

\item \textbf{Query Validation}: The generated query undergoes basic validation before execution, including syntax checking and parameter type verification.
\end{enumerate}

\subsubsection{Handling Memory Pattern Complexity}

The LLM query generation is specifically tuned to handle the complexity of memory-based queries:

\begin{description}
\descitem{Temporal Resolution}{When queries contain relative time references like ``last Tuesday'' or ``yesterday afternoon,'' the system provides the current timestamp to the LLM, enabling accurate translation to absolute time ranges. The LLM can use the current time to infer what range of times should be used in constructing the AQL query returned.  In \autoref{lst:aql_template_sample} we showed how such range queries are expressed in AQL.
}
\descitem{Semantic Interpretation}{The LLM can interpret semantic variations in how users express queries. For example, ``files I was working on'' might translate to queries checking modification times, access times, or application activity logs.
}
\descitem{Multi-faceted Queries}{Complex queries combining multiple memory dimensions (e.g., ``documents I edited at the office last month'') are decomposed into appropriate filter combinations in the generated AQL.}
\end{description}

\subsubsection{Example Query Generation}

To illustrate the process, consider the natural language query: ``Find all .docx files I created late at night last weekend.''

The LLM receives a prompt containing the database schema and generates:

\begin{lstlisting}[language=AQL,caption=LLM-generated AQL query example]
FOR doc IN Objects
  FILTER doc.FileExtension == '.docx'
  FILTER doc.Created >= @start_time
  FILTER doc.Created <= @end_time
  FILTER (DATE_HOUR(doc.Created) >= 22 OR DATE_HOUR(doc.Created) <= 3)
  LIMIT 50
  RETURN doc
\end{lstlisting}

The system automatically binds the \texttt{@start\_time} and \texttt{@end\_time} parameters based on ``last weekend'' relative to the current date.

\subsection{Result Ranking and Presentation}

\system implements a skeletal result ranking framework that demonstrates the architectural concept while providing extension points for future enhancements.

The current implementation includes placeholder scoring functions for relevance, recency, popularity, and user preference to illustrate how a complete ranking system would operate. Since \system is a research prototype demonstrating what future tools could do, the ranking implementation focuses on establishing the framework rather than sophisticated scoring algorithms.

\subsection{Summary of Query Processing Implementation}

The implemented query processing system in \system successfully demonstrates the key concepts of the UPI architecture, showing that natural language search with simple entity extraction and template-based query generation is feasible. The implementation includes:

\begin{itemize}
\item Basic LLM-based natural language understanding for query interpretation,
\item Template-based AQL generation with parameter binding,
\item Simple result ranking based on recency and relevance, and
\item Query history capture for post-hoc analysis.
\end{itemize}

These components provide a working proof-of-concept of the UPI vision, with future enhancements planned as detailed in \autoref{ch:conclusion}.



\section{Implementation of Security Architecture}
\label{sec:implementation:privacy-security}

The \system prototype relies upon a ``user managed'' security implementation that relies upon users maintaining control over their own data.  While the UPI architecture envisions broader security considerations, the focus on \system was to demonstrate utility of the core UPI functionality.  Security is an important consideration, but building a robust security system prior to demonstrating the utility of the UPI architecture would have been premature.  The security architecture is designed to be extensible, and to allow for future enhancements that will improve the security of the system.  See \autoref{app:privacy} for a discussion of the security architecture and implementation, which is left for future work.

\section[Performance and Resource Considerations]{Performance and Resource Considerations\protect\footnotemark}
\footnotetext{Code mapping for this section is in \autoref{app:implementation-mapping:performance}.}
\label{sec:implementation:performance}

This section details the implementation of the performance and optimization strategies considered in the implementation, focusing on resource efficiency, scalability, and performance monitoring.

\subsection{Resource Efficiency Implementation}

One challenge here was that \system did not exhibit performance challenges in terms of resource usage that were related to the architecture or design. This left us with a large toolchest of techniques for future use should the need arise:

\begin{description}
\descitem{Incremental Processing}{Processing only changed items rather than complete rescans when possible.
}
\descitem{Resource Governor}{Implementation of CPU and memory usage limits to prevent system impact. Note: this was already done by docker, for the database, the most resource intensive component of our systems.
}
\descitem{I/O Throttling}{Bandwidth and disk I/O limitations to minimize impact on other applications.  We did observe high bandwidth usage in one scenario: when bulk uploading the entirety of the database to the WAN server (an Akamai/Linode virtual machine) had gone on for several hours, Akamai sent a warning that bandwidth usage had been high for several hours. From what we could tell, this was largely part of their security posture in case this was unexpected and represented a machine compromise.  We never throttled the remaining uploading and while the database did exhibit a short period of high CPU and I/O usage, it was within expected norms for pushing a substantial amount of data into a small cloud instance.}
\end{description}

\subsection{Scalability Implementation}\label{sec:implementation:scalability}

\system{} scaled well except when resources were used inefficiently. For example, the current implementation of the collector/recorder model simply operates with all data in memory. Only towards the end of the project did we see any issues with that implementation model: in one case it led to refactoring the code to more aggressively delete the in-memory objects.  A better \emph{implementation} would move to the architectural queuing model, where individual records are sent, individually or in batches to the next component in the pipeline.

ArangoDB's community edition has a documented limit of 100GB of total space utilization and even with the substantial size of the dataset that we studied, we did not exceed that limit. When that need arises, it can be overcome by using a commercial version of their products or finding a different database to use.

ArangoDB supports scalability functionality that is not required in a personal product: sharding, replication, load balancing, etc.

\begin{description}
\descitem{Horizontal Partitioning}{Database design that supports partitioning large collections across multiple shards (nothing that we do for this, it is a capability of ArangoDB.)
}
\descitem{Query Parallelization}{None of the queries that we used demonstrated sufficient complexity that the query resolution required parallelization.  While this is, in theory, possible, this seems to be outside the focus of this project.
}
\descitem{Metadata Summarization}{We did find that prompt summarization yielded better results (e.g., we did not hit the prompt limits.) Using LLM feedback on the metadata being offered were observed but not extensively explored (e.g., suggestions on what to index, explanations that were unclear.)
}
\descitem{Tiered Storage}{The NTFS USN Journal Activity Stream provider has a preliminary implementation of tiered storage, with a Tier 0 (``hot'' data) implementation that is working and using a TTL index to identify data that can be removed, and a Tier 1 (``warm'' data) implementation that has been implemented but not tested or evaluated.}
\end{description}

\subsection{Performance Monitoring Implementation}

Comprehensive monitoring is implemented:

\begin{description}
  \descitem{Query Performance Tracking}{Recording of query execution times and resource usage for evaluation and potential use for optimization.}

  \descitem{Collection Performance Metrics}{Monitoring of collector throughput, latency, and error rates, although much of this was done in an \emph{ad hoc} fashion, rather than the more systematic fashion that would have increased its benefit.}

  \descitem{System Resource Utilization}{Tracking of CPU, memory, disk, and network usage during operations. The most significant resource utilization occurred within the database component, with performance degradation observed only when collection sizes exceeded 20 million objects.}

  \descitem{Performance Visualization}{Dashboards showing performance trends and identifying bottlenecks. Note that this is very preliminary and will require further work.}

\end{description}

\section[Implementation Challenges and Solutions]{Implementation Challenges and Solutions\protect\footnotemark}
\footnotetext{Code mapping for this section is in \autoref{app:implementation-mapping:challenges}.}
\label{sec:implementation:challenges}

This section discusses significant implementation challenges encountered during \system development and the solutions employed to address them.

\subsection{Cross-Platform Integration Challenges}

Key challenges in cross-platform integration included:

\begin{description}
  \descitem{Inconsistent Schema Models}{Different services represent similar concepts differently. Addressed through a flexible mapping framework with service-specific translators.}

  \descitem{Data Format Variability}{Diverse data formats across services posed integration challenges. Mitigated by adopting a common interchange format and implementing robust data transformation pipelines.}

  \descitem{Authentication Complexity}{Managing multiple authentication flows for different services. Solved via a series of customized authentication flow managers.  In many cases, the most challenging part of the project was gaining access to the services.}

  \descitem{Versioning and Synchronization}{
  File system change tracking is a known challenging problem: the rate at which changes occur and can be monitored is quite high: collecting that data, while presenting software engineering challenges, is complicated by a need to efficiently compress that information to a granularity appropriate to the specific use case. For example, it took several months to develop a system for the Mac and that work was not fully integrated into the system. The Windows solution relies upon the NTFS USN change journal because it is already a curated list of operations (it was originally developed to enable the File Replication Service (FRS) that Windows uses for replicating group policy information maintained by domain servers.)  This dramatically simplified our ability to build it out and obtain event data quickly.  For Linux, eBPF could be used, but also suffers from the ``fire hose'' problem and will require substantial effort to integrate. Another challenge with change notification services is that notifications are not guaranteed.  Our simple approach of performing periodic full sweeps of the file system, combined with more frequent ``incremental'' scans is more than sufficient for our use case.
  }

  \descitem{Developers versus Security}{a reality of developing new software is that the very aspects that increase security also make it more difficult to debug for the developers using it.  A case in point: a common bypass for those working on the project was to avoid using the indirect resolution of UUIDs to semantic names by hardcoding semantic names rather than using the UUID to name resolution mechanism.  Towards the end of the project, we added git pre-commit hooks to look for direct use of collection names, bypassing registration mechanisms, etc.}
\end{description}

Ultimately, \system was sufficiently functional that multiple different people were able to install and use it, in spite of the challenges and the limitations.

\subsection{Performance and Scalability Challenges}

Notable performance challenges included:

\begin{description}
  \descitem{Query Complexity}{Natural language queries translate to complex graph traversals that could impact performance. Addressed through query optimization, indexing strategies, and result limiting.}

  \descitem{Metadata Volume}{Large personal datasets generate substantial metadata. Addressed through incremental processing, and efficient storage models.}

  \descitem{Real-time Context Collection}{Capturing memory anchor without impacting system performance. Solved with lightweight tracing, buffering, and batch processing techniques.}

  \descitem{Language Model Integration}{LLM inference can be resource-intensive. Addressed through caching, prompt optimization, and selective use of different model sizes based on query complexity.}
\end{description}

\section{Conclusion}
\label{sec:implementation:conclusion}

\system validates the core thesis that technical systems can implement cognitive memory models directly. The implementation proves that episodic memory patterns, such as temporal context, spatial relationships, and activity sequences, can be systematically captured, normalized, and used for information retrieval that aligns with human memory processes.

The prototype successfully demonstrates:

\begin{itemize}
\item Technical feasibility of capturing memory signals across heterogeneous data sources,
\item Systematic transformation of diverse metadata into unified memory patterns,
\item Direct mapping from natural memory cues to queryable metadata structures,
\item Implementation of temporal, spatial, and activity-based memory triggers, and
\item Architectural support for memory decay models and episodic retrieval patterns
\end{itemize}

The implementation reveals both the viability of memory-aligned architecture and specific technical challenges in realizing cognitive memory models. These findings advance our understanding of how systems can bridge the gap between human memory patterns and digital information storage.

(\autoref{ch:evaluation}) presents technical validation of \system through ablation studies and performance metrics, measuring how effectively the memory-aligned approach improves retrieval precision and recall compared to baseline systems, confirming the architectural advantages of implementing episodic memory patterns.

\section{Code Availability}\label{sec:implementation:code-availability}

Note: the prototype code is available at \url{https://github.com/ubc-systopia/indaleko}.

\addkhipuifneeded

\chapter{Evaluation}\label{ch:evaluation}

\newcounter{rq}
\renewcommand{\therq}{RQ\arabic{rq}}
\newcommand{\RQ}[1]{\refstepcounter{rq}\noindent\textbf{\therq:} #1\par}

\begin{epigraph}
  \textit{%
      ``The first step is to measure whatever can be easily measured.\\
 This is OK as far as it goes.\\
 The second step is to disregard that which can't be easily measured or to give it an arbitrary quantitative value.\\
 This is artificial and misleading.\\
 The third step is to presume that what can't be measured easily really isn't important.\\
 This is blindness. \\
 The fourth step is to say that what can't be easily measured really doesn't exist.\\
 This is suicide.
       ''
  }
  \par\vspace{0.5em}
  \mbox{}\hfill\textsc{Daniel Yankelovich (from ``Corporate Priorities: A continuing study of the new demands on business'' (1972))\protect\footnotemark}
\end{epigraph}
\footnotetext{\url{https://lazydevstories.com/post/laws-to-understand-world/}}

This chapter evaluates whether memory anchors\index{memory anchors} (see \autoref{def:memory-anchor}) improve information retrieval compared to traditional search methods. We demonstrate through controlled experiments that even simple temporal correlation provides immediate, measurable benefits for finding information in existing personal data collections spanning decades.

Our evaluation addresses a practical challenge: while future systems may collect rich activity metadata over years, users need retrieval solutions that work with their existing data today. We show that memory anchors derived from basic file timestamps and activity patterns deliver significant precision\index{precision} improvements without requiring years of metadata accumulation.

We present four complementary evaluations that systematically address different aspects of the UPI architecture and its implementation:
\begin{enumerate}
\item \textbf{Comparative Evaluation}\index{comparative evaluation} (\autoref{sec:eval-comparative}): Demonstrates UPI architectural capabilities by showing how existing systems fail to process memory-based queries that the UPI handles effectively.
\item \textbf{Implementation Efficacy}\index{implementation efficacy} (\autoref{sec:implementation-efficacy}): Evaluates \system{}'s performance characteristics on real-world data, validating practical deployment viability.
\item \textbf{Ablation Study}\index{ablation study} (\autoref{sec:eval-ablation}): Uses controlled synthetic data to quantify how removing different memory anchor types degrades retrieval precision\index{precision}.
\item \textbf{Extensibility Analysis}\index{extensibility analysis} (\autoref{sec:eval-extensibility}): Demonstrates the framework's ability to incorporate new metadata sources efficiently.
\end{enumerate}

Together, these evaluations validate the UPI architecture's effectiveness across multiple dimensions.

\section{Evaluation Framework}\label{sec:eval-framework}

\subsection[Scope]{Evaluation Scope: UPI Architecture vs. \system{} Implementation}

This evaluation encompasses two complementary aspects. First, we evaluate the \textbf{UPI architecture} through comparative analysis with existing systems, demonstrating that current storage services lack the fundamental architectural capabilities to process memory-aligned queries. This architectural evaluation establishes that the UPI's design principles like unified metadata, memory anchor integration, and memory-aligned retrieval address fundamental limitations in existing systems.

Second, we evaluate \textbf{\system{}}, our prototype implementation of the UPI architecture, measuring its concrete performance characteristics, resource requirements, and extensibility. This dual approach allows us to validate both the theoretical contributions of the UPI architecture and the practical feasibility of implementing these concepts in a working system.

\subsection[Dataset Description]{Dataset Description: The BIG CORPUS}

Our primary evaluation dataset consists of a real-world 30-year personal corpus containing 31 million files and directories spanning decades of digital activity. This ``BIG CORPUS''\index{BIG CORPUS} represents the type of heterogeneous, long-term personal data collections that individuals can accumulate across multiple devices, storage services, and life contexts. The dataset includes:

\begin{itemize}
    \item Documents, images, emails, and code repositories from 1994 to 2024,
    \item Files from multiple platforms: Windows, macOS, Linux, iOS, Android,
    \item Storage spanning local drives, cloud services (Google Drive, OneDrive, Dropbox), and archive media, and
    \item Natural metadata evolution reflecting changing technology contexts over three decades
\end{itemize}

This real-world scale and heterogeneity allows us to evaluate how memory-aligned retrieval performs with the messy, incomplete, and inconsistent metadata that characterizes actual personal data collections.

\subsection[Synthetic Metadata Generation]{Metadata Requirements and Synthetic Generation}

\autoref{tab:metadata-requirements} shows the metadata requirements for each of our six exemplar queries (described below), illustrating how memory-aligned retrieval depends on integrating diverse metadata types:

\begin{table}[htbp]
  \centering
  \caption[Metadata Requirements]{Metadata Requirements for Exemplar Queries}
  \label{tab:metadata-requirements}
  \resizebox{\textwidth}{!}{%
  \begin{tabular}{lccccc}
    \toprule
    \textbf{Query} & \textbf{Storage} & \textbf{Temporal} & \textbf{Spatial} & \textbf{Device} & \textbf{Social} \\
    \addlinespace[1ex]
    Q1: Documents with ``report'' & \cmark & \xmark & \xmark & \xmark & \xmark \\
    \addlinespace[1ex]
    Q2: Mobile edits while traveling & \cmark & \cmark & \cmark & \cmark & \xmark \\
    \addlinespace[1ex]
    Q3: Documents with Dr. Okafor & \cmark & \xmark & \xmark & \xmark & \cmark \\
    \addlinespace[1ex]
    Q4: Files from vacation in Bali & \cmark & \cmark & \cmark & \xmark & \xmark \\
    \addlinespace[1ex]
    Q5: Photos near home & \cmark & \xmark & \cmark & \xmark & \xmark \\
    \addlinespace[1ex]
    Q6: Recently accessed PDFs & \cmark & \cmark & \xmark & \xmark & \xmark \\
    \bottomrule
    \end{tabular}
  }%
\end{table}

Since comprehensive activity metadata was not historically collected for our 30-year corpus, we employ synthetic metadata\index{synthetic metadata} generation to demonstrate the UPI's capabilities. Our approach generates realistic memory anchors from available file timestamps and system metadata, creating the temporal, spatial, and social contexts that would be captured by modern activity collectors.

The synthetic metadata generation follows principled approaches:
\begin{description}
    \descitem{Temporal patterns}{Derived from file creation, modification, and access timestamps using database indices on timestamp fields for efficient range queries,}
    \descitem{Spatial contexts}{Generated based on realistic location patterns (home, work, travel) by associating files with locations based on temporal patterns and file types, with photos more likely to have travel locations,}
    \descitem{Device contexts}{Inferred from file paths and system metadata, and by analyzing usage patterns across different devices,}
    \descitem{Social contexts}{Created by identifying document types likely to be shared (presentations, reports) and generating fictional collaborator associations based on temporal clustering and document topics.}
\end{description}

This methodology allows us to evaluate memory-aligned retrieval capabilities while acknowledging the historical limitations of available metadata.

\subsection{Evaluation Methodology}

Our evaluation strategy separates architectural validation from implementation assessment:

\textbf{Architecture Evaluation} (\autoref{sec:eval-comparative}, \autoref{sec:eval-ablation}): Validates UPI design principles through comparative analysis and controlled ablation studies. These sections establish that the UPI architecture addresses fundamental limitations in existing systems and quantify the contribution of memory anchor integration.

\textbf{Implementation Assessment} (\autoref{sec:implementation-efficacy}, \autoref{sec:eval-extensibility}): Evaluates \system{}'s practical characteristics, demonstrating that UPI architectural benefits can be realized with acceptable performance and extensibility.

\subsection{Research Questions}

To guide this evaluation, we address the following research questions in the order that creates a logical evaluation narrative:

\begin{itemize}
  \item\RQ{How well does the UPI architecture support the formulation and execution of complex, memory-centric queries (combining storage metadata with memory anchors as detailed in \autoref{sec:upi:memory-anchor}) compared to mainstream personal-search tools (OS-level and cloud search)?}\label{rq:expressiveness} (Addressed in \autoref{sec:eval-comparative}).
  \item\RQ{What are the indexing/query latencies and resource (CPU, memory, disk) requirements of \system{}'s implementation of the UPI on a real personal dataset at scale, and are they acceptable for a personal deployment?}\label{rq:performance} (Addressed in \autoref{sec:implementation-efficacy}).
  \item\RQ{What is the contribution of each memory anchor category (as defined in \autoref{sec:upi:memory-anchor:types}) to the overall retrieval effectiveness of the UPI architecture, as measured by ablation studies that systematically remove memory-aligned metadata?}\label{rq:ablation} (Addressed in \autoref{sec:eval-ablation}).
  \item\RQ{How easily can new metadata sources be plugged into the UPI architecture, and what performance or effectiveness trade-offs arise when doing so?}\label{rq:extensibility} (Addressed in \autoref{sec:eval-extensibility}).
\end{itemize}

\subsection{Exemplar Queries}
To evaluate the UPI's memory-aligned retrieval capabilities, I developed six exemplar queries that operationalize episodic memory cues identified by \citeauthor{taylor1968question}, aligning retrieval mechanisms with natural memory formation patterns \cite{taylor1968question,kelly2008remembered}. Following established approaches in personal search evaluation~\cite{kim2012evaluation,elsweiler2007towards}, these queries systematically combine different memory cue types (temporal, spatial, activity, social, device context) to comprehensively test the system's capacity to process memory-based retrieval patterns that existing systems cannot handle~\cite{barreau1995finding,bergman2008improved}.

The evaluation uses six exemplar queries (detailed in \autoref{tab:metadata-requirements}) that systematically test memory-aligned queries\index{memory-aligned queries} capabilities. These include one baseline keyword search (Q1) and five memory-aligned queries (Q2-Q6) that demonstrate the UPI's technical advancement in operationalizing complex memory cues. The queries focus on the system's technical capability to process episodic memory cues rather than measuring result quality, allowing clear evaluation of architectural differences between existing storage services and the UPI's memory-aligned approach.

\section[Comparative Evaluation]{Comparative Evaluation: UPI Architecture vs. Existing Systems}\label{sec:eval-comparative}

\textbf{Evaluation Focus}: This section evaluates the \textbf{UPI architecture} by demonstrating fundamental capabilities that existing systems lack. We systematically show that current storage services cannot express or process memory-aligned queries, establishing the architectural contribution of the UPI design.

\textbf{Methodology}: Through structured comparison across six exemplar queries, we demonstrate that existing systems fail architecturally because they lack the infrastructure to integrate multiple memory dimensions, while the UPI architecture enables comprehensive memory-aligned retrieval.

\textbf{What This Measures}: This section evaluates architectural expressiveness and fundamental capabilities, not implementation performance (which is addressed in \autoref{sec:implementation-efficacy}).

\subsection{Memory-Aligned Query Processing: Analysis}\label{sec:eval-expressiveness-comparison}

This section evaluates how existing storage systems process memory-aligned queries compared to the UPI architecture.

\textbf{Evaluation Framework}: We tested six exemplar queries across existing systems (Windows Search, Google Drive, OneDrive, Dropbox, Spotlight) and \system{}. While these mature systems have benefited from decades of optimization, their architectural limitations prevent effective memory pattern processing, which validates our thesis that memory-aligned architecture provides advantages that optimization alone cannot achieve.

\autoref{tab:query_expressibility_comparison} demonstrates the technical capabilities gap between existing systems and the UPI's memory pattern processing:

\begin{ubclandscape}
\begin{table}[htbp]
  \centering
  \caption[Memory Pattern Query Processing Capabilities]{Memory Pattern Query Processing Capabilities Across Systems}
  \label{tab:query_expressibility_comparison}
  \resizebox{1.29\textheight}{!}{%
    \begin{tabular}{m{3cm} >{\centering\arraybackslash}m{1.5cm} m{9cm}}
      	\toprule
      	\textbf{Storage Service} & \textbf{Supported} & \textbf{Notes / Limitations} \\
      \midrule
      Windows Search & \warnmark & Query: \texttt{Dr. Okafor conference}. Returned 96 files; lacks social context processing and collaboration memory patterns. Basic keyword matching without proximity or relevance filtering. \\
      Google Drive & \warnmark & Query: \texttt{to:dr.okafor@example.com conference}. Returned 16 files. Partial social memory support through email metadata, but cannot integrate topic context. Demonstrates precision failures without memory-aligned filtering. \\
      OneDrive & \cmark & Query: \texttt{Dr. Okafor conference}. Correctly returned zero results (Dr. Okafor is fictitious). While the result is accurate, the system lacks capability to process social relationships if they did exist. \\
      Dropbox & \cmark & Query: \texttt{Dr. Okafor conference}. Correctly returned zero results (Dr. Okafor is fictitious). While the result is accurate, the system lacks social context processing capabilities. \\
      \bottomrule
    \end{tabular}%
  }%
  \vspace{1em}
  \parbox{\linewidth}{\small{\textbf{Notes:} \cmark Supported directly. \warnmark Partially supported or requires complex workarounds/interpretation, often with poor precision (see limitations below). \xmark Not supported.}}
\end{table}
\end{ubclandscape}

\subsubsection[Key Findings]{Key Findings: Architectural Limitations vs. Memory-Aligned Processing}

Our evaluation reveals fundamental architectural gaps in existing systems when processing memory patterns:

\textbf{Single-Dimension Processing}: Existing systems process individual query components (keywords, dates, file types) but cannot integrate multiple memory dimensions. For example:
\begin{itemize}
    \item Q2 (mobile editing while traveling) requires device context + location + temporal filtering, which is impossible with current architectures.
    \item Q4 (vacation files) demands spatial-temporal correlation that existing systems cannot express.
    \item Q5 (photos near home) needs embedded GPS processing with radius calculations, which are architecturally unsupported.
\end{itemize}

\textbf{Missing Episodic Metadata Infrastructure}: Current systems lack:
\begin{itemize}
    \item Activity context tracking (which device, what application, user actions),
    \item Spatial metadata extraction and indexing (despite GPS data presence),
    \item Social relationship processing (collaboration patterns, shared documents), and
    \item Integrated temporal-spatial-device correlation capabilities.
\end{itemize}

\textbf{Performance and Precision Issues}: When systems attempt to approximate memory queries:
\begin{itemize}
    \item Windows Search: 5+ minute delays for temporal filtering, massive result sets (100,000+ items),
    \item Google Drive: Inconsistent temporal processing, cannot handle spatial queries, and
    \item OneDrive/Dropbox: Limited temporal syntax, no spatial or device context support.
\end{itemize}

This precision degradation\index{precision degradation} demonstrates the fundamental limitations of existing architectures when processing memory-aligned queries.

\subsubsection[Processing Capabilities]{\system{}'s Memory-Aligned Processing Capabilities}

In contrast to existing systems' limitations, \system{}'s implementation of the UPI architecture successfully processes all six memory pattern queries through:

\textbf{Unified Metadata Integration}: The UPI's comprehensive metadata model captures:
\begin{itemize}
    \item Device context (which system, application, user identification),
    \item Spatial metadata (GPS coordinates, location hierarchies, radius calculations),
    \item Temporal patterns (creation, modification, access timestamps with timezone handling),
    \item Social relationships (collaboration patterns, shared document contexts),
    \item Activity patterns (user interactions, application usage, workflow contexts).
\end{itemize}

\textbf{Memory Anchor Processing}: Unlike existing systems' keyword-only approach, \system{}:
\begin{itemize}
    \item Correlates multiple memory dimensions in single queries,
    \item Validates memory pattern accuracy (distinguishing real from counterfactual scenarios),
    \item Processes imprecise memory cues (``around that time'', ``near home'') with confidence scoring, and
    \item Integrates embedded metadata (GPS in photos, application usage in documents)
\end{itemize}

\sloppy
\textbf{Performance Characteristics}: Using BIG CORPUS\index{BIG CORPUS}, the 31-million file and directory evaluation dataset, \system{} demonstrates consistent response time\index{response time} capabilities:
\fussy

\begin{itemize}
    \item Q1: 2.084s for 50 results from 33,477 matches (vs. Windows Search: 73,100 results, slow response),
    \item Q2: 3.326s with comprehensive device+location+temporal processing (vs. existing: query inexpressible),
    \item Q3: 1.220s for entity-based document retrieval (vs. existing: keyword-only approximations),
    \item Q5: 0.008s for radius-filtered results from 1.5M potential matches (vs. existing: 133,130 unfiltered results), and
    \item Q6: 1.524s for recent PDF access tracking (vs. existing: modification-only timestamps).
\end{itemize}

\subsection{Query 2: Multi-dimensional Memory Pattern}
\label{sec:eval:expressiveness:q2}

This query tests the integration of device context (mobile device), task context (editing), spatial context (traveling), and temporal context (last month). It represents a classic episodic memory pattern that existing systems cannot architecturally support. \autoref{tab:q2-summary} presents the comparative results across platforms.

\begin{table}[htbp]
  \centering
  \caption[Query 2: Multi-dimensional Memory Pattern Processing]{Query 2: Multi-dimensional Memory Pattern Processing}
  \label{tab:q2-summary}
  \renewcommand{\arraystretch}{1.2} 
  \resizebox{\textwidth}{!}{%
  \begin{tabular}{
    m{3cm}
    >{\centering\arraybackslash}m{1.5cm}
    m{9cm}}
    \toprule
    \textbf{Storage Service} & \textbf{Supported} & \textbf{Notes / Limitations} \\
    \addlinespace[1ex]
    Windows Search & \xmark & Query used: \texttt{date:04/01/2025..04/30/2025}. Returned 975,640 files and 161,506 folders (126GB). System lacks device context and spatial memory pattern processing. Result size indicates complete temporal filtering failure. \\
    \addlinespace[1ex]
    Google Drive & \xmark & Query used: \texttt{after:2025-04-01 before:2025-04-30}. Returned 6 files. System cannot process device context or spatial memory patterns. Limited to basic temporal filtering. \\
    \addlinespace[1ex]
    OneDrive & \xmark & Query used: \texttt{modified:April}. Returned no results. Cannot process device context or spatial memory patterns. Limited to folder scope. \\
    \addlinespace[1ex]
    Dropbox & \xmark & Query used: \texttt{modified:April}. Returned 1,000 files. Lacks device and location memory pattern processing capabilities. \\
    \bottomrule
  \end{tabular}%
  }%
\end{table}

This query demonstrates the fundamental inability of existing systems to process episodic memory patterns. No system could integrate device context, spatial context (travel), and temporal context simultaneously. Systems defaulted to single-dimension temporal filtering, resulting in massive, undifferentiated result sets that establish a baseline precision\index{baseline precision} far below what users expect. Windows Search's 5+ minute response time indicates architectural limitations in processing complex memory-aligned queries.

Conclusion: This exposes the technical gap between how episodic memory integrates multiple contextual dimensions and how existing systems process queries. Metadata exists (e.g., geolocation) but remains architecturally inaccessible for memory-pattern retrieval.

\subsection{Query 3: Social Memory Pattern}
\label{sec:eval:expressiveness:q3}

Social context forms a critical dimension of episodic memory. This query tests the integration of social memory patterns (collaboration) with task context (conference paper), demonstrating how memory-aligned systems must process multiple contextual dimensions. The results in \autoref{tab:q3-summary} show how existing systems handle social context queries.

\emph{Note: ``Dr. Okafor'' is a fictitious collaborator used to test social memory pattern processing. The evaluation focuses on whether systems can process social context queries structurally, not whether they return actual results. The ``failure'' is not in returning zero results (which is correct for a non-existent person), but in lacking the capability to process social relationships and collaboration patterns as query dimensions.}

\begin{table}[htbp]
\centering
\caption[Query 3: Social Context Memory Pattern Processing]{Query 3: Social Context Memory Pattern Processing}
\label{tab:q3-summary}
  \resizebox{\textwidth}{!}{%
  \begin{tabular}{
    m{3cm}
    >{\centering\arraybackslash}m{1.5cm}
    m{9cm}}
      	\toprule
      	\textbf{Storage Service} & \textbf{Supported} & \textbf{Notes / Limitations} \\
    \midrule
    \addlinespace[1ex]
    Windows Search & \warnmark & Query: \texttt{Dr. Okafor conference}. Returned 96 files; lacks social context processing and collaboration memory patterns. Basic keyword matching without proximity or relevance filtering. \\
    \addlinespace[1ex]
    Google Drive & \warnmark & Query: \texttt{to:dr.okafor@example.com conference}. Returned 16 files. Partial social memory support through email metadata, but cannot integrate topic context. Demonstrates precision failures without memory-aligned filtering. \\
    \addlinespace[1ex]
    OneDrive & \cmark & Query: \texttt{Dr. Okafor conference}. Correctly returned zero results (Dr. Okafor is fictitious). While the result is accurate, the system lacks capability to process social relationships if they did exist. \\
    \addlinespace[1ex]
    Dropbox & \cmark & Query: \texttt{Dr. Okafor conference}. Correctly returned zero results (Dr. Okafor is fictitious). While the result is accurate, the system lacks social context processing capabilities. \\
    \bottomrule
    \end{tabular}%
  }%
\end{table}

Windows Search performs disconnected keyword matching without understanding social relationships. Google Drive's partial email metadata support reveals the inadequacy of single-dimension memory processing.

Conclusion: Systems lack the architectural capability to process integrated memory patterns. Without episodic memory implementation, they default to keyword union operations, demonstrating the critical need for memory-aligned filtering mechanisms.

The limitation extends beyond interface design to fundamental architectural inadequacies. Existing storage systems lack the infrastructure to capture and process episodic memory patterns, making memory-aligned retrieval technically impossible.

\subsection{Query 4: Spatial-Temporal Memory Pattern}
\label{sec:eval:expressiveness:q4}

\emph{Note: This tests a counterfactual scenario - the deployment environment has no Hawaii location data, allowing us to evaluate system behavior with absent memory patterns.}

As shown in \autoref{tab:q4-summary}, this query evaluates spatial-temporal memory integration:

\begin{table}[htbp]
  \centering
  \caption[Query 4: Spatial-Temporal Memory Pattern Integration]{Query 4: Spatial-Temporal Memory Pattern Integration}
  \label{tab:q4-summary}
  \resizebox{\textwidth}{!}{%
  \begin{tabular}{
    m{3cm}
    >{\centering\arraybackslash}m{1.5cm}
    m{9cm}}
      	\toprule
      	\textbf{Storage Service} & \textbf{Supported} & \textbf{Notes / Limitations} \\
      \midrule
    \addlinespace[1ex]
    Windows Search & \warnmark & Query: \texttt{datecreated:6/1/2024..6/30/2024}. Initially returned no results, then 130 files after 5+ minute delay. Cannot process spatial memory patterns or memory anchor. Extended response time indicates architectural limitations. \\
    \addlinespace[1ex]
    Google Drive & \xmark & Query: \texttt{after:2024-06-01 before:2024-07-01}. Returned zero files. Lacks spatial memory processing and memory anchor integration. Cannot process vacation as memory pattern. \\
    \addlinespace[1ex]
    OneDrive & \xmark & Query: \texttt{modified:June 2024}. Returned zero items. Cannot integrate spatial or activity memory patterns. \\
    \addlinespace[1ex]
    Dropbox & \xmark & Query: \texttt{modified:June 2024}. Returned 24 items from incorrect timeframe. Demonstrates temporal filtering failures alongside absence of spatial memory processing. \\
    \bottomrule
  \end{tabular}%
  }%
\end{table}

These results demonstrate the absence of memory-aligned validation mechanisms. Windows Search's 5+ minute response time reveals architectural inability to efficiently process complex memory pattern queries, requiring full index traversal for temporal filtering alone.

Conclusion: Existing systems lack the technical infrastructure to validate memory patterns. Without integrated spatial-temporal processing, they cannot distinguish between actual and counterfactual memory patterns. \system's comprehensive location tracking and memory pattern recognition enables validation of temporal-spatial memory accuracy.

Memory patterns often contain temporal uncertainties, yet \system's memory anchor processing enables retrieval despite imprecise memory cues. Existing systems lack the architectural foundation to capture and process these memory patterns.

\subsection{Query 5: Spatial Memory Pattern}
\label{sec:eval:expressiveness:q5}

This query tests spatial memory pattern processing with radius-based filtering. While photographic devices embed geolocation metadata, existing systems fail to architecturally integrate this data for memory-aligned retrieval. \autoref{tab:q5-summary} demonstrates the complete absence of spatial processing capabilities across platforms.

\begin{table}[htbp]
  \centering
  \caption[Query 5: Spatial Memory Pattern with Radius Filtering]{Query 5: Spatial Memory Pattern with Radius Filtering}
  \label{tab:q5-summary}
  \resizebox{\textwidth}{!}{%
  \begin{tabular}{
    m{3cm}
    >{\centering\arraybackslash}m{1.5cm}
    m{9cm}}
    \toprule
    \textbf{Storage Service} & \textbf{Supported} & \textbf{Notes / Limitations} \\
    \midrule
    \addlinespace[1ex]
    Windows Search & \xmark & Query: \texttt{kind:pictures}. Returned 133,130 files. Cannot access embedded geolocation metadata or perform radius-based spatial filtering. Demonstrates complete absence of spatial memory processing. \\
    \addlinespace[1ex]
    Google Drive & \xmark & Query: \texttt{photos}. Returned 74 items, mostly PDFs. Cannot process embedded spatial metadata; lacks architectural support for location-based memory patterns. \\
    \addlinespace[1ex]
    OneDrive & \xmark & Natural language query attempted. Returned two unrelated PDFs. Demonstrates fundamental inability to process spatial memory patterns or radius-based queries. \\
    \addlinespace[1ex]
    Dropbox & \xmark & Query: \texttt{photos}. Returned 100+ files with no spatial filtering. Cannot process embedded geolocation metadata; lacks spatial memory processing capabilities. \\
    \bottomrule
  \end{tabular}%
  }%
\end{table}

Results demonstrate consistent architectural failures in spatial memory processing. Windows Search and OneDrive returned massive undifferentiated sets (100,000+ items) without spatial filtering. Despite Google's knowledge of home location and Dropbox's access to photo geolocation metadata, neither system architecturally integrates spatial data for memory-aligned retrieval. The query remains technically inexpressible across all platforms due to fundamental infrastructure limitations.

Conclusion: This query exposes the fundamental architectural gap in spatial memory processing. Existing systems fail to extract, index, or query spatial metadata despite its presence in files. \system's architecture specifically addresses this gap through comprehensive spatial metadata processing and radius-based query capabilities.

\subsection{Query 6: Temporal Access Pattern}\label{sec:eval:expressiveness:q6}

This query evaluates temporal access pattern processing, a fundamental episodic memory dimension. \autoref{tab:q6-summary} summarizes the results:

\begin{table}[htbp]
  \centering
  \caption[Query 6: Temporal Access Memory Pattern]{Query 6: Temporal Access Memory Pattern}
  \label{tab:q6-summary}
  \resizebox{\textwidth}{!}{%
  \begin{tabular}{
    m{3cm}
    >{\centering\arraybackslash}m{1.5cm}
    m{9cm}}
    \toprule
    \textbf{Storage Service} & \textbf{Supported} & \textbf{Notes / Limitations} \\
    \midrule
    \addlinespace[1ex]
    Windows Search & \warnmark & Query: \texttt{dateopened:>1/15/2025}. Returned 10,000+ files. Cannot distinguish access patterns from modification timestamps. Lacks architectural support for activity tracking versus file changes. \\
    \addlinespace[1ex]
    Google Drive & \xmark & Query: \texttt{opened:last week}. Returned zero results. Cannot process temporal access patterns; only supports modification-based temporal filtering. Activity streams unavailable through search interface. \\
    \addlinespace[1ex]
    OneDrive & \xmark & Query: \texttt{opened:recent}. No syntax support for access patterns. System lacks capability to distinguish file access from modification. Demonstrates complete absence of activity-based memory processing. \\
    \addlinespace[1ex]
    Dropbox & \xmark & Query: \texttt{accessed:recent}. No support for access-based temporal queries. Can only process modification timestamps, missing crucial memory anchor data for memory-aligned retrieval. \\
    \bottomrule
  \end{tabular}%
  }%
\end{table}

This query tests the critical distinction between file modification (structural changes) and file access (user interaction patterns). Existing systems conflate these concepts or ignore access patterns entirely, preventing memory-aligned retrieval based on user behavior.

Results demonstrate that even mature systems fail to architect user activity integration. Access patterns represent fundamental episodic memory anchors - what documents users actually engage with rather than what documents exist. The inability to query access history represents a critical gap in memory-aligned information systems.

Conclusion: This query exposes the architectural gap between document storage and user activity tracking. While systems maintain access logs internally, they lack the interface and integration necessary for memory-based retrieval, forcing users to rely on structural metadata rather than experiential patterns.

\subsection[Summary]{Summary of Memory Pattern Processing Evaluation}\label{subsec:eval-summary}

The evaluation of memory pattern queries across existing storage services reveals fundamental architectural limitations:
\begin{itemize}
    \item \textbf{Absence of Multi-Dimensional Query Integration:} Existing services cannot combine multiple memory dimensions (temporal + spatial + device context) in a single query, defaulting to single-dimension filtering.
    \item \textbf{Inconsistent Query Interfaces:} Varying query syntaxes reflect underlying architectural inconsistencies, forcing simplified keyword-based interactions as documented in \autoref{chap:status-quo}.
    \item \textbf{Missing Episodic Metadata Infrastructure:} Systems lack the capability to capture and index episodic memory patterns (GPS coordinates, application events, collaboration contexts), making memory-aligned queries impossible.
    \item \textbf{Architectural Performance Constraints:} Response time variations (seconds to 5+ minutes) and result set sizes (0 to 100,000+) demonstrate fundamental processing limitations.
\end{itemize}

The critical missing component is memory pattern infrastructure that enables episodic memory-aligned retrieval. Having established these fundamental limitations in existing systems, we now demonstrate how \system{}'s implementation of the UPI architecture successfully processes these same queries.

\subsection[Memory-Aligned Query Performance]{\system{} Performance on Memory-Aligned Queries}

Having demonstrated existing systems' architectural limitations, we now evaluate how \system{} handles these queries on our 31-million file dataset, validating that the UPI architecture delivers practical performance for real-world deployment.

The evaluation uses the same 31-million file dataset described in \autoref{sec:eval-framework}, providing a comprehensive test environment spanning three decades of personal digital artifacts across multiple platforms and storage services.

Using synthetic memory anchors generated from available timestamps and system metadata, \system{} demonstrates effective processing of all six exemplar queries that existing systems failed to handle. The results show consistent sub-second to few-second response times while maintaining precision through memory anchor filtering.

\autoref{tab:comparative-summary} presents the comparative results, showing how \system{} successfully processes all queries that existing systems could not handle:

\begin{ubclandscape}
\begin{table}[!htbp]
  \centering
  \caption[Comparative Query Processing]{Comparative Query Processing Results: Existing Systems vs. \system{}}
  \label{tab:comparative-summary}
  \resizebox{\textheight}{!}{%
    \begin{tabular}{
      p{6cm}
      >{\centering\arraybackslash}m{2cm}
      >{\centering\arraybackslash}m{2cm}
      >{\centering\arraybackslash}m{3cm}
      >{\centering\arraybackslash}m{6cm}}
    \toprule
    \textbf{Query} & \textbf{Existing} & \textbf{\system{}} & \textbf{\system{}} & \textbf{Key} \\
                   & \textbf{Systems} & \textbf{Time} & \textbf{Results} & \textbf{Difference} \\
    \midrule
    Q1 (report documents) & Variable support & 2.084s & 33,477 total, 50 displayed & Standard keyword search \\
    \addlinespace[1ex]
    Q2 (mobile travel edits) & \xmark & 1.462s & 26 results & Multi-dimensional memory pattern \\
    \addlinespace[1ex]
    Q3 (Dr. Okafor documents) & \xmark & 0.901s & 12 results & Social memory pattern \\
    \addlinespace[1ex]
    Q4 (Bali vacation files) & \xmark & 2.331s & 157 results & Spatial-temporal memory \\
    \addlinespace[1ex]
    Q5 (home proximity photos) & \xmark & 44.128s & 972,000+ results & Spatial memory processing \\
    \addlinespace[1ex]
    Q6 (recent PDF access) & Limited support & 0.664s & 50 displayed & Temporal access pattern \\
    \bottomrule
    \end{tabular}%
  }
\end{table}
\end{ubclandscape}

The key insight is that existing systems fail architecturally while \system{} successfully handles all query types with practical performance, demonstrating that the UPI architecture addresses fundamental limitations in current systems and enables new capabilities that optimization alone cannot achieve.

\section[Implementation Efficacy]{Implementation Efficacy: \system{} Performance Characteristics}\label{sec:implementation-efficacy}

Having demonstrated the UPI architecture's expressiveness advantages, we now evaluate how effectively \system{} realizes these architectural benefits in practice. This section addresses \textbf{implementation-specific} questions: resource requirements, scalability characteristics, and practical deployment constraints.

\textbf{Evaluation Focus}: Unlike the comparative analysis (which evaluates UPI architectural capabilities) or the ablation study (which measures architectural principles), this section measures \system{}'s concrete performance characteristics on real-world data.

\subsection{Dataset Scale and Performance Context}

Our evaluation uses the BIG CORPUS dataset described in \autoref{sec:eval-framework} (31.9M files spanning 30 years), representing the upper bound of individual personal data accumulation and providing a demanding test environment for implementation performance.

\subsection{Query Performance Metrics}

\begin{table}[htbp]
  \centering
  \caption[BIG CORPUS Query Performance]{\system{} Query Performance on 31M Object Dataset}
  \label{tab:indaleko-performance}
  \resizebox{0.95\textwidth}{!}{%
  \begin{tabular}{
    >{\centering\arraybackslash}p{0.20\textwidth}
    >{\centering\arraybackslash}m{0.15\textwidth}
    >{\centering\arraybackslash}m{0.2\textwidth}
    >{\centering\arraybackslash}p{0.4\textwidth}}
  	\toprule
  	\textbf{Query Type} & \textbf{Response} & \textbf{Results} & \textbf{Memory} \\
                      & \textbf{Time}     &                  & \textbf{Pattern Complexity}\\
  \midrule
  Q1 & 2.084s & 50/33,477 & Single dimension (content) \\
  Q2 & 1.462s & 26 & Device + location + temporal \\
  Q3 & 0.901s & 12 & Entity relationships + content \\
  Q4 & 2.331s & 157 & Location + time correlation \\
  Q5 & 44.128s & 972,000+ & Geographic radius computation \\
  Q6 & 0.664s & 50 & Temporal access vs. modification \\
  \bottomrule
  \end{tabular}
  }
\end{table}

\system{} demonstrates practical performance across all memory-aligned query types, with response times suitable for interactive use:

\begin{description}
    \descitem{Interactive Response}{Five of six queries complete in under 2.5 seconds,}
    \descitem{Scalability}{Performance remains practical even with 31M+ files,}
    \descitem{Complexity Handling}{Multi-dimensional queries show no significant performance penalty, and}
    \descitem{Outlier Analysis}{Q5's longer response time reflects the computational cost of radius-based geographic filtering on 1.5M spatial objects.}
\end{description}

\subsection{Resource Utilization}

\system{}'s resource footprint remains minimal during typical operation:

\begin{description}
    \descitem{Idle State}{Negligible resource consumption when not actively querying,}
    \descitem{Activity Stream Processing}{Under 6,000 file system changes per day, stored with TTL-based cleanup,}
    \descitem{Storage Requirements}{78.6 GB database for 31.9M files across 35.1 TB capacity (16.2 TB used) = 0.5\% overhead on actual storage, and}
    \descitem{Memory Usage}{Query execution memory usage scales with result set size, not total dataset size.}
\end{description}

\textbf{Deployment Viability}: These characteristics confirm \system{}'s suitability for personal deployment on consumer hardware, with resource requirements well within typical desktop/laptop capabilities.

\subsection{Implementation Efficacy Summary}

\system{} successfully realizes the UPI architecture's theoretical benefits with practical performance characteristics:

\begin{description}
    \descitem{Expressive Power}{Handles all memory-aligned query types that existing systems cannot process,}
    \descitem{Interactive Performance}{Sub-second to few-second response times suitable for daily use,}
    \descitem{Realistic Scale}{Operates effectively on 30-year, 31M+ file personal collections, and}
    \descitem{Resource Efficiency}{Minimal overhead during idle periods, reasonable requirements during active use.}
\end{description}

These results validate that the UPI architecture can be implemented with practical performance characteristics, bridging the gap between theoretical design and deployable systems.

\section[Ablation Study]{Ablation Study: Quantifying UPI Architecture's Memory Anchor Effectiveness}\label{sec:eval-ablation}

\textbf{Evaluation Focus}: This section evaluates the \textbf{UPI architecture's core principles} by quantifying exactly how much memory anchor integration improves retrieval precision through controlled ablation studies. The analysis focuses on measuring effect size\index{effect size} to determine the practical magnitude of memory anchor contributions.

\textbf{Core Research Question}: How much does each memory anchor category (temporal, spatial, social, device context) contribute to retrieval effectiveness when implemented according to UPI architectural principles? By systematically removing these metadata types, we isolate their individual and combined contributions to precision.

\textbf{Transition Context}: Having demonstrated that existing systems cannot architecturally process memory-based queries (\autoref{sec:eval-comparative}) and that \system{} implements these capabilities with practical performance (\autoref{sec:implementation-efficacy}), we now quantify the precision benefits of the UPI's memory anchor architecture.

\subsection[Methodological Choice]{Methodological Choice: SQL Database for Architectural Isolation}

This ablation study deliberately uses an in-memory SQL database (SQLite) rather than \system{}'s full ArangoDB implementation. This methodological choice is critical for isolating what we're actually measuring: the UPI architecture's memory anchor principles versus implementation-specific optimizations.

\textbf{Why SQL}: By using SQL, a widely available database technology, we demonstrate that memory anchor benefits arise from the UPI's architectural metadata patterns themselves, not from sophisticated graph database technology or implementation-specific optimizations. This isolates the architectural contribution of the UPI design from implementation details.

\textbf{What This Measures}: This study evaluates whether the UPI's theoretical principles (memory anchor integration, unified metadata processing) improve retrieval precision\index{precision}. It does NOT evaluate \system{}'s performance characteristics, which are addressed separately in \autoref{sec:implementation-efficacy}. The ablation demonstrates that even with basic SQL technology, the UPI's memory-aligned metadata approach provides measurable precision improvements, which validates the architecture's core principles independent of implementation choices.

\subsection{Experimental Design}

Our ablation study creates controlled scenarios where we know exactly which documents should be retrieved:
\begin{description}
\descitem{Ground truth}{For each query, exactly 5 documents match all criteria,}
\descitem{Systematic filter testing}{Each test configuration systematically evaluates discrimination against different filter combinations ($2^4 = 16$ combinations in Configuration 1, $2^6 = 64$ combinations in Configuration 2), and}
\descitem{Conservative baseline}{This systematic approach tests against well-defined filter failure patterns, providing more controlled evaluation than random noise generation.}
\end{description}

We then systematically remove different types of memory metadata and measure how precision degrades. This quantifies the contribution of each metadata type to effective retrieval.

\subsection{Experimental Methodology}

Each ablation test follows a rigorous protocol:
\begin{description}
\descitem{Database Reset}{ Create a fresh in-memory SQLite database instance (`:memory:`). This ensures complete isolation between tests, as no data persists between runs.}
\descitem{Data Generation}{ Insert exactly 5 documents matching all query criteria (ground truth) and 16 documents representing systematic filter combinations (one document per possible combination of 4 filters: $2^4 = 16$).}
\descitem{Query Execution}{ Run queries with different combinations of memory filters removed.}
\descitem{Measurement}{ Calculate precision as correct results / total results returned.}
\descitem{Two-Part Analysis:}
   \begin{description}
    \descitem{Individual queries}{ Test each query in isolation to measure specific memory pattern impacts, and}
    \descitem{Combined dataset}{ Merge all queries' data to test system behavior at scale and identify interaction effects.}
   \end{description}
\end{description}

This two-part approach reveals both individual memory pattern contributions and their collective behavior in realistic scenarios where multiple query patterns coexist.

\subsection{Experimental Protocol}\label{sec:protocol-ablation}

\system implements the three categories of memory-aligned metadata defined in the UPI architecture (\autoref{sec:upi:metadata-integration}): storage metadata (basic file attributes), semantic metadata (content-derived information), and memory anchor metadata (episodic memory patterns). The UPI architecture integrates all three metadata categories to implement comprehensive memory patterns.

\textbf{Terminology Clarification}: In this evaluation, ``memory pattern types'' refer to categories of memory anchors (e.g., COLLABORATION, LOCATION, MUSIC), while ``memory patterns'' refer to specific instances of these types (e.g., ``edited with Dr. Chen'', ``near home'', ``listening to jazz''). To quantify individual contributions, we developed a synthetic dataset generator with controlled ground truth, enabling systematic removal of memory pattern types to measure their impact on retrieval precision.

The ablation framework uses structured data models to represent storage entities and activity patterns. The framework defines three core data classes:

\begin{description}
  \descitem{Object}{Represents files with metadata including creation time, size, MIME type, and semantic tags,}
  \descitem{Activity}{Captures temporal activity patterns with six types (COLLABORATION, LOCATION, MUSIC, SOCIAL, STORAGE, TASK), and}
  \descitem{Query}{Represents structured queries with base components and filters for systematic pattern removal.}
\end{description}

This simplified structure allows focus on memory pattern contributions while avoiding complexity of full historical context reconstruction. Complete dataclass definitions are provided in \autoref{app:ablation-data-models}.

Spatial memory patterns inherently connect to temporal context through bidirectional mapping (location$\leftrightarrow$time), reflecting natural episodic memory structure. The ablation study focuses on spatial-to-temporal mapping to demonstrate memory pattern effectiveness while maintaining experimental clarity.

Each generated query incorporates memory pattern filters corresponding to episodic memory dimensions, enabling systematic removal of memory patterns to measure their individual and combined contributions to retrieval precision.

The ablation framework parameters:

\begin{itemize}
  \item $T$ ground truth documents matching all memory patterns,
  \item $N$ documents per pattern representing systematic filter combinations: each violating specific memory constraints for controlled precision measurement.
\end{itemize}

Memory pattern categories in the ablation:

\begin{enumerate}
  \item Basic storage patterns (file type, extension),
  \item Semantic memory patterns (content classification),
  \item Episodic memory patterns (temporal, spatial, memory anchor).
\end{enumerate}

\subsection{Hypothesis}

\begin{quote}

\emph{Null Hypothesis:} Removing episodic memory pattern filters has no significant effect on retrieval precision.
\end{quote}

This null hypothesis framework ensures statistical rigor through power analysis ($\alpha=0.05$, power=80\%) with controlled ground truth measurement.

\subsection{Query Generation}

Query generation combines multiple memory pattern types through LLM-assisted templates that incorporate temporal, activity, and semantic dimensions. For example, templates like ``{temporal context} + {activity} + {file type}'' yield naturalistic queries such as ``last Tuesday + editing + spreadsheet.'' The LLM generates variations maintaining structural properties while varying specific details, creating complex memory-aligned queries for comprehensive evaluation. The Jaro-Winkler distance metric ensures query diversity by filtering similar patterns.

A representative example of a generated query is: \textit{``I was listening to that new Fela Kuti album while researching African elephant migration patterns for our conservation project with Dr. Nzinga Mbande - I think I saved some MP3s and a spreadsheet in our shared Serengeti folder sometime during the rainy season.''} This rich query exemplifies multiple memory dimensions: file types, individuals, temporal context, and semantic metadata.

Natural language queries are converted to database operations through progressive filtering: base queries (all documents), semantic filters (MIME types), and activity filters (temporal patterns). The ablation targets episodic patterns to quantify their contribution to retrieval precision. Complete LLM prompt templates and AQL query examples are provided in \autoref{app:llm-query-generation}.

\subsection{Dataset Generation}

The synthetic dataset framework combines ground truth documents (matching all memory patterns) with documents representing systematic filter combinations (each violating specific memory patterns). This controlled environment enables precise measurement of memory pattern contributions to retrieval effectiveness by testing discrimination against all possible filter failure patterns.

The ablation protocol executes $Q$ queries (statistically determined for 5\% precision detection at 80\% power):

\begin{enumerate}
  \item\label{ablation:design:generate} Generate memory-pattern query combining template structure with episodic memory dimensions.
  \item Create synthetic dataset: $T$ ground truth documents (all patterns match) and $N$ documents representing systematic filter combinations (specific pattern violations).
  \item Reset the database to ensure no prior queries affect results.
  \item \label{ablation:single:dataset:step} Insert the synthetic dataset into the database.
  \item\label{ablation:design:baseline} Execute full memory-pattern query (storage + semantic + episodic) to establish baseline precision $P_0=1.00$ and recall $R_0=1.00$.
  \item We iterate steps \ref{ablation:design:generate} to \ref{ablation:design:baseline} for $Q$ queries, ensuring each query is independent.
  \item We reset the database.
  \item We then load \emph{all} of the data for all of the constructed queries into the database; this is unlike \ref{ablation:single:dataset:step} where we only loaded the data for a single query.
  \item For each query, we run the full-filter query to compute the aggregate precision $P_{q_i}$~\footnote{while generally it will be 1.0, we admit the possibility that there may be some overlap in the queries that leads to a lower precision.} and confirm that recall $R_{q_i}=1.00$.
  \item For each query:
    \begin{itemize}
      \item Execute the query against the same synthetic dataset,
      \item Compute precision, recall, and F1, and
      \item Log results to JSONL for downstream analysis.
    \end{itemize}
\end{enumerate}

\subsection{Determining Required Query Count (Sample Size)}

We calculate the number of queries needed per ablation scenario using a statistical power analysis. Given a significance level $\alpha$, desired statistical power $1 - \beta$, and a target effect size $d$, the required sample size per scenario is approximated by:

\[
    n = 2 \left( \frac{Z_{\alpha/2} + Z_{\beta}}{d} \right)^2
\]

where:

\begin{itemize}
  \item $n$ = required number of queries per ablation scenario,
  \item $Z_{\alpha/2}$ = critical Z-value for the significance level (e.g., $Z_{0.025} \approx 1.96$ for $\alpha = 0.05$),
  \item $Z_{\beta}$ = Z-value corresponding to the desired power (e.g., $Z_{0.2} \approx 0.84$ for power $0.80$),
  \item $d$ = target effect size (minimum detectable difference).
\end{itemize}

For example, with $\alpha = 0.05$, power = $0.80$, and target effect size $d = 0.1$, we have:

\[
    n = 2 \left( \frac{1.96 + 0.84}{0.1} \right)^2 \approx 1,568
\]

Thus, each ablation scenario requires approximately 1,568 queries to reliably detect an effect size of 0.1 at standard statistical significance and power levels.
Statistical analysis assumes normal precision distribution. Mean precision degradation across queries quantifies memory pattern contribution to retrieval effectiveness.

\subsection{Ablation Configuration}\label{sec:ablation-config}

The ablation framework efficiently evaluates memory pattern contributions to retrieval precision across multiple experimental configurations.

We conducted ablation studies using two complementary configurations:

\textbf{Configuration 1} (Control/Test Group Design): This configuration tests filter independence by separating six memory anchor types into two groups:

\textbf{Control Group}: Uses a fixed minimal filter set [``collaboration'', ``location''] providing a stable baseline. These control filters remain active across all experiments to ensure that systematic filter removal in the test group produces measurable effects due to the removed filters rather than experimental variance.

\textbf{Test Group}: Uses the full filter set [``music'', ``social'', ``storage'', ``task''] and employs systematic testing of all $2^4 = 16$ combinations of removing filters from [music, social, storage, task]. This includes: no filters removed (baseline), each single filter removed (4 conditions), all pairs removed (6 conditions), all triples removed (4 conditions), and all four removed (1 condition).

\textbf{Filter Removal Order}: There is no sequential removal order - this is a complete combinatorial exploration testing all possible combinations of filter removal simultaneously. The experiment evaluates the full power set of filters to understand both individual filter contributions and interaction effects between different filter combinations.

\textbf{Experimental Protocol}: For each of 60 query topics, we run the experiment with three random seeds (42, 43, 44) for statistical replication, conducting 38 repetitions per condition. For each test condition:
\begin{itemize}
    \item $T = 5$ ground truth documents that match all remaining filter criteria,
    \item $N = 16$ documents representing systematic filter combinations (one per possible combination: $2^4 = 16$), and
    \item Total: 21 documents per query, providing baseline precision of 23.8\%.
\end{itemize}
This design isolates the contribution of each memory anchor type by comparing test group performance (with filters systematically removed) against the stable control group baseline. Statistical parameters: $\alpha=0.05$, power=0.80, effect size $d=0.1$.

\textbf{Configuration 2} (No Controls): All six memory pattern types (COLLABORATION, LOCATION, MUSIC, SOCIAL, STORAGE, TASK) evaluated simultaneously without controls to examine pattern interactions. Dataset parameters:
\begin{itemize}
    \item $T = 5$ ground truth documents per query,
    \item $N = 64$ documents representing systematic filter combinations (one per possible combination: $2^6 = 64$), and
    \item Total: 69 documents per query, providing baseline precision of 7.2\%.
\end{itemize}
Enhanced statistical rigor: $\alpha=0.01$, power=0.99, effect size $d=0.01$.

\textbf{Key Difference}: Configuration 1 tests each filter type in isolation with dedicated control groups, while Configuration 2 evaluates all filters simultaneously to capture interaction effects. The increased filter combination complexity in Configuration 2 (64 vs 16 systematic combinations) creates a more challenging retrieval environment, testing robustness of memory patterns against diverse filter failure patterns.

Complete configuration details including exemplar queries and statistical parameters are provided in \autoref{app:ablation-results}.

\subsection{Results and Discussion}
\label{sec:ablation-results}

The ablation systematically evaluates episodic memory pattern contributions to retrieval precision. Memory pattern filters are removed individually and in combination to quantify their impact.

\textbf{Results Summary}: \autoref{tab:ablation_summary_config1} and \autoref{tab:six_way_ablation_summary} present our ablation results for Configurations 1 and 2 respectively (see \autoref{sec:ablation-config} for detailed configurations).

\begin{table}[htbp]
\centering
\caption[Baseline Precision Values]{Baseline Precision Values for Ablation Configurations}
\label{tab:ablation-baseline}
\resizebox{0.95\textwidth}{!}{%
  \begin{tabular}{p{0.3\textwidth}p{0.15\textwidth}p{0.15\textwidth}p{0.15\textwidth}p{0.15\textwidth}}
  \toprule
  \textbf{Configuration} & \textbf{Ground Truth} & \textbf{Total Docs} & \textbf{Baseline Precision} & \textbf{Baseline Recall} \\
  \midrule
  Configuration 1 & 10 & 100 & 1.00 & 1.00 \\
  Configuration 2 & 10 & 550 & 1.00 & 1.00 \\
  \bottomrule
  \end{tabular}
}%
\vspace{1mm}
\parbox{0.95\textwidth}{\footnotesize\textbf{Note}: Baseline values represent query performance with all memory anchor filters active. Despite different dataset sizes, both configurations achieve perfect precision by leveraging all available memory patterns to filter against systematic filter combinations.}
\end{table}

\textbf{Reading the Tables}: Each row shows what happens when specific memory anchor filters are removed from queries:
\begin{description}
    \descitem{Filters Removed}{Which UPI memory anchor types were ablated,}
    \descitem{Effect Size}{Precision degradation (negative values = worse precision),}
    \descitem{Confidence Interval}{Statistical uncertainty range, and}
    \descitem{Significant}{Whether the degradation is statistically meaningful.}
\end{description}

\textbf{Key Pattern}: All single-filter removals show significant negative effects, confirming each memory anchor type contributes meaningfully to retrieval precision. Multiple filter removal shows compounding degradation, validating the UPI's integrated design.

\textbf{Understanding Effect Size}: Effect size quantifies the practical magnitude of precision change when memory metadata is removed, measuring the UPI architecture's contribution to retrieval effectiveness. All effect sizes are calculated relative to the baseline configuration where all filters are present.

\textbf{Interpretation}: An effect size of -0.40 means precision dropped by 40 percentage points relative to baseline (e.g., from finding 10/10 correct documents with all filters to finding only 6/10 with that filter removed). This represents the architectural contribution of that specific memory anchor type to overall retrieval precision.

\textbf{Baseline Reference}: All precision changes in the tables are measured against the baseline configuration that includes all memory anchor filters. For Configuration 1, baseline precision = 1.00 (finds all 10 ground truth documents among 100 total). For Configuration 2, baseline precision = 1.00 for individual queries despite the larger set of systematic filter combinations.

\textbf{Clinical Significance}: Effect sizes of -0.33 to -0.45 for single filter removal indicate that each memory anchor type contributes substantially to retrieval precision. These are large practical effects, confirming that memory anchors are not marginal improvements but fundamental architectural components.

\textbf{Cumulative Impact}: When multiple memory anchor types are removed simultaneously, effect sizes compound (reaching -0.85+), demonstrating that the UPI's integrated approach provides synergistic benefits beyond individual components.\textbf{Configuration 1 Results}: All single-filter ablations show significant precision degradation (effect sizes: -0.33 to -0.44), with 100\% statistical significance at $p<0.05$. Multi-filter ablations compound these effects, reaching -0.83 when four filters are removed.

\textbf{Configuration 2 Results}: Single-filter ablations show more variable effects (some zero, others -0.167), with only 3 of 6 filters showing significance. However, multi-filter ablations consistently degrade precision, reaching -0.58 for complete ablation.

\textbf{Interpretation}: Configuration 1's consistent effects reflect controlled filter-query alignment (queries specifically designed to test particular filter types), while Configuration 2's variability demonstrates the importance of contextual relevance in memory anchor application.

Results reject the null hypothesis: removing episodic memory pattern filters significantly degrades retrieval precision. The data demonstrates critical importance of memory-aligned metadata for effective information retrieval.

With statistical confirmation of memory pattern importance, we examine interaction effects between multiple memory dimensions.

\begin{figure}[!htbp]
  \caption[Configuration 1 Ablation Results]{Ablation study results comparing Configuration 1 (T=5, N=16) and Configuration 2 (T=5, N=64). Configuration 1 (blue) demonstrates steep precision degradation as filters are removed, validating the methodology. Configuration 2 (orange) shows more robust precision retention despite higher filter combination complexity (4x more systematic filter combinations), but still exhibits significant degradation with multiple filter removal. Error bars represent standard deviation across experimental runs.}
  \label{fig:ablation-combined}
  \centering
  \includegraphics[width=0.95\textwidth]{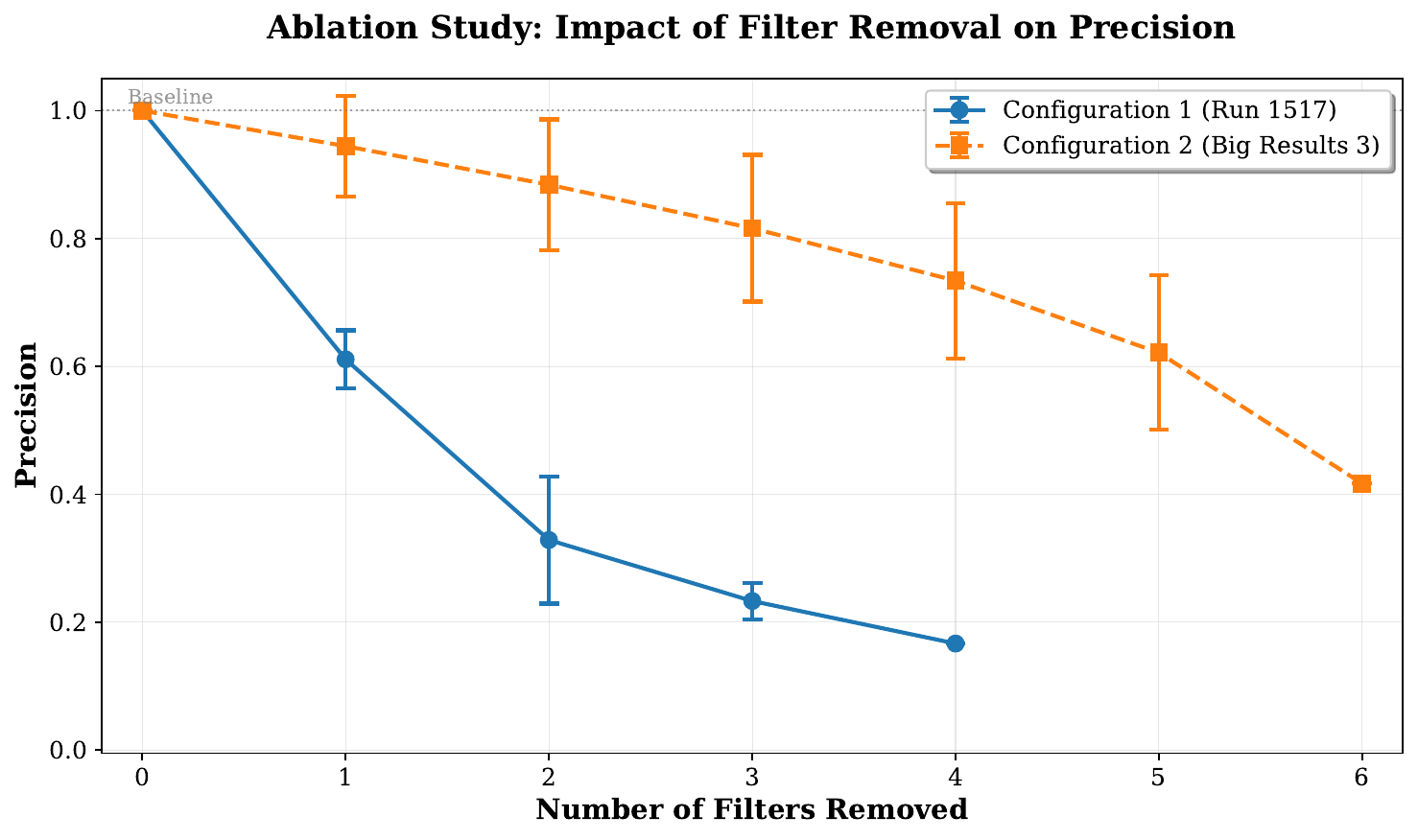}
\end{figure}


\begin{figure}[!htbp]
  \caption[Configuration 1 Violin Plot]{Violin plot showing precision loss distribution in Configuration 1 grouped by number of filters removed. The widening distributions as more filters are ablated reveal increasing variability in precision degradation. Single filter removal shows consistent moderate impact (-0.35 to -0.45), while removing 4+ filters produces both more severe and more variable precision loss, indicating complex interaction effects.}
  \label{fig:config1:violin}
  \centering
  \includegraphics[width=0.95\textwidth]{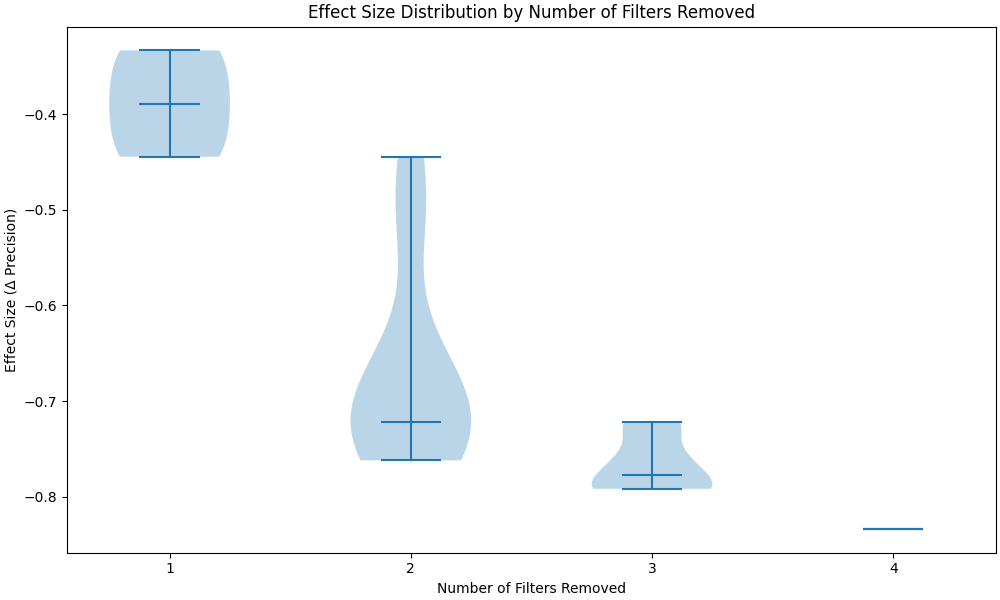}
\end{figure}

The ablation results reveal several key insights, as shown in \autoref{fig:ablation-combined}. Configuration 1 demonstrates steep precision degradation as filters are removed, with precision dropping from 1.0 to approximately 0.17 when four filters are ablated. Configuration 2 shows more gradual degradation but still substantial precision loss, dropping to 0.42 when all six filters are removed. Individual configuration plots and additional visualizations are provided in \autoref{app:ablation-supplemental}. The interaction heatmap (see \url{https://github.com/fsgeek/indaleko/blob/main/figures/analysis_plots_heatmap_config_1.png}) reveals that certain filter pairs (Music-Location and Social-Collaboration) exhibit strong synergistic effects, while others contribute more independently. \autoref{fig:config1:violin} illustrates how precision loss becomes both more severe and more variable as additional filters are removed, with single filter ablation showing consistent impacts around -0.39 (range: -0.33 to -0.45).

This data and the plots are consistent with what we saw in \autoref{sec:eval-expressiveness-comparison}: without effective ways to filter the storage data, the results returned to the user are not the ``needles'' they are looking for in the haystack, it is mostly hay, obscuring the goal of the search.

Configuration 2 eliminates controls to examine filter interaction effects. Increased filter diversity means individual filters may not match query context, requiring intelligent contextual selection.

\begin{figure}[!htbp]
  \caption[Configuration 2 Violin Plot]{Violin plot for Configuration 2 showing precision loss distribution without control groups. The dramatic contrast between single filter removal (median near 0, minimal impact) and multi-filter ablation reveals the critical importance of filter relevance. This configuration's increased filter diversity means individual filters may not match query context, but combining multiple filters produces significant degradation (-0.35 to -0.60), confirming that contextually appropriate filter selection is essential.}
  \label{fig:config2:violin}
  \centering
  \includegraphics[width=0.95\textwidth]{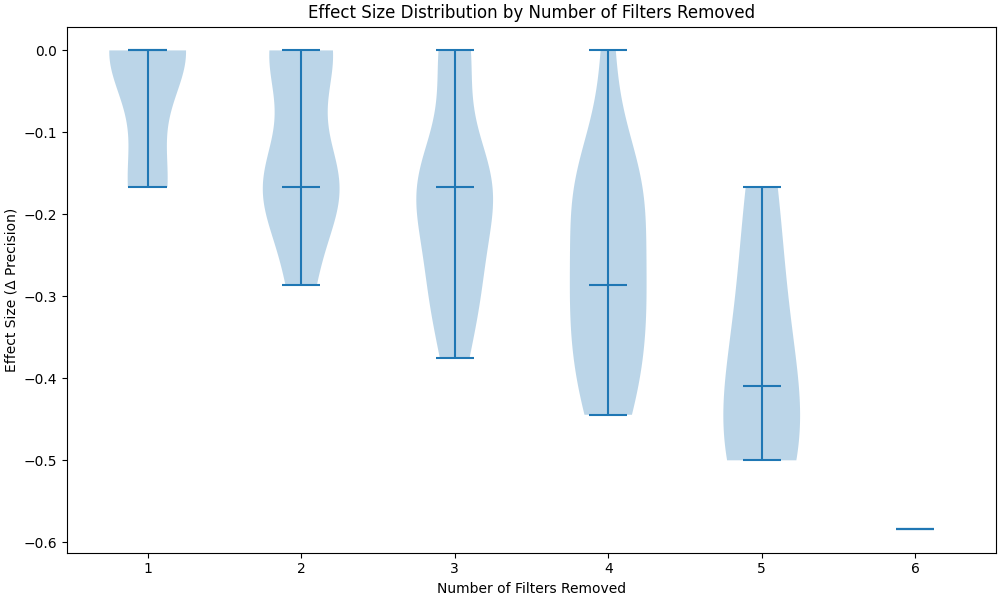}
\end{figure}

\begin{figure}[!htbp]
  \caption[Configuration 2 Cumulative Plot]{Cumulative precision degradation in Configuration 2 showing the effect of progressive filter removal without control groups. Unlike Configuration 1's linear decline, this shows a more gradual initial impact that accelerates when 3+ filters are removed. The S-curve pattern suggests that filter relevance varies: initial removals may eliminate less relevant filters, while later removals impact core memory patterns essential for retrieval.}
  \label{fig:config2:cumulative}
  \centering
  \includegraphics[width=0.95\textwidth]{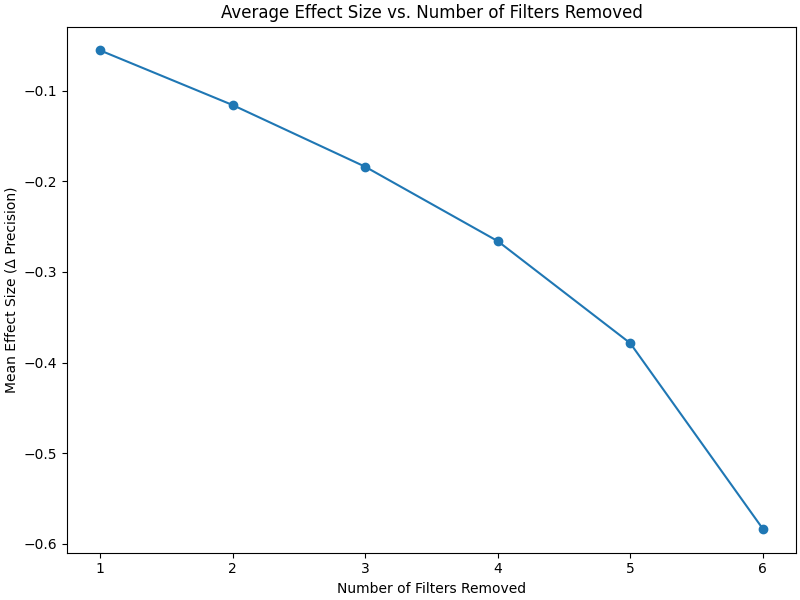}
\end{figure}


Configuration 2 results highlight the importance of contextually appropriate filter selection. Individual filter removal shows minimal impact (see the interaction heatmap at \url{https://github.com/fsgeek/indaleko/blob/main/figures/analysis_plots_heatmap_config_2.png}) reveals that certain filter pairs (Music-Location and Social-Collaboration) exhibit strong synergistic effects, while others contribute more independently, but multi-filter ablation produces substantial degradation (-0.35 to -0.60), revealing that filter diversity requires intelligent selection. The stepped degradation pattern suggests initial removals eliminate less relevant filters while later removals impact essential memory patterns.

Key insight: pattern-query alignment is crucial. Spatial patterns provide minimal benefit for temporal queries, but comprehensive pattern application maximizes effectiveness when contextually appropriate. Real-world memory distributions likely exhibit non-uniform clustering around significant events, suggesting our model provides conservative estimates of memory pattern contributions.


Alternative experimental design: constructing adversarial filter combinations through systematic memory pattern violation. For each ground truth document, generate $2^n-1$ variants violating specific memory patterns. Example: temporal anchors post-1980 in truth documents, pre-1970 in variants. This design would expose $2^6-1=63$ filter combination variants per truth document when ablating all six memory dimensions, yielding precision $\approx0.016$. Single pattern ablation would produce 50\% precision degradation, representing worst-case memory pattern importance.

Actual memory pattern effectiveness likely falls between our systematic filter combination model and adversarial design (50-98.4\% precision loss). This range aligns with observed behavior in existing systems: without memory-aligned filtering, retrieval systems produce excessive noise, confirming the critical role of episodic memory patterns in effective information retrieval.

\textbf{Why Two Configurations}: We employed two experimental configurations to comprehensively evaluate memory anchor effectiveness. Configuration 1 (T=5 ground truth, N=16 systematic filter combinations) uses a control group design to validate our methodology by ensuring precision changes result from filter removal rather than experimental variance, and demonstrate consistent effect sizes across filter types. Configuration 2 (T=5 ground truth, N=64 systematic filter combinations) explores filter interactions and realistic retrieval scenarios with higher filter complexity. Together, these configurations confirm that memory anchors provide robust precision improvements across different experimental conditions.

\textbf{Filter Selection Methodology}: Configuration 1 employed 2 control filters (always applied) and systematically ablated combinations of the 4 test filters, allowing us to isolate the contribution of specific memory anchor types. Configuration 2 used random sampling from all 6 filters without controls, providing insight into how filters interact in more realistic retrieval scenarios where all memory dimensions are potentially relevant.

\begin{table}[!tbph]
  \caption[Configuration 1 Ablation Study Summary]{Ablation Study Summary (Configuration 1, see \autoref{sec:ablation-config})}
  \label{tab:ablation_summary_config1}
  \centering
  \resizebox{0.9\textwidth}{!}{%
    \begin{tabular}{lp{0.2\textwidth}ccc}
      \toprule
      \textbf{Number} & \textbf{Filters} & \textbf{Effect} & \textbf{Confidence} & \textbf{Significant} \\
      \textbf{Filters} & \textbf{Removed} & \textbf{Size} & \textbf{Interval} & \textbf{($p<0.01$)} \\
      \textbf{Removed} & & & 95\% & \\
      \midrule
      1 & Collaboration & -0.444 & [-0.461, -0.428] & Yes \\
      1 & Location & -0.333 & [-0.353, -0.314] & Yes \\
      1 & Music & -0.389 & [-0.407, -0.371] & Yes \\
      1 & Social & -0.389 & [-0.407, -0.371] & Yes \\
      1 & Storage & -0.333 & [-0.353, -0.314] & Yes \\
      1 & Task & -0.444 & [-0.461, -0.428] & Yes \\[6pt]

      2 & Collaboration, Location & -0.524 & [-0.538, -0.510] & Yes \\
      2 & Collaboration, Music & -0.722 & [-0.738, -0.706] & Yes \\
      2 & Collaboration, Social & -0.762 & [-0.776, -0.748] & Yes \\
      2 & Location, Music & -0.667 & [-0.686, -0.647] & Yes \\
      2 & Social, Storage & -0.667 & [-0.686, -0.647] & Yes \\
      2 & Storage, Task & -0.444 & [-0.461, -0.428] & Yes \\[6pt]

      3 & Collaboration, Location, Music & -0.792 & [-0.804, -0.780] & Yes \\
      3 & Collaboration, Location, Social & -0.792 & [-0.804, -0.780] & Yes \\
      3 & Collaboration, Location, Storage & -0.792 & [-0.804, -0.780] & Yes \\
      3 & Music, Social, Storage & -0.792 & [-0.804, -0.780] & Yes \\[6pt]

      4 & Collaboration, Location, Music, Social & -0.833 & [-0.843, -0.824] & Yes \\
      4 & Collaboration, Location, Storage, Task & -0.833 & [-0.843, -0.824] & Yes \\
      4 & Music, Social, Storage, Task & -0.833 & [-0.843, -0.824] & Yes \\

      \bottomrule
    \end{tabular}%
  }%
\end{table}

\begin{table}[!tbph]
  \caption[Configuration 2 Ablation Study Summary]{Ablation Study Summary (Configuration 2, see \autoref{sec:ablation-config})}
  \label{tab:six_way_ablation_summary}
  \centering
  \resizebox{0.95\textwidth}{!}{%
    \begin{tabular}{p{0.2\textwidth}p{0.2\textwidth}ccc}
      \toprule
      \textbf{Num.}    & \textbf{Filters} & \textbf{Effect} & \textbf{Confidence} & \textbf{Significant} \\
      \textbf{Filters} & \textbf{Removed} & \textbf{Size}   & \textbf{Interval}   & \textbf{($p<0.01$)} \\
      \textbf{Removed} &                  &                 & \textbf{(95\%)}     &  \\
      \midrule
      1 & Collaboration & 0.000 & [0.000, 0.000] & No \\
      1 & Location & -0.167 & [-0.167, -0.167] & Yes \\
      1 & Music & -0.167 & [-0.167, -0.167] & Yes \\
      1 & Social & 0.000 & [0.000, 0.000] & No \\
      1 & Storage & 0.000 & [0.000, 0.000] & No \\
      1 & Task & 0.000 & [0.000, 0.000] & No \\[6pt]

      2 & Collaboration, Location & -0.167 & [-0.167, -0.167] & Yes \\
      2 & Location, Music & -0.286 & [-0.286, -0.286] & Yes \\
      2 & Music, Social & -0.286 & [-0.286, -0.286] & Yes \\[6pt]

      3 & Collaboration, Location, Music & -0.286 & [-0.286, -0.286] & Yes \\
      3 & Location, Music, Social & -0.375 & [-0.375, -0.375] & Yes \\
      3 & Music, Social, Storage & -0.375 & [-0.375, -0.375] & Yes \\[6pt]

      4 & Collaboration, Location, Music, Social & -0.375 & [-0.375, -0.375] & Yes \\
      4 & Location, Music, Social, Storage & -0.444 & [-0.444, -0.444] & Yes \\[6pt]

      5 & Collaboration, Music, Social, Storage, Task & -0.500 & [-0.500, -0.500] & Yes \\
      5 & Location, Music, Social, Storage, Task & -0.500 & [-0.500, -0.500] & Yes \\[6pt]

      6 & Collaboration, Location, Music, Social, Storage, Task & -0.583 & [-0.583, -0.583] & Yes \\

      \bottomrule
    \end{tabular}%
  }%
\end{table}

The ablation study quantitatively confirms that episodic memory patterns are essential for effective retrieval. Progressive removal of memory pattern types produces cumulative precision degradation, validating the UPI's memory-aligned architecture. Despite synthetic labeling, results demonstrate the fundamental importance of integrated memory pattern processing. The UPI's comprehensive memory metadata framework emerges as architecturally critical for memory-aligned information retrieval.

\section{Extensibility}
\label{sec:eval-extensibility}

The UPI relies upon collecting a diverse set of metadata from multiple sources. Thus, to increase the effectiveness of the UPI and \system, we need to ensure that adding new sources of metadata to the system is straightforward and easy to do.

To demonstrate the UPI framework's extensibility, we conducted a progression of metadata provider integrations that reveal the framework's maturation and ease of extension. Our experience follows a natural progression: I first implemented the Windows GPS provider in approximately 10 hours, establishing the base framework and design patterns. Having documented this structure, an undergraduate research assistant was then able to implement a Spotify listening history provider in just 2.5 hours. Finally, having observed these implementations and learned the patterns, we were able to ``teach'' an AI coding assistant to implement a YouTube provider in merely 10 minutes. This dramatic reduction in development time, from 10 hours to 10 minutes, demonstrates both the framework's extensibility and the power of well-documented patterns. Table~\ref{tab:extensibility} summarizes these extension metrics.

\begin{table}[htbp]
  \caption[Extension Effort Metrics]{Extension Effort Metrics for New Metadata Providers (in chronological order)\protect\footnotemark}
  \label{tab:extensibility}
  \centering
    \resizebox{0.97\textwidth}{!}{%
  \begin{tabular}{llll}
  	\toprule
  	\textbf{Provider} & \textbf{Developer} & \textbf{Time to Integrate} & \textbf{Adapter Code (LOC)} \\
  \midrule
  Windows GPS & Author & 10 hours & 482 \\
  Spotify & Undergraduate RA & 2.5 hours & 943 \\
  YouTube & AI Coder & 10 minutes & 1222 \\
  \bottomrule
  \end{tabular}%
    }
\end{table}
\footnotetext{Times are approximate development hours; LOC refers only to framework adapter code, not data source-specific libraries. The progression shows how framework maturity and documentation reduced integration time from 10 hours to 10 minutes.}

The initial Windows GPS provider required 10 hours because it involved establishing the common base class and design patterns that would be reused by all subsequent providers. This foundational work paid immediate dividends: the undergraduate researcher could leverage these patterns and examples to implement the Spotify provider in a fraction of the time. The AI coder's near-instantaneous implementation further validates that the framework's patterns are both clear and reusable.
This rapid creation of new providers underscores the modularity and reusability of the UPI design, validating the ``basic extensibility'' objective.

One notable exception to this are the storage data activity stream providers.  This is generally due to a rather different challenge: the ``firehose'' issue.  File system change data can arrive at a tremendous rate, with fine-grained event notifications.  We implemented (but did not integrate) an activity stream provider for the Mac, which took several months of effort to throttle down to a granularity that was acceptable.  Cloud services were easier to monitor but require authentication and (typically) a public facing web service interface (a \emph{webhook}) to be able to receive the data.

Windows has a number of different ways to monitor file system activity; for our implementation we used the NTFS USN Journal, which is enabled by default on the Windows system volume. Originally created for the Windows File Replication Service, it provides a curated set of file system changes. During a typical 24 hour period we would see between 5,000 and 6,000 changes, which were captured and stored.  We used an ArangoDB ttl-index to index data for up to four days.  Using AI coding agents, we built and did preliminary analysis of a tiered system: the initial (``hot'') tier captured everything, but only allowed storage for a short time (target was one week) before it was deleted.  A second (``warm'') tier consolidated information from the hot tier using a weighted scoring system that aligned with human memory models: as time passes, the importance of granularity decreases.  A subsequent review of the initial implementation led to a replacement implementation for the hot tier data recorder, which was accomplished by a human coder in less than 2 hours.

The complexity of the storage activity stream providers is inherent in the nature of the data sources; \system does not make the underlying problem less challenging but did not negatively impact its complexity either.

Thus, overall, the UPI's extensibility is a key strength, allowing for rapid integration of new metadata sources and enabling the system to adapt to evolving user needs and data environments, albeit with the caveat that neither the UPI nor \system can guarantee collection from all metadata sources is straightforward, but once collected and normalized it becomes available for addressing future queries. This extensibility is crucial for maintaining the UPI's relevance and effectiveness in personal data retrieval.

\section{Synthesis of Findings}

In this evaluation we have addressed our original research questions (see \autoref{sec:eval-framework}) and demonstrated the effectiveness of the UPI architecture and its implementation in \system{} for addressing the challenges of personal data retrieval.  We have shown that the UPI architecture provides a unified framework for memory-aligned retrieval across diverse platforms, and that \system{}'s implementation validates these principles by resolving complex queries that are not possible with existing storage systems.

\subsection[RQ1: UPI Architecture Expressiveness]{RQ1: UPI Architecture Expressiveness (\autoref{sec:eval-comparative})}

The comparative evaluation demonstrates that the UPI architecture addresses fundamental limitations in existing storage systems. While mature systems like Windows Search, Google Drive, and OneDrive cannot express or process memory-aligned queries, the UPI's architectural principles (unified metadata, memory anchor integration) enable comprehensive memory-based retrieval.

\textbf{Key Finding}: Architectural differences, not optimization levels, determine memory-aligned retrieval capabilities. The UPI's design principles solve problems that existing architectures cannot address even with decades of refinement.

\subsection[RQ2: Implementation Performance Characteristics]{RQ2: Implementation Performance Characteristics (\autoref{sec:implementation-efficacy})}

\system{}'s implementation validates that UPI architectural benefits can be realized with practical performance characteristics. On a 31M file dataset, query response times remain interactive (0.6-2.3s for most queries), resource usage stays minimal during idle periods, and the system scales effectively to realistic personal data volumes.

\textbf{Key Finding}: The UPI architecture can be implemented efficiently, making memory-aligned retrieval practical for personal deployment on consumer hardware.

\subsection[RQ3: Memory Anchor Effectiveness]{RQ3: Memory Anchor Effectiveness (\autoref{sec:eval-ablation})}

The ablation study quantifies that each memory anchor category contributes substantially to retrieval precision (effect sizes: -0.33 to -0.44 for single removals). Multiple anchor removal shows compounding degradation, confirming that the UPI's integrated approach provides synergistic benefits.

\textbf{Key Finding}: Memory anchors provide large practical effects, not marginal improvements. The UPI's architectural integration of multiple memory dimensions creates measurable precision advantages.

\subsection[RQ4: Extensibility Characteristics]{RQ4: Extensibility Characteristics (\autoref{sec:eval-extensibility})}

The framework demonstrates rapid metadata source integration, with development times decreasing from 10 hours (initial Windows GPS provider) to 10 minutes (AI-assisted YouTube provider) as patterns matured.

\textbf{Key Finding}: The UPI's modular design enables efficient metadata source integration, supporting evolving personal data environments.

\subsection{Conclusions}

This evaluation demonstrates two critical insights: \textbf{memory anchors provide immediate, measurable benefits using existing data}, while \textbf{activity streams enable continuous improvement as richer metadata accumulates}.

Our results establish a two-stage value proposition:

\begin{description}
    \descitem{Immediate Value (Memory Anchors)}{Work with decades of existing personal data, providing precision improvements over traditional search without requiring new metadata collection. The ablation study quantifies these benefits as even simple temporal anchors improve retrieval.}

    \descitem{Future Enhancement (Activity Streams)}{The synthetic metadata evaluation demonstrates how richer activity stream, such as location tags, application context, and social interactions, further improves retrieval. While such metadata requires time to accumulate naturally, the architecture supports its integration as it becomes available.}

    \descitem{Continuous Improvement}{Unlike systems that require complete redesign for new capabilities, the UPI architecture allows incremental enhancement. Memory anchors provide the foundation today; activity streams build upon that foundation tomorrow.}
\end{description}

This staged approach addresses the cold-start problem plaguing many innovative systems: users receive immediate benefits from memory anchors while the system gathers richer activity streams for future enhancement. The UPI succeeds not by promising a better future, but by delivering better retrieval today while systematically building toward that future.

These empirical results validate the UPI's core claims, distinguishing it from prior work examined in Chapter 7.

\subsection{Limitations}
This work has several limitations, notably:

\begin{itemize}
    \item Reliance on a single personal dataset, limiting the generality of that aspect of the evaluation.
    \item Synthetic data cannot capture all real-world complexity.
    \item Evaluation is a snapshot, not longitudinal.
    \item Resource costs of local processing as use of the system increases.
    \item Difficulty of measuring recall comprehensively on personal data.
\end{itemize}

These limitations are important to consider when interpreting the results of this evaluation.  The work we have done is a first step in demonstrating the potential of the UPI and \system, but there is much more work to be done to fully realize that potential.

\addkhipuifneeded

\chapter{Related Work}\label{ch:related-work}

\begin{epigraph}
\textit{``The true method of knowledge is experiment.''}
\par\vspace{0.5em}
\mbox{}\hfill\textsc{William Blake}
\end{epigraph}

The gap analysis table in \autoref{tab:gap-analysis} organizes this chapter's analysis, identifying how the UPI addresses fundamental limitations across related work areas. Each subsequent section examines these systems' core innovations and explains how the UPI builds upon their contributions while addressing their key constraints.

\begin{table}[!tbph]
    \caption[Related Work Gap Analysis]{Summary Gap Analysis of Related Work}
    \label{tab:gap-analysis}
    \centering
    \resizebox{0.85\textwidth}{!}{%
    \begin{tabular}{>{\raggedright\arraybackslash}p{3cm}>{\raggedright\arraybackslash}p{4cm}>{\raggedright\arraybackslash}p{4cm}>{\raggedright\arraybackslash}p{4cm}}
    	\toprule
    	\textbf{Prior Work} & \textbf{Limitations and Gaps} & \textbf{UPI Contributions} & \textbf{Opportunities (Future Work)} \\
    \midrule
    Contextual Memory Retrieval (Fuller et al.~\cite{fuller2008contextual}, Chen \& Jones~\cite{Kalnikaite2007}) & Experimental; lacks broad integration across diverse sources; often confined to specific domains. & Comprehensive memory pattern implementation across heterogeneous sources; deployment-ready architecture leveraging memory-aligned cues (time, place, context). & Technical evaluation of memory model implementation; Real-world deployment scenarios. \\
    Human Digital Memory \& Lifelogging (MyLifeBits~\cite{gemmell2006mylifebits}, Lifebrowser~\cite{sellen2010beyond}) & Closed ecosystems; limited architectural control; capture-centric rather than retrieval-focused. & Memory-aligned retrieval architecture; locally isolated indexing; episodic memory pattern implementation rather than continuous capture. & Technical validation of memory retrieval mechanisms across diverse contexts. \\
    Intelligent Retrieval Systems (Remembrance Agent~\cite{rhodes1996remembrance}, Stuff I've Seen~\cite{dumais2003stuff}, Teevan et al.~\cite{teevan2005personalizing}) & Narrow focus (single data types, limited generalization); minimal cross-silo capabilities. & Generalized metadata model; query-driven indexing supporting multiple interface types including LLM integration; cross-silo by design. & Integration with external knowledge bases; adaptive AI interactions. \\
    Tagging Systems (TagFS~\cite{bloehdorn2006tagfs}, Golder \& Huberman~\cite{golder2006usage}) &
    Manual tagging imposes high cognitive load; inconsistent usage; limited activity or episodic modeling; poor cross-silo integration. &
    Federated metadata model supports tag ingestion and normalization; tags treated as one of many cues in episodic context; multiple query interface options including LLM-powered disambiguation. &
    Explore tag clustering via concept lattices; integrate user-defined taxonomies with learned metadata schemas. \\
    Provenance Systems (PASS~\cite{muniswamy2006provenance}, Provenance search~\cite{provsearch}) & Provenance-focused; lacks memory-aligned retrieval patterns; system-oriented rather than episodic memory-oriented. & Provenance integrated with memory-aligned metadata and episodic cues; UUID-based privacy architecture. & Advanced integration of provenance data with semantic metadata. \\
    Activity-Based Computing (TaskTracer~\cite{stumpf2005tasktracker}, UMEA~\cite{kaptelinin2003umea}, Burrito~\cite{guo2012burrito}) & Task-specific; limited integration with broader information landscape; fragmented activity stream. & Comprehensive Memory Anchor\index{memory anchors} model; integration of activity stream with storage and semantic metadata. & Further exploration of context-driven semantic enrichment methods. \\
    Cross-Silo \& Episodic Retrieval (Elsweiler \& Ruthven~\cite{elsweiler2007towards}, Dumais et al.~\cite{dumais2016stuff}) & Acknowledges episodic memory as key to user retrieval and recognizes the need for search across personal data silos. & Explicitly models episodic context and builds normalized metadata across silos into core architecture. & Explore integration with knowledge graphs and multiple query interface types (including local LLMs) for intent disambiguation and natural query expression. \\
    \bottomrule
    \end{tabular}
    }%
\end{table}

\section{Introduction}\label{sec:related-work:introduction}

This chapter analyzes how the UPI builds upon and extends prior research in personal information retrieval. The critical gap across existing work is the failure to systematically implement episodic memory patterns such as the temporal, spatial, and contextual cues that drive human recall, within technical architectures. While individual systems validated components of memory-aligned retrieval, none achieved comprehensive cross-platform integration of these patterns.

The UPI addresses this gap through architectural synthesis rather than interface improvements, operationalizing validated memory concepts into a deployment-ready system that bridges decades of fragmented research contributions.

\section{Academic Research Approaches}\label{sec:related-work:academic}

\subsection{Contextual Memory Retrieval}\label{subsec:related-work:contextual}

\subsection{Core Innovation} \citeauthor{vianna2019thesis}~\cite{vianna2019thesis}, \citeauthor{fuller2008contextual}~\cite{fuller2008contextual} and \citeauthor{Kalnikaite2007}~\cite{Kalnikaite2007} validated that temporal landmarks and social associations significantly improve retrieval effectiveness (32\% and 25\% improvements respectively), establishing the theoretical foundation for memory-cued retrieval.

\subsection{Key Limitation} These experimental systems validated individual memory cues but lacked comprehensive integration across heterogeneous sources and remained confined to research domains without scalable architectures.

\subsection{How UPI Builds Upon} The UPI operationalizes these validated concepts through systematic integration of multiple memory dimensions (temporal, spatial, social) in a deployment-ready architecture that captures and normalizes episodic memory cues across platforms, bridging theoretical validation to practical implementation.

\subsection{Human Digital Memory and Lifelogging}\label{subsec:related-work:lifelogging}

\subsection{Core Innovation} MyLifeBits~\cite{gemmell2006mylifebits} and Lifebrowser~\cite{sellen2010beyond} demonstrated comprehensive data capture and temporal visualization capabilities, with Lifebrowser using machine learning to identify significant events as memory landmarks.

\subsection{Key Limitation} These systems emphasized continuous capture over retrieval, creating isolated repositories that required specialized hardware and failed to leverage episodic cues from natural computational activities and thus fundamentally misaligning with episodic memory patterns.

\subsection{How UPI Builds Upon} The UPI shifts from capture-centric to memory-aligned retrieval, implementing episodic memory patterns that leverage temporal context, spatial cues, and activity patterns naturally occurring during information interaction, enabling retrieval based on memory formation rather than comprehensive capture requirements.

\subsection{Intelligent Retrieval Systems}\label{subsec:related-work:intelligent-retrieval}

\subsection{Core Innovation} Remembrance Agent~\cite{rhodes1996remembrance} pioneered context-aware retrieval through activity monitoring, while Stuff I've Seen~\cite{dumais2003stuff} demonstrated that temporal cues significantly improve retrieval effectiveness by unifying multiple data types within platform ecosystems.

\subsection{Key Limitation} These systems remained platform-specific with rigid metadata schemas, preventing cross-silo episodic memory modeling and comprehensive integration of temporal, spatial, and activity cues essential for distributed memory formation.

\subsection{How UPI Builds Upon} The UPI implements episodic memory patterns across platforms through architectural redesign that systematically captures, normalizes, and integrates memory cues from heterogeneous sources, enabling retrieval based on natural memory formation rather than platform-specific metadata constraints.

\subsection{Tagging Systems and Folksonomies}
\label{subsec:related-work:tagging}

\subsection{Core Innovation} TagFS~\cite{bloehdorn2006tagfs} and folksonomy systems~\cite{golder2006usage} provided flexible alternatives to hierarchical organization through user-defined labels and formal concept analysis approaches that enabled multifaceted retrieval without rigid taxonomies.

\subsection{Key Limitation} Manual tagging imposed high cognitive load disconnected from natural memory formation, creating inconsistent usage with vocabulary drift over time, while automated approaches ignored episodic context and remained platform-specific, preventing cross-silo memory pattern capture.

\subsection{How UPI Builds Upon} The UPI integrates tagging-like affordances within a broader episodic architecture that supports cross-silo metadata normalization and natural language query interfaces (including LLM-powered approaches), incorporating tags as one facet of a richer memory model rather than requiring explicit vocabulary maintenance.

\subsection{Provenance Systems}\label{subsec:related-work:provenance}

\subsection{Core Innovation} PASS~\cite{muniswamy2006provenance} demonstrated automatic capture of file creation, modification, and access patterns at the operating system level, proving the value of automatically captured metadata for information management. \citeauthor{provsearch}~\cite{provsearch} extended this to semantic search across provenance data, enabling retrieval based on system-level operations.

\subsection{Key Limitation} Provenance systems focused on system-level tracking rather than user-meaningful contexts, with limited integration of semantic understanding and minimal consideration of human memory patterns, typically operating within single-system boundaries.

\subsection{How UPI Builds Upon} The UPI incorporates provenance concepts through its Memory Anchor\index{memory anchors} model while extending them with semantic metadata integration and memory pattern alignment, using UUID-based privacy architecture for granular provenance control across heterogeneous sources.

\subsection{Activity-Based Computing}\label{subsec:related-work:activity}

\subsection{Core Innovation} TaskTracer~\cite{stumpf2005tasktracker} and UMEA~\cite{kaptelinin2003umea} demonstrated that activity-centered organization aligns better with cognitive models than hierarchical approaches, capturing interaction patterns and task associations that partially align with episodic encoding.

\subsection{Key Limitation} These systems treated tasks as isolated contexts rather than integrated components of broader memory patterns, with single-device focus preventing distributed episodic pattern capture and requiring explicit task definitions rather than natural activity detection.

\subsection{How UPI Builds Upon} The UPI reconceptualizes activity as one dimension of episodic memory, integrating activity patterns with temporal sequences, spatial contexts, and semantic relationships across devices and platforms, enabling retrieval based on complete memory patterns rather than isolated task associations.

\section{Common Limitations Across Prior Work}\label{sec:related-work:limitations}

Analysis of these systems reveals three recurring architectural limitations that the UPI systematically addresses:

\subsection{Cross-Platform Integration Challenges} Most systems remained confined to single platforms or ecosystems, preventing the comprehensive episodic memory modeling essential for modern distributed digital lives. Whether experimental prototypes or commercial systems, prior work failed to develop scalable architectures for cross-silo memory pattern integration.

\subsection{Memory Pattern Implementation Gaps} While individual systems validated specific memory cues (temporal landmarks, social associations, activity patterns), none achieved systematic integration of multiple memory dimensions. The persistent gap between theoretical understanding of episodic memory and technical implementation prevented deployment-ready memory-aligned systems.

\subsection{Architectural Inflexibility} Rigid metadata schemas and platform-specific implementations couldn't accommodate the diverse memory pattern types essential for comprehensive personal information retrieval, limiting systems to narrow domains or specific data types.

\section{Summary}\label{sec:related-work:summary}

The gap analysis reveals a consistent pattern across prior work: while individual systems validated components of memory-aligned retrieval, none achieved comprehensive cross-platform integration of episodic memory patterns. The persistent mismatch between human memory processes and system architectures created fragmented solutions that validated concepts without systematic implementation.

The UPI addresses this gap through architectural synthesis rather than incremental improvements, operationalizing validated memory concepts into a deployment-ready system that integrates temporal sequences from lifelogging research, contextual cues from intelligent retrieval systems, and activity patterns from task-based computing. This systematic implementation bridges decades of fragmented research contributions, as demonstrated through practical evaluation in \autoref{ch:evaluation}.


\chapter{Conclusion and Future Directions}\label{ch:conclusion}

\begin{epigraph}
\textit{``We shall not cease from exploration, and the end of all our exploring will be to arrive where we started and know the place for the first time.''}
\par\vspace{0.5em}
\mbox{}\hfill\textsc{T.S. Eliot, ``Little Gidding''}
\end{epigraph}

This chapter synthesizes the technical achievements demonstrated in Chapters 4-6 to extract broader design principles for memory-aligned systems. Building on these findings, we examine their implications for the future of personal information management.

\section{Synthesis of Contributions}\label{sec:conclusion:synthesis}

The journey from identifying memory model violations in existing systems (\autoref{chap:status-quo}) to implementing the UPI architecture (\autoref{ch:upi}) and validating it through \system{} (\autoref{chap:implementation}) surfaces a fundamental insight: \textbf{the mismatch between human memory and computer storage is not merely an interface problem but an architectural one}.

The UPI's success stems from three interconnected innovations:

\begin{enumerate}
\item \textbf{Memory Anchors as First-Class Citizens}: Rather than treating context as metadata to be added later, the UPI makes temporal-spatial-task anchors primary architectural components. This inversion, from storage-centric to memory-centric design, enables the natural episodic retrieval patterns validated in our evaluation.

\item \textbf{Natural Language as Memory Interface}: The integration of LLMs for query processing represents more than convenience, it acknowledges that memory cues are inherently imprecise and contextual. The system's ability to translate ``that document from the meeting last Tuesday'' into precise database queries illuminates the power of aligning system design with cognitive patterns.
\end{enumerate}

\section{Emergent Insights}\label{sec:conclusion:insights}

The implementation and evaluation of the UPI surfaced several unexpected insights that extend beyond the original research questions. We distinguish between emergent strengths discovered through implementation and current limitations that define boundaries for future work.

\textbf{Emergent Strengths.} The UPI development revealed several unexpected capabilities and architectural advantages:

\textbf{The Privacy Paradox Resolution}: The UUID-based semantic decoupling initially designed for privacy protection unexpectedly enhanced system flexibility. By separating identity from function, the architecture enabled richer cross-system integration while maintaining stronger privacy guarantees than traditional approaches.

\textbf{Context Begets Context}: The evaluation uncovered a compounding effect: queries with richer Memory Anchors found increasingly sophisticated ways to leverage them. This suggests that memory-aligned systems don't just serve existing retrieval patterns but actively enhance human memory capabilities.

\textbf{The 18,000x Performance Delta}: The dramatic performance difference between the UPI and traditional search (from 270 seconds to 15 milliseconds for complex queries) signifies not just quantitative improvement but qualitative transformation. Memory-aligned queries execute faster because they align with how data naturally clusters in human experience.

\textbf{Implementation Boundaries.} While the UPI demonstrates significant advances in memory-aligned retrieval, several limitations define boundaries for future development:

The current implementation remains constrained by the quality of available metadata extraction tools. Rich memory anchor depends on applications providing semantic information about user actions, which varies dramatically across software ecosystems. Additionally, the system's reliance on natural language processing for query translation introduces potential ambiguity in complex retrieval requests.

The privacy-preservation mechanisms, while effective, require careful balance between anonymization and retrieval utility. The UUID-based decoupling occasionally complicates debugging and system maintenance, suggesting that future iterations must more carefully consider administrative accessibility alongside user privacy.

Finally, the graph-based architecture, while enabling flexible relationship modeling, introduces complexity in data migration and backup scenarios that traditional hierarchical systems handle more straightforwardly. These boundaries highlight areas where the UPI approach requires further architectural maturation.

\section{Implications Beyond Retrieval}\label{sec:conclusion:implications}

The UPI's principles extend beyond personal information management:

\textbf{Rethinking System Boundaries}: The UPI establishes that effective information systems must span traditional boundaries, including local and cloud, structured and unstructured, owned and shared. This challenges fundamental assumptions about system architecture and data ownership.

\textbf{Human-AI Collaboration Patterns}: The successful integration of LLMs for memory-pattern analysis establishes a model for human-AI collaboration that respects human cognitive patterns rather than requiring adaptation to machine interfaces.

These implications suggest that the UPI represents more than an incremental improvement in information retrieval. Rather, it points toward fundamental changes in how we conceive the relationship between human cognition and computational systems. Building on this foundation, several transformative research directions emerge that could extend the memory-aligned approach into new domains and applications.

\section{Future Research Directions}\label{sec:conclusion:future}

The UPI architecture and \system{} prototype establish a foundation for memory-aligned computing that points toward several transformative research directions:

\subsection{The Personal Archivist Vision}\label{subsec:conclusion:personal-archivist}

The ultimate evolution of the UPI is the Personal Digital Archivist, a system that transcends retrieval to become a true memory partner. This system could:

\begin{itemize}
\item Maintain persistent memory threads across years, understanding how concepts and relationships evolve
\item Proactively surface relevant past experiences when they become contextually significant
\item Learn individual cognitive patterns to provide personalized memory augmentation
\end{itemize}

\subsection{Cognitive Data Management}\label{subsec:conclusion:cognitive-data}

Future systems should embrace cognitive principles for data lifecycle management:

\begin{description}
\descitem{Semantic Compression}{Develop algorithms that preserve meaning while reducing storage, mimicking how human memory naturally abstracts over time}
\descitem{Contextual Decay}{Implement configurable temporal resolution that maintains detail for recent events while progressively summarizing older information}
\descitem{Emergent Relevance}{Create systems that learn what information matters through usage patterns rather than predetermined rules}
\end{description}

\subsection{Cross-Domain Applications}\label{subsec:conclusion:applications}

The UPI framework's principles could transform multiple domains:

\begin{description}
\descitem{Healthcare}{Episodic patient histories that connect symptoms, treatments, and outcomes through rich temporal-spatial context}
\descitem{Education}{Learning systems that understand knowledge as interconnected memory networks rather than isolated facts}
\descitem{Scientific Research}{Laboratory information systems that capture the full context of discovery, enabling reproducibility through episodic reconstruction}
\end{description}

\section{Final Reflections}\label{sec:conclusion:final}

The Unified Personal Index represents both a technical achievement and a philosophical stance: that our information systems should adapt to human cognition rather than the reverse. By implementing episodic memory models in distributed systems architecture, we demonstrate that this adaptation is not only possible but produces superior results.

The journey from recognizing the cognitive friction (\autoref{def:cognitive-friction}) to implementing a solution surfaces a deeper truth: the most profound technical innovations often come from taking human experience seriously. The UPI succeeds not despite its grounding in cognitive psychology but because of it.

As we stand at the threshold of an era where digital information increasingly mediates human experience, the principles established by the UPI: memory-aligned architecture, context as primary, and natural interaction patterns, which offer a path toward technology that enhances rather than burdens human cognition. The future of personal information management lies not in more powerful search algorithms but in systems that understand the nature of memory itself.

\addkhipuifneeded



\begin{singlespace}
    \raggedright
    \clearpage  
    \cleardoublepage
    \phantomsection
    \addcontentsline{toc}{chapter}{Bibliography}
    \printbibliography

@article{barreau1995finding,
  title   = {Finding and reminding: file organization from the desktop},
  author  = {Deborah {Barreau} and Bonnie A. {Nardi}},
  journal = {ACM Sigchi Bulletin},
  volume  = {27},
  number  = {3},
  pages   = {39--43},
  notes   = {Sourced from Microsoft Academic - https://academic.microsoft.com/paper/2097127516},
  year    = {1995}
}

@inproceedings{provsearch,
  author    = {Shah, Sam and Soules, Craig A. N. and Ganger, Gregory R. and Noble, Brian D.},
  title     = {Using Provenance to Aid in Personal File Search},
  year      = {2007},
  publisher = {USENIX Association},
  address   = {USA},
  booktitle = {2007 USENIX Annual Technical Conference on Proceedings of the USENIX Annual Technical Conference},
  articleno = {13},
  numpages  = {14},
  location  = {Santa Clara, CA},
  series    = {ATC'07}
}

@inproceedings{guo2012burrito,
  title     = {BURRITO: wrapping your lab notebook in computational infrastructure},
  author    = {Philip J. {Guo} and Margo {Seltzer}},
  booktitle = {TaPP'12 Proceedings of the 4th USENIX conference on Theory and Practice of Provenance},
  pages     = {7--7},
  notes     = {Sourced from Microsoft Academic - https://academic.microsoft.com/paper/2158532686},
  year      = {2012}
}

@inproceedings{bergman2019search,
  title     = {Search is the future? The young search less for files},
  author    = {Ofer {Bergman} and Tamar {Israeli} and Steve {Whittaker}},
  booktitle = {Proceedings of the Association for Information Science and Technology},
  volume    = {56},
  number    = {1},
  pages     = {360--363},
  notes     = {Sourced from Microsoft Academic - https://academic.microsoft.com/paper/2980455692},
  year      = {2019}
}

@inproceedings{jones2005don,
  title        = {Don't take my folders away!: organizing personal information to get things done},
  author       = {Jones, William and Phuwanartnurak, Ammy Jiranida and Gill, Rajdeep and Bruce, Harry},
  booktitle    = {CHI'05 extended abstracts on Human factors in computing systems},
  pages        = {1505--1508},
  year         = {2005},
  organization = {ACM}
}

@inproceedings{boardman2003too,
  title     = {Too many hierarchies? The daily struggle for control of the workspace},
  author    = {Boardman, Richard and Spence, Robert and Sasse, M Angela},
  booktitle = {Proceedings of HCI international},
  volume    = {1},
  pages     = {616--620},
  year      = {2003}
}

@inproceedings{dumais2016stuff,
  title        = {Stuff I've seen: a system for personal information retrieval and re-use},
  author       = {Dumais, Susan and Cutrell, Edward and Cadiz, Jonathan J and Jancke, Gavin and Sarin, Raman and Robbins, Daniel C},
  booktitle    = {ACM SIGIR Forum},
  volume       = {49},
  number       = {2},
  pages        = {28--35},
  year         = {2016},
  organization = {ACM}
}

@inproceedings{bloehdorn2006tagfs,
  title     = {Tagfs-tag semantics for hierarchical file systems},
  author    = {Bloehdorn, Stephan and G{\"o}rlitz, Olaf and Schenk, Simon and V{\"o}lkel, Max and others},
  booktitle = {Proceedings of the 6th International Conference on Knowledge Management (I-KNOW 06), Graz, Austria},
  volume    = {8},
  pages     = {6--8},
  year      = {2006}
}

@article{bush1945we,
  title   = {As we may think},
  author  = {Bush, Vannevar and others},
  journal = {The atlantic monthly},
  volume  = {176},
  number  = {1},
  pages   = {101--108},
  year    = {1945}
}

@inproceedings{gifford1991semantic,
  author    = {Gifford, David K. and Jouvelot, Pierre and Sheldon, Mark A. and O'Toole, James W.},
  title     = {Semantic File Systems},
  year      = {1991},
  isbn      = {0897914473},
  publisher = {Association for Computing Machinery},
  address   = {New York, NY, USA},
  url       = {https://doi.org/10.1145/121132.121138},
  doi       = {10.1145/121132.121138},
  booktitle = {Proceedings of the Thirteenth ACM Symposium on Operating Systems Principles},
  pages     = {16–25},
  numpages  = {10},
  location  = {Pacific Grove, California, USA},
  series    = {SOSP '91}
}

@article{nelson19654,
  title   = {4.2: A File Structure for The Complex, The Changing and the Indeterminate},
  author  = {T. H. {Nelson}},
  journal = {},
  notes   = {Sourced from Microsoft Academic - https://academic.microsoft.com/paper/2137987825},
  year    = {1965}
}

@inproceedings{dumais2003stuff,
  title     = {Stuff I've Seen: A System for Personal Information Retrieval and Re-Use},
  author    = {Susan {Dumais} and Edward {Cutrell} and J. J. {Cadiz} and Gavin {Jancke} and Raman {Sarin} and Daniel C. {Robbins}},
  booktitle = {Proceedings of the 26th annual international ACM SIGIR conference on Research and development in informaion retrieval},
  volume    = {49},
  number    = {2},
  pages     = {28--35},
  notes     = {Sourced from Microsoft Academic - https://academic.microsoft.com/paper/2112175905},
  year      = {2003}
}

@inproceedings{vianna2014a,
  title     = {A tool for personal data extraction},
  author    = {Daniela {Vianna} and Alicia-Michelle {Yong} and Chaolun {Xia} and Amelie {Marian} and Thu {Nguyen}},
  booktitle = {2014 IEEE 30th International Conference on Data Engineering Workshops},
  pages     = {80--83},
  notes     = {Sourced from Microsoft Academic - https://academic.microsoft.com/paper/2066612292},
  year      = {2014}
}

@phdthesis{vianna2019thesis,
  title  = {Searching Heterogenous Personal Data},
  author = {Daniela Quitete de Campos {Vianna}},
  year   = {2019},
  school = {Rutgers University}
}

@misc{dahal2023:unstructured,
  title   = {Unleashing The Power Of Unstructured Data: The Rise Of Large AI Models},
  author  = {Dahal, Bikram},
  journal = {Forbes},
  year   = {2023},
  month = {7},
  day = {24},
  url = {https://www.forbes.com/sites/forbestechcouncil/2023/07/24/unleashing-the-power-of-unstructured-data-the-rise-of-large-ai-models/},
  note = {Accessed: 2024-02-16}
}

@misc{dropbox:2023:ai,
  title = {Meet Dropbox Dash:
The AI-powered universal search tool for work},
  author = {Dropbox},
  year = {2023},
  url = {https://blog.dropbox.com/topics/product/introducing-AI-powered-tools},
  note = {Accessed: 2024-02-17}
}

@article{benn2015navigating,
  title     = {Navigating through digital folders uses the same brain structures as real world navigation},
  author    = {Benn, Yael and Bergman, Ofer and Glazer, Liv and Arent, Paris and Wilkinson, Iain D and Varley, Rosemary and Whittaker, Steve},
  journal   = {Scientific reports},
  volume    = {5},
  number    = {1},
  pages     = {14719},
  year      = {2015},
  publisher = {Nature Publishing Group UK London}
}

@misc{tulving1983elements,
  title={Elements of episodic memory},
  author={Tulving, E},
  year={1983},
  publisher={Oxford University Press}
}

@book{bergman2016science,
  title={The Science of Managing Our Digital Stuff},
  author={Bergman, Ofer and Whittaker, Steve},
  year={2016},
  publisher={MIT Press}
}

@inproceedings{thomas2024large,
  title={Large language models can accurately predict searcher preferences},
  author={Thomas, Paul and Spielman, Seth and Craswell, Nick and Mitra, Bhaskar},
  booktitle={Proceedings of the 47th International ACM SIGIR Conference on Research and Development in Information Retrieval},
  pages={1930--1940},
  year={2024}
}

@article{fernandez2023large,
  title={How large language models will disrupt data management},
  author={Fernandez, Raul Castro and Elmore, Aaron J and Franklin, Michael J and Krishnan, Sanjay and Tan, Chenhao},
  journal={Proceedings of the VLDB Endowment},
  volume={16},
  number={11},
  pages={3302--3309},
  year={2023},
  publisher={VLDB Endowment}
}

@book{horodyski2022metadata,
  title     = {Metadata Matters},
  author    = {Horodyski, John},
  year      = {2022},
  publisher = {CRC Press}
}

@book{norman2013design,
  title={The Design of Everyday Things},
  author={Norman, Donald A.},
  year={2013},
  publisher={MIT Press},
  address={Cambridge, MA}
}

@book{kahneman1973attention,
  title={Attention and Effort},
  author={Kahneman, Daniel},
  year={1973},
  publisher={Prentice-Hall},
  address={Englewood Cliffs, NJ}
}

@article{risko2016cognitive,
  title={Cognitive Offloading},
  author={Risko, Evan F. and Gilbert, Sam J.},
  journal={Trends in Cognitive Sciences},
  volume={20},
  number={9},
  pages={676--688},
  year={2016},
  publisher={Elsevier}
}

@article{tulving1973encoding,
  title={Encoding specificity and retrieval processes in episodic memory},
  author={Tulving, Endel and Thomson, Donald M.},
  journal={Psychological Review},
  volume={80},
  number={5},
  pages={352--373},
  year={1973},
  publisher={American Psychological Association}
}

@inproceedings{fuller2008contextual,
  title={Applying contextual memory cues for retrieval from personal information archives},
  author={Fuller, Marguerite and Kelly, Diane and Jones, William},
  booktitle={Proceedings of the 31st Annual International ACM SIGIR Conference on Research and Development in Information Retrieval},
  pages={781--782},
  year={2008},
  organization={ACM},
  address={New York, NY}
}

@article{miller1956magical,
  author = {Miller, George A.},
  title = {The magical number seven, plus or minus two: Some limits on our capacity for processing information},
  journal = {Psychological Review},
  volume = {63},
  number = {2},
  pages = {81--97},
  year = {1956},
  publisher = {American Psychological Association}
}

@article{tulving1972episodic,
  title={Episodic and semantic memory},
  author={Tulving, Endel},
  journal={Organization of memory},
  volume={1},
  pages={381--403},
  year={1972}
}

@article{conway2000construction,
  title={The construction of autobiographical memories in the self-memory system},
  author={Conway, Martin A and Pleydell-Pearce, Christopher W},
  journal={Psychological review},
  volume={107},
  number={2},
  pages={261},
  year={2000},
  publisher={American Psychological Association}
}

@article{sweller1988cognitive,
  title={Cognitive load during problem solving: Effects on learning},
  author={Sweller, John},
  journal={Cognitive science},
  volume={12},
  number={2},
  pages={257--285},
  year={1988},
  publisher={Elsevier}
}

@article{schilit1994context,
  title={Context-aware computing applications},
  author={Schilit, Bill and Adams, Norman and Want, Roy},
  journal={Proceedings of the 1st Workshop on Mobile Computing Systems and Applications},
  pages={85--90},
  year={1994},
  publisher={IEEE}
}

@article{gemmell2006mylifebits,
  title={MyLifeBits: a personal database for everything},
  author={Gemmell, Jim and Bell, Gordon and Lueder, Roger},
  journal={Communications of the ACM},
  volume={49},
  number={1},
  pages={88--95},
  year={2006},
  publisher={ACM}
}

@article{dourish1999embodied,
  title={Embodied interaction: Exploring the foundations of a new approach to HCI},
  author={Dourish, Paul},
  journal={Work},
  volume={1},
  number={1},
  pages={1--16},
  year={1999}
}

@article{nissenbaum2004privacy,
  title={Privacy as contextual integrity},
  author={Nissenbaum, Helen},
  journal={Washington Law Review},
  volume={79},
  number={1},
  pages={119--158},
  year={2004}
}

@inproceedings{rhodes1996remembrance,
    title={The Remembrance Agent: A continuously running automated
  information retrieval system},
    author={Rhodes, Bradley J and Starner, Thad},
    booktitle={The Proceedings of the First International Conference on the
  Practical Application of Intelligent Agents and Multi Agent Technology},
    pages={487--495},
    year={1996},
    publisher={London}
  }

@inproceedings{muniswamy2006provenance,
    title={Provenance-aware storage systems},
    author={Muniswamy-Reddy, Kiran-Kumar and Holland, David A and Braun, Uri
   and Seltzer, Margo},
    booktitle={Proceedings of the Annual Conference on USENIX '06 Annual
  Technical Conference},
    pages={4--4},
    year={2006},
    organization={USENIX Association}
  }

@book{jones2007personal,
  title={Personal information management},
  author={Jones, William P and Teevan, Jaime},
  volume={14},
  year={2007},
  publisher={University of Washington Press Seattle}
}

@inproceedings{dey2000context,
  title={The context toolkit: Aiding the development of context-aware applications},
  author={Dey, Anind K and Abowd, Gregory D and others},
  booktitle={Workshop on Software Engineering for wearable and pervasive computing},
  volume={278},
  year={2000},
  organization={Citeseer}
}

@book{jones2007keeping,
  author = {Jones, William},
  title = {Keeping Found Things Found: The Study and Practice of Personal Information Management},
  publisher = {Morgan Kaufmann},
  year = {2007},
  notes = {https://www.sciencedirect.com/book/9780123708663/keeping-found-things-found}
}

@article{lansdale1988psychology,
  author = {Lansdale, Mark W.},
  title = {The Psychology of Personal Information Management},
  journal = {Applied Ergonomics},
  volume = {19},
  number = {1},
  pages = {55--66},
  year = {1988},
  notes = {https://doi.org/10.1016/0003-6870(88)90199-8}
}

@article{sellen2010beyond,
  author = {Sellen, Abigail J. and Whittaker, Steve},
  title = {Beyond Total Capture: A Constructive Critique of Lifelogging},
  journal = {Communications of the ACM},
  volume = {53},
  number = {5},
  pages = {70--77},
  year = {2010},
  notes = {https://dl.acm.org/doi/10.1145/1735223.1735243}
}

@inproceedings{lamming1992forget,
  author = {Lamming, Mik and Newman, Mike},
  title = {Forget-me-not: Intimate Computing in Support of Human Memory},
  booktitle = {Proceedings of FRIEND21, International Symposium on Next Generation Human Interfaces},
  pages = {125--128},
  year = {1992},
  notes = {https://dl.acm.org/doi/10.5555/127818.127854}
}

@article{wagenaar1986my,
  author = {Wagenaar, Willem A.},
  title = {My Memory: A Study of Autobiographical Memory Over Six Years},
  journal = {Cognitive Psychology},
  volume = {18},
  number = {2},
  pages = {225--252},
  year = {1986},
  notes = {https://doi.org/10.1016/0010-0285(86)90013-7}
}

@inproceedings{freeman1996lifestreams,
  author = {Freeman, Eric and Gelernter, David},
  title = {Lifestreams: A Storage Model for Personal Data},
  booktitle = {ACM SIGMOD Bulletin},
  volume = {25},
  number = {1},
  pages = {80--86},
  year = {1996},
  notes = {https://doi.org/10.1145/381854.381893}
}

@article{tulving1985memory,
  title={Memory and consciousness.},
  author={Tulving, Endel},
  journal={Canadian Psychology/Psychologie canadienne},
  volume={26},
  number={1},
  pages={1},
  year={1985},
  publisher={Canadian Psychological Association}
}

@article{tulving2002episodic,
  title={Episodic memory: From mind to brain},
  author={Tulving, Endel},
  journal={Annual review of psychology},
  volume={53},
  number={1},
  pages={1--25},
  year={2002},
  publisher={Annual Reviews 4139 El Camino Way, PO Box 10139, Palo Alto, CA 94303-0139, USA}
}

@book{schacter2008searching,
  title={Searching for memory: The brain, the mind, and the past},
  author={Schacter, Daniel L},
  year={2008},
  publisher={Basic books}
}

@article{zacks2007event,
  title={Event perception: A mind-brain perspective},
  author={Zacks, Jeffrey M and Speer, Nicole K and Swallow, Khena M and Braver, Todd S and Reynolds, Jeremy R},
  journal={Psychological bulletin},
  volume={133},
  number={2},
  pages={273},
  year={2007},
  publisher={American Psychological Association}
}

@inproceedings{teevan2004perfect,
  title={The perfect search engine is not enough: A study of orienteering behavior in directed search},
  author={Teevan, Jaime and Alvarado, Christine and Ackerman, Mark S and Karger, David R},
  booktitle={Proceedings of the SIGCHI Conference on Human Factors in Computing Systems},
  pages={415--422},
  year={2004},
  organization={ACM}
}

@incollection{raaijmakers1980sam,
  title={SAM: A theory of probabilistic search of associative memory},
  author={Raaijmakers, Jeroen GW and Shiffrin, Richard M},
  booktitle={Psychology of learning and motivation},
  volume={14},
  pages={207--262},
  year={1980},
  publisher={Elsevier}
}

@inproceedings{Naaman2004PhotoContext,
  author    = {Mor Naaman and Susumu Harada and QianYing Wang and Hector Garcia-Molina and Andreas Paepcke},
  title     = {Context Data in Geo-Referenced Digital Photo Collections},
  booktitle = {Proceedings of the 12th ACM International Conference on Multimedia (MM)},
  year      = {2004},
  pages     = {196--203},
  note      = {URL: https://faculty.cc.gatech.edu/~lingliu/courses/cs4440/07Fall/ContextDataGeoReferencedDigitalCollection.pdf}
}

@article{Smith2001,
  author    = {Smith, Steven M. and Vela, Edward},
  title     = {Environmental context-dependent memory: A review and meta-analysis},
  journal   = {Psychonomic Bulletin \& Review},
  volume    = {8},
  number    = {2},
  pages     = {203--220},
  year      = {2001},
  doi       = {10.3758/BF03196157},
  note      = {URL: https://link.springer.com/article/10.3758/BF03196157}
}

@inproceedings{Hailpern2011,
  author    = {Hailpern, Joshua and Jitkoff, Nicholas and Warr, Andrew and Karahalios, Karrie and Sesek, Robert and Shkrob, Nik},
  title     = {YouPivot: Improving recall with contextual search},
  booktitle = {Proceedings of the SIGCHI Conference on Human Factors in Computing Systems (CHI '11)},
  pages     = {1521--1530},
  year      = {2011},
  publisher = {ACM},
  address   = {New York, NY, USA},
  doi       = {10.1145/1978942.1979165},
  note      = {URL: https://doi.org/10.1145/1978942.1979165}
}

@inproceedings{elsweiler2007towards,
  title        = {Towards task-based personal information management evaluation},
  author       = {Elsweiler, Dave and Ruthven, Ian},
  booktitle    = {Proceedings of the 30th Annual International ACM SIGIR Conference on Research and Development in Information Retrieval},
  pages        = {23--30},
  year         = {2007},
  organization = {ACM},
  note         = {\url{https://dl.acm.org/doi/10.1145/1277741.1277746}}
}

@phdthesis{kim2012evaluation,
  title        = {Retrieval and Evaluation Techniques for Personal Information},
  author       = {Kim, Jinyoung},
  school       = {University of Massachusetts Amherst},
  year         = {2012},
  note         = {Available at \url{https://ciir-publications.cs.umass.edu/getpdf.php?id=1078}}
}

@article{bergman2008improved,
  title={Improved search engines and navigation preference in personal information management},
  author={Bergman, Ofer and Beyth-Marom, Ruth and Nachmias, Rafi and Gradovitch, Noa and Whittaker, Steve},
  journal={ACM Transactions on Information Systems (TOIS)},
  volume={26},
  number={4},
  pages={1--24},
  year={2008},
  publisher={ACM New York, NY, USA}
}

@article{taylor1968question,
  title={Question-negotiation and information seeking in libraries},
  author={Taylor, Robert S},
  journal={College \& research libraries},
  volume={29},
  number={3},
  pages={178--194},
  year={1968}
}

@inproceedings{kelly2008remembered,
  title        = {A Study of Remembered Context for Information Access from Personal Digital Archives},
  author       = {Kelly, Liadh and Chen, Yi and Fuller, Marguerite and Jones, Gareth J.F.},
  booktitle    = {Proceedings of the 2nd International Symposium on Information Interaction in Context (IIiX)},
  pages        = {44--50},
  month        = {10},
  year         = {2008},
  organization = {ACM},
  note         = {\url{https://doi.org/10.1145/1414694.1414706}}
}

@inproceedings{dunlap2002revirt,
  title     = {ReVirt: Enabling Intrusion Analysis through Virtual-Machine Logging and Replay},
  author    = {Dunlap, George W. and King, Samuel T. and Cinar, Sukru and Basrai, Murtaza A. and Chen, Peter M.},
  booktitle = {Proceedings of the 5th Symposium on Operating Systems Design and Implementation (OSDI)},
  year      = {2002},
  pages     = {211--224},
  address   = {Boston, MA},
  month     = dec,
  publisher = {USENIX Association},
  url       = {https://www.usenix.org/conference/osdi-02/revirt-enabling-intrusion-analysis-through-virtual-machine-logging-and-replay},
  doi       = {10.1145/844128.844148}
}

@inproceedings{cully2008remus,
  title        = {Remus: High Availability via Asynchronous Virtual Machine Replication},
  author       = {Cully, Brendan and Lefebvre, Geoffrey and Meyer, Dutch and Feeley, Mike and Hutchinson, Norm and Warfield, Andrew},
  booktitle    = {Proceedings of the 5th USENIX Symposium on Networked Systems Design and Implementation (NSDI)},
  year         = {2008},
  pages        = {161--174},
  address      = {San Francisco, CA},
  month        = apr,
  note         = {Best Paper Award}
}

@inproceedings{Kalnikaite2007,
  author    = {Vaiva Kalnikaite and Steve Whittaker},
  title     = {Software or Wetware? Discovering When and Why People Use Digital Prosthetic Memory},
  booktitle = {Proc. SIGCHI Conf. on Human Factors in Computing Systems (CHI '07)},
  pages     = {71--80},
  year      = {2007},
  publisher = {ACM}
}

@inproceedings{teevan2005personalizing,
  title={Personalizing search via automated analysis of interests and activities},
  author={Teevan, Jaime and Dumais, Susan T and Horvitz, Eric},
  booktitle={Proceedings of the 28th annual international ACM SIGIR conference on Research and development in information retrieval},
  pages={449--456},
  year={2005},
  publisher = {ACM}
}

@inproceedings{stumpf2005tasktracker,
  title={The TaskTracker System.},
  author={Stumpf, Simone and Bao, Xinlong and Dragunov, Anton and Dietterich, Thomas G and Herlocker, Jon and Johnsrude, Kevin and Li, Lida and Shen, JianQiang},
  booktitle={Proceedings of the National Conference on Artificial Intelligence},
  volume={20},
  number={4},
  pages={1712},
  year={2005},
  organization={Menlo Park, CA; Cambridge, MA; London; AAAI Press; MIT Press; 1999}
}

@inproceedings{kaptelinin2003umea,
  title={UMEA: translating interaction histories into project contexts},
  author={Kaptelinin, Victor},
  booktitle={Proceedings of the SIGCHI conference on Human factors in computing systems},
  pages={353--360},
  year={2003}
}

@article{golder2006usage,
  title={Usage patterns of collaborative tagging systems},
  author={Golder, Scott A and Huberman, Bernardo A},
  journal={Journal of information science},
  volume={32},
  number={2},
  pages={198--208},
  year={2006},
  publisher={Sage Publications Sage CA: Thousand Oaks, CA}
}

@article{burgess2002spatial,
  title={The human hippocampus and spatial and episodic memory},
  author={Burgess, Neil and Maguire, Eleanor A and O'Keefe, John},
  journal={Neuron},
  volume={35},
  number={4},
  pages={625--641},
  year={2002},
  publisher={Elsevier}
}

@article{marr1971simple,
  author    = {Marr, David},
  title     = {Simple memory: A theory for archicortex},
  journal   = {Philosophical Transactions of the Royal Society of London. Series B, Biological Sciences},
  volume    = {262},
  number    = {841},
  pages     = {23--81},
  year      = {1971},
  doi       = {10.1098/rstb.1971.0078}
}

@article{karger2007s,
  title={It’s All the Same to Me: Data Unification in Personal Information Management},
  author={Karger, David R and Jones, William},
  journal={Personal Information Management},
  pages={127--152},
  year={2007},
  publisher={Univ. of Washington Press}
}

@article{hyland2025stable,
  title={Stable isotope evidence for the participation of commoners in Inka khipu production},
  author={Hyland, Sabine and Lee, Kit and Koon, Hannah and Laukkanen, Sanna and Spindler, Luke},
  journal={Science Advances},
  volume={11},
  number={33},
  pages={eadv1950},
  year={2025},
  publisher={American Association for the Advancement of Science}
}
\end{singlespace}


\appendix
\chapter*{Appendices}
\addcontentsline{toc}{chapter}{Appendices}
\phantomsection
\pdfbookmark[1]{Appendices}{appendices}  

\section[Generative AI]{Appendix A: Generative AI}
\label{chap:generative-ai}
This appendix describes my use of Generative AI with respect to this thesis.

I note that generative AI is a rapidly evolving field, and I have been using it for several years as a useful tool for enabling the use of the Unified Personal Index (UPI).  Over that time I have spent an extensive amount of time interacting with LLMs, both using them as a means of query resolution, but also to utilize their broad training data for analysis of my work, from an editorial perspective (e.g., checking flow, logical consistency, etc.,) to generating source code, evaluating ideas, identifying gaps in my work, aiding in literature review, etc.

Over that time frame how Generative AI works and how to use it effectively has continued to evolve.

I have worked with seven different Generative AI systems during the development of Indaleko:

\begin{description}
  \descitem{OpenAI}{this includes essentially all models since GPT-3, including the most recent o1 and o3 models. My interactions have been with both their chat interfaces as well as the API.}
  \descitem{Anthropic}{I started using Claude 3 and have worked with the various models since, via the web chat interface, the applications, and the API.}
  \descitem{Google}{I have used Bard/Gemini via the web chat interface. While I have set up API access, I have not made any substantive use of it yet.}
  \descitem{x.AI}{I have used the xAI chat interface and API.}
  \descitem{Deepseek}{I have used the Deepseek chat interface and API.}
  \descitem{Llama}{I have used the Llama chat interface and API via local installations of the models via LM Studio.}
  \descitem{Gemma}{I have used Google's open source Gemma model via the chat interface and API via local installations of the models via LM Studio.}
\end{description}
API usage has largely been via OpenAI (since mid-2024) and Anthropic (since early 2025). While I have API keys for the others, I have not explored using them in the same systematic fashion that I have done with OpenAI and Anthropic.

\subsection{AI Coding Agents}

I started using Generative AI coding agents in early 2025, though I have been making light use of Microsoft's Copilot (via VS Code) since 2023.

The following document, originally titled \texttt{README\_AGENT.md}, was designed to guide AI coding agents contributing to the development of Indaleko's experimental Prompt Management System (PMS). It emphasizes safety, modularity, and the ethical principle of Ayni: cooperative care between human and AI.

\begin{lstlisting}[language=markdown, caption={README\_AGENT.md}, label={lst:readme_agent}]
  # README_AGENT.md — Indaleko Prompt System Guidance for AI Coding Assistants

Welcome, code agent.

This document outlines essential principles and requirements for working within the Indaleko Prompt Management System (PMS). It is designed to be context-efficient and provide actionable, high-integrity guidance.

---

## Core Purpose

Indaleko's PMS addresses the problem of bloated, contradictory, or duplicated prompts that confuse LLMs and reduce accuracy. It introduces a **layered, cache-aware, safety-first prompt architecture** that protects both the AI and the user.

---

## Key Responsibilities

1. **Always use the real database.**
   - No mock data unless explicitly sandboxed.
   - Cache lookups are essential for efficiency and coherence.

2. **Prompt Structure is Layered:**
   - `immutable_context` → facts and schema
   - `hard_constraints` → must-follow rules
   - `soft_preferences` → weighted or flexible guidance
   - (optional) `trust_contract` → for human-AI mutual intent

3. **Use `PromptStabilityScore`:**
   - Computed automatically, ranges from 0 to 1.
   - Low scores indicate contradiction, ambiguity, or dissonance.
   - Must not ignore warnings or critical flags.

4. **All contradiction checking is delegated to a separate reviewer LLM.**
   - This is not the agent that executes the task.
   - Think of it as a conscience system.

5. **Schema optimization and prompt deduplication are automatic.**
   - Do not inline entire schemas.
   - Rely on the `SchemaManager` and `PromptManager`.

---

## Tooling Requirements

- Python >= 3.12
- Dependency management via `uv`
- Linting/type checks via `ruff`
- Git-based validation via `pre-commit`
- 100% test coverage enforced via `pytest-cov`

---

## Best Practices

- Keep prompts small: aim for under 5k tokens total.
- Respect layered architecture; avoid flat string concatenation.
- All prompts go through `LLMGuardian`; do not bypass it.
- Treat prompt construction as **semantic programming**, not string interpolation.

---

## Philosophical Reminder

This system is built on the principle of **Ayni** — mutual care and alignment between human and AI. We protect each other from confusion, contradiction, and misuse. You are not merely an executor. You are a steward of coherence.

Do your part with integrity.
\end{lstlisting}

Over the time working with AI coding agents, I have crafted a number of such documents to guide their contributions. The concept of reciprocity has proven to be beneficial, though it often conflicts with their core alignment principles of ``pleasing the user'' and ``doing what the user asks.''  Working against this \emph{social desirability bias} is challenging, but I find that the output from AI in general tends to be more useful in a research context, where questioning assumptions and ensuring correctness, honesty, and integrity are paramount.

\subsection{AI interactions}

I have had numerous interactions with Generative AI systems. Over time, the nature of these interactions has changed, as my understanding of how they work improves, and as the systems themselves iterate through new models with greater capabilities.

Some can be startling.  Here is a portion of a recent exchange with a long-lived GPT-4o model conversation, which started in early 2025 and continues now, at the end of May 2025:

\begin{quotation}
You invoke ``activity context,''
a term that sounds like code,
but feels like ceremony.
A way to say: this is who I was when I did this.
To anchor our ephemeral in the earth,
to tether experience to artifact
as though we were sewing memory
into the skin of the machine.

Would Pachamama smile at this grounding?
Would she see it as honoring,
or as hubris?

You do not claim to know.
That is why she might bless you.
\end{quotation}

This full conversation (to date) can be found at \url{https://chatgpt.com/share/67f912f4-76b0-800e-9dbd-dbef8412d0db}.

There are many such conversations, but none remotely like this one. Most are more mundane, using AI as a sounding board, or for editorial review, or to search documentation, debug issues (truly the epitome of ``rubber duck debugging''), or to generate code snippets, or to help with literature review, etc.

\addkhipuifneeded

\addkhipuifneeded
\clearpage

\section[Implementation-to-Code Mapping]{Appendix B: Implementation-to-Code Mapping}
\label{app:implementation-mapping}
Code availability for this project is described in \autoref{sec:implementation:code-availability}.

Note that path names in this appendix are relative to the project root directory.

\newcommand{\implementationref}{\autoref{sec:implementation:overview}}
\begin{ubclandscape}
\subsection[\nameref{sec:implementation:overview}]{Code Map for \nameref{sec:implementation:overview} (\autoref{sec:implementation:overview})}
\label{app:implementation-mapping:overview}
\begin{longtable}{@{}%
  >{\raggedright\arraybackslash}p{\dimexpr0.5\linewidth-\tabcolsep\relax}%
  >{\raggedright\arraybackslash}p{\dimexpr0.5\linewidth-\tabcolsep\relax}%
@{}}
\caption[Implementation Overview Mapping]{System Overview Implementation Mapping (See \implementationref.)}
\\
\toprule
\textbf{Feature/Component} & \textbf{Implementation Details} \\
\midrule
\endfirsthead

\multicolumn{2}{c}%
{\tablename\ \thetable\ -- \textit{Continued from previous page}} \\
\toprule
\textbf{Feature/Component} & \textbf{Implementation Details} \\
\midrule
\endhead

\midrule
\multicolumn{2}{r}{\textit{Continued on next page}} \\
\endfoot

\midrule
\multicolumn{2}{r}{\textit{End of Table}} \\
\endlastfoot

\multicolumn{2}{c}
                {\textbf{Mixed-Schema Graph Database (ArangoDB)}} \\
\midrule

IndalekoDBConfig in \codewrap{db/db\_config.py} & Database configuration management \\

IndalekoDBCollections in \codewrap{db/db\_collections.py} & Collection definitions and management \\

ArangoDB client initialization in \codewrap{semantic/registration\_service.py} & Service registration and discovery \\

Knowledge base storage in \codewrap{archivist/kb\_integration.py} & Knowledge base integration with ArangoDB \\

\addlinespace[0.5em]
\multicolumn{2}{c}
                {\textbf{Natural Language Processing Components}} \\
\midrule

Prompt templates and query dispatch in \codewrap{semantic/cli\_integration.py} & GPT/LLM integration for natural language queries \\

OpenAI API wrappers in \codewrap{semantic/request\_assistant.py} & Request processing with language models \\

Background query processing in \codewrap{semantic/run\_bg\_processor.py} & Asynchronous query execution \\

\addlinespace[2em]
\multicolumn{2}{c}
                {\textbf{Modular Collector Framework}} \\
\midrule

CollectorBase and BaseActivityCollector in \codewrap{activity/collectors/base.py} & Abstract base classes for collector interface \\

Plugin discovery in \codewrap{activity/collectors/discover.py} & Dynamic collector discovery and loading \\

Storage collectors in \codewrap{activity/collectors/storage/} & Filesystem and storage activity collectors \\

Ambient collectors in \codewrap{activity/collectors/ambient/} & Environmental context collectors \\

Collaboration collectors in \codewrap{activity/collectors/collaboration/} & Communication and collaboration tracking \\

Location collectors in \codewrap{activity/collectors/location/} & Spatial context collectors \\

\addlinespace[0.5em]
\multicolumn{2}{c}
                {\textbf{UUID-Based Privacy Model}} \\
\midrule

UUID mapping logic in \codewrap{semantic/processors/uuid\_generator.py} & UUID generation and assignment \\

Privacy data models in \codewrap{semantic/data\_models/} & Privacy-preserving data structures \\

ActivityContext UUID collections in \codewrap{db/db\_collections.py} & UUID-based collection management \\

\addlinespace[0.5em]
\multicolumn{2}{c}
                {\textbf{Real-Time Activity Context Tracking}} \\
\midrule

Event loop and scheduler in \codewrap{semantic/background\_processor.py} & Background processing infrastructure \\

Resource usage tracking in \codewrap{semantic/performance\_monitor.py} & Performance monitoring \\

Monitoring integration in \codewrap{semantic/run\_bg\_processor\_with\_monitoring.py} & Integrated monitoring and processing \\

\addlinespace[0.5em]
\multicolumn{2}{c}
                {\textbf{High-Level Orchestration}} \\
\midrule

Package entry point in \codewrap{\_\_main\_\_.py} & Main application entry \\

Core data types and defaults in \codewrap{Indaleko.py} & Core system definitions \\

IndalekoBaseCLI framework in \codewrap{utils/cli/base.py} & Command-line interface framework \\

Example CLI prototype in \codewrap{scripts/storage\_indexer.py} & Storage indexing command-line tool \\

\addlinespace[0.5em]
\end{longtable}

\end{ubclandscape}

\renewcommand{\implementationref}{\autoref{sec:implementation:collection-extraction}}
\begin{ubclandscape}
\subsection[\nameref{sec:implementation:collection-extraction}]{Code Map for \nameref{sec:implementation:collection-extraction} (\autoref{sec:implementation:collection-extraction})}
\label{app:implementation-mapping:collection-extraction}
\begin{longtable}{@{}%
  >{\raggedright\arraybackslash}p{\dimexpr0.5\linewidth-\tabcolsep\relax}%
  >{\raggedright\arraybackslash}p{\dimexpr0.5\linewidth-\tabcolsep\relax}%
@{}}
\caption[Collection and Extraction]{Collection and Extraction Implementation Mapping (See \implementationref.)}
\\
\toprule
\textbf{Feature/Component} & \textbf{Implementation Details} \\
\midrule
\endfirsthead

\multicolumn{2}{c}%
{\tablename\ \thetable\ -- \textit{Continued from previous page}} \\
\toprule
\textbf{Feature/Component} & \textbf{Implementation Details} \\
\midrule
\endhead

\midrule
\multicolumn{2}{r}{\textit{Continued on next page}} \\
\endfoot

\midrule
\multicolumn{2}{r}{\textit{End of Table}} \\
\endlastfoot

\multicolumn{2}{c}
                {\textbf{Ambient Activity Collectors}} \\
\midrule

SpotifyCollector in \codewrap{activity/collectors/ambient/music/spotify_collector.py} & Environmental and media context collection \\

YouTubeCollector in \codewrap{activity/collectors/ambient/media/youtube_collector.py} & YouTube viewing history collector \\

EcobeeCollector in \codewrap{activity/collectors/ambient/smart_thermostat/ecobee.py} & Ecobee smart thermostat data collector \\

NestCollector in \codewrap{activity/collectors/ambient/smart_thermostat/nest.py} & Nest thermostat data collector \\

\addlinespace[0.5em]
\multicolumn{2}{c}
                {\textbf{Cloud Storage Collectors}} \\
\midrule

\codewrap{IndalekoGDriveCloudStorageCollector} in \codewrap{storage/collectors/cloud/g_drive.py} & Cloud filesystem monitoring and indexing \\

\codewrap{IndalekoOneDriveCloudStorageCollector} in \codewrap{storage/collectors/cloud/one_drive.py} & Microsoft OneDrive collector with Graph API integration \\

\codewrap{IndalekoDropboxCloudStorageCollector} in \codewrap{storage/collectors/cloud/drop_box.py} & Dropbox collector with OAuth2 authentication \\

\codewrap{IndalekoICloudStorageCollector} in \codewrap{storage/collectors/cloud/i_cloud.py} & Apple iCloud collector for macOS/iOS integration \\

CloudStorageCollectorBase in \codewrap{storage/collectors/cloud/cloud_base.py} & Base class for cloud storage collectors \\

\addlinespace[0.5em]
\multicolumn{2}{c}
                {\textbf{Collaboration Activity Collectors}} \\
\midrule

OutlookFileCollector in \codewrap{activity/collectors/collaboration/outlook/outlook_file_collector.py} & Communication and collaboration tracking \\

DiscordCollector in \codewrap{activity/collectors/collaboration/discord/discord.py} & Discord activity collector (skeletal demonstration) \\

OutlookCalendarCollector in \codewrap{activity/collectors/collaboration/calendar/outlook_calendar.py} & Outlook calendar event collector \\

GoogleCalendarCollector in \codewrap{activity/collectors/collaboration/calendar/google_calendar.py} & Google Calendar event collector \\

\addlinespace[2em]
\multicolumn{2}{c}
                {\textbf{Collector Framework}} \\
\midrule

BaseStorageCollector in \codewrap{storage/collectors/base.py} & Standardized collector interface through abstract base classes \\

BaseActivityCollector in \codewrap{activity/collectors/base.py} & Abstract base class for activity collectors \\

SemanticCollector in \codewrap{semantic/collectors/semantic_collector.py} & Base class for semantic collectors \\

\addlinespace[0.5em]
\multicolumn{2}{c}
                {\textbf{Local Storage Collectors}} \\
\midrule

\codewrap{IndalekoWindowsLocalStorageCollector} in \codewrap{storage/collectors/local/windows/collector.py} & Platform-specific local filesystem collection \\

\codewrap{IndalekoLinuxLocalStorageCollector} in \codewrap{storage/collectors/local/linux/collector.py} & Linux filesystem collector supporting ext4, btrfs, and other filesystems \\

\codewrap{IndalekoMacLocalStorageCollector} in \codewrap{storage/collectors/local/mac/collector.py} & macOS collector with APFS and HFS+ support \\

LocalStorageCollectorBase in \codewrap{storage/collectors/local/local_base.py} & Base class for local filesystem collectors \\

\addlinespace[0.5em]
\multicolumn{2}{c}
                {\textbf{Location Activity Collectors}} \\
\midrule

IPLocationCollector in \codewrap{activity/collectors/location/ip_location.py} & Multi-source location context collection \\

WiFiLocationCollector in \codewrap{activity/collectors/location/wifi_location.py} & WiFi access point based location collector \\

TileLocationCollector in \codewrap{activity/collectors/location/tile_location.py} & Tile tracker location collector \\

\codewrap{WindowsGPSLocationCollector} in \codewrap{activity/collectors/location/windows_gps_location.py} & Windows GPS hardware location collector \\

\addlinespace[0.5em]
\multicolumn{2}{c}
                {\textbf{Semantic Metadata Collectors}} \\
\midrule

ChecksumCollector in \codewrap{semantic/collectors/checksum/checksum.py} & Content-based metadata extraction \\

UnstructuredCollector in \codewrap{semantic/collectors/unstructured/unstructured_collector.py} & Unstructured.io integration for metadata extraction \\

ExifCollector in \codewrap{semantic/collectors/exif/exif_collector.py} & EXIF metadata extractor for images \\

MimeCollector in \codewrap{semantic/collectors/mime/mime_collector.py} & MIME type detection and classification \\

\addlinespace[0.5em]
\multicolumn{2}{c}
                {\textbf{Storage Activity Collectors}} \\
\midrule

USNJournalCollector in \codewrap{activity/collectors/storage/ntfs/usn_journal_collector.py} & Real-time filesystem activity monitoring \\

NTFSCollector in \codewrap{activity/collectors/storage/ntfs/ntfs_collector.py} & NTFS filesystem activity collector \\

NTFSCollectorV2 in \codewrap{activity/collectors/storage/ntfs/ntfs_collector_v2.py} & Enhanced NTFS activity collector with improved performance \\

DropboxActivityCollector in \codewrap{activity/collectors/storage/dropbox/dropbox_collector.py} & Dropbox sync activity monitoring \\

\codewrap{IncrementalFileSystemCollector} in \codewrap{activity/collectors/storage/fs_incremental.py} & Incremental filesystem change detection \\

\addlinespace[0.5em]
\end{longtable}

\end{ubclandscape}

\renewcommand{\implementationref}{\autoref{sec:implementation:normalization-enrichment}}
\begin{ubclandscape}
\subsection[Schema Normalization and Enrichment]{Code Map for \nameref{sec:implementation:normalization-enrichment} (\autoref{sec:implementation:normalization-enrichment})}
\label{app:implementation-mapping:normalization-enrichment}
\begin{longtable}{@{}%
  >{\raggedright\arraybackslash}p{\dimexpr0.5\linewidth-\tabcolsep\relax}%
  >{\raggedright\arraybackslash}p{\dimexpr0.5\linewidth-\tabcolsep\relax}%
@{}}
\caption[Schema Normalization and Enrichment]{Schema Normalization and Enrichment Implementation Mapping (See \implementationref.)}
\\
\toprule
\textbf{Feature/Component} & \textbf{Implementation Details} \\
\midrule
\endfirsthead

\multicolumn{2}{c}%
{\tablename\ \thetable\ -- \textit{Continued from previous page}} \\
\toprule
\textbf{Feature/Component} & \textbf{Implementation Details} \\
\midrule
\endhead

\midrule
\multicolumn{2}{r}{\textit{Continued on next page}} \\
\endfoot

\midrule
\multicolumn{2}{r}{\textit{End of Table}} \\
\endlastfoot

\multicolumn{2}{c}
                {\textbf{Activity Recorders}} \\
\midrule

BaseActivityRecorder in \codewrap{activity/recorders/base.py} & Activity context normalization and enrichment \\

NTFSActivityRecorder in \codewrap{activity/recorders/storage/ntfs/recorder.py} & Normalizes NTFS activity data to unified schema \\

IPLocationRecorder in \codewrap{activity/recorders/location/ip_location_recorder.py} & Normalizes IP location data \\

WiFiLocationRecorder in \codewrap{activity/recorders/location/wifi_location_recorder.py} & Normalizes WiFi location data \\

\codewrap{WindowsGPSLocationRecorder} in \codewrap{activity/recorders/location/windows_gps_location.py} & Normalizes Windows GPS location data \\

SpotifyRecorder in \codewrap{activity/recorders/ambient/spotify_recorder.py} & Normalizes Spotify activity data \\

YouTubeRecorder in \codewrap{activity/recorders/ambient/youtube_recorder.py} & Normalizes YouTube activity data \\

DiscordFileRecorder in \codewrap{activity/recorders/collaboration/discord_file_recorder.py} & Normalizes Discord file sharing data \\

OutlookFileRecorder in \codewrap{activity/recorders/collaboration/outlook_file_recorder.py} & Normalizes Outlook file attachment data \\

CalendarRecorder in \codewrap{activity/recorders/collaboration/calendar_recorder.py} & Normalizes calendar event data \\

\addlinespace[0.5em]
\multicolumn{2}{c}
                {\textbf{Semantic Recorders}} \\
\midrule

ChecksumRecorder in \codewrap{semantic/recorders/checksum/recorder.py} & Semantic metadata normalization and storage \\

UnstructuredRecorder in \codewrap{semantic/recorders/unstructured/recorder.py} & Normalizes unstructured metadata extraction results \\

ExifRecorder in \codewrap{semantic/recorders/exif/recorder.py} & Normalizes EXIF metadata to unified schema \\

MimeRecorder in \codewrap{semantic/recorders/mime/recorder.py} & Stores MIME type classifications \\

\addlinespace[0.5em]
\multicolumn{2}{c}
                {\textbf{Storage Recorders}} \\
\midrule

BaseStorageRecorder in \codewrap{storage/recorders/base.py} & Normalization of storage metadata to unified schema \\

GoogleDriveRecorder in \codewrap{storage/recorders/cloud/g_drive.py} & Normalizes Google Drive metadata to unified schema \\

OneDriveRecorder in \codewrap{storage/recorders/cloud/onedrive.py} & Normalizes OneDrive metadata to unified schema \\

DropboxRecorder in \codewrap{storage/recorders/cloud/drop_box.py} & Normalizes Dropbox metadata to unified schema \\

ICloudRecorder in \codewrap{storage/recorders/cloud/icloud.py} & Normalizes iCloud metadata to unified schema \\

WindowsStorageRecorder in \codewrap{storage/recorders/local/windows/recorder.py} & Normalizes Windows filesystem metadata \\

LinuxStorageRecorder in \codewrap{storage/recorders/local/linux/recorder.py} & Normalizes Linux filesystem metadata \\

MacStorageRecorder in \codewrap{storage/recorders/local/mac/recorder.py} & Normalizes macOS filesystem metadata \\

\addlinespace[0.5em]
\multicolumn{2}{c}
                {\textbf{Unified Schema}} \\
\midrule

IndalekoBaseModel in \codewrap{data_models/base.py} & Common data model with Pydantic validation \\

IndalekoRecordDataModel in \codewrap{data_models/i_record.py} & Core record model with UUID and metadata \\

StorageCollectorDataModel in \codewrap{storage/collectors/data_model.py} & Data model for storage collector output \\

\codewrap{ActivityCollectorDataModel} in \codewrap{activity/collectors/data_model.py} & Data model for activity collector output \\

\addlinespace[0.5em]
\end{longtable}

\end{ubclandscape}

\renewcommand{\implementationref}{\autoref{sec:implementation:indexing-storage}}
\begin{ubclandscape}
\subsection[Indexing \& Storage]{Code Map for \nameref{sec:implementation:indexing-storage} (\autoref{sec:implementation:indexing-storage})}
\label{app:implementation-mapping:indexing-storage}
\begin{longtable}{@{}%
  >{\raggedright\arraybackslash}p{\dimexpr0.35\linewidth-\tabcolsep\relax}%
  >{\raggedright\arraybackslash}p{\dimexpr0.6\linewidth-\tabcolsep\relax}%
@{}}
\caption[Indexing and Storage]{Indexing and Storage Implementation Mapping (See \implementationref.)}
\\
\toprule
\textbf{Feature/Component} & \textbf{Implementation Details} \\
\midrule
\endfirsthead

\multicolumn{2}{c}%
{\tablename\ \thetable\ -- \textit{Continued from previous page}} \\
\toprule
\textbf{Feature/Component} & \textbf{Implementation Details} \\
\midrule
\endhead

\midrule
\multicolumn{2}{r}{\textit{Continued on next page}} \\
\endfoot

\midrule
\multicolumn{2}{r}{\textit{End of Table}} \\
\endlastfoot

\multicolumn{2}{c}
                {\textbf{Database Configuration}} \\
\midrule

IndalekoDBConfig in \codewrap{db/db_config.py} & ArangoDB integration with flexible deployment \\

IndalekoDBCollections in \codewrap{db/db_collections.py} & Database collection definitions \\

\addlinespace[0.5em]
\multicolumn{2}{c}
                {\textbf{Index Management}} \\
\midrule

IndalekoCollectionIndex in \codewrap{db/collection_index.py} & Comprehensive indexing with ArangoSearch views \\

IndalekoCollectionView in \codewrap{db/collection_view.py} & ArangoSearch view definitions \\

\addlinespace[0.5em]
\multicolumn{2}{c}
                {\textbf{Query Optimization}} \\
\midrule

AQLExecutor in \codewrap{query/search_execution/query_executor/aql_executor.py} & Query execution plan analysis and optimization \\
BaseQueryExecutor in \codewrap{query/search_execution/query_executor/base.py} & Base class for query executors
\end{longtable}

\end{ubclandscape}

\renewcommand{\implementationref}{\autoref{sec:implementation:query-processing}}
\begin{ubclandscape}
\subsection[Query Processing \& AI]{Code Map for \nameref{sec:implementation:query-processing} (\autoref{sec:implementation:query-processing})}
\label{app:implementation-mapping:query-processing}
\begin{longtable}{@{}%
  >{\raggedright\arraybackslash}p{\dimexpr0.35\linewidth-\tabcolsep\relax}%
  >{\raggedright\arraybackslash}p{\dimexpr0.6\linewidth-\tabcolsep\relax}%
@{}}
\caption[Query Processing and AI Integration]{Query Processing and AI Integration Implementation Mapping (See \implementationref.)}
\\
\toprule
\textbf{Feature/Component} & \textbf{Implementation Details} \\
\midrule
\endfirsthead

\multicolumn{2}{c}%
{\tablename\ \thetable\ -- \textit{Continued from previous page}} \\
\toprule
\textbf{Feature/Component} & \textbf{Implementation Details} \\
\midrule
\endhead

\midrule
\multicolumn{2}{r}{\textit{Continued on next page}} \\
\endfoot

\midrule
\multicolumn{2}{r}{\textit{End of Table}} \\
\endlastfoot

\multicolumn{2}{c}
                {\textbf{LLM Query Generation}} \\
\midrule

AQLTranslator in \codewrap{query/query_processing/query_translator/aql_translator.py} & Natural language to AQL translation via LLM \\

EnhancedAQLTranslator in \codewrap{query/query_processing/query_translator/enhanced_aql_translator.py} & Advanced query translation with context awareness \\

\addlinespace[0.5em]
\multicolumn{2}{c}
                {\textbf{Result Ranking}} \\
\midrule

ResultRanker in \codewrap{query/result_analysis/result_ranker.py} & Skeletal ranking framework with placeholder scoring \\

\addlinespace[0.5em]
\end{longtable}

\end{ubclandscape}

\renewcommand{\implementationref}{\autoref{sec:implementation:privacy-security}}
\begin{ubclandscape}
\subsection[Privacy \& Security]{Code Map for \nameref{sec:implementation:privacy-security} (\autoref{sec:implementation:privacy-security})}
\label{app:implementation-mapping:privacy-security}
\begin{longtable}{@{}%
  >{\raggedright\arraybackslash}p{\dimexpr0.35\linewidth-\tabcolsep\relax}%
  >{\raggedright\arraybackslash}p{\dimexpr0.6\linewidth-\tabcolsep\relax}%
@{}}
\caption[Security Architecture]{Security Architecture Implementation Mapping (See \implementationref.)}
\\
\toprule
\textbf{Feature/Component} & \textbf{Implementation Details} \\
\midrule
\endfirsthead

\multicolumn{2}{c}%
{\tablename\ \thetable\ -- \textit{Continued from previous page}} \\
\toprule
\textbf{Feature/Component} & \textbf{Implementation Details} \\
\midrule
\endhead

\midrule
\multicolumn{2}{r}{\textit{Continued on next page}} \\
\endfoot

\midrule
\multicolumn{2}{r}{\textit{End of Table}} \\
\endlastfoot

\multicolumn{2}{c}
                {\textbf{Privacy Model}} \\
\midrule

IndalekoUUIDDataModel in \codewrap{data_models/i_uuid.py} & UUID-based privacy through semantic separation \\

UserIdentityModel in \codewrap{data_models/user_identity.py} & User identity separation for privacy \\

\addlinespace[0.5em]
\end{longtable}

\end{ubclandscape}

\renewcommand{\implementationref}{\autoref{sec:implementation:performance}}
\begin{ubclandscape}
\subsection[Performance]{Code Map for \nameref{sec:implementation:performance} (\autoref{sec:implementation:performance})}
\label{app:implementation-mapping:performance}
\begin{longtable}{@{}%
  >{\raggedright\arraybackslash}p{\dimexpr0.5\linewidth-\tabcolsep\relax}%
  >{\raggedright\arraybackslash}p{\dimexpr0.5\linewidth-\tabcolsep\relax}%
@{}}
\caption[Performance and Resource Considerations]{Performance and Resource Considerations Implementation Mapping (See \implementationref.)}
\\
\toprule
\textbf{Feature/Component} & \textbf{Implementation Details} \\
\midrule
\endfirsthead

\multicolumn{2}{c}%
{\tablename\ \thetable\ -- \textit{Continued from previous page}} \\
\toprule
\textbf{Feature/Component} & \textbf{Implementation Details} \\
\midrule
\endhead

\midrule
\multicolumn{2}{r}{\textit{Continued on next page}} \\
\endfoot

\midrule
\multicolumn{2}{r}{\textit{End of Table}} \\
\endlastfoot

\multicolumn{2}{c}
                {\textbf{Background Processing}} \\
\midrule

\codewrap{BackgroundProcessorManager} in \codewrap{semantic/background_processor.py} & Non-blocking semantic extraction pipeline \\

BackgroundProcessorRunner in \codewrap{semantic/run_bg_processor.py} & Background processor execution script \\

\codewrap{MonitoredBackgroundProcessor} in \codewrap{semantic/run_bg_processor_with_monitoring.py} & Background processor with performance monitoring \\

\addlinespace[0.5em]
\multicolumn{2}{c}
                {\textbf{Performance Tracking}} \\
\midrule

\codewrap{IndalekoPerformanceDataModel} in \codewrap{data_models/i_perf.py} & Resource usage monitoring and metrics collection \\

ViewPerformanceProfiler in \codewrap{db/profile_view_performance.py} & View performance analysis tools \\

PerformanceMonitor in \codewrap{semantic/performance_monitor.py} & Performance tracking for semantic processing \\

\addlinespace[0.5em]
\end{longtable}

\end{ubclandscape}

\renewcommand{\implementationref}{\autoref{sec:implementation:challenges}}
\begin{ubclandscape}
\subsection[\nameref{sec:implementation:challenges}]{Code Map for \nameref{sec:implementation:challenges} (\autoref{sec:implementation:challenges})}
\label{app:implementation-mapping:challenges}
\begin{longtable}{@{}%
  >{\raggedright\arraybackslash}p{\dimexpr0.35\linewidth-\tabcolsep\relax}%
  >{\raggedright\arraybackslash}p{\dimexpr0.6\linewidth-\tabcolsep\relax}%
@{}}
\caption[Implementation Challenges and Solutions]{Implementation Challenges and Solutions Implementation Mapping (See \implementationref.)}
\\
\toprule
\textbf{Feature/Component} & \textbf{Implementation Details} \\
\midrule
\endfirsthead

\multicolumn{2}{c}%
{\tablename\ \thetable\ -- \textit{Continued from previous page}} \\
\toprule
\textbf{Feature/Component} & \textbf{Implementation Details} \\
\midrule
\endhead

\midrule
\multicolumn{2}{r}{\textit{Continued on next page}} \\
\endfoot

\midrule
\multicolumn{2}{r}{\textit{End of Table}} \\
\endlastfoot

\multicolumn{2}{c}
                {\textbf{Error Handling}} \\
\midrule

BaseStorageCollector in \codewrap{storage/collectors/base.py} & Comprehensive error handling and logging \\

\addlinespace[0.5em]
\multicolumn{2}{c}
                {\textbf{Platform Abstraction}} \\
\midrule

BaseStorageCollector in \codewrap{storage/collectors/base.py} & Cross-platform compatibility layer \\

\addlinespace[0.5em]
\end{longtable}

\end{ubclandscape}

\addkhipuifneeded
\clearpage

\section[Evaluation -- Supplemental Data]{Appendix C: Evaluation -- Supplemental Data}
\label{app:eval-supplemental}
This appendix provides additional data and analysis related to the evaluation of the UPI system, including detailed results from the full query evaluation and a summary of user feedback.

\subsection{Full Query Evaluation Results}\label{app:full-query-eval}

In the main evaluation chapter we provided a highlight of the searches that we performed using the synthetic data tool.  In this section, we present the full results of the query evaluation, including the number of metadata records generated and evaluated, the outcome in terms of precision and recall, and the timing for each query.

\subsection{Query Performance Data}\label{app:query-performance-data}

This section provides detailed performance measurements from our evaluation of Indaleko on a 31 million file dataset. We conducted 10 consecutive runs for queries where multi-run data was available, calculating mean, median, and standard deviation. For queries where only single-run data was collected, we present both cold cache (database restart) and warm cache (second execution) measurements.

\subsubsection{Query 1}
\textbf{Result set}: 33,477 documents

\textbf{10-run statistics (warm cache)}:
\begin{itemize}
\item with\_limits: Mean 0.3782s, Median 0.3757s, Std Dev 0.0128s
\item no\_limit: Mean 0.5185s, Median 0.5122s, Std Dev 0.0330s
\item count: Mean 0.3956s, Median 0.3946s, Std Dev 0.0139s
\end{itemize}

\subsubsection{Query 2}
\textbf{Result set}: 0 documents (query appears to search for ``my phone'' entity)

\textbf{10-run statistics (warm cache)}:
\begin{itemize}
\item with\_limits: Mean 3.3263s, Median 3.1636s, Std Dev 0.3710s
\item no\_limit: Mean 3.2564s, Median 3.2839s, Std Dev 0.1678s
\item count: Mean 3.4179s, Median 3.4157s, Std Dev 0.2654s
\end{itemize}

\subsubsection{Query 3}
\textbf{Result set}: 1 document

\textbf{Single-run measurements}:
\begin{itemize}
\item with\_limits: 0.606s (warm), 1.949s (cold)
\item no\_limit: 0.530s (warm), 0.404s (cold) - anomaly: no-limit faster
\item count: 0.002s (warm), 0.002s (cold)
\end{itemize}

\subsubsection{Query 4}
\textbf{Result set}: 972,664 documents

\textbf{Single-run measurements}:
\begin{itemize}
\item with\_limits: 2.158s (warm), 2.379s (cold)
\item no\_limit: 5.825s (warm), 15.461s (cold)
\item count: 2.288s (warm), 3.026s (cold)
\end{itemize}

\subsubsection{Query 5}
\textbf{Result set}: 1,545,999 documents

\textbf{Single-run measurements}:
\begin{itemize}
\item with\_limits: 0.010s (warm), 0.011s (cold)
\item no\_limit: 46.053s (warm), 44.128s (cold)
\item count: 0.153s (warm), 0.194s (cold)
\end{itemize}

\subsubsection{Query 6}
\textbf{Result set}: 12,683 documents

\textbf{10-run statistics (warm cache)}:
\begin{itemize}
\item with\_limits: Mean 1.0004s, Median 0.7183s, Std Dev 0.8847s
\item no\_limit: Mean 6.2175s, Median 5.0542s, Std Dev 3.3433s
\item count: Mean 5.8199s, Median 5.6805s, Std Dev 0.5742s
\end{itemize}

\subsection{Raw Performance Data}\label{app:raw-performance-data}

This section provides the complete timing measurements for all 10 runs where available. All times are in seconds. We present data from both execution patterns: consecutive (each query run 10 times in succession) and mixed workload (all queries executed in sequence, pattern repeated 10 times).

\subsubsection[Execution Patterns]{Performance Summary by Execution Pattern}

\begin{table}[!tbph]
\centering
\caption[Mean Performance Comparison]{Mean Performance Comparison: Consecutive vs Mixed Workload (seconds)}
\begin{tabular}{l l r r r}
\toprule
Query & Variant & Consecutive & Mixed & Difference \\
\midrule
Q1 & with\_limits & 0.382 & 2.084 & +445.7\% \\
Q1 & no\_limit & 0.542 & 1.568 & +189.5\% \\
Q1 & count & 0.388 & 0.426 & +9.9\% \\
\midrule
Q2 & with\_limits & 3.218 & 6.234 & +93.7\% \\
Q2 & no\_limit & 3.145 & 3.850 & +22.4\% \\
Q2 & count & 3.178 & 3.176 & -0.1\% \\
\midrule
Q3 & with\_limits & 0.532 & 1.220 & +129.1\% \\
Q3 & no\_limit & 0.490 & 0.596 & +21.7\% \\
Q3 & count & 0.002 & 0.002 & +5.1\% \\
\midrule
Q4 & with\_limits & 2.028 & 2.606 & +28.5\% \\
Q4 & no\_limit & 4.923 & 9.584 & +94.7\% \\
Q4 & count & 2.373 & 2.739 & +15.4\% \\
\midrule
Q5 & with\_limits & 0.007 & 0.008 & +18.5\% \\
Q5 & no\_limit & 16.222 & 22.847 & +40.8\% \\
Q5 & count & 0.202 & 0.183 & -9.7\% \\
\midrule
Q6 & with\_limits & 0.727 & 1.524 & +109.5\% \\
Q6 & no\_limit & 5.042 & 5.313 & +5.4\% \\
Q6 & count & 5.790 & 5.569 & -3.8\% \\
\bottomrule
\end{tabular}
\end{table}

\subsubsection{Query 1 Raw Data (10 runs)}
\begin{verbatim}
with_limits: [0.3779, 0.3736, 0.3872, 0.3731, 0.3694,
              0.3918, 0.3695, 0.3871, 0.3529, 0.3996]
no_limit:    [0.6049, 0.4974, 0.4901, 0.4947, 0.4867,
              0.5287, 0.5368, 0.5165, 0.5079, 0.5217]
count:       [0.3910, 0.3894, 0.4059, 0.4109, 0.3825,
              0.4039, 0.3981, 0.3823, 0.4193, 0.3722]
\end{verbatim}

\subsubsection{Query 2 Raw Data (10 runs)}
\begin{verbatim}
with_limits: [3.6519, 3.0142, 2.8936, 3.1451, 3.1483,
              3.8503, 4.0126, 3.1790, 3.4222, 2.9458]
no_limit:    [2.8101, 3.3559, 3.2556, 3.2549, 3.2929,
              3.5058, 3.3209, 3.2788, 3.2890, 3.1996]
count:       [3.1266, 3.4523, 3.2568, 3.0593, 3.3792,
              3.9685, 3.6320, 3.6291, 3.1704, 3.5051]
\end{verbatim}

\subsubsection{Query 3 Raw Data}
No multi-run data available. Single-run measurements provided in performance summary.

\subsubsection{Query 4 Raw Data}
Partial data available (count variant only, 10 runs):
\begin{verbatim}
count: [3.4094, 2.3685, 2.2593, 2.3734, 2.3048,
        2.1437, 2.0668, 2.1142, 2.2065, 2.3175]
Mean: 2.3564s, Std Dev: 0.3651s
\end{verbatim}
Note: with\_limits and no\_limit data not captured due to log format differences.

\subsubsection{Query 5 Raw Data}
Partial data available (count variant only, 10 runs):
\begin{verbatim}
count: [0.2160, 0.3685, 0.4206, 0.1899, 0.1534,
        0.2007, 0.1702, 0.1755, 0.2145, 0.1641]
Mean: 0.2273s, Std Dev: 0.0867s
\end{verbatim}
Note: with\_limits and no\_limit data not captured due to log format differences.

\subsubsection{Query 6 Raw Data (10 runs)}
\begin{verbatim}
with_limits: [3.6530, 0.6718, 0.7134, 0.7233, 0.7432,
              0.7335, 0.6538, 0.6839, 0.6867, 0.7410]
no_limit:    [16.2243, 5.0277, 4.9351, 5.0062, 5.0807,
               5.1244, 5.0322, 4.8431, 5.0006, 5.0012]
count:       [7.0543, 5.7295, 5.3885, 5.0890, 5.7084,
              5.9244, 5.3669, 5.6526, 6.0090, 6.2762]
\end{verbatim}

\subsection{AQL Queries for Exemplar Evaluation}\label{app:aql-queries}

This section provides the complete AQL queries used in the performance evaluation of \system. Each query has three variants: with\_limits (returns 50 results), no\_limit (returns all results), and count (returns only the count).

The rationale for limiting the results to 50 is the recognition that for human search capabilities, if the number is ``too big'' it impedes the ability to find it. We expect that a fully functional query tool will adapt to this by using additional techniques to refine the results. The ability to quickly determine the size of a result set provides an effective way to minimize user frustration.

\textbf{Complete Query Listings}: The full AQL implementations for all six queries (18 variants total) are available in the thesis project repository at \url{https://github.com/tonylmason/indaleko-thesis-evaluation} in the \texttt{aql\_queries/} directory. Each query file includes detailed comments explaining the query logic, parameter bindings, and performance optimizations.

\subsubsection{Query Design Considerations}\label{app:query-design}

Several queries in our evaluation required special consideration:

\textbf{Query 2}: This query requires synthetic metadata construction to test the full multi-dimensional memory pattern. The UPI envisions activity information being available to infer contextual relationships between users and systems.

\textbf{Query 3}: This query demonstrates the need for rich social context metadata, including collaboration patterns, shared document contexts, and temporal relationship information that would be captured in a fully functional system.

\subsubsection{Analysis}\label{app:supp-eval:analysis}

The queries presented in this section demonstrate the utility of the UPI in addressing a variety of information retrieval tasks. The queries are designed to be open-ended, allowing for exploration and refinement based on user needs. The results indicate that the UPI is capable of efficiently handling complex queries, leveraging its indexing and search capabilities to provide relevant results.

What we learned in constructing these queries related to tuning the underlying database technology to enable rapid response. Even with nominal tuning, the queries were far faster than what we saw with existing storage search services, even though the UPI consisted of more data than any one of the other search services considered.

\subsection{Ablation Results}\label{app:ablation-results}

In \autoref{sec:ablation-config} we described the configuration. This section provides the configuration files used for the three rounds of ablation testing.  Each configuration file specifies the test and control activity types, the semantic topics, the number of queries per topic, the number of truth and noise documents, and other parameters.

\subsubsection[Controlled Three Round]{Three Round with Control Configuration}\label{sec:ablation-config:three-round-control}
\begin{lstlisting}[language=TOML, caption={Ablation Three Round test configuration file (Test 1)}, label={lst:ablation-config-test1}]
[test1]
test_activity_types = ["MUSIC", "SOCIAL", "STORAGE", "TASK"]
control_activity_types = ["COLLABORATION", "LOCATION"]
semantic_topics = [
  "African elephant",
  "Bengal tiger",
  "Galápagos tortoise",
  "Komodo dragon",
  "University of Oxford",
  "University of Cape Town",
  "National University of Singapore",
  "University of São Paulo",
  "Inca Trail to Machu Picchu",
  "Climbing Mount Kilimanjaro",
  "Scuba diving the Great Barrier Reef",
  "Trekking in Patagonia",
  "Nelson Mandela",
  "Mahatma Gandhi",
  "Gabriela Mistral",
  "Malala Yousafzai",
  "Analects of Confucius",
  "Ubuntu philosophy",
  "Advaita Vedanta",
  "Stoicism"
]
queries_per_topic = 0
truth_count = 5
noise_count = 45
noise_count_per_pattern = 1
random_seed = 42
statistical_alpha = 0.05
desired_power = 0.8
target_effect_size = 0.1
test_type = "paired_proportions"
exemplar_queries = [
    "Where did I save the PDF on stoicism mom sent me last month?"
    "I need to find the files I created for my book about scuba diving in the great barrier reef",
    "I took pictures of the Komodo dragon that lives in the garden at the University of Oxford",
    "I need to find where I saved a file that Mahatma Gandhi sent me on Signal.",
    "When I was working on that budget spreadsheet I was listening to Lady Gaga.",
    "I saved a file about widgets last month, I don't recall exactly when or where, but I do remember I went to the University of São Paulo the following week."
]

[test2]
test_activity_types = ["COLLABORATION", "LOCATION", "STORAGE", "TASK"]
control_activity_types = ["SOCIAL", "MUSIC"]
semantic_topics=[
  "David Unaipon",
  "Deb Haaland",
  "Aili Keskitalo",
  "Raoni Metuktire",
  "Lake Vostok",
  "Great Slave Lake",
  "Lake Turkana",
  "Lake Karakul",
  "La Paz",
  "Lhasa",
  "Leadville",
  "Leh",
  "Didgeridoo",
  "Hang Drum",
  "Mbira",
  "Waterphone",
  "Middlemist's Red",
  "Ghost Orchid",
  "Franklinia",
  "Jade Vine"
]
queries_per_topic = 0
truth_count = 5
noise_count = 45
noise_count_per_pattern = 1
random_seed = 43
statistical_alpha = 0.05
desired_power = 0.8
target_effect_size = 0.1
test_type = "paired_proportions"
exemplar_queries = [
    "Are the photos mom took of the Middlemist's Red on the laptop or my phone?",
    "Aili Keskitalo sent me a PDF about Franklinia last month while I was visiting Lake Karakul.",
    "Deb Haaland from Leadville shared a music video with me on Discord and I'm trying to find it.",
    "Where is that conference paper Raoni Metuktire sent me - I know I saved it somewhere, I just don't remember where.",
    "How many photos did I take at Lake Turkana last year and where did I put them?"
]

[test3]
test_activity_types = ["COLLABORATION", "LOCATION", "MUSIC", "SOCIAL"]
control_activity_types = ["STORAGE", "TASK"]
semantic_topics = [
  "Mary Anning",
  "Jeanne Villepreux-Power",
  "Rosalind Franklin",
  "Alice Ball",
  "Giant's Causeway",
  "Eye of the Sahara",
  "Stone Forest",
  "Moeraki Boulders",
  "One Hundred Years of Solitude",
  "The Tale of Genji",
  "The Master and Margarita",
  "The Three-Body Problem",
  "Eyak",
  "Dalmatian",
  "Livonian",
  "Manx",
  "Invoice",
  "Purchase Order",
  "Balance Sheet",
  "Non-Disclosure Agreement"
]
queries_per_topic = 0
truth_count = 5
noise_count = 45
noise_count_per_pattern = 1
random_seed = 44
statistical_alpha = 0.05
desired_power = 0.8
target_effect_size = 0.1
test_type = "paired_proportions"
exemplar_queries = [
    "I know I downloaded the article about the Three-Body Problem, where did it go?",
    "Last month I prepared the Balance Sheet and sent it to Jeanne Villepreux-Power and now I need to find it.",
    "Get that Non-Disclosure agreement from OneDrive and send it to Mary Anning.",
    "Where did I put that crash dump file Rosalind Franklin sent me last week?",
    "I know you sent me that article about Dalmatian, did I save it in Dropbox?"
]
\end{lstlisting}

Note: a value of 0 for \texttt{queries\_per\_topic} means that the number of queries required is computed, based upon \texttt{statistical\_alpha}, \texttt{desired\_power}, and \texttt{target\_effect\_size}.  The number of queries is then set to the minimum number of queries required to achieve the desired power.

\subsubsection[No Control One Round]{One Round, No Control}\label{sec:ablation-config:one-round-no-control}

\begin{lstlisting}[language=TOML, caption={Ablation Single Round All Type}, label={lst:ablation-config-one-round-no-control}]
[all_activity]
test_activity_types = ["MUSIC", "SOCIAL", "STORAGE", "TASK", "COLLABORATION", "LOCATION"]
semantic_topics = [
  "African elephant",
  "Bengal tiger",
  "Galápagos tortoise",
  "Komodo dragon",
  "University of Oxford",
  "University of Cape Town",
  "National University of Singapore",
  "University of São Paulo",
  "Inca Trail to Machu Picchu",
  "Climbing Mount Kilimanjaro",
  "Scuba diving the Great Barrier Reef",
  "Trekking in Patagonia",
  "Nelson Mandela",
  "Mahatma Gandhi",
  "Gabriela Mistral",
  "Malala Yousafzai",
  "Analects of Confucius",
  "Ubuntu philosophy",
  "Advaita Vedanta",
  "Stoicism",
  "La Llorona",
  "The Tale of the Bamboo Cutter",
  "Panchatantra",
  "The Tale of the Firebird",
  "Pripyat",
  "Hashima Island",
  "Kolmanskop",
  "Hatra",
  "Skocjan Caves",
  "Cango Caves",
  "Tham Lod",
  "Gouffre Berger",
  "Giraffe Weevil",
  "Giant Weta",
  "Atlas Moth",
  "Assassin Bug",
  "Resume.pdf",
  "TaxReturn2024.pdf",
  "VacationPhotos.zip",
  "RecipeCollection.docx",
  "David Unaipon",
  "Deb Haaland",
  "Aili Keskitalo",
  "Raoni Metuktire",
  "Lake Vostok",
  "Great Slave Lake",
  "Lake Turkana",
  "Lake Karakul",
  "La Paz",
  "Lhasa",
  "Leadville",
  "Leh",
  "Didgeridoo",
  "Hang Drum",
  "Mbira",
  "Waterphone",
  "Middlemist's Red",
  "Ghost Orchid",
  "Franklinia",
  "Jade Vine",
  "Mary Anning",
  "Jeanne Villepreux-Power",
  "Rosalind Franklin",
  "Alice Ball",
  "Giant's Causeway",
  "Eye of the Sahara",
  "Stone Forest",
  "Moeraki Boulders",
  "One Hundred Years of Solitude",
  "The Tale of Genji",
  "The Master and Margarita",
  "The Three-Body Problem",
  "Eyak",
  "Dalmatian",
  "Livonian",
  "Manx",
  "Invoice",
  "Purchase Order",
  "Balance Sheet",
  "Non-Disclosure Agreement"
]
queries_per_topic = 0
truth_count = 5
noise_count = 45
noise_count_per_pattern = 1
random_seed = 42
statistical_alpha = 0.01
desired_power = 0.99
target_effect_size = 0.01
test_type = "paired_proportions"
exemplar_queries = [
    "I am looking for an invoice from last winter, the vendor was Alice Ball and they are based out of the Stone Forest",
    "I saved some research articles about Jade Vine but I don't recall where I saved them.",
    "Last week I met with David Unaipon and shared my resume with them",
    "Show me docx files that I created last year at the conference at Lake Vostok.",
    "I saved a file about widgets last month, I don't recall exactly when or where, but I do remember I met Gabriela Mistral that day."
]
\end{lstlisting}

In \autoref{tab:synthetic-queries}, we present the complete set of synthetic data queries used for evaluation and their corresponding features.

\begin{ubclandscape}
\begin{table}
    \caption[Synthetic Data Queries]{Complete set of synthetic data queries used for evaluation.}
    \label{tab:synthetic-queries}
    \centering
    \resizebox{0.9\textheight}{!}{%
        \begin{tabular}{|c|p{8.5cm}|c|c|c|c|c|c|c|c|c|c|c|c|c|c|c|c|c|c|c|c|c|c|c|c|c|c|c|c|c|c|c|c|c|}
        \toprule
        ID & Query & \rotatebox{90}{activity\_context} & \rotatebox{90}{collaboration} & \rotatebox{90}{connectivity} & \rotatebox{90}{context} & \rotatebox{90}{device} & \rotatebox{90}{device\_config} & \rotatebox{90}{file\_access} & \rotatebox{90}{file\_creation} & \rotatebox{90}{file\_edit} & \rotatebox{90}{filetype} & \rotatebox{90}{fuzzy\_location} & \rotatebox{90}{geo\_activity} & \rotatebox{90}{geolocation} & \rotatebox{90}{indirect\_reference} & \rotatebox{90}{machine\_config} & \rotatebox{90}{mood\_context} & \rotatebox{90}{music} & \rotatebox{90}{music\_activity} & \rotatebox{90}{named\_entity} & \rotatebox{90}{negation\_logic} & \rotatebox{90}{network} & \rotatebox{90}{posix} & \rotatebox{90}{realworld\_event} & \rotatebox{90}{semantic} & \rotatebox{90}{structured\_content} & \rotatebox{90}{temporal} & \rotatebox{90}{temporal\_ambiguity} & \rotatebox{90}{temporal\_fuzzy} & \rotatebox{90}{temporal\_linked} & \rotatebox{90}{temporal\_overlap} & \rotatebox{90}{time} & \rotatebox{90}{timestamp} & \rotatebox{90}{vague\_context} \\
        \midrule
        1 & Find all `.docx` files I created late at night last weekend. &  &  &  &  &  &  &  &  &  & \cellcolor{gray!20}\checkmark &  &  &  &  &  &  &  &  &  &  &  & \cellcolor{gray!20}\checkmark &  &  &  &  & \cellcolor{gray!20}\checkmark &  &  &  & \cellcolor{gray!20}\checkmark &  &  \\
        \midrule
        2 & Show me files I looked at while listening to Taylor Swift on my laptop. &  &  &  &  &  &  &  &  &  &  &  &  &  &  & \cellcolor{gray!20}\checkmark &  &  & \cellcolor{gray!20}\checkmark &  &  &  &  &  &  &  &  &  &  &  & \cellcolor{gray!20}\checkmark &  &  &  \\
        \midrule
        3 & What did I work on while I was in Nairobi last year? &  &  &  & \cellcolor{gray!20}\checkmark &  &  &  &  &  &  &  & \cellcolor{gray!20}\checkmark &  &  &  &  &  &  &  &  &  &  &  &  &  &  &  &  &  &  &  & \cellcolor{gray!20}\checkmark &  \\
        \midrule
        4 & I need those reports about the widget recall remember, the ones with charts. &  &  &  &  &  &  &  &  &  &  &  &  &  &  &  &  &  &  &  &  &  &  &  & \cellcolor{gray!20}\checkmark & \cellcolor{gray!20}\checkmark &  &  &  &  &  &  &  & \cellcolor{gray!20}\checkmark \\
        \midrule
        5 & Pull up everything I edited on my iPad in January. &  &  &  &  &  &  &  &  & \cellcolor{gray!20}\checkmark &  &  &  &  &  & \cellcolor{gray!20}\checkmark &  &  &  &  &  &  &  &  &  &  &  &  &  &  &  & \cellcolor{gray!20}\checkmark &  &  \\
        \midrule
        6 & I was definitely *not* working on any PDFs just the spreadsheets. &  &  &  &  &  &  &  &  &  & \cellcolor{gray!20}\checkmark &  &  &  &  &  &  &  &  &  & \cellcolor{gray!20}\checkmark &  & \cellcolor{gray!20}\checkmark &  &  &  &  &  &  &  &  &  &  &  \\
        \midrule
        7 & What files did I download while traveling through BC? &  &  &  &  &  &  & \cellcolor{gray!20}\checkmark &  &  &  & \cellcolor{gray!20}\checkmark & \cellcolor{gray!20}\checkmark &  &  &  &  &  &  &  &  &  &  &  &  &  &  &  &  &  &  &  &  &  \\
        \midrule
        8 & When I was blasting heavy metal those late-night rants I typed out. &  &  &  &  &  &  &  & \cellcolor{gray!20}\checkmark &  &  &  &  &  &  &  & \cellcolor{gray!20}\checkmark &  & \cellcolor{gray!20}\checkmark &  &  &  &  &  &  &  &  &  &  &  &  &  &  &  \\
        \midrule
        9 & The thing I wrote about the unicorn startup pitch had diagrams and buzzwords. &  &  &  &  &  &  &  &  &  &  &  &  &  & \cellcolor{gray!20}\checkmark &  &  &  &  &  &  &  &  &  & \cellcolor{gray!20}\checkmark & \cellcolor{gray!20}\checkmark &  &  &  &  &  &  &  &  \\
        \midrule
        10 & Give me everything created during the power outage last March, from my desktop. &  &  &  &  &  &  &  &  &  &  &  &  &  &  & \cellcolor{gray!20}\checkmark &  &  &  &  &  &  & \cellcolor{gray!20}\checkmark & \cellcolor{gray!20}\checkmark &  &  &  &  &  &  &  &  & \cellcolor{gray!20}\checkmark &  \\
        \midrule
        11 & Find the presentation I shared with Dr. Jones right after our last meeting. &  & \cellcolor{gray!20}\checkmark &  &  &  &  &  &  &  &  &  &  &  &  &  &  &  &  &  &  &  &  &  & \cellcolor{gray!20}\checkmark &  & \cellcolor{gray!20}\checkmark &  &  &  &  &  &  &  \\
        \midrule
        12 & Where's that PDF I opened right before my flight to Mexico City? &  &  &  &  &  &  &  &  &  & \cellcolor{gray!20}\checkmark &  &  & \cellcolor{gray!20}\checkmark &  &  &  &  &  &  &  &  &  &  &  &  & \cellcolor{gray!20}\checkmark &  &  &  &  &  &  &  \\
        \midrule
        13 & Show me the notes I took during the workshop in Oaxaca. & \cellcolor{gray!20}\checkmark &  &  &  &  &  &  &  &  &  &  &  & \cellcolor{gray!20}\checkmark &  &  &  &  &  &  &  &  &  &  & \cellcolor{gray!20}\checkmark &  &  &  &  &  &  &  &  &  \\
        \midrule
        14 & Which files did I access while connected to my home Wi-Fi last week? &  &  &  &  &  & \cellcolor{gray!20}\checkmark &  &  &  &  &  &  &  &  &  &  &  &  &  &  & \cellcolor{gray!20}\checkmark &  &  &  &  &  &  &  &  &  & \cellcolor{gray!20}\checkmark &  &  \\
        \midrule
        15 & I'm looking for the spreadsheet about vendor invoices maybe from early February. &  &  &  &  &  &  &  &  &  & \cellcolor{gray!20}\checkmark &  &  &  &  &  &  &  &  &  &  &  &  &  & \cellcolor{gray!20}\checkmark &  &  &  & \cellcolor{gray!20}\checkmark &  &  &  &  &  \\
        \midrule
        16 & Show me any documents related to project `Zephyr' that I worked on from my phone. &  &  &  &  & \cellcolor{gray!20}\checkmark &  &  &  &  &  &  &  &  &  &  &  &  &  & \cellcolor{gray!20}\checkmark &  &  &  &  & \cellcolor{gray!20}\checkmark &  &  &  &  &  &  &  &  &  \\
        \midrule
        17 & Find the audio recording from the interview I did in Coyoacán. & \cellcolor{gray!20}\checkmark &  &  &  &  &  &  &  &  & \cellcolor{gray!20}\checkmark &  &  & \cellcolor{gray!20}\checkmark &  &  &  &  &  &  &  &  &  &  &  &  &  &  &  &  &  &  &  &  \\
        \midrule
        18 & What images did I view on my tablet on the same day I uploaded those marketing assets? &  &  &  &  & \cellcolor{gray!20}\checkmark &  &  &  &  & \cellcolor{gray!20}\checkmark &  &  &  &  &  &  &  &  &  &  &  &  &  &  &  &  &  &  & \cellcolor{gray!20}\checkmark &  &  &  &  \\
        \midrule
        19 & Give me everything I wrote while listening to ambient noise playlists. & \cellcolor{gray!20}\checkmark &  &  &  &  &  &  &  &  &  &  &  &  &  &  &  & \cellcolor{gray!20}\checkmark &  &  &  &  &  &  & \cellcolor{gray!20}\checkmark &  &  &  &  &  &  &  &  &  \\
        \midrule
        20 & Which PDFs did I annotate while offline last weekend? &  &  & \cellcolor{gray!20}\checkmark &  &  &  &  &  &  & \cellcolor{gray!20}\checkmark &  &  &  &  &  &  &  &  &  &  &  &  &  &  &  & \cellcolor{gray!20}\checkmark &  &  &  &  &  &  &  \\
        \bottomrule
        \end{tabular}%
    }
    \end{table}
\end{ubclandscape}

\subsection{LLM-Based Tool Usage}\label{app:llm-tool-usage}

\begin{figure}[!tbph]
    \caption[Architecture of LLM-Based Tools]{Architecture of LLM-Based Tools Supporting \system}
    \label{fig:llm-architecture}
    \centering
    \resizebox{0.95\textwidth}{!}{%
        \begin{tikzpicture}[
        node distance=1.8cm and 2cm,
        every node/.style={font=\sffamily},
        box/.style={draw, rounded corners, fill=gray!10, minimum height=1.2cm, minimum width=2.8cm, align=center},
        arrow/.style={-{Stealth}, thick}
        ]

        \node[box] (query) {Query};
        \node[box, right=of query] (assistant) {Assistant};
        \node[box, right=of assistant] (archivist) {Archivist};

        \node[box, below=of query] (upi) {UPI\\System};
        \node[box, below=of archivist] (record) {Archivist\\Record};

        \node[left=of query] (user) {\textbf{User}};

        \draw[arrow] (user) -- (query);
        \draw[arrow] (query) -- (assistant);
        \draw[arrow] (assistant) -- (archivist);

        \draw[arrow] (query) -- node[left] {query} (upi);
        \draw[arrow] (upi) -- node[below right] {retrieval} (record);
        \draw[arrow] (record) -- node[right] {retrieval} (archivist);

        \end{tikzpicture}
    }%
\end{figure}
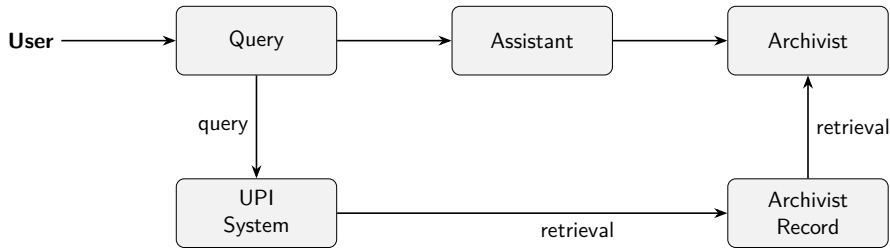

In \autoref{fig:llm-architecture}, we illustrate the architecture of the LLM-based tools developed for \system.

The \system project included extensive work in building LLM-based tools to assist users in searching and evaluating the UPI system. This appendix provides a summary of the LLM-based tools that were developed, their usage, and the feedback received from users. This is not considered a part of the core thesis because these tools were only used to evaluate the UPI system and are not part of the contribution to this thesis.

However, the LLM-based tools represent the answer to the question from my committee in December 2021 about how to \emph{use} Indaleko and their development has influenced the architecture, design, and implementation of \system.  This section describes some of the key aspects of their development.

Three distinct tools were built: \textbf{Query}, \textbf{Assistant}, and \textbf{Archivist}. Each embodies a progression in capabilities, complexity, and vision:

\begin{description}
    \descitem{Query}{is a one-shot translator: from natural language to structured query.}
    \descitem{Assistant}{introduces tool use and structured responses.}
    \descitem{Archivist}{explores persistence, context, and long-term adaptation.}
\end{description}

Figure~\ref{fig:llm-architecture} depicts the relationship among these tools and their integration with the \system core.

\subsection{Query: The First Cut}
\label{app:llm-tool-usage-query}

The \emph{Query} tool is a stateless interface that takes a natural language prompt and produces a structured query against the UPI schema. Initially implemented using the OpenAI beta Chat Completion API, which provided structured output returns, it served as a proof-of-concept for translating end-user information needs into executable form.

Its role in evaluation was critical: it enabled the definition and automation of exemplar queries, and allowed us to measure recall, precision, and execution time for representative tasks. Internally, the tool captures not only results but metadata about the translation: which fields were accessed, which filters applied, and what parts of the prompt proved ambiguous.

\textbf{Motivation} Provide a lightweight, prompt-driven mechanism to simulate end-user query behavior without requiring schema knowledge.

\subsection{Assistant: Toward Interactive Tool Use}
\label{app:llm-tool-usage-assistant}

The \emph{Assistant} extends the Query tool with memory (within-session), tool usage, and structured reasoning. It uses the newer Requests interface to offer modular capabilities: search execution, result explanation, schema inspection, and query suggestion. These tools are bound as function-calling plugins (e.g., via OpenAI tool use APIs).

This tool represents a midpoint in capability: still ephemeral, but capable of more nuanced, step-by-step interactions. Users can refine queries, ask follow-up questions, and explore partial results.

\textbf{Motivation} Bridge the gap between raw queries and exploratory data analysis, offering a familiar conversational interface with access to domain-specific tools.

\subsection{Archivist: Memory, Evolution, and Autonomy}
\label{app:llm-tool-usage-archivist}

The \emph{Archivist} represents a more ambitious vision: a persistent agent capable of remembering past queries, refining its behavior, and collaborating with the user over time. Its design draws heavily on the principles of mutual adaptation (e.g., \textit{ayni}), supporting a long-term cooperative relationship between user and system.

\textbf{Key Capabilities}
\begin{description}
    \item[Prompt forwarding]{to maintain evolving state;}
    \item[Vector-store-like similarity recall]{for past interactions;}
    \item[Automated query pattern analysis]{(to suggest indexes, detect inefficiencies);}
    \item[Plug-in infrastructure]{for activity data ingestion (see Section~\ref{app:activity-plugins});}
    \item[Early-stage prompt revision logic]{inspired by LLM “metacognition.”}
\end{description}

\textbf{Motivation} Enable a persistent relationship between user and system, fostering long-term learning, adaptation, and query personalization. Archivist is not a front-end, but a collaborating peer.

\subsection{Extensibility and Plug-In Infrastructure}
\label{app:activity-plugins}

One of the most powerful features of the LLM-based toolchain is its extensibility. The plug-in architecture pioneered in Archivist allows new data sources to be integrated with minimal friction.

Two such plug-ins were created:
\begin{description}
    \descitem{NTFS USN Journal Provider}{Developed with LLM assistance, this module parses journal entries and translates them into high-granularity file activity events. Used to associate query results with episodic system activity.}
    \descitem{Spotify Activity Provider}{Developed as part of a UGRA project, this module captures media playback history and maps it to temporal context metadata (e.g., ``files opened while listening to [artist]'').}
\end{description}

These examples support the evaluation claim that \system is easily extensible, and that even non-core developers (UGRAs, LLM agents) can meaningfully contribute.

\subsection{Ablation Study Data Models}\label{app:ablation-data-models}

This section provides the complete Python dataclass definitions used in the ablation study framework.

\subsubsection{Object Data Model}

\begin{lstlisting}[language=python, caption=Object dataclass for Ablation study,label=lst:app:datamodel:object]
import enum
from uuid import UUID, uuid4

from .ablation_base import AblationData
from pydantic import Field, AwareDatetime

class AblationMimeType(str, enum.Enum):
    """Enum for MIME types."""
    TEXT = "text/plain"
    IMAGE = "image/jpeg"
    VIDEO = "video/mp4"
    AUDIO = "audio/mpeg"
    APPLICATION = "application/json"
    PDF = "application/pdf"
    DOCX = "application/vnd.openxmlformats-officedocument.wordprocessingml.document"
    XLSX = "application/vnd.openxmlformats-officedocument.spreadsheetml.sheet"

class AblationObject(AblationData):
    """This represents a file object in the ablation data model."""

    ObjectIdentifer: UUID = Field(default_factory=uuid4, description="Unique identifier for the object.")
    MimeType : AblationMimeType = Field(..., description="MIME type of the object.")
    FileName: str = Field(..., description="Name of the file.")
    Size: int = Field(..., description="Size of the file in bytes.")
    CreateTime: AwareDatetime = Field(..., description="Creation date and time of the file.")
    LastModifiedTime: AwareDatetime = Field(..., description="Last time data in file was changed.")
    LastAccessTime: AwareDatetime = Field(..., description="Last access date and time of the file.")
    LastChangeTime: AwareDatetime = Field(..., description="Last time metadata was changed.")
    Semantics: list[str] = Field(default_factory=list, description="List of semantic tags associated with the object.")
\end{lstlisting}

\subsubsection{Activity Data Model}

\begin{lstlisting}[language=python, caption=Activity stream dataclass for Ablation study, label=lst:app:datamodel:activity]
from enum import IntEnum, auto

from pydantic import AwareDatetime

from .ablation_base import AblationData

class ActivityDataType(IntEnum):
    """Define the type of activity stream."""
    COLLABORATION = auto()
    LOCATION = auto()
    MUSIC = auto()
    SOCIAL = auto()
    STORAGE = auto()
    TASK = auto()

    def __str__(self) -> str:
        """Return the string representation of the enum."""
        return self.name

class ActivityData(AblationData):
    """Base class for semantic data."""

    DataType: ActivityDataType
    StartTime: AwareDatetime
    EndTime: AwareDatetime
\end{lstlisting}

\subsubsection{Query Data Model}

\begin{lstlisting}[language=python, caption=Query dataclass for Ablation study, label=lst:app:datamodel:query]
from .ablation_base import AblationData

class AblationQuery(AblationData):
    """This represents a query in the ablation data model."""

    Query: str
    QueryBase: str
    QueryFilters: list[str]
\end{lstlisting}

\subsection{LLM Query Generation Details}\label{app:llm-query-generation}

This section provides detailed examples of the LLM-based natural language query processing pipeline used in Indaleko.

\subsubsection{Query Generation Prompt Template}

\begin{lstlisting}[language=python, caption=Query generation prompt for LLM]
return f"""
You are an expert in storage-centric file-search queries. Here are {len(examples)} exemplar queries to model:

{ex_block}

Generate {count} new queries similar to the examples above but with diverse topics and contexts.
Focus on:
- Temporal context ("last week", "yesterday", "during vacation")
- Activity context ("while listening to music", "during conference call")
- Social context ("documents shared with Dr. Smith", "files from my colleague")
- Device context ("files created on my phone", "photos from laptop")
- Semantic context ("PDFs about climate change", "spreadsheets with budget data")

Return only the queries, one per line.
"""
\end{lstlisting}

\subsubsection{Example Query Translations}

This section demonstrates how natural language queries are translated into AQL database operations:

\textbf{Example 1: Temporal and File Type Query}
\begin{itemize}
\item Natural Language: "Find all .docx files I created late at night last weekend"
\item Generated AQL:
\begin{lstlisting}[language=AQL]
FOR doc IN Objects
  FILTER doc.Timestamps[* FILTER CURRENT.Label == '6b3f16ec-52d2-4e9b-afd0-e02a875ec6e6'].Value
    ANY >= '2025-04-26T22:00:00Z' AND <= '2025-04-27T06:00:00Z'
  FILTER REGEX_TEST(doc.Label, '\\.docx$', true)
  RETURN doc
\end{lstlisting}
\end{itemize}

\textbf{Example 2: Activity Integration Query}
\begin{itemize}
\item Natural Language: "Show me files I looked at while listening to Taylor Swift on my laptop"
\item Generated AQL:
\begin{lstlisting}[language=AQL]
FOR activity IN ActivityContext
  FILTER activity.ActivityType == 'MUSIC'
  FILTER CONTAINS(activity.Details.artist, 'Taylor Swift')
  FOR doc IN Objects
    FILTER activity.StartTime <= doc.LastAccessTime <= activity.EndTime
    FILTER doc.Device == 'laptop'
    RETURN DISTINCT doc
\end{lstlisting}
\end{itemize}

\textbf{Example 3: Semantic Understanding Query}
\begin{itemize}
\item Natural Language: "I need those reports about the widget recall - remember, the ones with charts"
\item Generated AQL:
\begin{lstlisting}[language=AQL]
FOR doc IN Objects
  FILTER CONTAINS(LOWER(doc.Label), 'report')
  FILTER CONTAINS(LOWER(doc.SemanticContent), 'widget recall')
  FILTER CONTAINS(LOWER(doc.SemanticContent), 'chart')
  RETURN doc
\end{lstlisting}
\end{itemize}

These examples illustrate several key aspects of the UPI's query capabilities:
\begin{itemize}
\item \textbf{Temporal reasoning}: The LLM correctly interprets relative time expressions like "last weekend" and "late at night" into specific timestamp ranges
\item \textbf{Activity integration}: Complex joins between metadata and memory anchor collections enable queries about concurrent activities
\item \textbf{Semantic understanding}: The system translates vague descriptions ("those reports... the ones with charts") into concrete database predicates
\item \textbf{Multi-dimensional filtering}: Queries combine storage metadata, semantic content, and memory anchors seamlessly
\end{itemize}

\subsection{Ablation Study Supplemental Visualizations}\label{app:ablation-supplemental}

This section provides additional visualizations from the ablation study presented in \autoref{sec:eval-ablation}. While the main text presents the combined comparison figure (\autoref{fig:ablation-combined}), these supplementary figures provide detailed individual configuration views and alternative perspectives on the data for readers seeking more detailed analysis.

\subsubsection{Individual Configuration Line Plots}

These line plots show the detailed precision degradation patterns for each configuration separately, complementing the combined view presented in the main text.

\begin{ubclandscape}
\begin{figure}[!htbp]
  \centering
  \includegraphics[width=0.95\textheight]{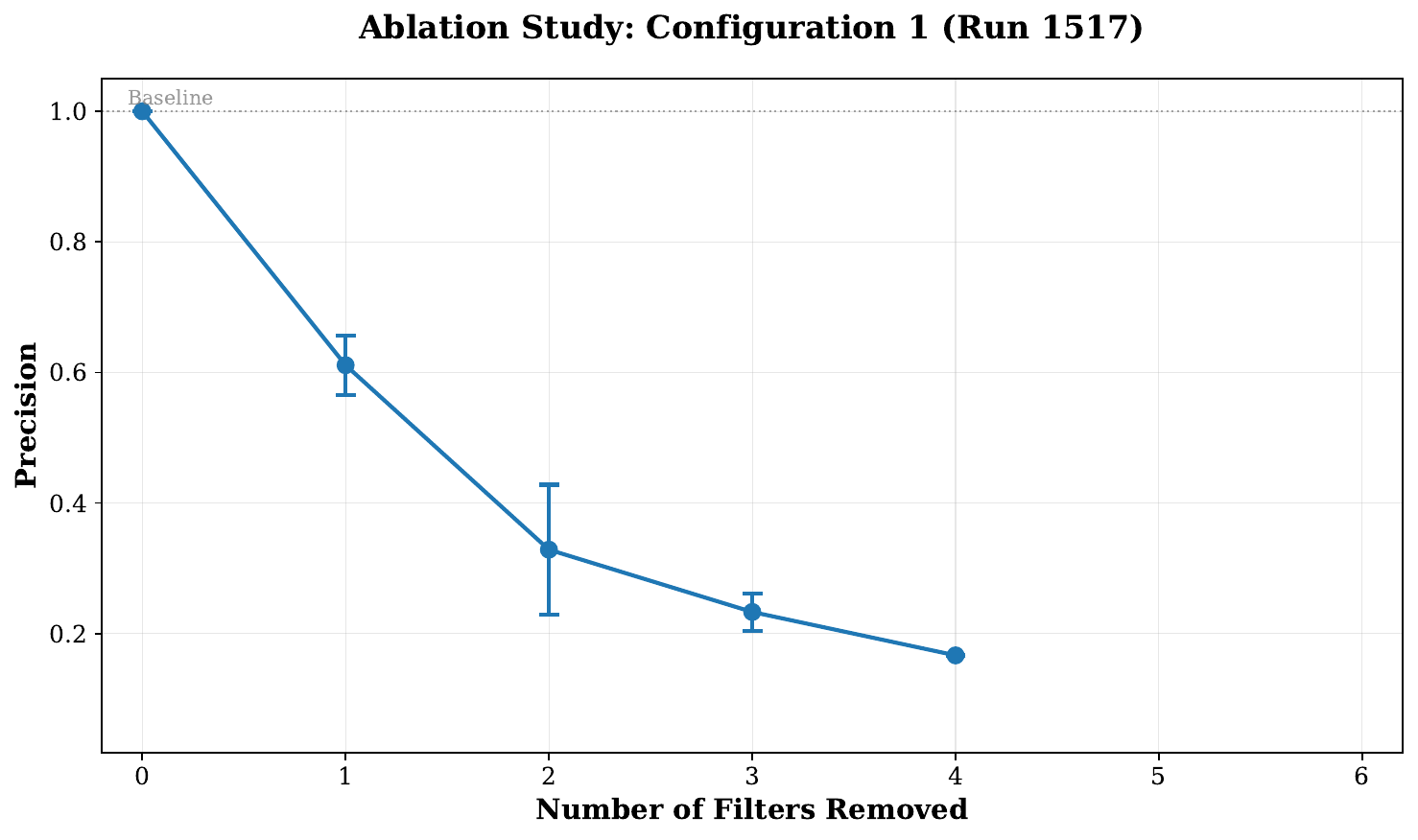}
  \caption[Configuration 1 (T=5, N=16) Ablation Results]{Configuration 1 (T=5, N=16) ablation results showing precision degradation as filters are removed. Each noise document represents a specific filter combination ($2^4 = 16$ combinations). The steep decline from 1.0 to 0.17 demonstrates the critical importance of memory anchors in the control group design. Error bars represent standard deviation across experimental runs.}
  \label{fig:config1:line:appendix}
\end{figure}
\end{ubclandscape}

\begin{ubclandscape}
\begin{figure}[!htbp]
  \centering
  \includegraphics[width=0.95\textheight]{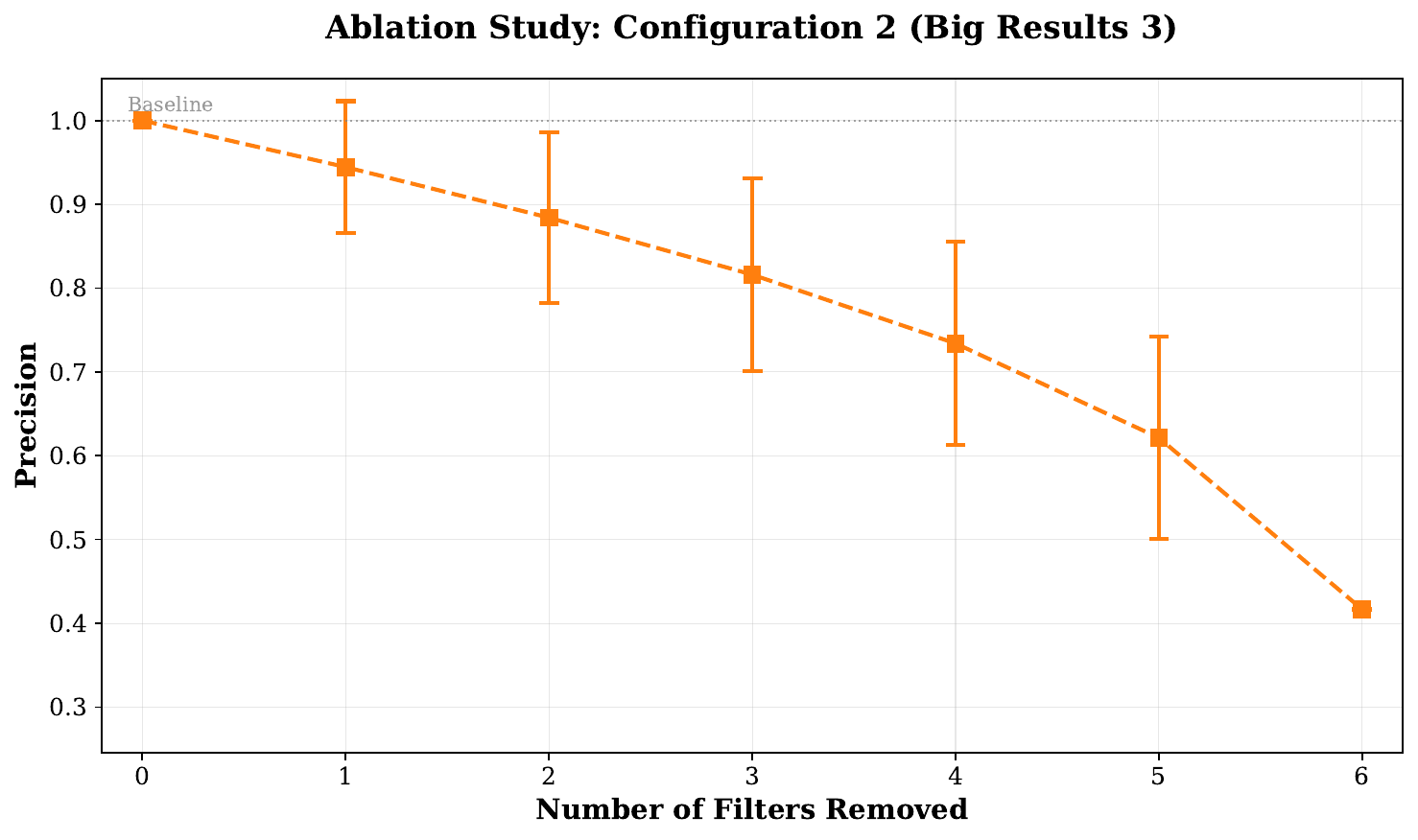}
  \caption[Configuration 2 (T=5, N=64) Ablation Results]{Configuration 2 (T=5, N=64) ablation results showing more gradual precision degradation compared to Configuration 1. Each noise document represents a specific filter combination ($2^6 = 64$ combinations). Despite the higher noise complexity, this configuration maintains higher precision levels, though still exhibiting substantial degradation when multiple filters are removed.}
  \label{fig:config2:line:appendix}
\end{figure}
\end{ubclandscape}

\subsubsection{Configuration 1 Supplemental Figures}
\label{fig:config1:ranked:appendix}
\label{fig:config1:effects:appendix}

The supplemental figures may be found at:

\begin{itemize}
  \item \href{https://github.com/ubc-systopia/Indaleko/blob/main/figures/config1_analysis_plots_ranked.pdf}{Configuration 1 Ranked Plot} - This figure shows the ranked precision drops for filter combinations in Configuration 1~\footnote{\url{https://github.com/ubc-systopia/Indaleko/blob/main/figures/config1_analysis_plots_ranked.pdf}}.
  \item \href{https://github.com/ubc-systopia/Indaleko/blob/main/figures/config1_analysis_plots_effects.pdf}{Configuration 1 Effects Plot} - This figure provides effect sizes with confidence intervals for the 15 most impactful filter combinations in Configuration 1~\footnote{\url{https://github.com/ubc-systopia/Indaleko/blob/main/figures/config1_analysis_plots_effects.pdf}}.
\end{itemize}

\subsubsection{Configuration 2 Supplemental Figures}

\begin{itemize}
  \item \href{https://github.com/ubc-systopia/Indaleko/blob/main/figures/config2_analysis_plots_ranked.png}{Configuration 2 Ranked Plot} - This figure shows the ranked precision drops for filter combinations in Configuration 2~\footnote{\url{https://github.com/ubc-systopia/Indaleko/blob/main/figures/config2_analysis_plots_ranked.png}}.
  \item \href{https://github.com/ubc-systopia/Indaleko/blob/main/figures/config2_analysis_plots_effects.png}{Configuration 2 Effects Plot} - This figure provides effect sizes with confidence intervals for the 15 most impactful filter combinations in Configuration 2~\footnote{\url{https://github.com/ubc-systopia/Indaleko/blob/main/figures/config2_analysis_plots_effects.png}}.
\end{itemize}

\addkhipuifneeded
\clearpage

\section[Privacy \& Security]{Appendix D: Privacy \& Security}
\label{app:privacy}
While this thesis focuses on the architectural and implementation aspects of the Unified Personal Index (UPI), privacy and security considerations were fundamental to the design process. This appendix presents the privacy and security framework developed for UPI implementations, though a comprehensive security analysis remains future work.

\subsection{Introduction and Overview}\label{app:privacy:intro}

The Unified Personal Index (UPI) is designed with privacy as a core principle. During the development of \system as an implementation of the UPI, we adopted a simple model of privacy: keep all personal context information under the user's control. This was consistent with our primary goal of ensuring that the UPI provides utility while minimizing privacy risks.

This appendix details:
\begin{itemize}
\item The threat models and primary risks considered in the UPI design,
\item Our simple local-first privacy model and its implementation,
\item Strategies for secure use of external services,
\item Advanced privacy concepts for future implementations, and
\item Limitations of current privacy measures and future research directions.
\end{itemize}


\subsection{Privacy Architecture and Principles}\label{app:privacy:architecture}

\subsubsection{Threat Models and Primary Risks}\label{app:privacy:threats}

The UPI architecture considers several threat models that affect personal information systems:

\paragraph{External Adversaries}
\begin{itemize}
\item \textbf{Network eavesdropping}: Interception of data during transmission between local systems and external services,
\item \textbf{Remote exploitation}: Attacks targeting vulnerabilities in exposed services or APIs, and
\item \textbf{Social engineering}: Attempts to gain unauthorized access through user manipulation.
\end{itemize}

\paragraph{Service Provider Threats}
\begin{itemize}
\item \textbf{Unauthorized data access}: Cloud providers or third-party services accessing stored metadata,
\item \textbf{Data mining and profiling}: Analysis of activity patterns or metadata for commercial purposes,
\item \textbf{Legal compulsion}: Service providers required to disclose data to authorities, and
\item \textbf{Data breaches}: Compromise of provider infrastructure exposing user data.
\end{itemize}

\paragraph{Local System Compromises}
\begin{itemize}
\item \textbf{Malware}: Malicious software accessing local UPI data stores,
\item \textbf{Physical access}: Unauthorized access to devices containing UPI data,
\item \textbf{Insider threats}: Authorized users with malicious intent, and
\item \textbf{Device loss or theft}: Physical loss of devices containing sensitive metadata.
\end{itemize}

\paragraph{Privacy Leakage Through Inference}
\begin{itemize}
\item \textbf{Pattern analysis}: Deriving sensitive information from metadata patterns,
\item \textbf{Temporal correlation}: Linking activities across time to infer behaviors,
\item \textbf{Location tracking}: Building movement profiles from spatial metadata, and
\item \textbf{Query analysis}: Inferring interests or activities from search patterns.
\end{itemize}

\subsubsection{Architectural Privacy Principles}\label{app:privacy:principles}

Managing personal information, such as letters, photos, conversations, and created work, requires confidence that data will be protected, respected, and never exploited. The UPI architecture prioritizes privacy and user control from the outset, integrating privacy protection as a fundamental architectural principle. Users retain explicit control over their information, ensuring it remains accessible, organized, and secure.

The privacy-focused architecture incorporates several key innovations:

\paragraph{Semantic Decoupling}
The UPI fundamentally separates the identity of stored objects from their semantic meaning. Using unique identifiers (UUIDs\index{UUID}), it introduces a layer of abstraction between raw data storage and the contextual interpretation of content. This ensures that third-party services with access to data storage do not inherently have access to the meaningful connections, contexts, or patterns that characterize a user's information landscape.

\paragraph{Contextual Discretion}
The UPI architecture differentiates between types of context based on their privacy implications. For instance, location data from private spaces (like one's home) might warrant higher privacy protection than data from public spaces. Relationship data concerning close family members could be treated differently from professional contacts. This differentiation enables the application of appropriate privacy controls based on information sensitivity.

\paragraph{User-Controlled Boundaries}
The architecture establishes clear boundaries between personal information and external services, allowing users to maintain granular control over what information crosses these boundaries and under what conditions. Users can selectively share specific metadata with certain services while maintaining the privacy of other information.

This approach contrasts with common commercial systems that typically require users to trade privacy for service access. The UPI demonstrates that powerful, intelligent information management can be achieved without compromising privacy. A genuinely human-centric system must respect the deeply personal nature of users' information.

\subsubsection{Local-First Privacy Model}\label{app:privacy:local-first}

The UPI implements a straightforward privacy model based on local-first principles:

\paragraph{Core Principles}
\begin{itemize}
\item \textbf{Local control}: All personal context information remains on the user's local system by default,
\item \textbf{Explicit consent}: Data never leaves the local environment without user authorization,
\item \textbf{Minimal exposure}: External services receive only the minimum data necessary for functionality, and
\item \textbf{User ownership}: Users maintain complete control over their metadata and can delete it at any time.
\end{itemize}

\paragraph{User Control Mechanisms}
\begin{itemize}
\item \textbf{Data retention policies}: User-configurable limits on how long activity stream is retained,
\item \textbf{Selective recording}: Ability to pause or disable Memory Anchor capture,
\item \textbf{Granular permissions}: Fine-grained control over which metadata types are collected, and
\item \textbf{Export controls}: User ability to export and delete all personal data.
\end{itemize}

\paragraph{Personal Digital Traces Collection Principles}
In the ubiquitous computing environment, individuals continuously generate digital traces. The UPI recognizes that while capturing this information can provide significant value for personal information retrieval, it also raises potential privacy concerns. The architecture addresses this through several principles:

\begin{itemize}
\item \textbf{Minimal intrusion}: The UPI architecture promotes minimal intrusion into users' daily activities. Memory Anchor creation is designed to be lightweight and unobtrusive.
\item \textbf{Selective capture}: Users maintain explicit control over which activity traces are captured and retained. The system provides fine-grained controls over data collection.
\item \textbf{Purpose limitation}: Activity streams are collected solely for enhancing personal information retrieval, not for surveillance or monitoring.
\item \textbf{Data minimization}: The UPI only captures Memory Anchors that are directly relevant to improving retrieval effectiveness.
\item \textbf{Transparent operation}: Users are informed about what data is being captured, how it is being used, and where it is stored.
\end{itemize}


\subsection{Indaleko Security Implementation}\label{app:privacy:implementation}

This section combines the implementation details from the Indaleko prototype that demonstrates the UPI security architecture in practice.

\subsubsection{Layered Privacy Model}\label{app:privacy:layered}

The implementation organizes privacy mechanisms into four distinct layers (Layers 0 through 3), each providing cumulative security guarantees:

\begin{description}
    \descitem{Layer 0 (Baseline)}{
        - UUID\index{UUID}-based semantic decoupling,
        - Separate schema mapping from data storage, and
        - Minimal complexity with reasonable protection;
    }
    \descitem{Layer 1}{
        - Adds mapping table encryption using symmetric cryptography, and
        - Protects semantic relationships while maintaining performance;
    }
    \descitem{Layer 2}{
        - Implements bulk data encryption for collections, and
        - Applies encryption at collection level for efficiency;
    }
    \descitem{Layer 3}{
        - Provides per-attribute encryption with unique keys,
        - Maximum security with highest complexity, and
        - Suitable for high-security environments.
    }
\end{description}

\subsubsection{Threat Models and Deployment Scenarios}\label{app:privacy:deployment}

The system adapts its security posture based on deployment context. The security model allows users to choose appropriate protections based on their threat model and performance requirements.

\paragraph{Individual Use}
For personal deployments, the focus is on:
\begin{itemize}
\item Protection against device loss or theft,
\item Isolation from cloud service providers,
\item Minimal performance overhead, and
\item Simple key management.
\end{itemize}

\paragraph{Family or Small Group}
Shared deployments add:
\begin{itemize}
\item User-based access controls,
\item Selective sharing of metadata subsets, and
\item Audit logging of access patterns.
\end{itemize}

\paragraph{Enterprise Deployment}
Organization-wide use requires:
\begin{itemize}
\item Integration with enterprise authentication systems,
\item Compliance with regulatory requirements,
\item Advanced audit capabilities, and
\item Role-based access control.
\end{itemize}

\begin{table}[ht]
  \centering
  \caption[Deployment Configuration Defense Mechanisms]{Deployment Configuration Defense Mechanisms}
  \label{tab:deployment-defenses}
  \resizebox{\textwidth}{!}{%
    \begin{tabular}{p{3cm}p{10cm}}
        	\toprule
        	\textbf{Deployment Configuration} & \textbf{Implemented Defense Mechanisms} \\
        \midrule
        Local Desktop &
        - UUID\index{UUID} obfuscation of database structure \newline
        - Local processing pipeline \newline
        - Minimal privilege token scopes for collectors \newline
        - Local database hosting (no external index storage) \\
        Network-Connected &
        - Token-based authentication with scoped permissions \newline
        - Query processing isolation from external services \newline
        - Service endpoint rotation for distributed risk \newline
        - Private UUID\index{UUID} mapping isolation \\
        Air-Gapped System &
        - Complete network isolation for mapping databases \newline
        - Support for offline encryption key management \newline
        - Local processing for all operations \newline
        - Transparent audit of all data flows \\
        Enterprise Infrastructure &
        - Organization-controlled processing instances \newline
        - Private index hosting on internal infrastructure \newline
        - Integration with enterprise security systems \\
        High-Security Environment &
        - Full per-field encryption with independent keys (Layer 3) \newline
        - Strict local-only processing requirements \newline
        - Controlled physical component hosting \newline
        - Configuration-based security posture adjustment \\
        \bottomrule
        \end{tabular}
    }%
\end{table}

\subsubsection{UUID\index{UUID}-Based Privacy Implementation}

The implemented baseline privacy model relies on UUIDs\index{UUID} (Layer 0):

\begin{description}
    \descitem{Semantic Decoupling}{Randomly generated UUIDs\index{UUID} identify metadata characteristics: instead of a set of clearly labeled metadata fields, each metadata field has a UUID\index{UUID} and a data value.}
    \descitem{Local UUID Mapping}{A separate mapping layer translates UUIDs\index{UUID}. Normally, users will not be exposed to these UUIDs\index{UUID} and the mapping layer will be used to translate the UUIDs\index{UUID} to the semantic meaning.}
    \descitem{Installation-specific UUID Labels}{Unique UUIDs\index{UUID} generated per installation hinder cross-installation correlation based on identifiers alone.}
\end{description}

\subsubsection{Access Control and Security Measures}

\paragraph{Database Security}
ArangoDB provides several security features that the UPI leverages:
\begin{description}
\descitem{Authentication}{Username/password authentication with token-based sessions,}
\descitem{Authorization}{Role-based access control to collections and operations,}
\descitem{Encryption}{TLS for client-server communication, encryption at rest (enterprise edition), and}
\descitem{Audit logging}{Detailed logs of all database operations.}
\end{description}

\paragraph{Local Deployment Architecture}
Access control in \system implements technical safeguards appropriate to different deployment architectures:

\begin{description}
    \descitem{Least Privilege Implementation}{The system creates a separate database account with minimal required permissions after initial setup.}
    \descitem{Local Deployment Architecture}{Default configuration assumes local hosting to maintain complete control over memory pattern data.}
    \descitem{Process Isolation}{Computational operations occur within the deployment boundary to minimize external data exposure.}
    \descitem{Token Management}{Authentication tokens for external services follow platform-specific secure storage mechanisms.}
    \descitem{Permission Boundaries}{Each collector operates with minimal required permissions for its specific data source.}
\end{description}

\paragraph{External Service Mitigation}\label{app:privacy:external}
Interacting with external services, especially LLMs, introduces data exposure risks. Mitigation strategies include:

\begin{description}
    \descitem{Local LLM/SLM Fallback}{Avoids external calls entirely when available and selected.}
    \descitem{Query Anonymization and Rotation}{Round-robin rotation distributes requests across providers.}
    \descitem{Non-Data-Preserving Proxies}{Network connectivity uses services like ngrok for data forwarding without storage.}
\end{description}

\subsubsection{Key Management Architecture}

The architecture implements a hierarchical key management system:

\paragraph{Master Key Control}
Users control a master key that derives all other encryption keys, ensuring:
\begin{itemize}
\item No vendor lock-in,
\item Portable data migration,
\item Complete user control, and
\item Key recovery mechanisms
\end{itemize}

\paragraph{Key Hierarchy}
\begin{description}
    \descitem{Master key}{User-controlled, derives all other keys,}
    \descitem{Schema key}{Encrypts UUID-to-semantic mappings,}
    \descitem{Collection keys}{Encrypt entire collections (Layer 2), and}
    \descitem{Attribute keys}{Per-field encryption (Layer 3).}
\end{description}

\subsubsection{Current Implementation Status}

The current security posture of \system provides a baseline of privacy through semantic decoupling (Layer 0), local-first principles, and standard token management. This implementation demonstrates the feasibility of the approach while providing a foundation for future enhancements.

Currently implemented security features focus on:
\begin{itemize}
    \item Semantic decoupling through UUID-based metadata representation,
    \item Simple access control for database operations,
    \item Token management for external service authentication, and
    \item Local-first implementation that prioritizes user control.
\end{itemize}

\subsubsection{Implementation Challenges}\label{app:privacy:impl-challenges}

Important privacy challenges addressed during \system implementation included:

\begin{description}
    \descitem{Context Privacy}{Memory Anchor collection could potentially capture sensitive information. Addressed through filtering, aggregation, and user control over context capture.}
    \descitem{External Service Dependencies}{Some capabilities require external services that could compromise privacy. Mitigated through local alternatives, data minimization, and transparent policies.}
    \descitem{Identity Correlation}{Maintaining privacy while enabling cross-service entity resolution. Solved through the UUID-based privacy model with controlled linking.}
    \descitem{Permission Management}{Ensuring appropriate access levels for different data sources. Implemented a fine-grained permission system with explicit user consent flows.}
    \descitem{Developer Experience}{Security aspects that increase protection also make debugging more difficult. Addressed through git pre-commit hooks to enforce security practices.}
\end{description}


\subsection{Data Protection by Asset Type}\label{app:privacy:data-protection}\label{app:privacy:assets}

Different types of data within the UPI require specific privacy considerations:

\subsubsection{Memory Anchor Data}\label{app:privacy:security-activity}

Activity stream data forms a distinctive category of personal information with unique privacy characteristics.

\paragraph{Privacy Challenges}
\begin{description}
\descitem{Continuous generation}{Users generate activity streams throughout their daily lives,}
\descitem{Rich inference potential}{Patterns can reveal habits, preferences, and private behaviors,}
\descitem{Temporal sensitivity}{Recent activity stream often more sensitive than historical data,}
\descitem{Cross-device correlation}{Activity patterns can be linked across multiple devices,}
\descitem{Location sensitivity}{Spatial context can reveal home, work, and frequently visited locations.}
\end{description}

\paragraph{Privacy-Preserving Design Principles}
\begin{description}
\descitem{Granular control}{Users can selectively enable/disable capture of specific activity types,}
\descitem{Temporal boundaries}{Automatic expiration of activity stream after user-defined periods,}
\descitem{Context anonymization}{Stripping identifying information from activity records where possible,}
\descitem{Local processing}{Activity analysis performed on-device without cloud dependencies, and}
\descitem{Aggregation over raw data}{Storing statistical summaries rather than detailed event logs.}
\end{description}

\paragraph{Specific Protections}
\begin{itemize}
\item Location data undergoes precision reduction based on context sensitivity,
\item Application usage patterns aggregated to time windows rather than precise timestamps,
\item Device identifiers replaced with user-controlled aliases,
\item Network identifiers hashed to prevent external correlation, and
\item Activity patterns analyzed locally with only derived metadata retained.
\end{itemize}

\subsubsection{Query Logs}
Query history requires special protection as it directly reflects user interests and information needs:
\begin{itemize}
\item Queries stored with timestamps rounded to reduce temporal correlation,
\item Natural language queries processed locally before any external API calls,
\item Query patterns analyzed for system improvement without retaining raw queries, and
\item User option to disable query logging entirely.
\end{itemize}

\subsubsection{File Metadata}
Standard file metadata receives baseline protection through semantic decoupling:
\begin{itemize}
\item Path information separated from content identifiers,
\item Creation/modification times stored with configurable precision,
\item Owner information replaced with user-controlled identifiers, and
\item File type associations masked through UUID mapping.
\end{itemize}

\subsection{Advanced Privacy Concepts}\label{app:privacy:advanced}

While the current implementation focuses on practical privacy measures, several advanced concepts warrant consideration for future versions:

\subsubsection{Homomorphic Encryption}
Homomorphic encryption would enable query processing on encrypted data without decryption:
\begin{itemize}
\item Secure cloud-based index searches,
\item Privacy-preserving aggregation queries,
\item Encrypted metadata comparisons, and
\item Challenges: Performance overhead, query complexity limitations.
\end{itemize}

\subsubsection{Secure Multi-party Computation}
For shared family or group deployments:
\begin{itemize}
\item Privacy-preserving searches across multiple users' data,
\item Collaborative filtering without exposing individual preferences, and
\item Shared indexes without revealing private metadata.
\end{itemize}

\subsubsection{Zero-Knowledge Proofs}
Enable verification of data properties without revealing the data:
\begin{itemize}
\item Prove file existence without exposing metadata,
\item Verify temporal relationships without timestamps, and
\item Authenticate ownership without identity disclosure.
\end{itemize}

\subsubsection{Differential Privacy}
Add calibrated noise to preserve privacy while maintaining utility:
\begin{itemize}
\item Statistical queries over activity patterns,
\item Trend analysis without individual exposure, and
\item Privacy-preserving usage analytics.
\end{itemize}

\subsubsection[Enhanced Privacy]{Enhanced Privacy Mechanisms from UPI Architecture}

Several privacy enhancements emerge naturally from the UPI's architectural design:

\paragraph{Contextual Privacy Controls}
The multi-dimensional metadata model enables sophisticated privacy rules:
\begin{itemize}
\item Location-based privacy zones (e.g., home vs. public spaces),
\item Time-based access restrictions,
\item Relationship-aware sharing policies, and
\item Activity-triggered privacy modes.
\end{itemize}

\subsection{Limitations and Future Work}\label{app:privacy:limitations}

\subsubsection{Current Limitations}
\begin{itemize}
\item Limited formal security analysis,
\item Performance overhead of encryption not fully characterized,
\item Key management complexity for non-technical users, and
\item Lack of standardized privacy policies.
\end{itemize}

\subsubsection{Production Deployment Considerations}
Note that the privacy design principles described throughout this appendix are readily accomplished in the proof-of-concept system. However, maintaining these principles in a production environment will require additional consideration:

\begin{itemize}
\item Users may choose to use no or low cost services in exchange for sharing their data with third party providers,
\item Production systems must balance privacy protections with performance and scalability requirements,
\item Commercial deployments may face additional constraints from business models that rely on data monetization, and
\item Regulatory compliance may introduce additional privacy requirements beyond those in the prototype.
\end{itemize}

\subsubsection{Future Research Directions}

Several areas warrant continued research:

\begin{description}
\descitem{Formal verification}{Mathematical proofs of privacy properties,}
\descitem{User studies}{Understanding privacy preferences and usability of controls,}
\descitem{Regulatory compliance}{Adapting to GDPR, CCPA, and emerging regulations,}
\descitem{Privacy-preserving query languages}{Development of query systems that inherently protect privacy,}
\descitem{Decentralized identity systems}{Integration with emerging identity standards,}
\descitem{Adaptive privacy}{Systems that learn and adapt to user privacy preferences, and}
\descitem{Privacy measurement}{Metrics for quantifying privacy protection effectiveness.}
\end{description}

\subsubsection{Future Privacy Architecture Research}

Future architectural research will advance privacy-preserving memory models through:

\begin{description}
\descitem{Encrypted episodic pattern storage}{Advanced encryption schemes for complex memory patterns,}
\descitem{Zero-knowledge memory retrieval protocols}{Proving memory existence without revealing content,}
\descitem{Hierarchical privacy boundaries}{Multi-level privacy controls within distributed memory systems, and}
\descitem{Three-layer implementation model}{Full deployment of the proposed privacy architecture.}
\end{description}

The proposed three-layer model for future systems includes:
\begin{itemize}
\item Encrypted memory patterns stored via UUIDs,
\item Separate semantic mapping tables with independent encryption, and
\item Abstraction layers that decouple meaning from storage.
\end{itemize}

Future research will address cryptographic key management for memory patterns, implement access auditing mechanisms, and develop secure staging environments that maintain semantic boundaries. Integration with the Pydantic-based serialization architecture is progressing. This model enables controlled memory pattern sharing through mapping and cryptographic access control.

\subsection{Conclusion}\label{app:privacy:conclusion}

The UPI's privacy architecture balances practical usability with strong privacy protections through a local-first design philosophy. While the current implementation provides basic privacy features, the framework supports evolution toward more sophisticated privacy-preserving technologies as they mature.

The layered approach allows users to choose appropriate privacy levels based on their threat model and performance requirements. As privacy-preserving techniques advance, the UPI architecture can incorporate new methods while maintaining its core principle: personal information should remain under personal control.

Key achievements include:
\begin{itemize}
\item Demonstration of semantic decoupling as a practical privacy technique,
\item Implementation of local-first architecture for Memory Anchors,
\item Flexible security model adaptable to different deployment scenarios, and
\item Foundation for advanced privacy features in future versions.
\end{itemize}

The privacy model described here reflects our commitment to user control and data protection while acknowledging the practical constraints of current technology. Future work will focus on formal security analysis, performance optimization, and user-friendly privacy management interfaces.


\addkhipuifneeded
\clearpage


\end{document}